\documentclass[reprint,superscriptaddress,showpacs,preprintnumbers,amsmath,amssymb,aps,prc]{revtex4-1}

\usepackage{graphicx}
\usepackage{dcolumn}
\usepackage{bm}
\usepackage{hyperref}
\usepackage{multirow}
\usepackage{xfrac}

\begin{document}

\title{$\beta^-$ decay study of the $^{66}$Mn -- $^{66}$Fe -- $^{66}$Co -- $^{66}$Ni chain}

\author{M.~Stryjczyk}
\email{marek.stryjczyk@kuleuven.be}
\affiliation{KU Leuven, Instituut voor Kern- en Stralingsfysica, Celestijnenlaan 200D, 3001 Leuven, Belgium}
\author{Y.~Tsunoda}
\affiliation{Center for Nuclear Study, University of Tokyo, Hongo, Bunkyo-ku, Tokyo 113-0033, Japan}
\author{I.~G.~Darby}
\affiliation{KU Leuven, Instituut voor Kern- en Stralingsfysica, Celestijnenlaan 200D, 3001 Leuven, Belgium}
\author{H. De~Witte}
\affiliation{KU Leuven, Instituut voor Kern- en Stralingsfysica, Celestijnenlaan 200D, 3001 Leuven, Belgium}
\author{J.~Diriken}
\affiliation{KU Leuven, Instituut voor Kern- en Stralingsfysica, Celestijnenlaan 200D, 3001 Leuven, Belgium}
\affiliation{Belgian National Science Centre SCKCEN, Boeretang 200, B-2004 Mol, Belgium}
\author{D.~V.~Fedorov}
\affiliation{Petersburg Nuclear Physics Institute, NRC Kurchatov Institute, 188300 Gatchina, Russia}
\author{V.~N.~Fedosseev}
\affiliation{EN Department, CERN, CH-1211 Geneva 23, Switzerland}
\author{L.~M.~Fraile}
\affiliation{Grupo de F\'{i}sica Nuclea \& UPARCOS, Universidad Complutense, CEI Moncloa, 28040 Madrid, Spain}
\author{M.~Huyse}
\affiliation{KU Leuven, Instituut voor Kern- en Stralingsfysica, Celestijnenlaan 200D, 3001 Leuven, Belgium}
\author{U.~K\"oster}
\affiliation{Institut Laue-Langevin, 71 avenue des Martyrs, 38042 Grenoble, France}
\author{B.~A.~Marsh}
\affiliation{EN Department, CERN, CH-1211 Geneva 23, Switzerland}
\author{T.~Otsuka}
\affiliation{Center for Nuclear Study, University of Tokyo, Hongo, Bunkyo-ku, Tokyo 113-0033, Japan}
\affiliation{Department of Physics, University of Tokyo, Hongo, Bunkyo-ku, Tokyo 113-0033, Japan}
\affiliation{RIKEN Nishina Center, 2-1 Hirosawa, Wako, Saitama 351-0198, Japan}
\affiliation{National Superconducting Cyclotron Laboratory, Michigan State University, East Lansing, MI 48824, USA}
\affiliation{KU Leuven, Instituut voor Kern- en Stralingsfysica, Celestijnenlaan 200D, 3001 Leuven, Belgium}
\author{D.~Pauwels}
\affiliation{KU Leuven, Instituut voor Kern- en Stralingsfysica, Celestijnenlaan 200D, 3001 Leuven, Belgium}
\author{L.~Popescu}
\affiliation{Belgian National Science Centre SCKCEN, Boeretang 200, B-2004 Mol, Belgium}
\author{D.~Radulov}
\thanks{Deceased}
\affiliation{KU Leuven, Instituut voor Kern- en Stralingsfysica, Celestijnenlaan 200D, 3001 Leuven, Belgium}
\author{M.~D.~Seliverstov}
\affiliation{KU Leuven, Instituut voor Kern- en Stralingsfysica, Celestijnenlaan 200D, 3001 Leuven, Belgium}
\affiliation{Petersburg Nuclear Physics Institute, NRC Kurchatov Institute, 188300 Gatchina, Russia}
\affiliation{Department of Physics, University of York, York YO10 5DD, United Kingdom}
\author{A.~M.~Sj\"odin}
\affiliation{EN Department, CERN, CH-1211 Geneva 23, Switzerland}
\author{P.~Van~den~Bergh}
\affiliation{KU Leuven, Instituut voor Kern- en Stralingsfysica, Celestijnenlaan 200D, 3001 Leuven, Belgium}
\author{P.~Van~Duppen}
\affiliation{KU Leuven, Instituut voor Kern- en Stralingsfysica, Celestijnenlaan 200D, 3001 Leuven, Belgium}
\author{M.~Venhart}
\affiliation{KU Leuven, Instituut voor Kern- en Stralingsfysica, Celestijnenlaan 200D, 3001 Leuven, Belgium}
\affiliation{Institute of Physics, Slovak Academy of Sciences, SK-84511 Bratislava, Slovakia}
\author{W.~B.~Walters}
\affiliation{Department of Chemistry and Biochemistry, University of Maryland, College Park, Maryland 20742, USA}
\author{K.~Wimmer}
\affiliation{Department of Physics, University of Tokyo, Hongo, Bunkyo-ku, Tokyo 113-0033, Japan}
\affiliation{Physik Department E12, Technische Universit\"at M\"unchen, D-85748 Garching, Germany}

\date{\today}

\begin{abstract}
\begin{description}

\item[Background] Shell evolution can impact the structure of the nuclei and lead to effects such as shape coexistence. The nuclei around $^{68}$Ni represent an excellent study case, however, spectroscopic information of the neutron-rich, $Z<28$ nuclei is limited.

\item[Purpose] The goal is to measure $\gamma$-ray transitions in $^{66}$Fe, $^{66}$Co and $^{66}$Ni populated in the $\beta^-$ decay of $^{66}$Mn, to determine absolute $\beta$-feedings and relative $\gamma$-decay probabilities and to compare the results with Monte Carlo Shell Model calculations in order to study the influence of the relevant single neutron and proton orbitals occupancies around $Z=28$ and $N=40$.

\item[Method] The low-energy structures of $^{65,66}$Fe, $^{66}$Co and $^{66}$Ni were studied in the $\beta^-$ decay of $^{66}$Mn produced at ISOLDE, CERN. The beam was purified by means of laser resonance ionization and mass separation. The $\beta$ and $\gamma$ events detected by three plastic scintillators and two MiniBall cluster germanium detectors, respectively, were correlated in time to build the low-energy excitation schemes and to determine the $\beta$-decay half-lifes of the nuclei.

\item[Results] The relative small $\beta$-decay ground state feeding of $^{66}$Fe obtained in this work is at variant to the earlier studies. Spin and parity $1^+$ was assigned to the $^{66}$Co ground state based on the strong ground state feeding in the decay of $^{66}$Fe as well as in the decay of $^{66}$Co. Experimental log(\textit{ft}) values, $\gamma$-ray deexcitation patterns and energies of excited states were compared to Monte Carlo Shell Model calculations. Based on this comparison, spin and parity assignments for the selected number of low-lying states in the $^{66}$Mn to $^{66}$Ni chain were proposed.

\item[Conclusions] The $\beta$-decay chain starting $^{66}$Mn towards $^{66}$Ni, crossing $N=40$, evolves from deformed nuclei to sphericity. The $\beta$-decay population of a selected number of $0^+$ and $2^+$ states in $^{66}$Ni, which is understood within shape coexistence framework of Monte Carlo Shell Model calculations, reveals the crucial role of the neutron $0g_{9/2}$ shell and proton excitations across the Z=28 gap.

\end{description}
\end{abstract}


\maketitle

\section{\label{sec:introduction}Introduction}

The nickel isotopic chain with a magic number of protons ($Z=28$) is an excellent study case to test the nuclear shell model. It starts at $^{48}$Ni, which decays through a recently discovered $2p$ emission channel \cite{Pfutzner2002,Pomorski2011,Pomorski2014}, goes through the doubly-magic $N=Z=28$ $^{56}$Ni and ends beyond $^{78}$Ni, whose region has been studied extensively \cite{VandeWalle2009,Padgett2010,Xu2014,Shiga2016,Alshudifat2016,Olivier2017,Welker2017} to check the persistence of the magic numbers in nuclei with an extreme neutron-to-proton ratio. The region of $^{68}$Ni around $N=40$ is of particular interest. Some unexpected properties, such as a large excitation energy of the first $2^+$ state and a low B(E2,$2^+_1\rightarrow0^+_1$) \cite{Mueller2000,Sorlin2002,Bree2008}, have been measured in this nucleus and they might suggest an $N=40$ (Harmonic Oscillator magic number) subshell closure. On the other hand, adding or removing protons from the $N=40$ nuclei leads to an increase of the collectivity, which is manifested by a sudden decrease of the $2^+_1$-excitation energy and increase of B(E2) values \cite{Hannawald1999,Gade2010,Crawford2013,Celikovic2013,Louchart2013,Suchyta2014}. Several studies were performed to understand these properties \cite{Lenzi2010,Tsunoda2014,Santamaria2015,Togashi2015,Mougeot2018}. 

Recent developments in theoretical models suggest that in order to reproduce the structure of exotic nuclei, the tensor force has to be included into the nuclear shell model potential \cite{Otsuka2006}. Its monopole part influences the shell structure, which is known as shell evolution \cite{Otsuka2013}, and can lead to the erosion of the magic numbers \cite{Bastin2007,Steppenbeck2013,Utsuno2014} or changes in the single-particle shell ordering \cite{Franchoo1998,Flanagan2009,DeGroote2017}. In the $^{68}$Ni region, it is conjectured that shell evolution is responsible for a significant reduction of the energy gap between the $0f_{7/2}$ and the $0f_{5/2}$ proton shells, which gives a rise to creation of the different energy minima in the Potential Energy Surface (PES) \cite{Tsunoda2014,Walters2015,Otsuka2016,Leoni2017}. The occurrence of these phenomena were also discussed in Ref. \cite{Girod1988,Girod1989,Bonche1989,Moller2009}. 

In this work these phenomena are studied through the $\beta^-$ decay of $^{66}$Mn to three $A=66$ daughter nuclei: $^{66}$Fe, $^{66}$Co and $^{66}$Ni. The simultaneous analysis extended the amount of available experimental information \cite{Grzywacz1998,Recchia2012,Liddick2012,Liddick2013,Leoni2017,Olaizola2017,Olaizola2017a} and allowed us to address, at the same time, the increase of collectivity at $N=40$, shape coexistence in $^{68}$Ni region, and also the onset of deformation in the $A=66$ chain. State-of-the-art Monte Carlo Shell Model calculations were performed for the analyzed isotopes and allowed to make theory guided tentative spin assignments and to explain the selective population of states in the $\beta$-decay process. 

The paper is organized in the following way. The experimental setup and the analysis method are described in Section \ref{sec:setup}. The data analysis results for the decay of $^{66}$Mn, $^{66}$Fe and $^{66}$Co are presented in the subsections \ref{sec:results66Fe}, \ref{sec:results66Co} and \ref{sec:results66Ni}, respectively. In Subsection \ref{sec:resultsgsf} the details regarding the half-lifes of the analyzed nuclei and the direct feeding to the ground states are described. The discussion of the results and the interpretation are presented in Section \ref{sec:discussion}, and the conclusions are drawn in Section \ref{sec:conclusions}. 

\section{\label{sec:setup}Experimental setup and analysis}

The experiment was part of a campaign at ISOLDE, CERN to measure the $\beta$ decay of the neutron-rich $^{58,60-68}$Mn isotopes. The details on the experimental conditions are published in Ref. \cite{Pauwels2012,Radulov2013,Radulov2014,Flavigny2015}. Here we report only the essential information.

To produce pure beams of manganese, 1.4 GeV protons from the Proton Synchrotron Booster were impinged onto a UC$_{x}$ target (45 g/cm$^2$ thickness). Created fission products diffused from the hot target (about 2000 degrees Celsius \cite{Montano2013}) into the ion source. Manganese atoms were selectively ionized by the three Nd:YAG-pumped dye lasers provided by the RILIS laser system \cite{Fedosseev2012}. Subsequently, the ions were accelerated and separated with respect to their mass-over-charge ratio $A/Q$ using the High Resolution Separator (HRS). The slits between the two HRS dipole magnets were used to reduce isobaric contaminants. Finally, the ions were implanted on an aluminized mylar tape located inside a movable tape station \cite{Pauwels2008}. The implantation point was surrounded by three plastic $\Delta E$ $\beta$ detectors and two High Purity germanium (HPGe) MiniBall detector clusters for $\gamma$-rays \cite{Eberth2001}. Signals from the detectors were registered by the fully digital acquisition system, which was based on XIA-DGF4C modules \cite{DGFmanual} with an internal 40 MHz clock. More details about the detection system can be found in Ref. \cite{Pauwels2008,Radulov2013}.

\begin{figure*}
\includegraphics[width=\textwidth]{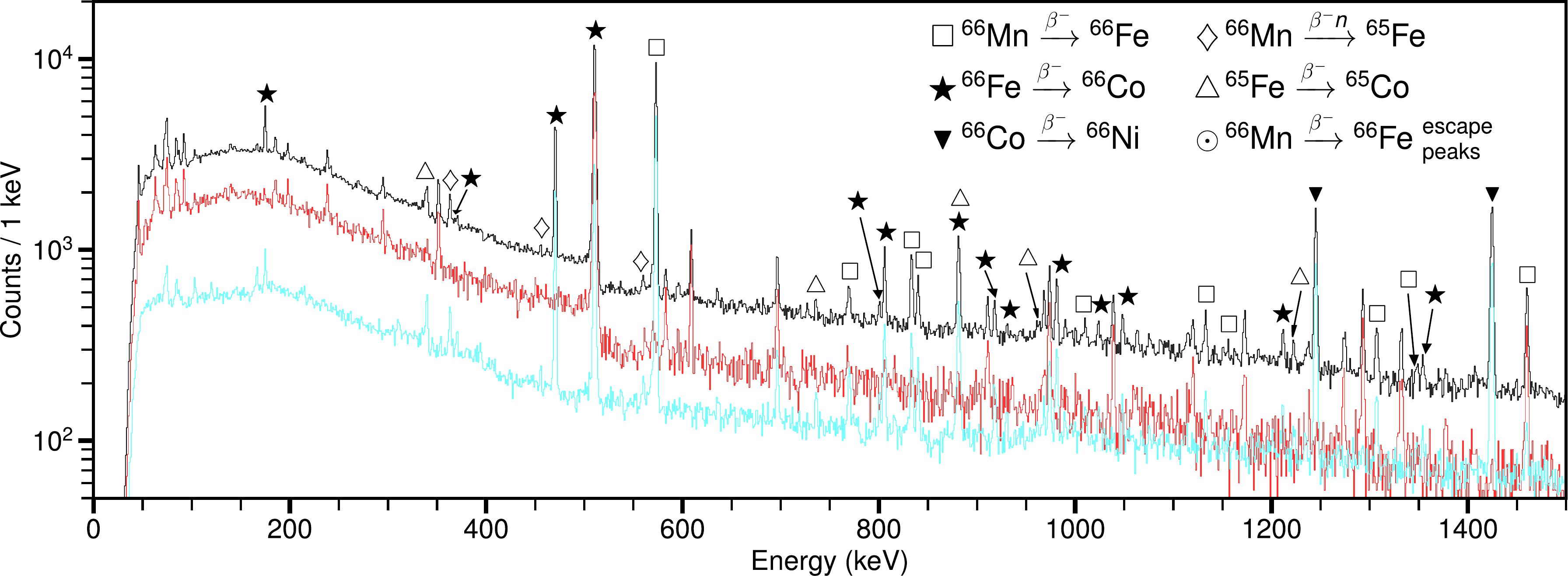}
\includegraphics[width=\textwidth]{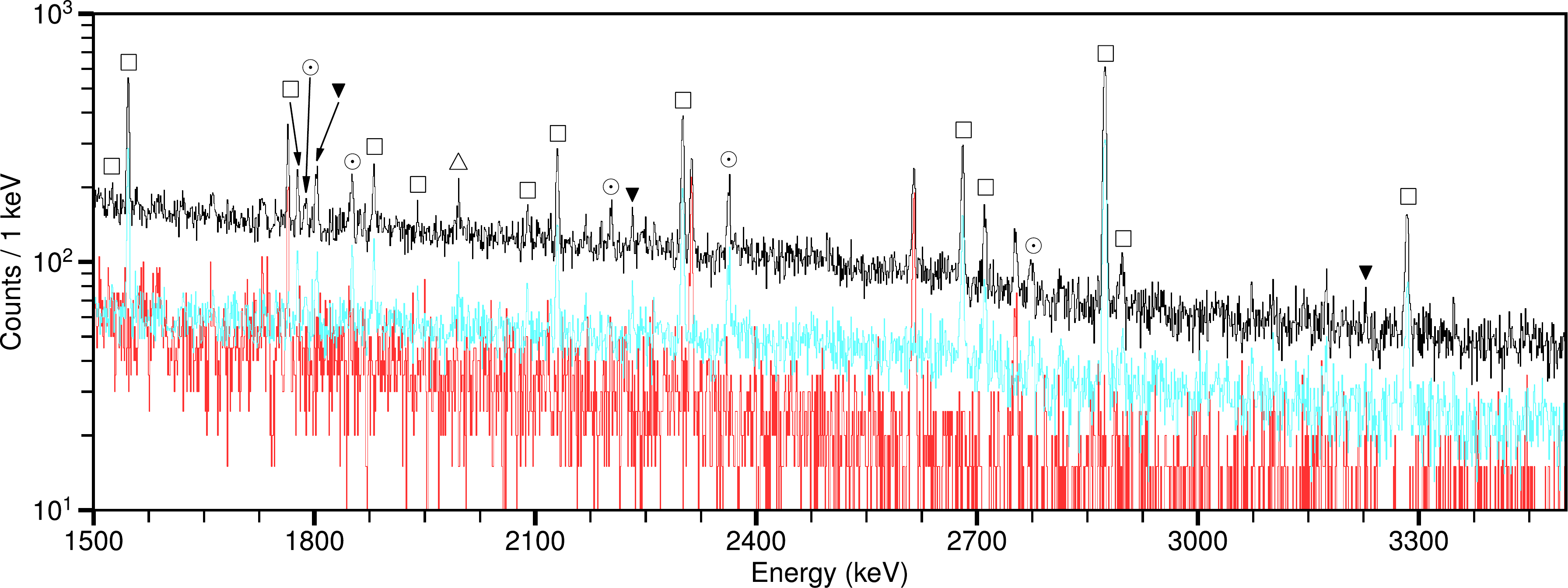}
\caption{\label{fig:singles}(Color online) Single-$\gamma$ spectrum collected in the \textit{laser-on} mode (black) and the \textit{laser-off} mode (red/medium grey, upscaled 5 times for better visual comparison), and $\beta$-$\gamma$ coincidence spectrum collected in the \textit{laser-on} mode (blue/light grey) from 0 to 1500 keV (top panel) and 1500 keV to 3500 keV (bottom panel) with the mass separator set to $A=66$. Peaks attributed to the decay of $^{66}$Mn and daughter activities are marked.}
\end{figure*}

The acquisition cycle was based on the CERN proton supercycle structure (SC). In our case, the SC contains 33 quasi-instantaneous proton pulses (PP), sent every 1.2 seconds. After registration of the PP signal, the acquisition was started and a beam gate was opened for 140 ms to allow the delivery of the manganese beam. Subsequently, the decay radiation was registered for 860 ms. One second after each PP, a forced read-out of 200 ms was performed to store the data. After the last PP in each SC, the tape was moved in order to remove long-lived daughter activities. In our experiments, 32 consecutive PP from each SC, from the 2$^{nd}$ to the 33$^{rd}$, were taken, while the first PP was skipped in order to move the tape.

The germanium detectors calibration was performed using standard sources of $^{60}$Co, $^{133}$Ba, $^{152}$Eu and $^{241}$Am. The measured photopeak $\gamma$-ray efficiency for the cobalt line at 1332 keV is 5.8(1) \% \cite{Pauwels2008,Radulov2013}.

The total measuring time at $A=66$ was 17696 s with the laser set on resonance for the ionization of the manganese (\textit{laser-on} mode) and 3062 s with one of the RILIS lasers blocked (\textit{laser-off} mode). The singles $\gamma$-spectra, which are presented together with the $\beta$-$\gamma$ coincidence spectrum in Fig. \ref{fig:singles}, indicate that the most important contaminants are singly-charged $^{66}$Ga$^{+}$ ions and doubly-charged $^{132}$Sb$^{2+}$. 

A statistical analysis of the data was performed by using the $SATLAS$ code \cite{Gins2017}, which allows to apply Bayesian approach. In this approach, the goal is to obtain the posterior probability density function (\textit{posterior pdf}), which shows the distribution of the model parameters given the data. By applying Bayes' theorem, the posterior pdf ($P(model|data)$) can be expressed as a normalized product of two factors (Eq. \ref{eq:bayes}): the likelihood function ($P(data|model)$), which represents the probability distribution of obtaining the data assuming the model, and the \textit{prior} ($P(model)$), which represents the knowledge about the parameters before the experiment.

\begin{equation}
P(model|data) \propto P(data|model)P(model)
\label{eq:bayes}
\end{equation}

To generate a representative posterior pdf, the Affine Invariant Markov chain Monte Carlo Ensemble sampler \cite{Foreman-Mackey2013}, which is an algorithm implementing a Markov chain Monte Carlo (MCMC) method, was used. A certain amount of samples generated at the beginning of each chain was discarded to accommodate for the tuning of the sampler parameters. This procedure is known as a burn-in. After sampling, a marginalization of the nuisance parameters were performed to obtain the distributions of the parameters of interest. The computed posterior probability density functions represent the entire knowledge of the parameters assuming the given data and the priors. In this work we adopted the 50 percentile of the posterior pdf as a Bayes estimator and the 16 and 84 percentiles as the limits of the 68\% credible interval (analogue of Gaussian $1\sigma$).

\section{\label{sec:results}Results}

\subsection{\label{sec:results66Fe}Decay of $^{66}$Mn}

\begin{figure*}
\includegraphics[width=\textwidth]{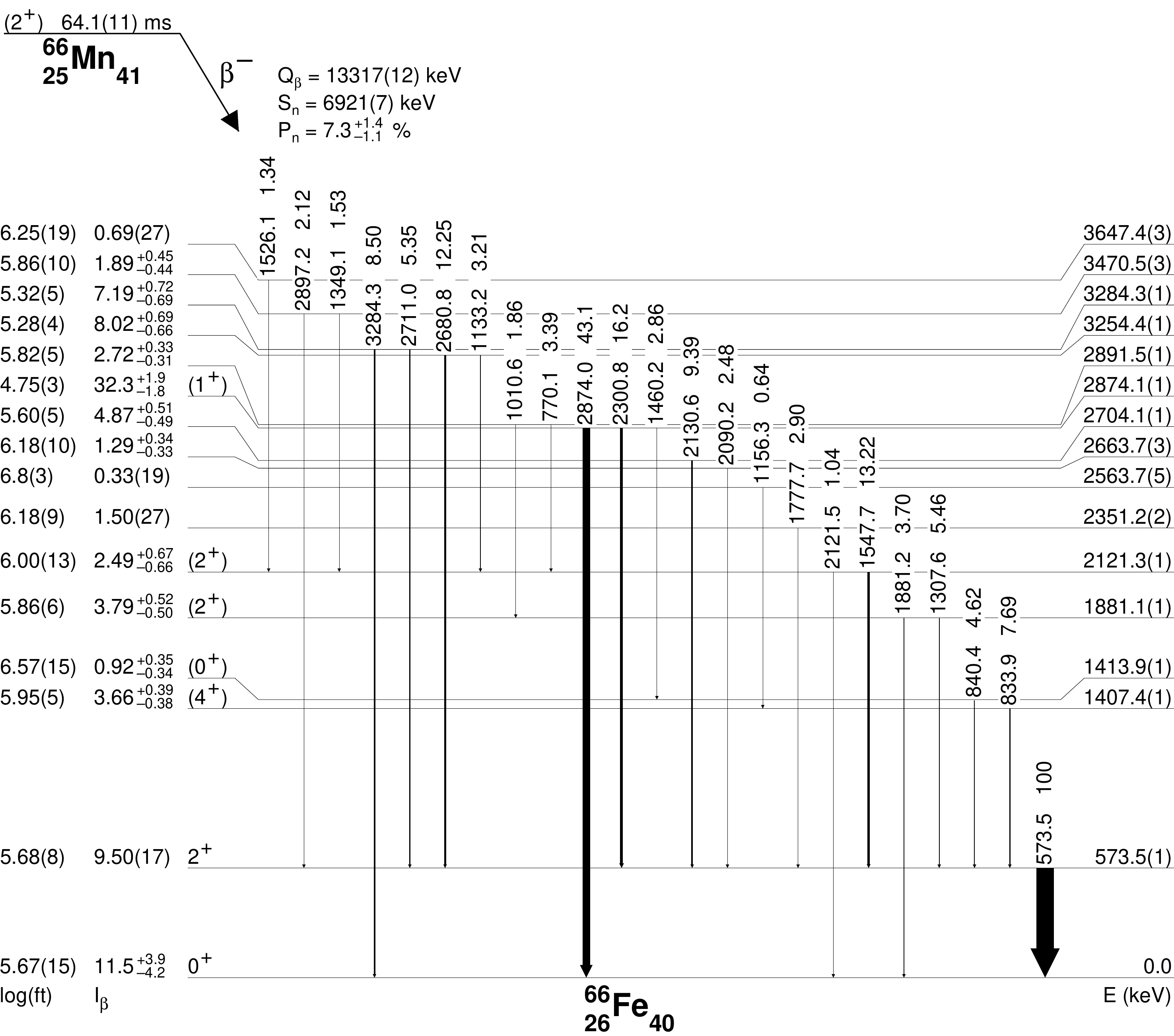}
\caption{\label{fig:66Mndecayscheme} The scheme of excited states in $^{66}$Fe populated in $\beta^-$ decay of $^{66}$Mn. $Q_{\beta^-}$ and $S_n$ values are taken from Ref. \cite{Wang2017}. The $\beta$-feeding of the states should be treated as upper limits and the log(\textit{ft}) values as lower limits due to the pandemonium effect. Half-life and $P_n$ are determined in the analysis. Spin assignments were made based on the experimental data and the Monte Carlo Shell Model calculations, see text for details. The level at 1413.9 keV is shifted 40 keV upwards on the scheme for better visual representation.}
\end{figure*}

\begin{figure}
\includegraphics[width=\columnwidth]{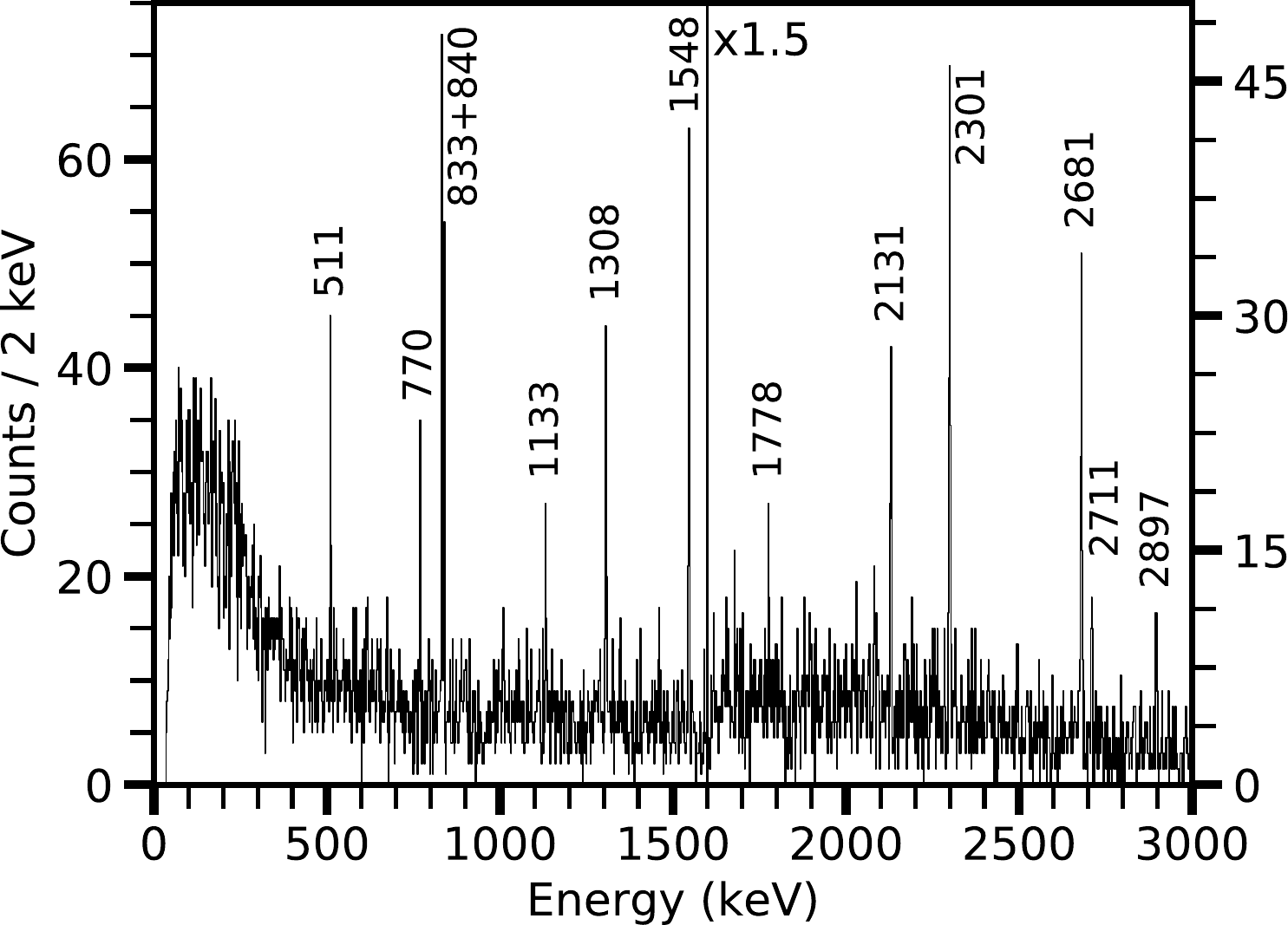}
\caption{\label{fig:gg573}A $\gamma$-$\gamma$ coincidence spectrum gated on the 574 keV transition. The most intense coincidences are labeled with the energy given in keV.}
\end{figure}

The decay scheme of $^{66}$Mn to the excited states in $^{66}$Fe was built by using $\gamma$-$\gamma$ and $\beta$-$\gamma$-$\gamma$ coincidence techniques (Fig. \ref{fig:66Mndecayscheme}). The energy gates were set on the intense $\gamma$-ray transitions known from the literature \cite{Hannawald1999,Lunardi2007,Adrich2008,Gade2010,Daugas2011,Rother2011,Crawford2013,Liddick2013,Radulov2014,Olaizola2017a} (Fig. \ref{fig:gg573}). In order to minimize the background level related to the decay of the daughter activities and in view of the short half-life of $^{66}$Mn (T$_{1/2}$ = 64.1(11) ms, details are presented in Section \ref{sec:resultsgsf}), for this part of the analysis only the data registered up to 400 ms after the PP were taken into account. The identified $\gamma$-ray transitions attributed to the decay of $^{66}$Mn to $^{66}$Fe are presented in the Table \ref{tab:gammas66Fe}.

\begin{table}
\caption{\label{tab:gammas66Fe}
The relative intensities of the $\gamma$-ray transitions assigned to the decay of $^{66}$Mn to $^{66}$Fe, normalized to the 574 keV transition. For the absolute intensity, multiply by 0.519$^{+0.027}_{-0.025}$.}
\begin{ruledtabular}
\begin{tabular}{ccccl}
E$_\gamma$		 	& I$_\gamma^{rel}$	& E$_{\mathrm{level}}^{\mathrm{initial}}$ & E$_{\mathrm{level}}^{\mathrm{final}}$ & Coincident lines (keV) \\
(keV) 				& 				&  (keV) 	&  (keV) 	& \\\hline
573.5(1) 			& 100.00 		& 573.5 		& 0.0		& 770.1, 833.9, 840.4, \\
					& 				& 			& 			& 1010.6, 1133.2, 1307.6, \\
					& 				& 			& 			& 1349.1, 1460.2, 1526.1, \\
					& 				& 			& 			& 1547.7, 1777.7, 2090.2, \\
					& 				& 			& 			& 2130.6, 2300.8, 2680.8, \\
					& 				& 			& 			& 2711.0, 2897.2 \\
770.1(1) 			& 3.39(34) 		& 2891.5 	& 2121.3	& 573.5, 1547.7 \\
833.9(1) 			& 7.69(54) 		& 1407.4 	& 573.5		& 573.5, 1156.3 \\
840.4(1) 			& 4.62(53) 		& 1413.9 	& 573.5		& 573.5, 1460.2 \\
1010.6(2) 			& 1.86(45) 		& 2891.5 	& 1881.1	& 573.5, 1307.6, 1881.2 \\
1133.2(1) 			& 3.21(42) 		& 3254.4 	& 2121.3	& 573.5, 1547.7 \\
1156.3(5) 			& 0.64(37)		& 2563.7 	& 1407.4	& 833.9 \\
1307.6(1) 			& 5.46(50) 		& 1881.1 	& 573.5		& 573.5, 1010.6 \\
1349.1(3) 			& 1.53(50) 		& 3470.5 	& 2121.3	& 573.5, 1547.7 \\
1460.2(2) 			& 2.86(40) 		& 2874.1 	& 1413.9	& 573.5, 840.4 \\
1526.1(3) 			& 1.34(52) 		& 3647.4 	& 2121.3	& 573.5, 1547.7 \\
1547.7(1) 			& 13.22(75) 		& 2121.3 	& 573.5		& 573.5, 770.1, 1133.2, \\
					& 				& 			& 			& 1349.1, 1526.1 \\
1777.7(2) 			& 2.90(50) 		& 2351.2 	& 573.5		& 573.5 \\
1881.2(1) 			& 3.70(66) 		& 1881.1 	& 0.0		& 1010.6 \\
2090.2(3) 			& 2.48(63) 		& 2663.7 	& 573.5		& 573.5 \\
2121.5(3) 			& 1.04(52) 		& 2121.3 	& 0.0		& \textendash \\
2130.6(1) 			& 9.39(84) 		& 2704.1 	& 573.5		& 573.5 \\
2300.8(1) 			& 16.2(13) 		& 2874.1 	& 573.5		& 573.5 \\
2680.8(1) 			& 12.3(10)		& 3254.4 	& 573.5		& 573.5 \\
2711.0(2) 			& 5.35(83) 		& 3284.3 	& 573.5		& 573.5 \\
2874.0(1) 			& 43.1(20) 		& 2874.1 	& 0.0		& \textendash \\
2897.2(4) 			& 2.12(68) 		& 3470.5 	& 573.5		& 573.5 \\
3284.3(1) 			& 8.50(93) 		& 3284.3 	& 0.0		& \textendash \\
\end{tabular}
\end{ruledtabular}
\end{table}

Most of the $\gamma$-ray transitions were placed on the scheme based on the coincidences. The transitions at 2122, 2874 and 3284 keV were placed based on the energy matching between the transition and already identified excited states. The missing coincidences between the 2122 keV line and the 770, 1133, 1349 and 1526 keV $\gamma$-ray transitions can be understood as due to the low intensity of the 2122 keV transition. 

The intensities of the $\gamma$-ray transitions were determined based on the $\beta$-$\gamma$ coincidence spectrum and were normalized to the strongest transition at 574 keV. In order to calculate the $\beta$ feeding to the excited states, the relative feeding of each state was normalized to the sum of all $\gamma$-ray transitions deexciting directly to the ground state (574, 1881, 2122, 2874 and 3284 keV). Then, the obtained values were corrected by the factor, which includes the direct feeding to the ground state ($I_{\beta gsf} = 11.5^{+3.9}_{-4.2} \%$) and the probability of $\beta^-$-delayed-neutron emission ($P_n = 7.3^{+1.4}_{-1.1}$). The asymmetric uncertainties of the $\beta$ feeding of excited states are reflecting the asymmetric uncertainty of the ground state feeding. The discussion regarding $I_{\beta gsf}$ and $P_n$ values is presented in Section \ref{sec:resultsgsf}. 

The log(\textit{ft}) values were calculated with the NNDC calculator \cite{logft}. The half-life of the parent nucleus (T$_{1/2}$ = 64.1(11) ms, see Section \ref{sec:resultsgsf} for details) was taken from our analysis and the $Q_{\beta^-} = 13317(12)$~keV from the AME2016 evaluation \cite{Wang2017}. In the case of asymmetric uncertainties of the input values, the larger value was taken. It should be noted that since the energy window for the decay is large, our $\beta$ feeding values should be treated as upper limits due to the \textit{pandemonium} effect \cite{Hardy1977}. 

Our analysis is extending the decay scheme presented in Ref. \cite{Liddick2013} and is consistent with the results presented recently \cite{Olaizola2017a}. As it was noted in \cite{Olaizola2017a}, the level at 2121 keV was mistakenly quoted at 2130 keV in \cite{Liddick2013}. We did not observe two weak $\gamma$-ray transitions reported at 2246 keV and 3074 keV, neither in coincidence with 574 keV transition nor in the single spectrum. The true summing effect was checked for the transition at 2122 keV, which was not observed in the previous $\beta^-$ decay studies, and it is included in the uncertainty of the intensity.

\subsubsection*{$\beta$-delayed-neutron channel}

Since the neutron separation energy in $^{66}$Fe ($S_n = 6921(7)$~keV) is much lower than $Q_{\beta^-} = 13317(12)$~keV \cite{Wang2017}, the emission of $\beta^-$-delayed-neutrons is possible. In our analysis we identified 4 $\gamma$-ray transitions assigned to this channel (Table \ref{tab:gammas65Fe}). The transitions at 364, 456 and 561 keV were already reported in the previous analysis of the $^{66}$Mn $\beta^-$ decay \cite{Olaizola2017a}.

\begin{table}[h]
\caption{\label{tab:gammas65Fe}
The relative intensities of the $\gamma$-ray transitions assigned to the $\beta$-delayed-neutron decay of $^{66}$Mn, normalized to 100 units of the 574 keV transition in $^{66}$Fe.}
\begin{ruledtabular}
\begin{tabular}{cccc}
E$_\gamma$ (keV) 	&  I$_\gamma^{rel}$ 	& E$_{\mathrm{level}}^{\mathrm{initial}}$ (keV) & E$_{\mathrm{level}}^{\mathrm{final}}$ (keV) 		\\\hline
363.7(1) 			& 5.24(37)			& 363.7		& 0.0 		\\
455.9(2)				& 0.89(21)			& 455.9		& 0.0 		\\
162.7(3)\footnotemark[1] 	& 0.30(8)\footnotemark[2]	& 560.6		& 397.9 		\\
560.6(2)				& 1.17(25)			& 560.6		& 0.0 		\\
\end{tabular}
\end{ruledtabular}
\footnotetext[1]{Seen only in a $\gamma$-delayed-$\gamma$ coincidence spectrum gated on the 364 keV transition.}
\footnotetext[2]{Calculated by multiplying the intensity of 561 keV transition by the ratio extracted from \cite{Olaizola2013}. See text for details.}
\end{table}

\begin{figure}
\includegraphics[width=\columnwidth]{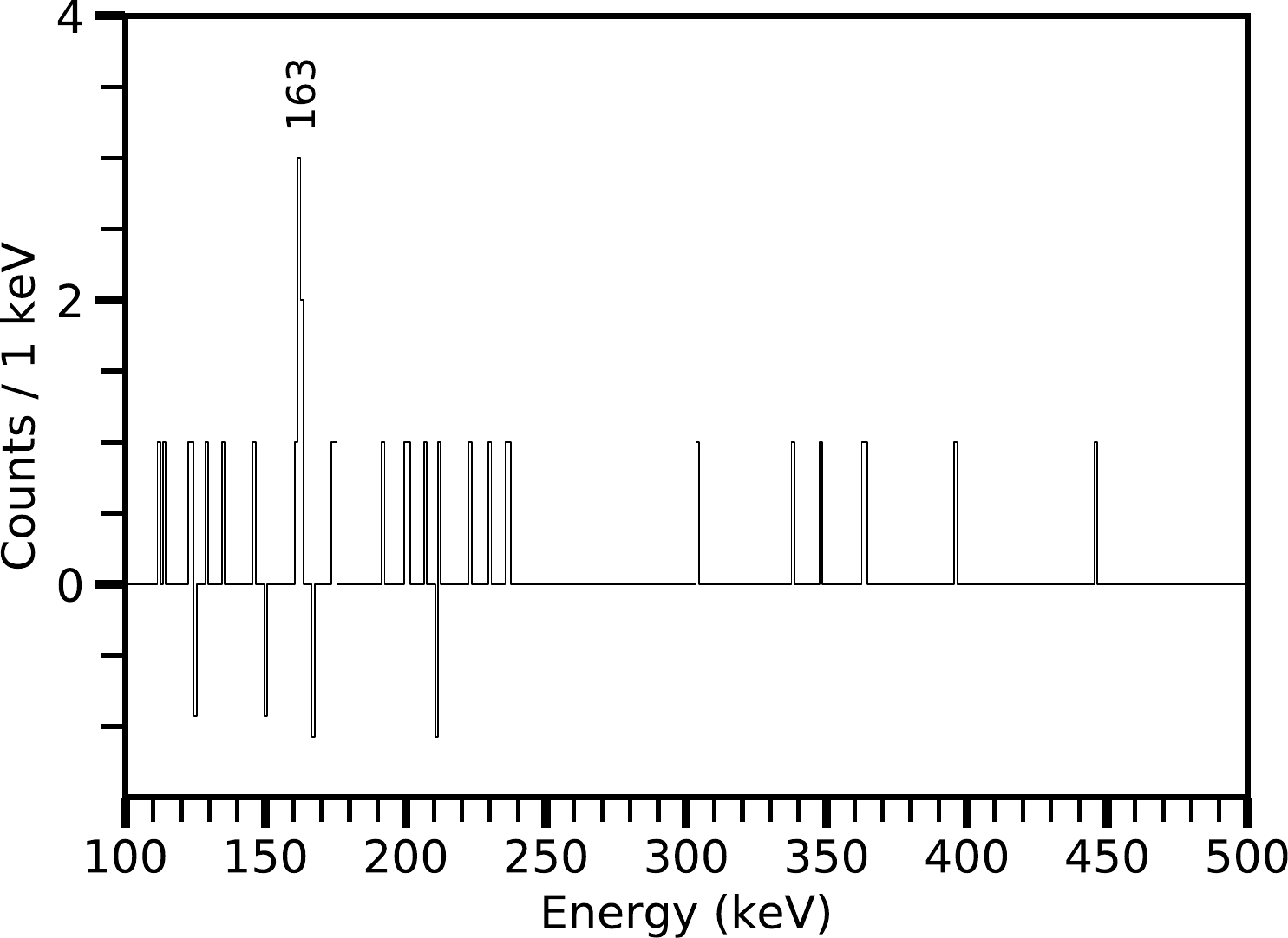}
\caption{\label{fig:gg363} A portion of the $\gamma$-delayed-$\gamma$ spectrum with background subtraction gated on the 364 keV transition (coincidence window: $-5$ to $-0.5$ $\mu$s). A peak at 163 keV is visible.}
\end{figure}

\begin{figure}
\includegraphics[width=\columnwidth]{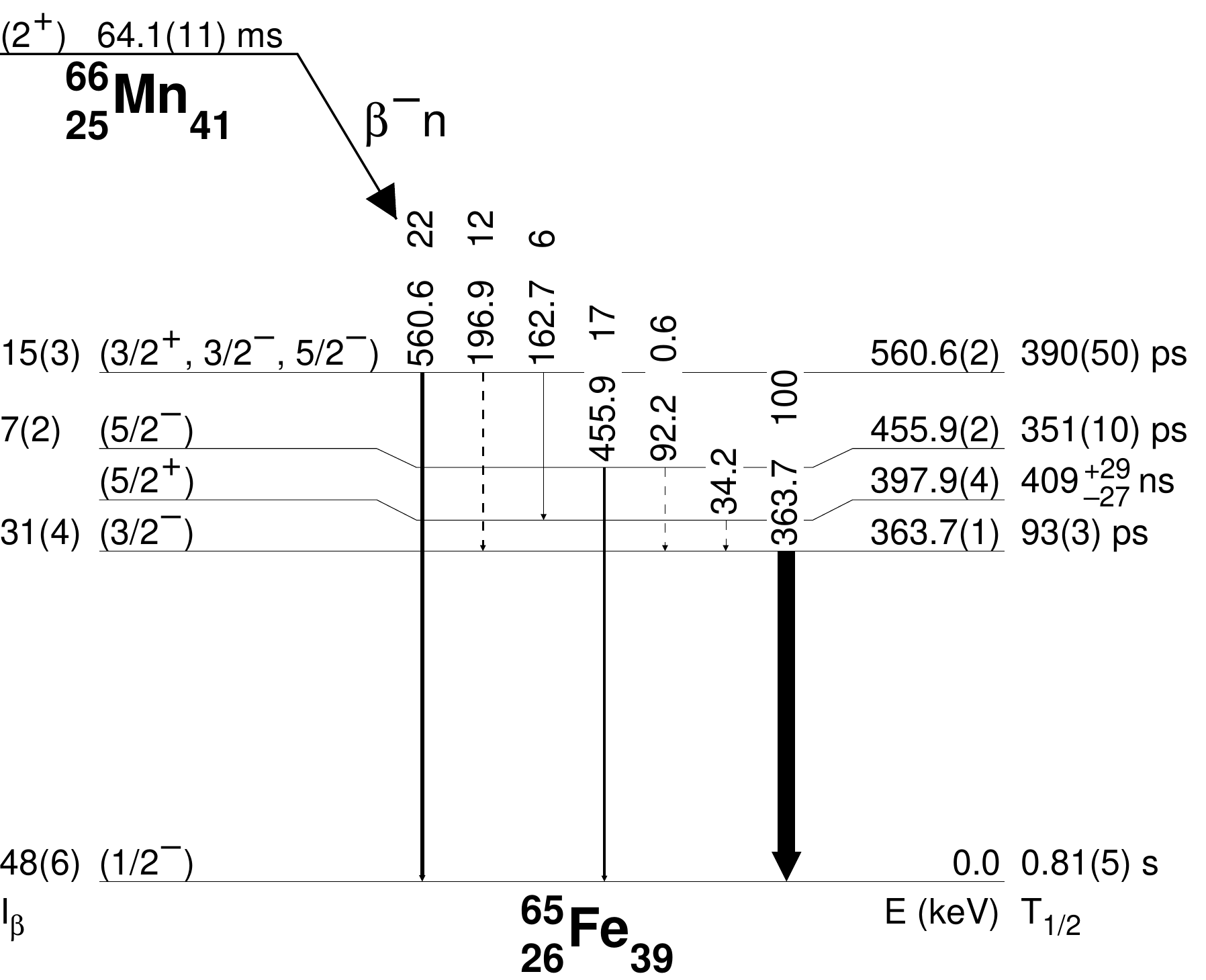}
\caption{\label{fig:66Mndecayscheme-bn} The excited states in $^{65}$Fe populated in $\beta^-$-delayed-neutron decay of $^{66}$Mn. Dotted lines represent transitions reported in Ref. \cite{Olaizola2013}, not observed in our analysis, and their energies are the differences between levels energies. Intensities are normalized to 100 units of the 364 keV transition. Spin assignments and half-lifes of the states in $^{65}$Fe, except the half-life of 398 keV level, are taken from Ref. \cite{Olaizola2013}.}
\end{figure}

The scheme of excited states based on the identified $\gamma$-rays is presented in Fig. \ref{fig:66Mndecayscheme-bn}. The intensities of the unobserved transitions, which are known from the $\beta^-$ decay studies of $^{65}$Mn \cite{Olaizola2013}, were included into apparent $\beta$ feeding calculations. They were obtained by multiplying the intensity of the observed $\gamma$-ray by the ratio of the intensities extracted from \cite{Olaizola2013}. The same procedure was performed for the 163 keV transition, which was observed only in the $\gamma$-delayed-$\gamma$ spectrum gated on the 364 keV transition (see Fig. \ref{fig:gg363}). Since the 34 keV $\gamma$-ray transition is below the measurable energy range of our setup, it was not possible to determine the feeding of the 398 keV level and, as a result,  the reported feeding of 364 keV transition should be treated as a sum of the 364 keV and 398 keV levels feedings. During the analysis we did not observe any transition which can be associated with a decay of the $^{65}$Fe high-spin isomeric state at 394 keV \cite{Olaizola2013}, hence, this level is not presented in our decay scheme. The details regarding the probability of the $\beta^-$-delayed-neutron decay ($P_n = 7.3^{+1.4}_{-1.1}$) are presented in Section \ref{sec:resultsgsf}.

To obtain the direct feeding to the $^{65}$Fe ground state, the $\gamma$-detection efficiency corrected counts of the 364, 456 and 561 keV ground state transitions were compared to the corrected counts of the 340, 736, 961, 1076 and 1223 keV observed in the $\beta^-$-decay of $^{65}$Fe. By using the absolute intensities reported in Ref. \cite{Radulov2014} and by making a cycle correction, we obtain 48(6)\% feeding of the $^{65}$Fe ground state in the $\beta^-$-delayed-neutron decay of $^{66}$Mn. This result is larger than 33(5)\% reported in \cite{Olaizola2017a} but consistent within $2\sigma$. 

\subsubsection*{Half-life of $^{65m2}$Fe}

The half-life of the isomeric state in $^{65}$Fe at 398 keV \cite{Grzywacz1998,Olaizola2013,Radulov2014} was deduced from the time difference between the signal from the $\beta$ detector (\textit{start}) and the 364 keV $\gamma$-ray transition deexciting the isomeric state (\textit{stop}). The fitting region was set from 600 ns up to 6 $\mu$s after the \textit{start} signal to remove the possible direct $\beta$ or intermediate $\gamma$-ray feeding to the 364 keV level.

\begin{figure}
\includegraphics[width=\columnwidth]{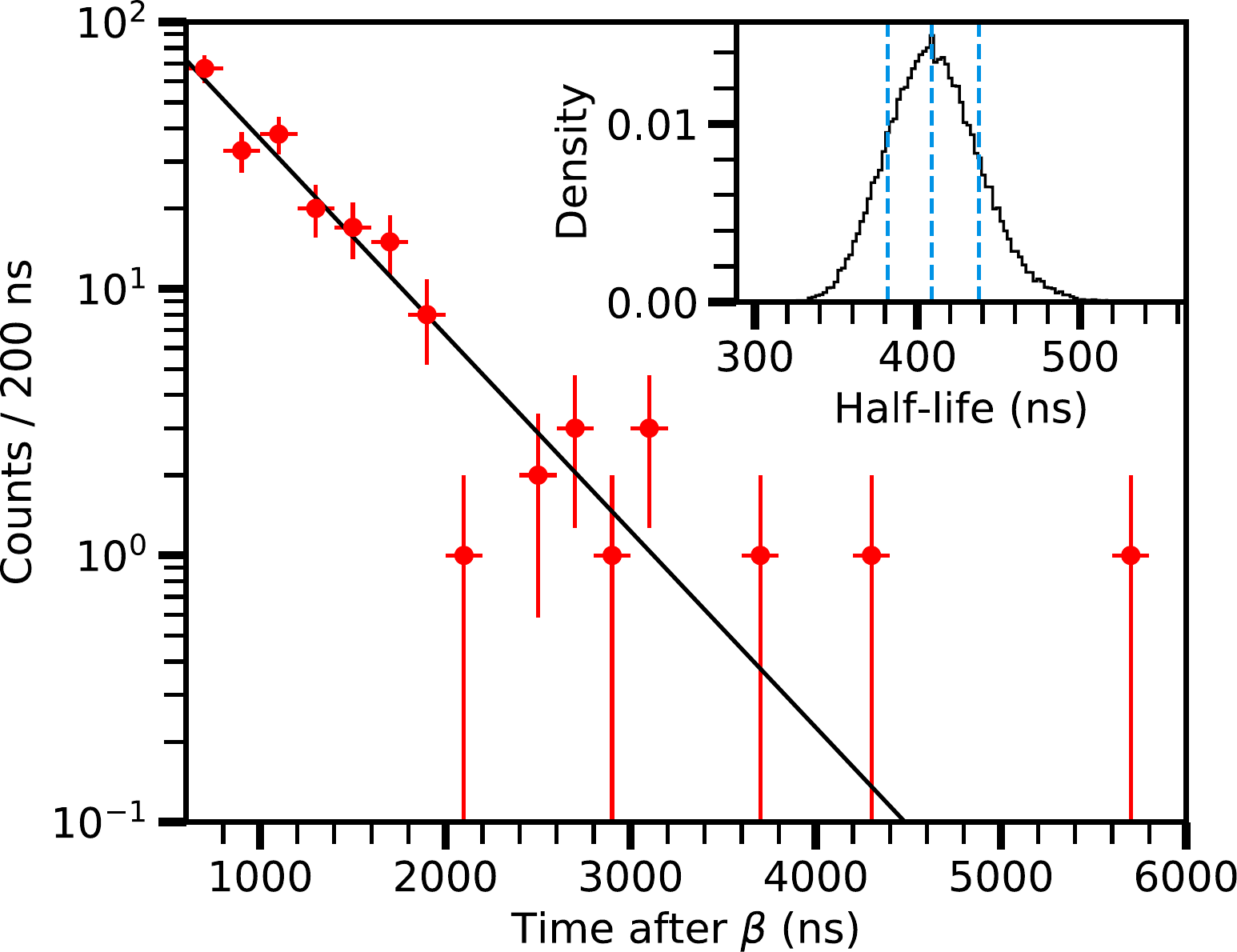}
\caption{\label{fig:65m2Fefit} Time behavior of the 364 keV $\gamma$-ray transition as a function of time after the $\beta$ signal with the fitted function. Insert: posterior probability density function of the half-life of the second isomeric state in $^{65}$Fe ($T_{1/2} = 409^{+29}_{-27}$~ns). The 16, 50 and 84 percentiles are indicated with vertical, dotted lines.}
\end{figure}

In our analysis the used model is the exponential decay $A(t) = A \cdot \mathrm{exp}(-\frac{ln(2)}{T_{1/2}}t)$. The likelihood function was built assuming that the number of counts in each bin is following the Poisson distribution while both free parameters, $A$ and $T_{1/2}$, were constrained by the prior to be non-negative. In total 100000 samples were taken from the posterior pdf (20 walkers with 5000 steps each), from which the first 15 \% were rejected as a burn-in. After the sampling, a marginalization of $A$, which is a nuisance parameter, was performed. The posterior pdf of the half-life and the fit to the data are presented in Fig. \ref{fig:65m2Fefit}. The obtained value, T$_{1/2} = 409^{+29}_{-27}$~ns, is in agreement with the previous experimental results reported by Grzywacz \textit{et al.} (430(130) ns \cite{Grzywacz1998}), Georgiev (434(35) ns \cite{Georgiev2001}), Daugas \textit{et al.} (420(13) ns \cite{Daugas2010}), Olaizola \textit{et al.} (437(55) ns \cite{Olaizola2013}) and Radulov (428(11) ns \cite{Radulov2014}).

\subsection{\label{sec:results66Co}Decay of $^{66}$Fe}

\begin{figure*}
\includegraphics[width=\textwidth]{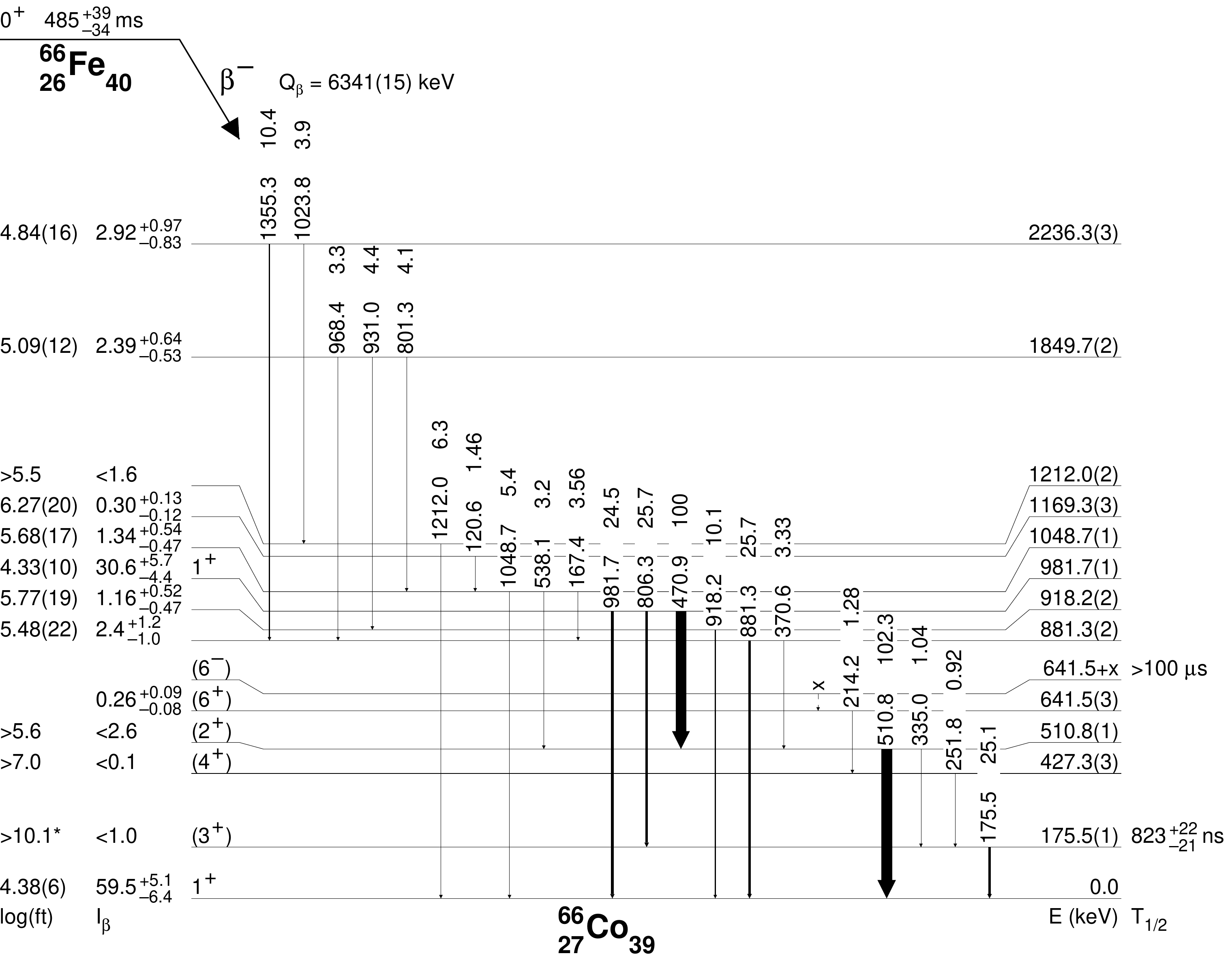}
\caption{\label{fig:66Fedecayscheme} The scheme of excited states in $^{66}$Co populated in $\beta^-$ decay of $^{66}$Fe. The $\beta$-feeding of the states should be treated as upper limits and the log(\textit{ft}) values as lower limits due to the pandemonium effect. The $Q_{\beta^-}$ value is taken from Ref. \cite{Wang2017} and the limit for the half-life of the second isomeric state from Ref. \cite{Grzywacz1998}. The half-life of the parent nucleus and the first excited state at 176 keV come from our analysis. The spin and parity assignments are made based on the experimental results and the Monte Carlo Shell Model calculations, see text for details. When indicated with an asterisk (*), the log(\textit{ft}) value was calculated assuming second-forbidden unique transition.}
\end{figure*}

\begin{table}[h]
\caption{\label{tab:gammas66Co}
Relative intensities (I$_\gamma^{rel}$) of the $\gamma$-ray transitions attributed to the decay of $^{66}$Fe to $^{66}$Co, normalized to 100 units of the 471 keV transition. For absolute intensity, multiply by 0.204$^{+0.037}_{-0.029}$.}
\begin{ruledtabular}
\begin{tabular}{ccccl}
E$_\gamma$		 	& I$_\gamma^{rel}$	& E$_{\mathrm{level}}^{\mathrm{initial}}$ & E$_{\mathrm{level}}^{\mathrm{final}}$ & Coincident lines (keV) \\
(keV) 				& 				&  (keV) 	&  (keV) 	& \\\hline
120.6(2) 			& 1.46(57)		& 1169.3 	& 1048.7	& 167.4 \\
167.4(1) 			& 3.56(69) 		& 1048.7 	& 881.3		& 120.6, 801.3, 881.3 \\
175.5(1) 			& 25.1(20) 		& 175.5 		& 0.0		& 214.2\footnotemark[1], 251.8\footnotemark[1], 335.0\footnotemark[1], \\
					&				&			&			& 470.9\footnotemark[1], 806.3 \\
214.2(2)\footnotemark[2]	& 1.28(36) 	& 641.5 		& 427.3		& 175.5\footnotemark[1] \\
251.8(3)\footnotemark[2]	& 0.92(24) 	& 427.3 		& 175.5		& 175.5\footnotemark[1] \\
335.0(3)\footnotemark[2]	& 1.04(32) 	& 510.8 		& 175.5		& 175.5\footnotemark[1] \\
370.6(3) 			& 3.33(89) 		& 881.3 		& 510.8		& 510.8 \\
470.9(1) 			& 100.0 			& 981.7 		& 510.8		& 510.8 \\
510.8(1)\footnotemark[3] 	& 102.3$^{+8.4}_{-10.0}$\footnotemark[4] 	& 510.8 		& 0.0		& 370.6, 470.9, 538.1 \\
538.1(2) 			& 3.2(10)		& 1048.7 	& 510.8		& 510.8 \\
801.3(2) 			& 4.1(13) 		& 1849.7 	& 1048.7	& 167.4, 1048.7 \\
806.3(1) 			& 25.7(25) 		& 981.7 		& 175.5		& 175.5 \\
881.3(2)\footnotemark[5] 	& 25.7(38)\footnotemark[6] 	& 881.3 		& 0.0		& 167.4, 968.4, 1355.3 \\
918.2(2) 			& 10.1(19) 		& 918.2		& 0.0		& 931.0 \\
931.0(3) 			& 4.4(12) 		& 1849.7 	& 918.2		& 918.2 \\
968.4(2) 			& 3.3(12) 		& 1849.7 	& 881.3		& 881.3 \\
981.7(1) 			& 24.5(30) 		& 981.7 		& 0.0		& \textendash\\
1023.8(4) 			& 3.9(22)		& 2236.3 	& 1212.0	& 1212.0 \\
1048.7(1) 			& 5.4(12) 		& 1048.7 	& 0.0		& 801.3 \\
1212.0(2) 			& 6.3(20) 		& 1212.0 	& 0.0		& 1023.8 \\
1355.3(3) 			& 10.4(30) 		& 2236.3 	& 881.3		& 881.3 \\
\end{tabular}
\end{ruledtabular}
\footnotetext[1]{Seen in $\gamma$-delayed-$\gamma$ coincidence.}
\footnotetext[2]{Energy and intensity obtained from $\gamma$-delayed-$\gamma$ coincidence spectrum gated on 176~keV transition.}
\footnotetext[3]{Energy obtained from $\gamma$-$\gamma$ coincidence spectrum gated on 471~keV line.}
\footnotetext[4]{Intensity obtained by analyzing the time behavior of the transition, see text for details.}
\footnotetext[5]{Energy obtained from $\gamma$-$\gamma$ coincidence spectrum gated on 167~keV line.}
\footnotetext[6]{Intensity obtained by subtracting intensity related to the decay of $^{65}$Fe to $^{65}$Co.}
\end{table}

\begin{figure}
\includegraphics[width=\columnwidth]{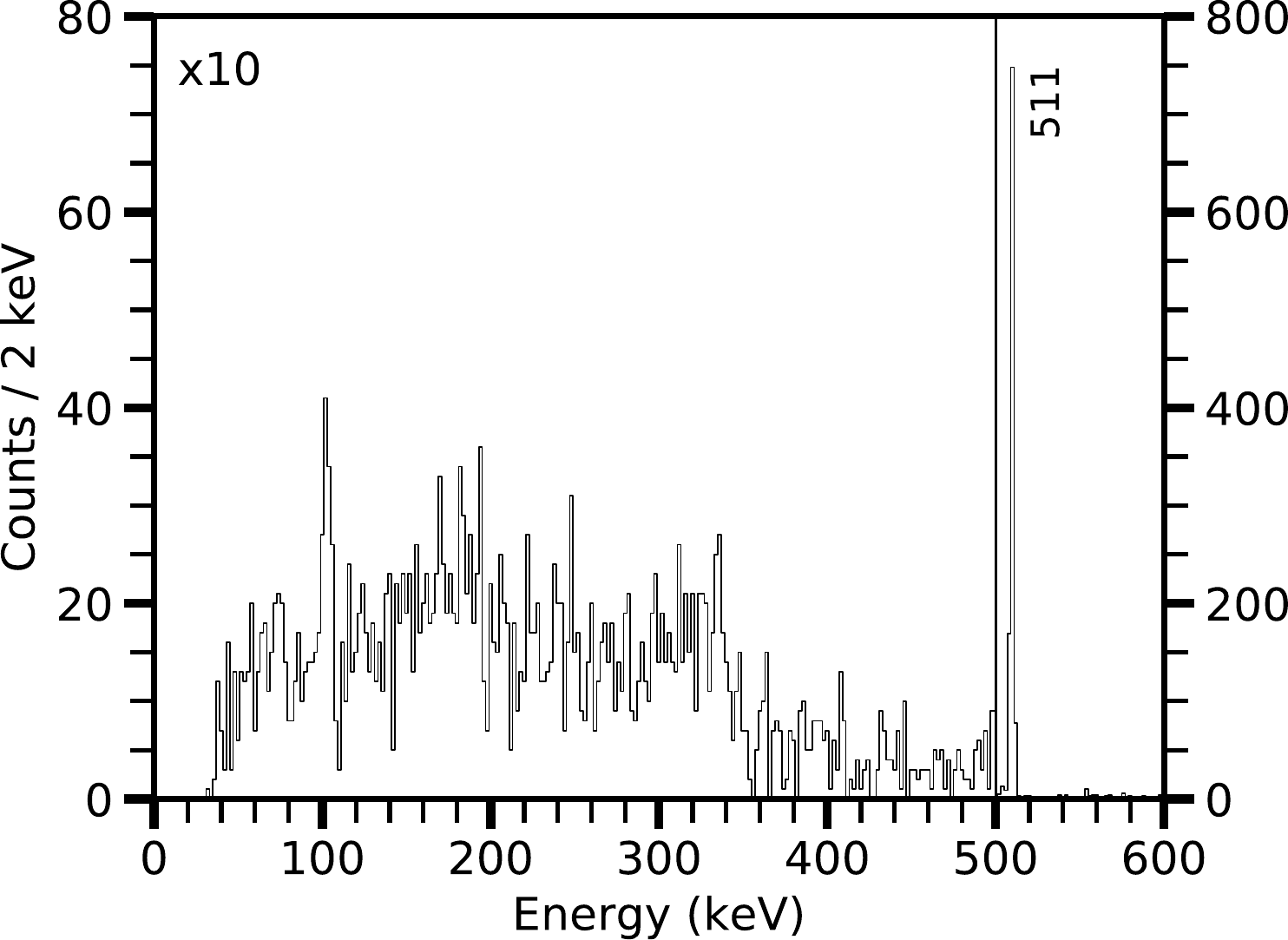}
\caption{\label{fig:gg470}A portion of the $\gamma$-$\gamma$ coincidence spectrum gated on the 471 keV transition. Beyond 600~keV no peaks were observed.}
\end{figure}

The scheme of excited states in $^{66}$Co was built using the techniques described in the previous section and is presented in Fig. \ref{fig:66Fedecayscheme}. The energy gates were set on the $\gamma$-ray transitions known from the previous  $\beta^-$ decay studies of $^{66}$Fe \cite{Ivanov2007,Liddick2012} (Fig. \ref{fig:gg470}). The list of transitions attributed to the decay of $^{66}$Fe is presented in the Table \ref{tab:gammas66Co}. 

To determine the energies and intensities of the transitions assigned to the decay of $^{66}$Fe, the data collected from 400 ms to 1 s after PP was used. This time condition allows to reduce the background coming from the $^{66}$Mn decay. The transitions for which determination of energy or intensity from $\beta$-$\gamma$ coincidence spectrum was not possible are described below.

\begin{figure}
\includegraphics[width=\columnwidth]{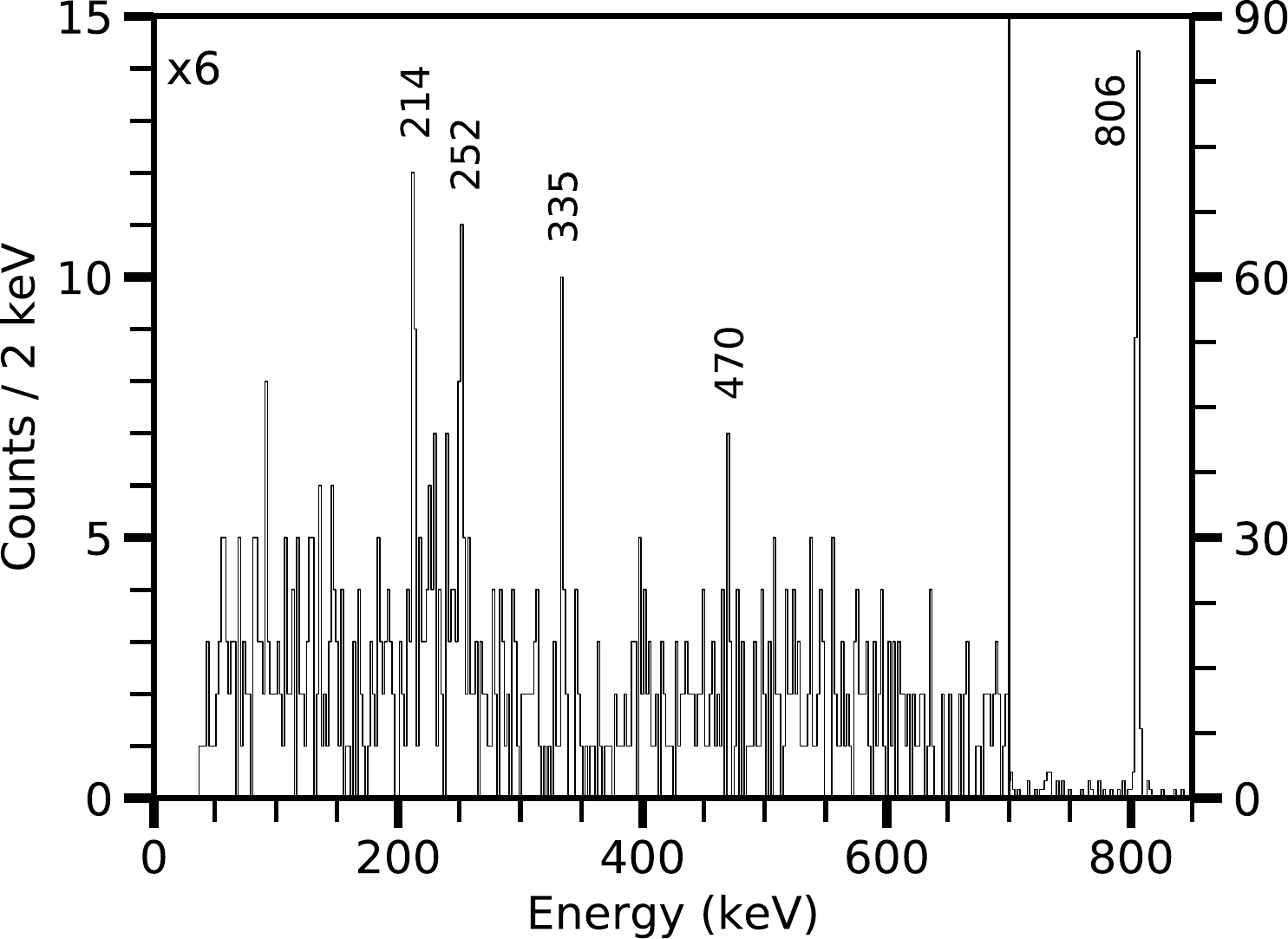}
\caption{\label{fig:gdg175}A portion of the $\gamma$-delayed-$\gamma$ coincidence spectrum gated on the 176 keV transition (coincidence time window $-2.5$ to $-0.2$ $\mu$s). The coincide transitions are labeled with the energy in keV.}
\end{figure}

The intensity of the transition at 176 keV deexciting an isomeric state was determined by using the single-$\gamma$ spectrum. The area of the $\gamma$-ray peak was corrected by the $\gamma$-ray detection efficiency and compared to the intensity of the 471 keV transition. Later, a $\gamma$-delayed-$\gamma$ spectrum gated on the 176 keV transition was used to obtain the intensities of the 214, 252 and 335 keV transitions (Fig. \ref{fig:gdg175}). Their peak areas were compared to the peak area of the 806 keV transition.

The $\gamma$-ray of 881 keV is emitted in the $\beta^-$ decay of $^{66}$Fe and $^{65}$Fe \cite{Pauwels2009,Radulov2014}, which is produced in the $\beta^-$-delayed-neutron decay of $^{66}$Mn. The intensity of this transition related solely to $^{65}$Co was extracted by taking the relative intensity of the 340 keV transition and multiplying it by the ratio of absolute intensities taken from \cite{Pauwels2009}. Then, the obtained value was subtracted from the total intensity of the 881 keV transition yielding to the intensity related to $^{66}$Co. The energy of this $\gamma$-ray was determined using the $\gamma$-$\gamma$ coincidence spectrum gated on the 167 keV transition.



\begin{figure}
\includegraphics[width=\columnwidth]{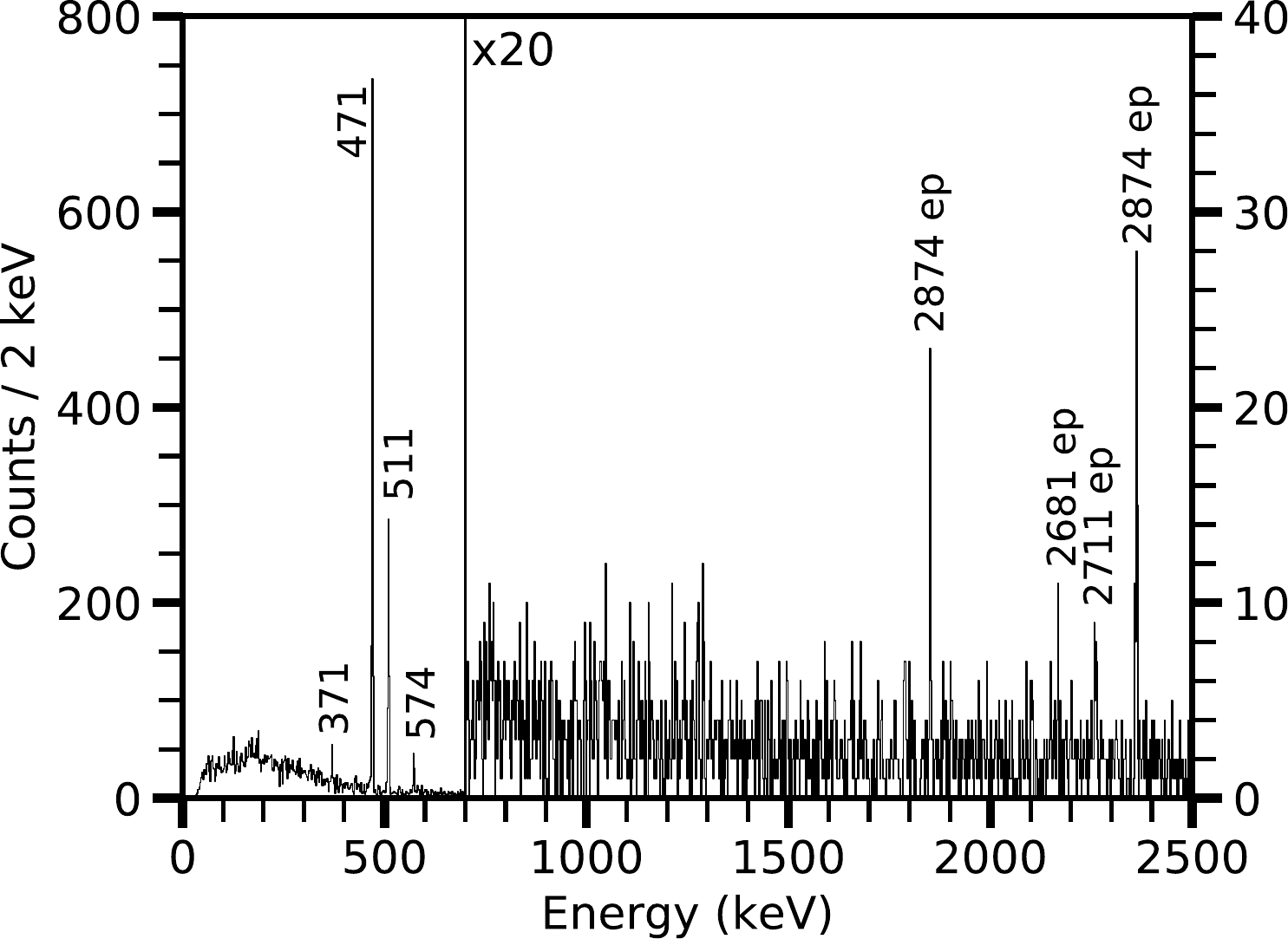}
\caption{\label{fig:gg511}A portion of the $\gamma$-$\gamma$ coincidence spectrum gated on the 511 keV transition. Transitions at 371 and 471 keV are assigned to the decay of $^{66}$Fe. There are also visible transitions assigned to the decay of $^{66}$Mn: escape peaks (ep) of the high energy $\gamma$-ray transition and their coincidences (the transition at 574 keV).}
\end{figure}

The intensity of the 511 keV transition was determined by analyzing the number of $\beta$-gated-$\gamma$ counts as a function of time after PP. We assumed there are four main sources of $\gamma$-rays with this energy: $\beta^-$ decay of $^{66}$Fe, escape peaks from the $^{66}$Mn decay high-energy $\gamma$-rays (Fig. \ref{fig:gg511}), Compton-scattered $\gamma$-rays and the environmental radiation. We also assumed there might be 511~keV $\gamma$-rays of different origin, for example the escape peaks of weak unobserved transitions from the decay of nuclei other than $^{66}$Mn. The intensity obtained from our analysis is equal $I_{511} = 102.3^{+8.4}_{-10.0}$. The details regarding the fitting procedure are presented in Appendix \ref{apx:511intensity}.

The presence of an isomeric state ($T_{1/2} > 100 \mu s$) in $^{66}$Co was first reported by Grzywacz et al. and it was proposed to be a high-spin state which deexcites through 252 keV and 214 keV $\gamma$-ray transitions \cite{Grzywacz1998}. These two transitions were observed in the $\gamma$-delayed-$\gamma$ spectrum gated on the 176 keV transition (Fig. \ref{fig:gdg175}). The results obtained in the multinucleon transfer studies of $^{70}$Zn beam on $^{238}$U target performed at Laboratori Nazionali di Legnaro suggest the order of the $\gamma$-rays in the cascade deexciting the 642 keV state should be reversed compared to Ref. \cite{Grzywacz1998} since only the 252 keV transition was observed \cite{Recchia2012}. These two transitions (214 keV and 252 keV) were also observed in two independent deep-inelastic reactions ($^{70}$Zn+$^{208}$Pb and $^{64}$Ni+$^{238}$U, see Ref. \cite{Pauwels2009,Broda2012,Chiara2012,Chiara2013} for experimental details) performed at Argonne National Laboratory \cite{ChiaraPrivate}. They were registered in a prompt coincidence window of 40~ns with the beam pulse and with each other, and in a delayed coincidence (outside the 40~ns window) with the 176~keV transition. These measurements contradict the isomeric nature of the 642 keV state, hence, we conclude that the isomeric state lies above the 642 keV level and the energy difference between them is below 50 keV which is the low-energy detection limit reported in \cite{Grzywacz1998}. Since both levels, 642 keV and $642+x$ keV, are proposed to be high-spin state (see discussion in Sec. \ref{sec:discussion}), the feeding of the 642 keV level reported in Fig. \ref{fig:66Fedecayscheme} should be treated as an unobserved feeding related to the pandemonium effect. 

For the states with apparent $\beta$ feeding consistent with zero, 95\% credible limits were calculated. The log(\textit{ft}) values were calculated as described in the previous section. The $Q_{\beta^-}$ was taken from AME2016 \cite{Wang2017} and the half-life (T$_{1/2} = 485^{+39}_{-34}$~ms, see following section for the details) from our analysis.

\subsubsection*{Half-life of $^{66m1}$Co}

The half-life of the first isomeric state was obtained by analyzing the time behavior of the 176 keV transition after the $\beta$ signal ($\beta$-$\gamma$ coincidence) and after the 806 keV transition ($\gamma$-$\gamma$ coincidence). To obtain the the $\gamma$-$\gamma$ coincidence data, the 806 keV transition was chosen as a \textit{start} signal and the 176 keV transition as a \textit{stop} signal. The data were fitted from 0.5 to 10 $\mu$s assuming an exponential decay model (Eq. \ref{eq:isomergg}). 

\begin{equation}
A_\gamma(t) = A_{\gamma0} e^{-\frac{ln(2)}{T_{1/2}}t}
\label{eq:isomergg}
\end{equation}

\noindent For the $\beta$-$\gamma$ coincidence, two separate sets of data were prepared. The first set contained the number of counts in the background area as a function of the time after the $\beta$ signal (light area in the insert of Fig. \ref{fig:66m1Cobg}) and it was described by an exponential decay model with a constant (Eq. \ref{eq:isomerbgbkg}), which reflects the existence of a time-dependent and a time-independent part of the background. 

\begin{equation}
A_{bkg}(t) = A_{bkg0} e^{-\frac{ln(2)}{T_{bkg}}t} + C
\label{eq:isomerbgbkg}
\end{equation}

\noindent The second set contained the counts in the peak area (dark area in the insert of Fig. \ref{fig:66m1Cobg}) which was described as an exponential decay function with the isomer half-life, and an exponential function and a constant to include the background (Eq. \ref{eq:isomerbg}).

\begin{equation}
A_{\beta}(t) = A_{\beta0} e^{-\frac{ln(2)}{T_{1/2}}t} + A_{bkg0} e^{-\frac{ln(2)}{T_{bkg}}t} + C
\label{eq:isomerbg}
\end{equation}

\noindent The background area was normalized to the number of channels in the peak area. It was assumed that the parameters in Eq. \ref{eq:isomerbgbkg} and background parameters in Eq. \ref{eq:isomerbg} ($A_{bkg0}$, $T_{bkg}$ and $C$) are identical. For both sets of data, the fitting region was set from 0.6 to 10 $\mu$s after $\beta$ signal. 

All datasets were fitted simultaneously assuming that the free parameters are non-negative, which was provided by using priors, and that counts in each bin are described by Poisson distribution. The random walk was performed with 20 walkers and 10000 steps, from which the first 15\% were rejected as a burn-in. The fits are presented in Fig. {\ref{fig:66m1Cobg} and \ref{fig:66m1Cogg} and the posterior probability density function of the half-life after the marginalization is presented as an insert of Fig. \ref{fig:66m1Cogg}. The value obtained in our analysis is equal to $823^{+22}_{-21}$~ns. It is in agreement with the half-life reported by Georgiev (830(10) ns \cite{Georgiev2001}) while there is a difference with the result reported by Grzywacz \textit{et al.} (1.21(1) $\mu$s \cite{Grzywacz1998}). Georgiev suggested that the value reported by Grzywacz \textit{et al.} is a mean lifetime since the difference between results is of about factor ln(2).

\begin{figure}
\includegraphics[width=\columnwidth]{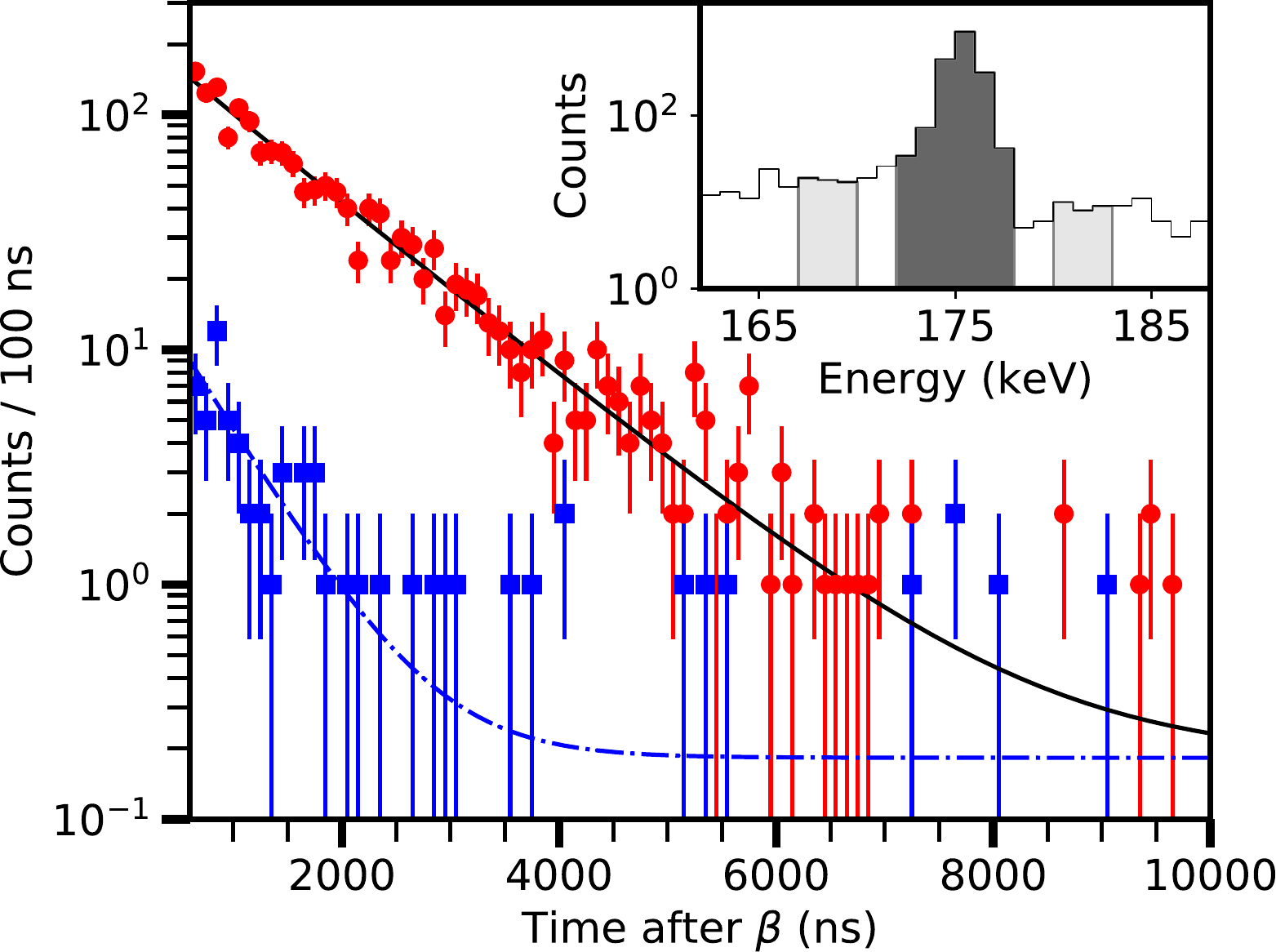}
\caption{\label{fig:66m1Cobg} (Color online) Counts in the peak area of the 176 keV $\gamma$-ray transition as a function of time after the $\beta$ particle (red circles) with the fitted function (black straight line) and counts in the background area (blue squares) with the fitted function (blue dash-dotted line). Insert: a portion of the $\beta$-delayed-$\gamma$ spectrum (coincidence time 0.5 to 10 $\mu$s) in the 176 keV $\gamma$-ray transition region with marked peak area (dark shade) and background areas (light shades). See text for details.}
\end{figure}

\begin{figure}
\includegraphics[width=\columnwidth]{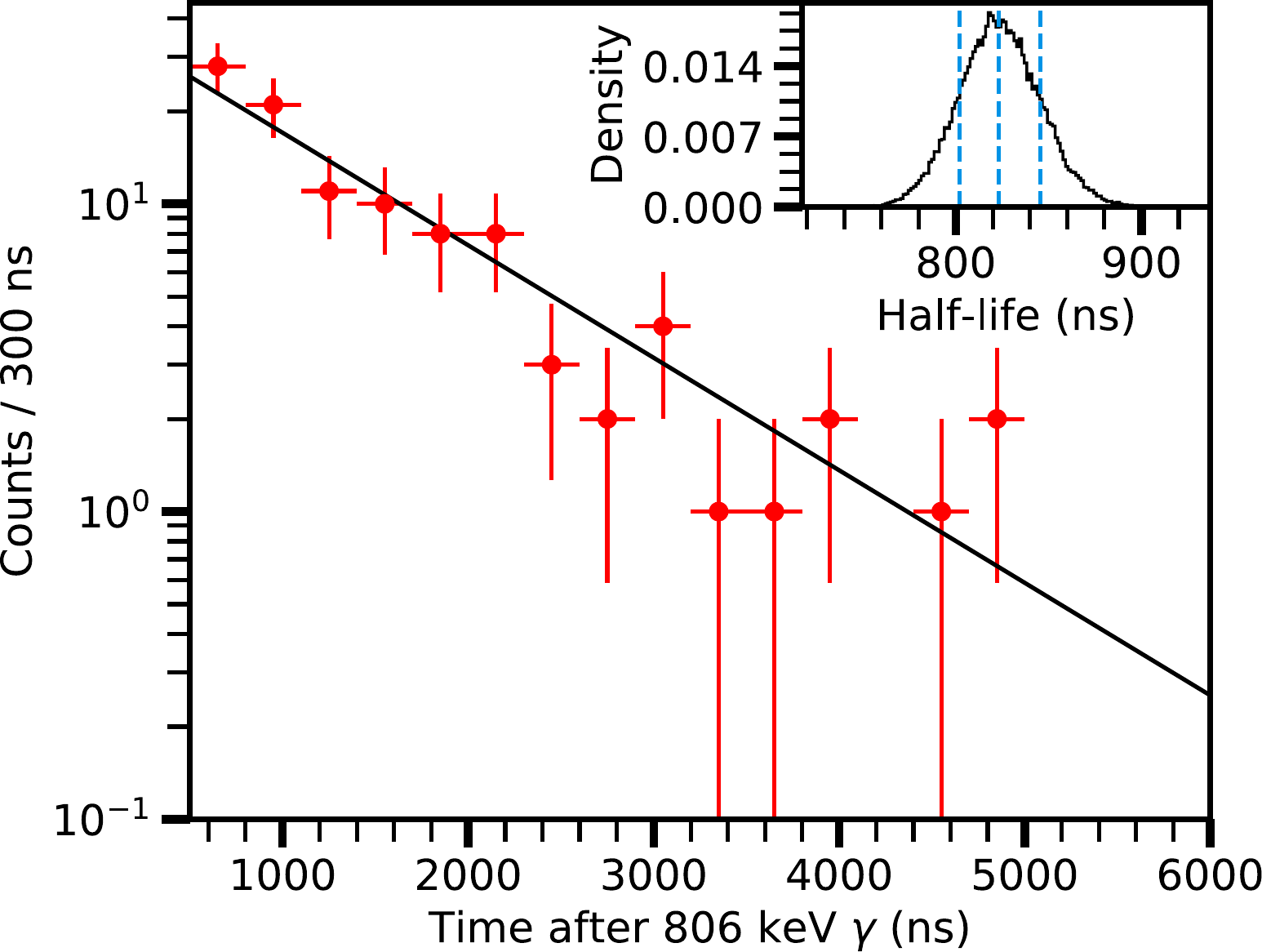}
\caption{\label{fig:66m1Cogg} Time behavior of the 176 keV $\gamma$-ray transition as a function of time after the 806 keV $\gamma$-ray transition. Insert: posterior probability density function of the half-life of the first isomeric state in $^{66}$Co ($T_{1/2} = 823^{+22}_{-21}$~ns). The 16, 50 and 84 percentiles are indicated with vertical, dotted lines.}
\end{figure}

\subsection{\label{sec:results66Ni}Decay of $^{66}$Co}

\begin{figure}
\includegraphics[width=\columnwidth]{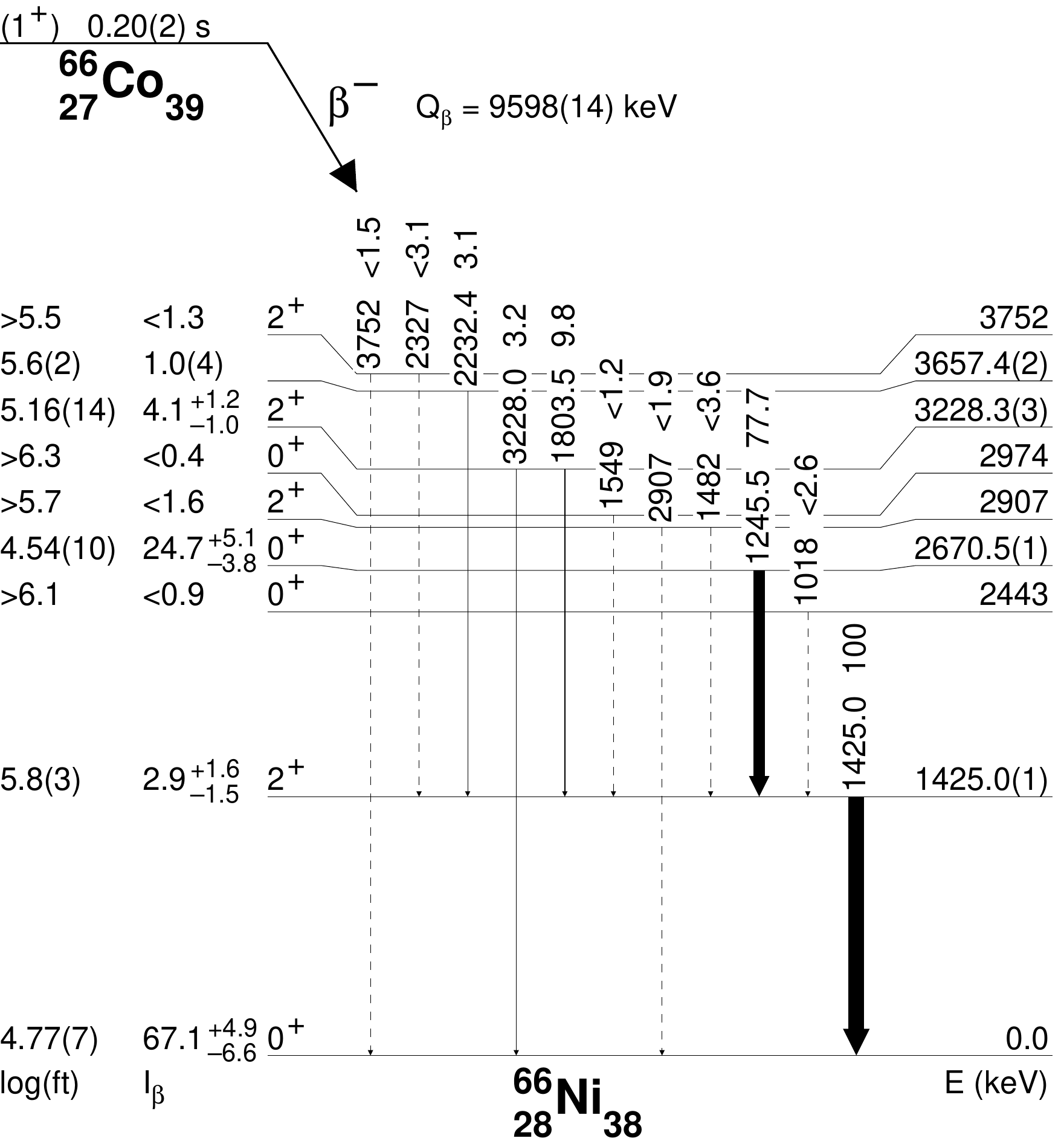}
\caption{\label{fig:66Codecayscheme} The scheme of excited states in $^{66}$Ni populated in $\beta^-$ decay of $^{66}$Co. The spin assignments and the energies of the unobserved states are taken from Ref. \cite{Darcey1971,Broda2012,Leoni2017}. The $\beta$-feeding of the states should be treated as upper limits and the log(\textit{ft}) values as lower limits due to the pandemonium effect.}
\end{figure}

The scheme of excited states in $^{66}$Ni (Fig. \ref{fig:66Codecayscheme}) was built by setting the energy gates on the previously known $\gamma$-ray transitions \cite{Darcey1971,Bernas1981,Bosch1988,Fister1990,Pawat1994,Ishii1997,Mueller2000,Recchia2012,Broda2012,Leoni2017,Olaizola2017} (Fig. \ref{fig:gg1425}). We confirmed a decay scheme recently published in \cite{Olaizola2017}. In addition we observed a $\gamma$-ray transition at 3228~keV, which was assigned to $^{66}$Ni based on the energy matching with the excited state. The list of transitions attributed to the decay of $^{66}$Co is presented in Table \ref{tab:gammas66Ni}. 

\begin{table}[h]
\caption{\label{tab:gammas66Ni}
Upper part: the relative intensities (I$_\gamma^{rel}$) of the $\gamma$-ray transitions assigned to the decay of $^{66}$Co, normalized to 100 units of the 1425 keV transition. For the absolute intensities, multiply by 0.319$^{+0.065}_{-0.048}$. Lower part: the unobserved transitions from the known $0^+$ and $2^+$ states in $^{66}$Ni with relative intensities given with 95\% credible limits.}
\begin{ruledtabular}
\begin{tabular}{cccc}
E$_\gamma$	 (keV)  	& I$_\gamma^{rel}$	& E$_{\mathrm{level}}^{\mathrm{initial}}$ (keV)  & E$_{\mathrm{level}}^{\mathrm{final}}$ (keV)   \\\hline
1245.5(1) 			& 77.7(31) 		& 2670.5 	& 1425.0 	\\
1425.0(1)			& 100.0			& 1425.0	& 0.0		\\
1803.5(3)			& 9.8(20)		& 3228.3	& 1425.0	\\
2232.4(2)			& 3.1(11)		& 3657.4	& 1425.0	\\
3228.0(6)			& 3.2(16)		& 3228.3	& 0.0		\\\\
1018\footnotemark[1]				& $<$2.6		& 2443		& 1425.0	\\
1482\footnotemark[1]				& $<$3.6		& 2907		& 1425.0	\\
1549\footnotemark[2]				& $<$1.2		& 2974		& 1425.0	\\
2327\footnotemark[3]				& $<$3.1		& 3752		& 1425.0	\\
2907\footnotemark[1]				& $<$1.9		& 2907		& 0.0		\\
3752\footnotemark[3]				& $<$1.5		& 3752		& 0.0		\\
\end{tabular}
\end{ruledtabular}
\footnotetext[1]{Energy taken from Ref.~\cite{Broda2012}.}
\footnotetext[2]{Energy taken from Ref.~\cite{Leoni2017}.}
\footnotetext[3]{Energy taken from Ref.~\cite{Darcey1971} accounting a systematic shift of $-6$~keV (Ref.~\cite{Broda2012}).}
\end{table}

Only selected states with spins and parities of $0^+$ and $2^+$ were observed in our analysis of the $^{66}$Co $\beta^-$ decay. The upper limits for the unobserved transitions from the known $0^+$ and $2^+$ states at 2443, 2907, 2974 and 3752 keV \cite{Darcey1971,Broda2012,Leoni2017} to $0^+_1$ and $2^+_1$ were determined with 95\% credible limits. For the state at 3746 keV, which was observed in the $(t,p)$ reaction \cite{Darcey1971}, a systematic shift of $-6$~keV proposed in Ref. \cite{Broda2012} was applied. The results are presented in Table \ref{tab:gammas66Ni}. 

The direct feeding to the ground state was obtained for the first time and is equal $67.1^{+4.9}_{-6.6}\%$. The log(\textit{ft}) values were calculated as described in the previous section. The Q$_{\beta^-}$ was taken from AME2016 \cite{Wang2017} and the half-life of $^{66}$Co (T$_{1/2} = 200(20)$~ms) from NNDC evaluation \cite{Browne2010}.

\begin{figure}
\includegraphics[width=\columnwidth]{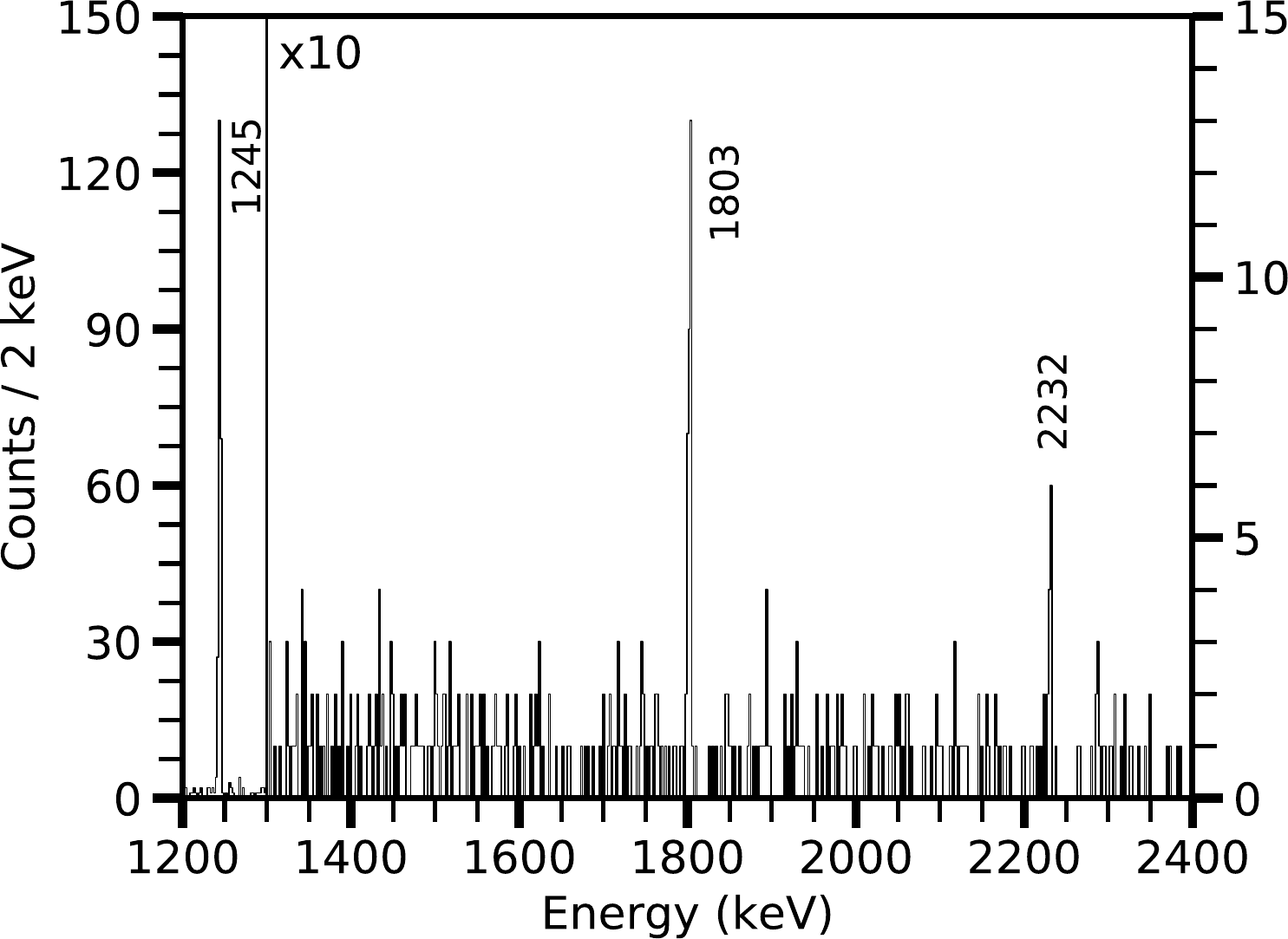}
\caption{\label{fig:gg1425}A portion of the $\gamma$-$\gamma$ coincidence spectrum gated on the 1425 keV transition. The coincide transitions are labeled with the energy in keV.}
\end{figure}

\subsection{\label{sec:resultsgsf}Half-lifes and ground state feedings}

To determine ground state feedings in the analyzed nuclei and their half-lifes, the numbers of registered $\gamma$-rays and $\beta$ particles were compared. It was assumed that after closing the beam gate, when there is no implantation, the number of registered $\gamma$-rays assigned to a particular decay channel as a function of time should be described by the $\gamma$-decay curve, which is an adequate Bateman's equation (see Appendix \ref{apx:bateman}). It was also assumed that the number of registered $\beta$ particles as a function of time ($\beta$-decay curve) can be described as a linear combination of all the $\gamma$-decay curves: 

\begin{equation} 
\beta(t) = \displaystyle\sum_{i} A_{i}\gamma_{i}^{sig}(t) \mathrm{.}
\label{eq:gsfassumptions}
\end{equation}

Due to the long half-lifes of nickel isotopes (2.5 h \cite{Browne2010a} and 54.6 h \cite{Browne2010} for $^{65}$Ni and $^{66}$Ni, respectively), parameters $A_{Cu65}$, $A_{Cu66}$ and $A_{Zn66}$ were set to zero. Because of the low statistics, the part of the equation $A_{Co65}\gamma_{Co65}(t) + A_{Ni65}\gamma_{Ni65}(t)$ was approximate by a constant value $C$. This parameter contains also the contribution of the beam contaminants to the $\beta$-decay curve. Since both $^{65}$Fe and $^{66}$Fe are produced in the $\beta^-$ decay of $^{66}$Mn, their $\gamma$-decay curves can be described by the same Bateman's equation. After applying these assumptions, Eq. \ref{eq:gsfassumptions} can be rewritten to Eq. \ref{eq:gsfassumptionsnew}, which constitutes the model for $\beta$-decay curve.

\begin{equation}
\beta(t) = A_{Fe}\gamma_{Fe}^{sig}(t) + A_{Co66}\gamma_{Co66}^{sig}(t) + A_{Ni66}\gamma_{Ni66}^{sig}(t) + C
\label{eq:gsfassumptionsnew}
\end{equation}

\begin{figure*}
\includegraphics[width=\textwidth]{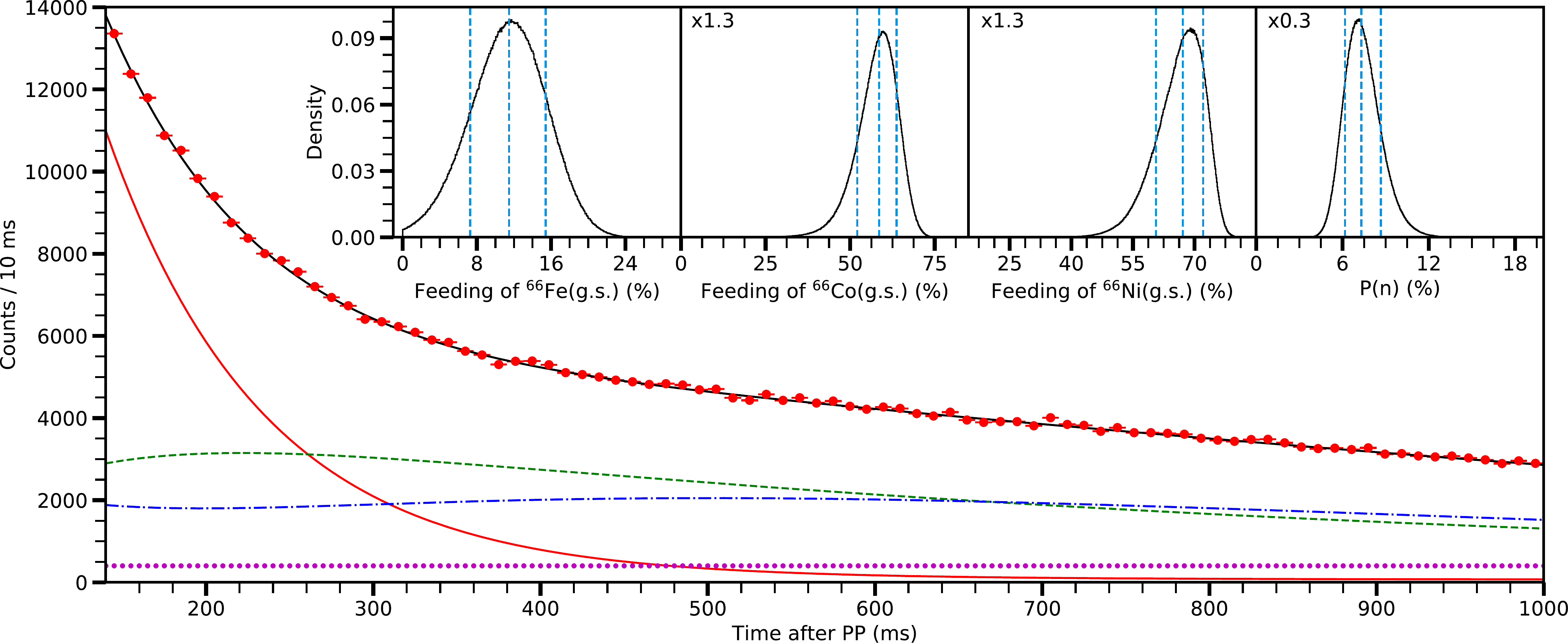}
\caption{\label{fig:gsffit} (Color online) The fit results of the $\gamma$-decay curves to the number of $\beta$ particles registered in time (red circles). The $\beta$-decay curve (Eq. \ref{eq:gsfassumptionsnew}) is plotted as a black straight line. The contribution of the $^{66}$Mn, $^{66}$Fe and $^{66}$Co decays is represented by the red straight line (Eq. \ref{eq:gsfbatemanfe}), green dashed line (Eq. \ref{eq:gsfbatemanco}) and blue dash-dotted line (Eq. \ref{eq:gsfbatemanni}), respectively. The purple dotted line represents a constant from Eq. \ref{eq:gsfassumptionsnew}. Insert: posterior probability density functions of the direct feeding to the ground state of (from left) $^{66}$Fe, $^{66}$Co and $^{66}$Ni, and the probability of $\beta$-delayed-neutron emission. The 16, 50 and 84 percentiles are indicated with vertical, dotted lines.}
\end{figure*}

The $\gamma$-decay curves were described by the most intense transitions from each decay (574, 471 and 1425 keV). The data were taken from the $\beta$-gated-$\gamma$ spectrum to include the efficiency of the $\beta$ detectors. To overcome the problem of the background in the $\gamma$ spectrum, for each transition two sets of data were prepared, one from the peak area and one from the background area, as it was discussed in the section related to the $^{66m1}$Co half-life. The assumed model for the background dataset was an exponential decay function and a constant while the peak area datasets were described by Eq. \ref{eq:gsfbkg}, where $\gamma^{sig}(t)$ is the relevant Bateman's equation and $\gamma^{bkg}(t)$ is the background model. Two additional datasets were prepared for the 364 keV transition in $^{65}$Fe. They were taken from the $\beta$-gated-$\gamma$ spectrum with longer coincidence time (from 0 to 4.5 $\mu$s) to account for the contribution of the isomeric state. 

\begin{equation}
\gamma(t) = \gamma^{sig}(t) + \gamma^{bkg}(t)
\label{eq:gsfbkg}
\end{equation}

The simultaneous fit of nine datasets (signal and background datasets for each of 574, 471, 1425 and 364 keV transitions and one dataset with $\beta$ particles) with 25 free parameters was performed with SATLAS. The likelihood function was built assuming that the number of counts in each bin in all datasets are following the Poisson distribution. The priors were set to constrain the half-life of $^{66}$Co to the literature value ($T_{1/2} = 200(20)$~ms \cite{Browne2010}), the $A$ parameters from Eq. \ref{eq:gsfassumptionsnew} to be equal or larger than 1, which reflects the fact that the number of decays through excited states cannot exceed the number of all decays, and the rest of free parameters to be non-negative. The random walk was performed with 60 walkers and 100000 steps, from which first 15\% were rejected as a burn-in. The results of the fit are presented in Fig. \ref{fig:gsffit}.

\begin{figure}
\includegraphics[width=\columnwidth]{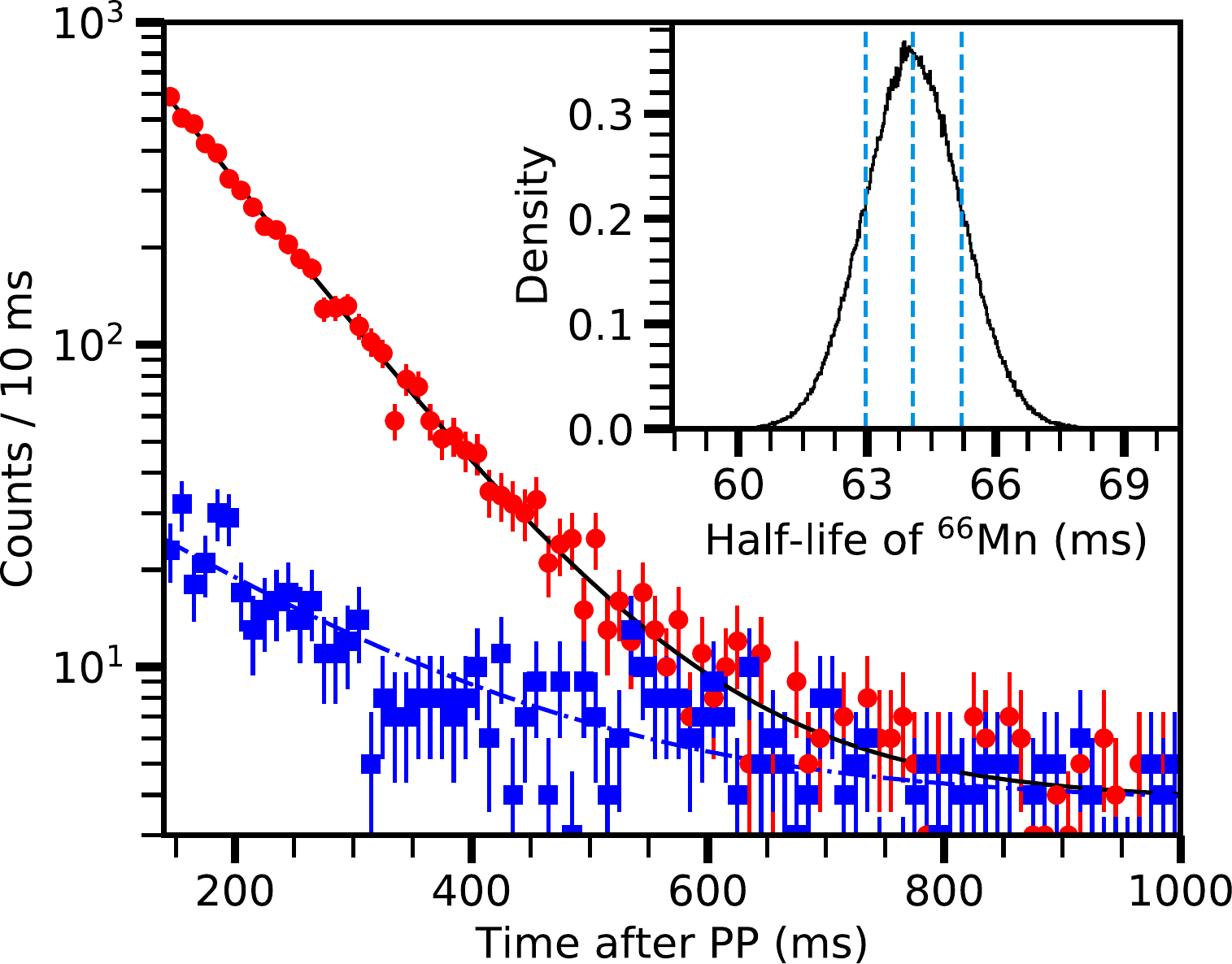}
\caption{\label{fig:t12mnfit}(Color online) Time behavior of the 574 keV transition as a function of time after PP (red circles) with the fitted function (black straight line) and the background area of the 574 keV transition (blue squares) with the fitted function (blue dash-dotted line). Insert: posterior probability density function of $^{66}$Mn half-life. The 16, 50 and 84 percentiles are indicated with vertical, dotted lines.}
\end{figure}

\begin{figure}
\includegraphics[width=\columnwidth]{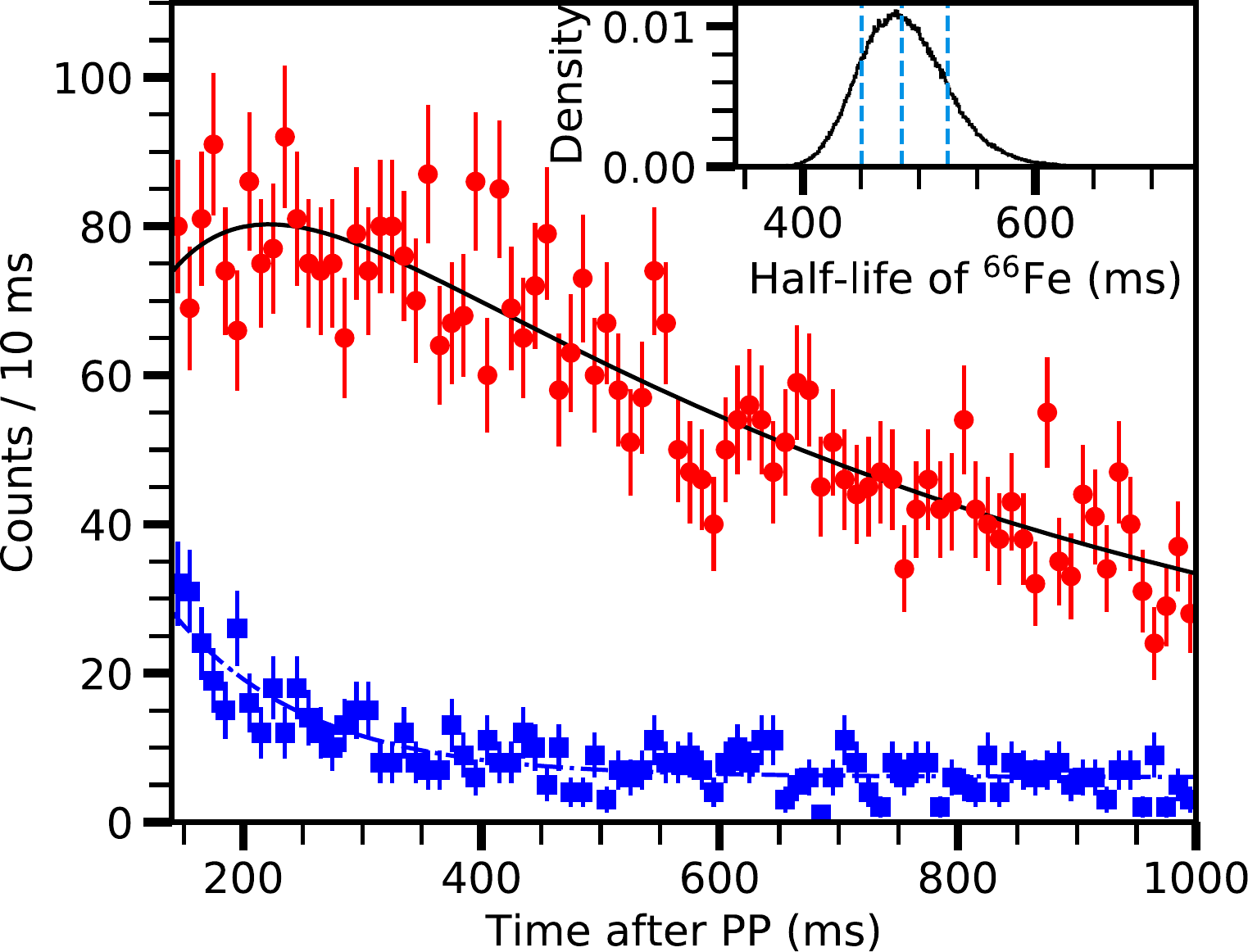}
\caption{\label{fig:t12fefit}(Color online) Time behavior of the 471 keV transition as a function of time after PP (red circles) with the fitted function (black straight line) and the background area of the 471 keV transition (blue squares) with the fitted function (blue dash-dotted line). Insert: posterior probability density function of $^{66}$Fe half-life. The 16, 50 and 84 percentiles are indicated with vertical, dotted lines.}
\end{figure}

From the fit, two half-lifes of the analyzed nuclei could be extracted. For $^{66}$Mn the obtained value is $T_{1/2} = 64.1(11)$~ms (Fig. \ref{fig:t12mnfit}), which is in an excellent agreement with the weighted average reported in NNDC (65(2) ms \cite{Browne2010}) and with newer experimental results reported by Daugas \textit{et al.} (65(5) ms \cite{Daugas2011}), Liddick \textit{et al.} (60(3) ms and 63(4)~ms \cite{Liddick2013}), and Olaizola \textit{et al.} (70(15) ms \cite{Olaizola2017a}). For $^{66}$Fe the obtained value is $\mathrm{T}_{1/2} = 485^{+39}_{-34}$~ms (Fig. \ref{fig:t12fefit}). This result is consistent with 440(60) ms coming from two separate experiments, reported in Ref. \cite{Ameil1998,Sorlin2000}, but it is significantly different from 351(6) ms reported in Ref. \cite{Liddick2012}.  

To extract ground state feedings, each $A$ parameter from Eq. \ref{eq:gsfassumptionsnew} was corrected by the $\gamma$-detection efficiency ($eff_\gamma$) and by the intensity factor defined as $f_I = I_\gamma \times (\sum I_{\gamma\mathrm{~to~g.s.}})^{-1}$, where $I_\gamma$ is the relative intensity of the transition used for fitting (574, 471 or 1425 keV) and $\sum I_{\gamma\mathrm{~to~g.s.}}$ is the sum of the relative intensities of the transitions deexciting directly to the ground state. The latter factor includes the information about the decays through excited states, which did not lead to the emission of the selected $\gamma$-ray transition. After applying the corrections, the $B$ parameters, defined as $B = A \times eff_\gamma \times f_I$, were computed.

The missing feeding ($mf$) is related to the $B$ parameter through Eq. \ref{eq:gsfequation}:

\begin{equation}
mf = 1 - \frac{1}{B} \mathrm{.}
\label{eq:gsfequation}
\end{equation}

\noindent In the case of $^{66}$Fe decay to $^{66}$Co and $^{66}$Co decay to $^{66}$Ni, the missing feedings are interpreted as ground state feedings (see the inserts of Fig. \ref{fig:gsffit} for the posterior probability density functions), while in the case of the $^{66}$Mn $\beta^-$ decay, the missing feeding is interpreted as a sum of the ground state feeding to $^{66}$Fe and the probability of the $\beta^-$-delayed-neutron decay. The latter one can be extracted by using Eq. \ref{eq:pn}:

\begin{equation}
P_n = \frac{1}{A_{Fe}} \times \frac{\alpha^{Fe65}}{\alpha^{Fe}} \times \frac{1}{1-gsf^{65}} \times \frac{1}{eff_\gamma^{364}f_I^{364}} \mathrm{,}
\label{eq:pn}
\end{equation}

\noindent where $A_{Fe}$ and $\alpha^{Fe}$ are parameters extracted from the fit of the 574 keV transition, $\alpha^{Fe65}$ is the parameter from the fit of the 364 keV transition, $gsf^{65}$ is the direct feeding of $^{65}$Fe from $^{66}$Mn decay, and $eff_\gamma^{364}$ and $f_I^{364}$ are the $\gamma$ detection efficiency and the intensity factor for the 364 keV transition, respectively. The derivation of this equation is presented in the Appendix \ref{apx:pneq}. The ground state feeding of $^{66}$Fe is defined as a difference between the missing feeding and the probability of $\beta$-delayed-neutron decay. The posterior probability density functions of both, the ground state feeding of $^{66}$Fe and the probability of the $\beta$-delayed-neutron, are presented as the insert in Fig. \ref{fig:gsffit}.

It should be noted that the presented method allows to estimate only the upper limits of the ground state feedings as the missing feeding consists of a \textit{true} ground state feeding, as well as an unobserved feeding from the higher-lying excited states (\textit{pandemonium} effect).

The $\beta$ feeding to the excited state is defined as a product of the apparent $\beta$ feeding to the selected state ($I_\beta^{app}$) and the total $\beta$ feeding to the excited states, which is linked to the missing feeding. After calculations, the formula used to obtain the $\beta$ feedings of the excited states in the analyzed nuclei is:

\begin{equation}
I_\beta = I_\beta^{app} \times (1-mf) = I_\beta^{app} \times \big(1 - (1 - \frac{1}{B})\big) = \frac{I_\beta^{app}}{B}\mathrm{.}
\label{eq:betafeedingn}
\end{equation}

The $11.5^{+3.9}_{-4.2}$\% of the direct feeding to the $^{66}$Fe ground state and the $7.3^{+1.4}_{-1.1}$\% probability of the $\beta^-$-delayed-neutron emission are not in agreement with previously reported $I_{\beta gsf} = 36(6)\%$ and $P_n = 4(1)\%$ in Ref. \cite{Liddick2013}, $I_{\beta gsf} = 47(8)\%$ and $P_n = 3.8(8)\%$ in Ref. \cite{Olaizola2017a} and $P_n = 9.5(5)\%$ in Ref. \cite{Hannawald2000}. Also the $59.5^{+5.1}_{-6.4}\%$ of the direct feeding to the $^{66}$Co ground state is not in agreement with 72(5)\% reported in Ref. \cite{Liddick2012,xundlLiddick}, which can be explained by the fact we identified three more $\gamma$-ray transitions at 982, 1049 and 1212~keV feeding the ground state. This discrepancy might also explain the difference between our analysis and the feeding to the $^{66}$Fe ground state reported in Ref. \cite{Olaizola2017a} since the latter one was based on the values from Ref. \cite{Liddick2012}.

\section{\label{sec:discussion}Discussion}

\begin{figure*}
\includegraphics[width=\textwidth]{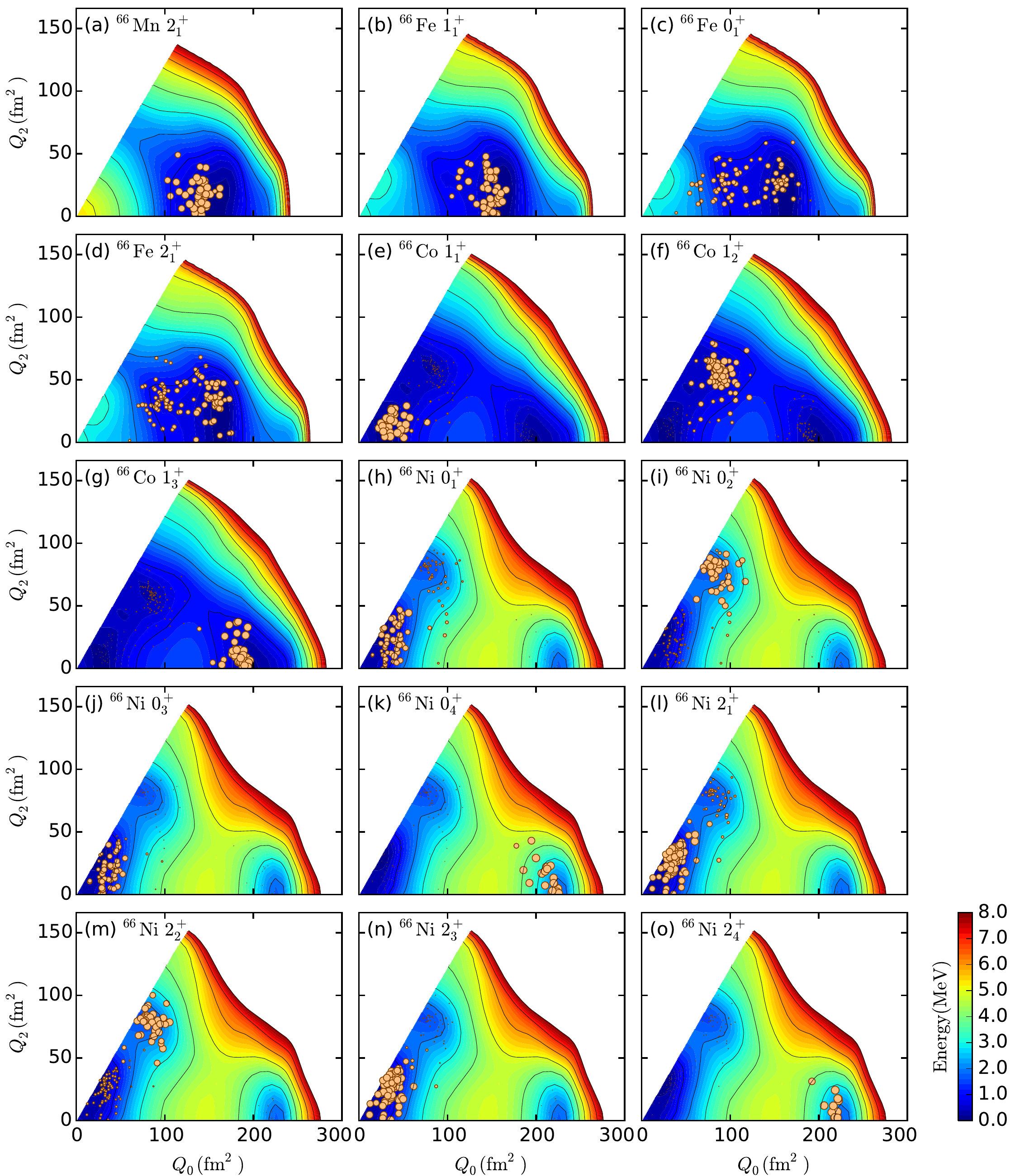}
\caption{\label{fig:tplots} (Color online) T-plots of the selected states in the $A=66$ chain.}
\end{figure*}

\begin{table*}
\caption{\label{tab:theoryBGT}
The overview of the B(GT) values obtained from MCSM calculations and associated log(\textit{ft}) values compared with the experimental results. $J^\pi_p$ and $J^\pi_d$ are spins and parities of parent and daughter nuclei, respectively. The log(\textit{ft}) values were calculated using a formula $\mathrm{\textit{ft}} = \kappa (g_A^2q^2B_{GT})^{-1}$, where $\sfrac{\kappa}{g_V^2} = 6147$~s, $\sfrac{g_A}{g_V} = -1.2772$ \cite{Brown2018} and $q = 0.744$ is a standard quenching factor \cite{Martinez-Pinedo1996}. The \textendash~symbol in the log(\textit{ft})$^{exp}$ column indicates that the calculated state was not linked with any of the experimentally observed levels.}
\begin{ruledtabular}
\begin{tabular}{ccccccccccccccc}
\multicolumn{5}{l}{$^{66}$Mn $\rightarrow$ $^{66}$Fe} & \multicolumn{5}{l}{$^{66}$Fe $\rightarrow$ $^{66}$Co} & \multicolumn{5}{l}{$^{66}$Co $\rightarrow$ $^{66}$Ni}\\[0.01cm]\cline{1-5}\cline{6-10}\cline{11-15}\\[0.01cm]
$J^\pi_p$ 	& $J^\pi_d$ 	& B(GT) 	& log(\textit{ft})$^{th}$ & log(\textit{ft})$^{exp}$ 	& $J^\pi_p$ 	& $J^\pi_d$ 	& B(GT) 	& log(\textit{ft})$^{th}$ & log(\textit{ft})$^{exp}$ 	& $J^\pi_p$ 	& $J^\pi_d$ 	& B(GT) 	& log(\textit{ft})$^{th}$ & log(\textit{ft})$^{exp}$ \\[0.1cm]\hline\\[0.01cm]
$2^+_1$ 	& $1^+_1$ 	& $1.8 \times 10^{-1}$ 	& 4.59 	& 4.75(3) 		& $0^+_1$ 	& $1^+_1$ 	& $9.8 \times 10^{-2}$ 	& 4.84 	& 4.38(6) 		& $1^+_1$ 		& $0^+_1$ & $1.9 \times 10^{-1}$ & 4.55 & 4.77(7)\\
			& $2^+_1$ 	& $2.8 \times 10^{-4}$ 	& 7.38 	& 5.68(8) 		& 			& $1^+_2$ 	& $2.0 \times 10^{-1}$ 	& 4.52 	& 4.33(10) 		&				& $0^+_2$ & $9.0 \times 10^{-3}$ & 5.88 & $>6.1$ \\
			& $2^+_2$ 	& $7.6 \times 10^{-4}$ 	& 6.95 	& 5.86(6) 		&			& $1^+_3$ 	& $7.8 \times 10^{-2}$ 	& 4.94 	& \textendash 	&				& $0^+_3$ & $3.2 \times 10^{-1}$ & 4.33 & 4.54(10) \\
			& $2^+_3$ 	& $1.1 \times 10^{-3}$ 	& 6.80 	& 6.00(13) 		&			& $1^+_4$ 	& $3.9 \times 10^{-3}$ 	& 6.25 	& \textendash 	&				& $0^+_4$ & $3.9 \times 10^{-5}$ & 8.25 & $>6.3$\\
			& $3^+_1$ 	& $1.6 \times 10^{-3}$ 	& 6.63 	& \textendash 	&			& 			&  						&  		& 				&				& $2^+_1$ & $6.7 \times 10^{-3}$ & 6.01 & 5.8(3)\\
			& 			&						&		& 				&			& 			&  						&  		& 				&				& $2^+_2$ & $3.8 \times 10^{-5}$ & 8.25 & $>5.7$\\
			& 			&						&		& 				&			& 			&  						&  		& 				&				& $2^+_3$ & $1.2 \times 10^{-1}$ & 4.75 & 5.16(14) \\
			& 			&						&		& 				&			& 			&  						&  		& 				&				& $2^+_4$ & $1.3 \times 10^{-7}$ & 10.73 & $>5.5$ \\

\end{tabular}
\end{ruledtabular}
\end{table*}

\begin{table*}
\caption{\label{tab:theoryoccupation}
The average occupation numbers of the selected states in the $A=66$ chain obtained from MCSM calculations.}
\begin{ruledtabular}
\begin{tabular}{cccccccccccccc}
Nucleus	& $J^\pi$	& \multicolumn{6}{l}{Proton occupation} & \multicolumn{6}{l}{Neutron occupation}\\[0.01cm]\cline{3-8}\cline{9-14}
				& 			& $0f_{7/2}$ & $1p_{3/2}$ & $0f_{5/2}$ & $1p_{1/2}$ & $0g_{9/2}$ & $1d_{5/2}$ & $0f_{7/2}$ & $1p_{3/2}$ & $0f_{5/2}$ & $1p_{1/2}$ & $0g_{9/2}$ & $1d_{5/2}$ \\[0.1cm]\hline
$^{66}$Mn 		& $2^+_1$ (g.s) 	& 4.39	 & 0.38	 & 0.17	 & 0.03	 & 0.02	 & 0.00	 & 7.89	 & 3.74	 & 3.93	 & 1.18	 & 3.93	 & 0.34	\\[0.2cm]
$^{66}$Fe 		& $0^+_1$ (g.s)		& 5.40	 & 0.34	 & 0.19	 & 0.03	 & 0.04	 & 0.01	 & 7.85	 & 3.52	 & 3.93	 & 1.26	 & 3.21	 & 0.23	\\
$^{66}$Fe 		& $0^+_2$ 			& 5.39	 & 0.35	 & 0.19	 & 0.03	 & 0.04	 & 0.01	 & 7.87	 & 3.65	 & 3.98	 & 1.33	 & 2.89	 & 0.28	\\
$^{66}$Fe 		& $1^+_1$ 			& 5.22	 & 0.46	 & 0.24	 & 0.04	 & 0.03	 & 0.00	 & 7.87	 & 3.46	 & 3.43	 & 1.15	 & 3.75	 & 0.35	\\
$^{66}$Fe 		& $2^+_1$ 	 		& 5.29 	 & 0.42 	 & 0.21 	 & 0.03 	 & 0.04 	 & 0.01 	 & 7.85 	 & 3.48 	 & 3.82 	 & 1.15 	 & 3.42 	 & 0.29 	\\
$^{66}$Fe 		& $2^+_2$ 			& 5.26 	 & 0.43 	 & 0.23 	 & 0.03   & 0.04 	 & 0.01 	 & 7.86 	 & 3.54 	 & 3.94 	 & 1.37 	 & 3.07 	 & 0.21 	\\
$^{66}$Fe 		& $2^+_3$ 			& 5.36 	 & 0.38 	 & 0.19 	 & 0.03 	 & 0.04 	 & 0.01 	 & 7.88 	 & 3.66 	 & 3.95 	 & 1.51 	 & 2.76 	 & 0.25 	\\
$^{66}$Fe 		& $4^+_1$ 			& 5.22	 & 0.47	 & 0.23	 & 0.04	 & 0.03	 & 0.00	 & 7.86	 & 3.46	 & 3.76	 & 0.97	 & 3.59	 & 0.36	\\[0.2cm]
$^{66}$Co 		& $1^+_1$ (g.s.) 	& 6.79	 & 0.07	 & 0.03	 & 0.01	 & 0.08	 & 0.01	 & 7.88	 & 3.75	 & 4.82	 & 1.81	 & 0.67	 & 0.06	\\
$^{66}$Co		& $1^+_2$ 			& 6.19	 & 0.52	 & 0.19	 & 0.03	 & 0.05	 & 0.01	 & 7.89	 & 3.61	 & 4.19	 & 0.98	 & 2.24	 & 0.08	\\
$^{66}$Co 		& $1^+_3$ 			& 5.29	 & 0.66	 & 0.74	 & 0.27	 & 0.03	 & 0.01	 & 7.83	 & 3.05	 & 3.54	 & 0.52	 & 3.58	 & 0.48	\\
$^{66}$Co		& $2^+_1$			& 6.75 	 & 0.12 	 & 0.04 	 & 0.01 	 & 0.08 	 & 0.01 	 & 7.90 	 & 3.75 	 & 4.72 	 & 1.82 	 & 0.75 	 & 0.06 	\\
$^{66}$Co		& $3^+_1$			& 6.63 	 & 0.20 	 & 0.08 	 & 0.01 	 & 0.07 	 & 0.01 	 & 7.88 	 & 3.70 	 & 4.83 	 & 1.13 	 & 1.38 	 & 0.08 	\\ 
$^{66}$Co		& $4^+_1$			& 6.69 	 & 0.16 	 & 0.06 	 & 0.01 	 & 0.07 	 & 0.01 	 & 7.89 	 & 3.75 	 & 5.03 	 & 1.08 	 & 1.18 	 & 0.07	\\
$^{66}$Co		& $6^+_1$		 	& 6.81 	 & 0.06 	 & 0.03 	 & 0.00 	 & 0.08 	 & 0.01 	 & 7.90 	 & 3.76 	 & 4.78 	 & 1.84 	 & 0.67 	 & 0.06	\\
$^{66}$Co		& $6^-_1$		 	& 6.61 	 & 0.23 	 & 0.08 	 & 0.01 	 & 0.06 	 & 0.01 	 & 7.89 	 & 3.69 	 & 4.41 	 & 1.54 	 & 1.41 	 & 0.06	\\[0.2cm]
$^{66}$Ni 		& $0^+_1$	(g.s.) 	& 7.59	 & 0.22	 & 0.06	 & 0.01	 & 0.10	 & 0.01	 & 7.86	 & 3.49	 & 4.74	 & 1.08	 & 0.77	 & 0.06	\\
$^{66}$Ni 		& $0^+_2$ 			& 6.61	 & 0.92	 & 0.30	 & 0.08	 & 0.07	 & 0.01	 & 7.83	 & 3.30	 & 3.44	 & 1.30	 & 2.07	 & 0.06	\\
$^{66}$Ni 		& $0^+_3$ 			& 7.71	 & 0.12	 & 0.05	 & 0.01	 & 0.10	 & 0.02	 & 7.89	 & 3.75	 & 4.83	 & 0.99	 & 0.50	 & 0.05	\\
$^{66}$Ni 		& $0^+_4$			& 5.34	 & 0.75	 & 1.37	 & 0.48	 & 0.06	 & 0.01	 & 7.77	 & 2.30	 & 3.35	 & 0.52	 & 3.51	 & 0.55	\\
$^{66}$Ni 		& $2^+_1$ 			& 7.52	 & 0.28	 & 0.07	 & 0.01	 & 0.10	 & 0.01	 & 7.88	 & 3.60	 & 4.63	 & 1.15	 & 0.69	 & 0.05	\\
$^{66}$Ni 		& $2^+_2$ 			& 6.59	 & 0.94	 & 0.30	 & 0.08	 & 0.07	 & 0.01	 & 7.83	 & 3.34	 & 3.46	 & 1.30	 & 2.02	 & 0.05	\\
$^{66}$Ni 		& $2^+_3$			& 7.68	 & 0.16	 & 0.04	 & 0.01	 & 0.10	 & 0.01	 & 7.90	 & 3.77	 & 4.08	 & 1.73	 & 0.46	 & 0.06	\\
$^{66}$Ni 		& $2^+_4$ 			& 5.33	 & 0.75	 & 1.38	 & 0.48	 & 0.05	 & 0.01	 & 7.77	 & 2.28	 & 3.36	 & 0.52	 & 3.51	 & 0.56	\\
\end{tabular}
\end{ruledtabular}
\end{table*}

In order to obtain better insight into the structure of analyzed nuclei, Monte Carlo Shell Model calculations (MCSM) were performed. These calculations were successfully used in this region to explain, next to excitation energies, a broad range of experimental observables, such as log(\textit{ft}) values \cite{Morales2017}, electromagnetic moments \cite{Wraith2017} and lifetimes of excited state \cite{Leoni2017}. We assumed $^{40}$Ca to be an inert core and we used the A3DA interaction, which covers $pfg_{9/2}d_{5/2}$ valance space for both protons and neutrons. The electromagnetic transitions were calculated with the effective charges and $g$ factors. Their values were set to 1.5$e$ and 0.5$e$ for protons and neutrons, respectively, and $g_l^\pi$ = 1.1, $g_l^\nu$ = 0.1 and $g_s$ = 0.7 $g_s^\mathrm{bare}$. The details of the MCSM technique can be found in Ref. \cite{Shimizu2012,Tsunoda2014,Shimizu2017}.

\subsection{Decay of $^{66}$Mn}

\begin{figure}
\includegraphics[width=\columnwidth]{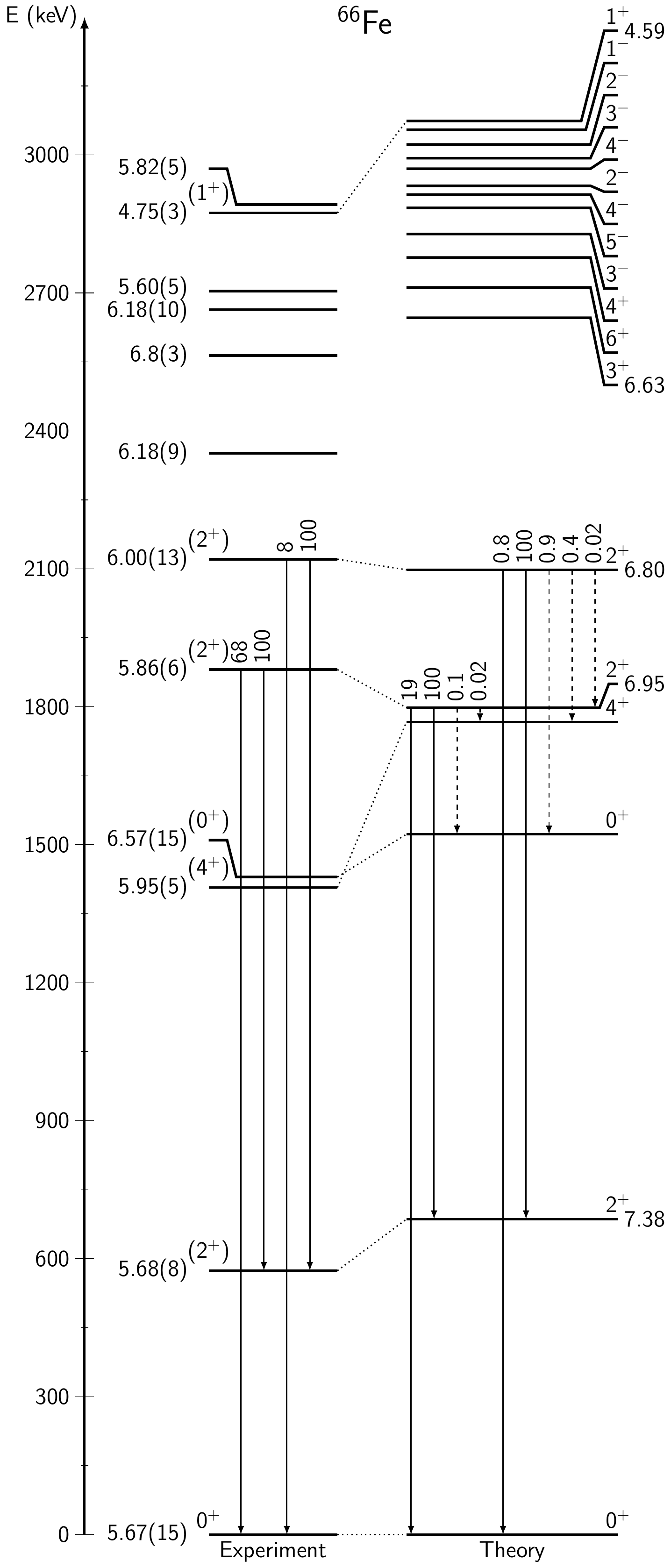}
\caption{\label{fig:66Feexpvsth} The comparison of the observed states in $^{66}$Fe, log(\textit{ft}) values and relative branching ratios from selected levels with the MCSM calculations. The states up to 3.1 MeV are presented. The theoretical intensities are calculated by taking experimental energies and B(M1) and B(E2) values from MCSM. Dashed lines represent transitions which were not observed experimentally. Experimental levels at 1407 and 1414 keV are shifted for the better visual representation. The calculated $4^-_2$ level is shifted $-20$~keV.}
\end{figure}

The $^{66}$Mn ground state has been previously tentatively assigned with a spin and parity $1^+$ based on the significant feeding to the ground state in the decay of $^{66}$Cr \cite{Liddick2011} and it was further supported by a strong direct $\beta$-feeding to the $^{66}$Fe ground state in the $^{66}$Mn decay, as reported in \cite{Liddick2013,Olaizola2017a}. However, the direct ground state feeding obtained in our work ($11.5^{+3.9}_{-4.2}$\%) is substantially lower than previously reported results (36(6)\% \cite{Liddick2013} and 47(8)\% \cite{Olaizola2017a}). Different spin and parity assignments have been investigated by computing the low-lying states in $^{66}$Mn in the Monte Carlo Shell Model calculations and the $2^+$ state was calculated as $^{66}$Mn ground state. Furthermore, the calculations were able to reproduce a strong direct $\beta$-feeding to the state at 2874 keV in $^{66}$Fe only assuming the $2^+$ state as a $^{66}$Mn ground state. There is a very good agreement between the theoretical and the experimental log(\textit{ft}) values (see Table \ref{tab:theoryBGT}) as well as the energy matching between the experimental level and the theoretical $1^+$ state (see Fig. \ref{fig:66Feexpvsth}). The $1^+$ assignment for the 2874 keV level is also consistent with a direct deexcitation to the $^{66}$Fe ground state and lack of $\gamma$-ray transition to the ($4^+$) state at 1407 keV. Thus, for the further discussion and interpretation, the $^{66}$Mn was tentatively assigned as ($2^+$) while the excited state at 2874 keV in $^{66}$Fe as ($1^+$).

Figure \ref{fig:66Feexpvsth} shows a comparison of the observed and calculated states in $^{66}$Fe together with the log(\textit{ft}) values and the intensities of the $\gamma$-ray transitions. There is a very good agreement between experimental and theoretical energies of the excited states, however, the calculated log(\textit{ft}) values are substantially larger than the experimental results. The $Q_{\beta^-}$ of about 13 MeV together with the fact that the highest observed state has about 3.6 MeV might suggest that part of the feeding from the high-lying states is not observed (\textit{pandemonium} effect). The low-efficiency experimental setups are known to be burdened with a systematic error related to the inability to detect high-energy and low-intensity $\gamma$-ray transitions. The Total Absorption Spectroscopy (TAS) measurements, which are not affected by this issue can significantly reduced the $\beta$-feedings of the low-lying states \cite{Jordan2013,Tain2015,Rasco2016,Rasco2017,Rice2017,Valencia2017,Fijakowska2017}. Consequently, the $\beta$-feedings presented in our work have to be treated as the upper limits while the associated log(\textit{ft}) values are the lower limits. 

From our work we confirm the spin sequence of the lowest lying $2^+_1$, $4^+_1$ and $0^+_2$ states. The theoretical half-life of the first excited state (T$_{1/2} = 26.5$~ps), computed by taking the calculated reduced transition probability and the experimental energy, is in very good agreement with the values reported by Rother \textit{et al.} (27.3(28) ps \cite{Rother2011}), Crawford \textit{et al.} ($31^{+3}_{-2}$ ps \cite{Crawford2013}) and Olaizola \textit{et al.} ($<$44 ps \cite{Olaizola2017a}). The half-lifes of the $4^+_1$ and $0^+_2$ states obtained from MCSM calculations assuming experimental transitions energies ($2.9$ and $9.2$~ps, respectively) are also in agreement with the recently reported limits of $<25$~ps and $<35$~ps \cite{Olaizola2017a}.

Based on a comparison with the MCSM calculations, we tentatively assign spin and parity of $2^+$ to the states at 1881 and 2121 keV. Next to the good energy matching between experimental results and theoretical calculations, the characteristic $\gamma$-decay pattern also agrees with the calculations. Using the calculated reduced transition probabilities and the experimental energies, the intensity ratios of transitions deexciting these two states were calculated and compared to the experimental values (see Fig. \ref{fig:66Feexpvsth}). The transitions from the $2^+_2$ and $2^+_3$ states to the $0^+_2$ and $4^+_1$ state are predicted to be two orders of magnitude weaker than the most intense transitions deexciting each state, which is below our detection limit. 

The Monte Carlo Shell Model calculations were used to understand the differences in the $\beta$-feedings of the $^{66}$Fe states. The $1^+_1$ state, which is calculated at 3074 keV, has a large B(GT) value corresponding to log(\textit{ft}) = 4.55 (see Table \ref{tab:theoryBGT}). The analysis of the $^{66}$Mn($2^+_1$) and $^{66}$Fe($1^+_1$) average occupation numbers (Table \ref{tab:theoryoccupation}) indicates that the $\beta^-$ decay is dominated by the $\nu 0f_{5/2} \rightarrow \pi 0f_{7/2}$ Gamow-Teller decay. The structure of the states can be also presented in the form of T-plots (Fig. \ref{fig:tplots}) which show the distribution of the MCSM basis vectors on the Potential Energy Surface (PES) \cite{Tsunoda2013,Tsunoda2014,Shimizu2017}. The area of the circles, which represent the basis vectors, is proportional to the overlap probability with the state wave function. For the ground state of $^{66}$Mn, the circles are located in the prolate deformation region of PES which might suggest, together with a significant occupation of the $\nu 0g_{9/2}$ shell, that this nucleus lies within the Island of Inversion located around $^{64}$Cr \cite{Lenzi2010}. A similar pattern of circles can be observed for the $1^+_1$ state in $^{66}$Fe which reflects the similarities in structure with the $^{66}$Mn ground state. The analysis of the $^{66}$Fe $2^+$ states average occupation numbers compared to the $2^+$ $^{66}$Mn ground state show an increase in the average occupation number of the $\pi 0f_{7/2}$ but a decrease in the $\nu 0g_{9/2}$ and $\nu 1p_{3/2}$ orbitals. A detailed analysis of the MCSM wave function indeed suggest that only minor components are relevant for the decay while the main components do not contribute to the process. The T-plot of the $2^+_1$ state (Fig. \ref{fig:tplots}) shows indeed that the wave function is fragmented and the overlap between this state wave function and the $^{66}$Mn ground state wave function is small.

\subsection{Decay of $^{66}$Fe}

\begin{figure}
\includegraphics[width=\columnwidth]{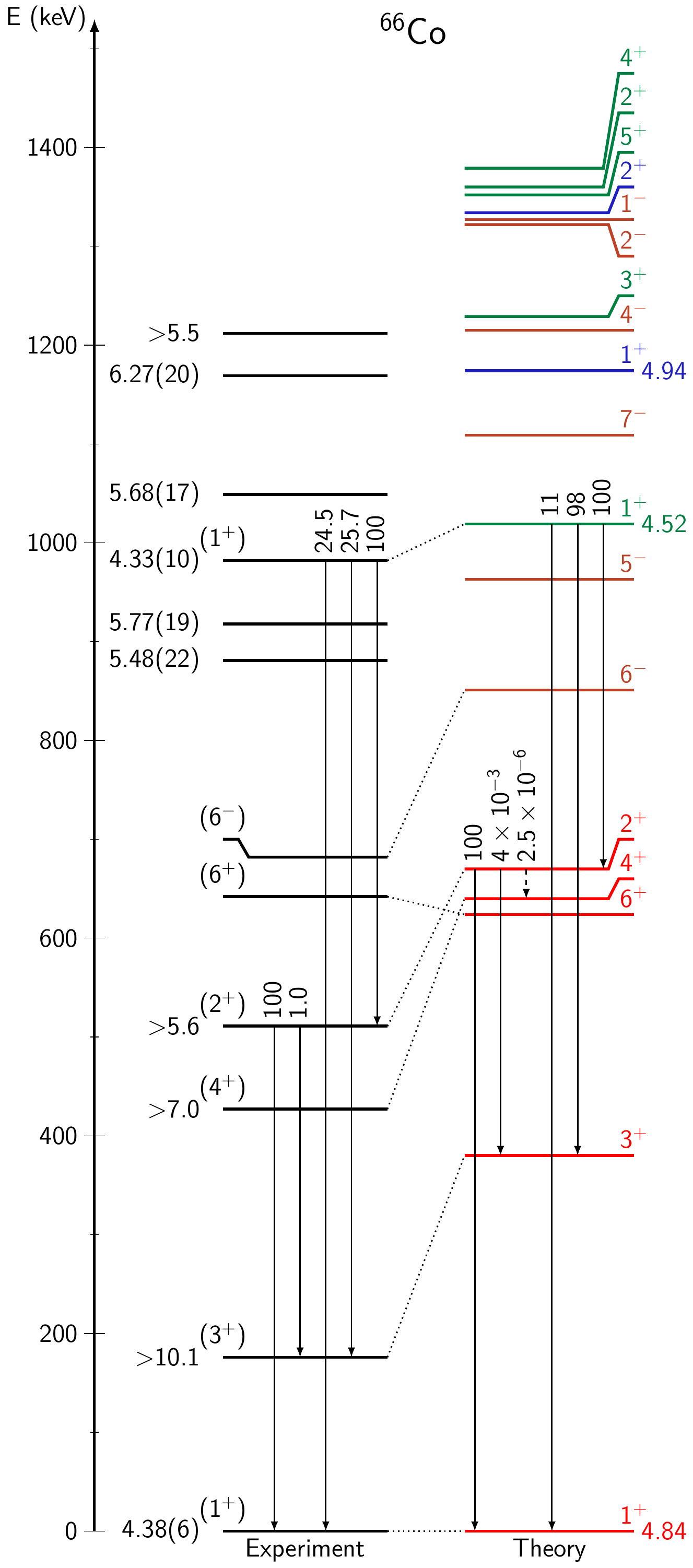}
\caption{\label{fig:66Coexpvsth} (Color online) The comparison of the observed states in $^{66}$Co, log(\textit{ft}) values and relative branching ratios from selected levels with the MCSM calculations. The states up to 1.2 MeV are presented. The theoretical intensities are calculated by taking experimental energies and B(M1) and B(E2) values from MCSM. Dashed lines represent transitions which were not observed experimentally. The spherical states are drawn in red, oblate in green, prolate in blue and the negative-parity states in brown. The calculated $4^+_1$ state is shifted $-10$~keV and the $6^+_1$ state is shifted $-20$~keV for better visual representation.}
\end{figure}

The strong population of the ground state and the excited state at 982 keV in $^{66}$Co which is resulting in the log(\textit{ft}) values of 4.38(6) and 4.33(10), respectively, suggests a spin and parity of $1^+$ for both of them. The ground state assignment is in contradiction with previously proposed ($3^+$) \cite{Bosch1988}, however, it is in agreement with the more recent experimental studies \cite{Liddick2012,Broda2012} as well as with the population of the $0^+$ and $2^+$ states in $^{66}$Ni in the $\beta^-$ decay of $^{66}$Co (see next section for details). The $1^+$ assignments are also supported by the Monte Carlo Shell Model calculations (Fig. \ref{fig:66Coexpvsth}). The $^{66}$Fe ground state wave function is predicted to be fragmented (Fig. \ref{fig:tplots}) which might suggest a transitional nature of this nucleus as it lies between the center of the Island of Inversion around $^{64}$Cr \cite{Lenzi2010} and the spherical $^{68}$Ni \cite{Tsunoda2014}, and allows the decay to three $1^+$ states with different shapes: spherical, oblate and prolate (Fig. \ref{fig:tplots}). The energies and the log(\textit{ft}) values are well reproduced for the first and the second $1^+$ states, however, we do not observed a state which can be assigned as the third $1^+$.

The half-life of the first excited state in $^{66}$Co at 176 keV ($T_{1/2} = 823^{+22}_{-21}$~ns) suggests a deexcitation through an E2 transition and, as a result, a spin and parity of $3^+$. This assignment is consistent with the low $\beta$-feeding of this state ($I_\beta < 1\%$) as well as with the Monte Carlo Shell Model calculations. The first excited state is predicted to be $3^+$ and its half-life, assuming the experimental energy and theoretical B(E2) value, is 536~ns. Hence, we propose a tentative spin and parity assignment of ($3^+$) for the state at 176 keV. 

The state at 511 keV was tentatively assigned spin and parity ($2^+$) based on the low $\beta$-feeding, the strong feeding from the $1^+_2$ state at 982 keV, the deexcitation to the $1^+_1$ ground state and ($3^+$) state, and based on the energy matching with the Monte Carlo Shell Model calculations (Fig. \ref{fig:66Coexpvsth}). 

The state at 642 keV was suggested to be an isomeric state ($T_{1/2} > 100$~$\mu$s) which deexcites through an M2 transition \cite{Grzywacz1998}. However, in the light of the results obtained in the deep-inelastic scattering \cite{ChiaraPrivate}, we propose that an isomeric state lies less than 50~keV above the 642 keV level. Based on the MCSM calculations, we tentatively assigned spins and parities of $6^-$, $6^+$ and $4^+$ for the levels $642+x$~keV, 642~keV and 427~keV, respectively. These assignments allow a deexcitation of the $642+x$ state through an E1 transition for which the Weisskopf estimate of the half-life, assuming the hindrance factor of $10^5$ \cite{Perdrisat1966} and energy of 5~keV, is about 330~$\mu s$ and it is consistent with the experimental limit. 

\subsection{Decay of $^{66}$Co}

\begin{figure}
\includegraphics[width=\columnwidth]{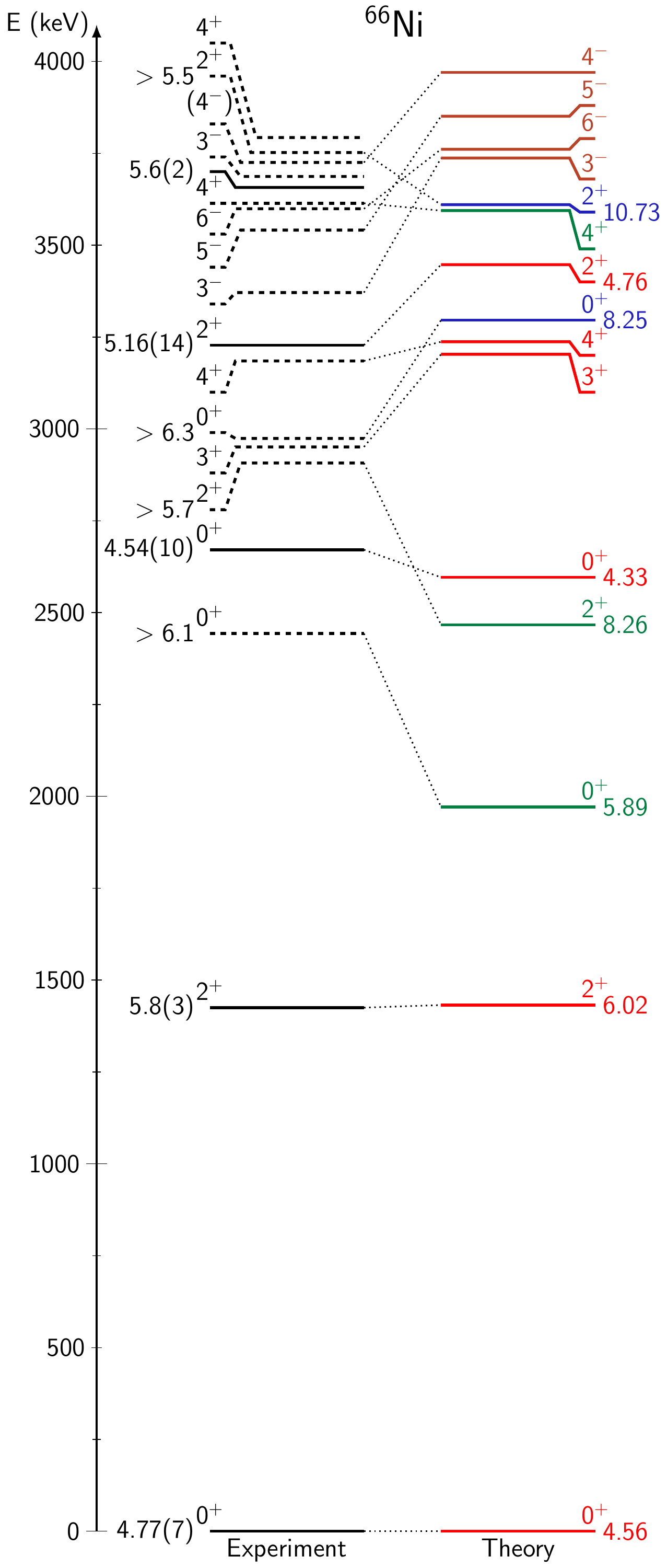}
\caption{\label{fig:66Niexpvsth} (Color online) The comparison of the states in $^{66}$Ni observed in the $\beta^-$ decay of $^{66}$Co (straight lines) and other experiments (dashed lines) \cite{Darcey1971,Broda2012,Leoni2017}, and the log(\textit{ft}) values with the MCSM calculations. The states up to 4 MeV are presented. The experimental $3^+_1$ level is shifted $-20$~keV and the theoretical $6^-_1$ level is shifted $20$~keV. The spherical states are drawn in red, oblate in green, prolate in blue and negative-parity states in brown.}
\end{figure}

Out of four known low-lying $0^+$ and $2^+$ states in $^{66}$Ni, only two of them are populated in the $\beta^-$ decay of $^{66}$Co. This selective behavior in the $\beta^-$ decay can be understood by looking at the MCSM calculations. A strong population of the $0^+_1$ and $0^+_3$ states can be understood as a single-particle Gamow-Teller decay of neutron at $0f_{5/2}$ shell to the proton at $0f_{7/2}$ shell. Although the average occupations of the $\nu 0f_{5/2}$ shell are similar to the ground state of $^{66}$Co  (Table \ref{tab:theoryoccupation}), the wave functions of the discussed nickel states have two strong components, $\nu f_{5/2}^6$ and $\nu f_{5/2}^4 + \nu p_{3/2}^2$, from which only the second one is participating in the $\beta^-$ decay process. Since the amplitudes of these two components in the $0^+_1$ and $0^+_3$ states are similar, the calculated B(GT) values are of the same order (Table \ref{tab:theoryBGT}). The same reasoning works for the $2^+_1$ and $2^+_3$ states, where the component of interest, $\nu f_{5/2}^4 + \nu p_{3/2}^2$, is coupled to $J=2$. The difference is that for the $2^+_1$ state, the amplitude of this component is much smaller than for $2^+_3$, which leads to the differences in the average occupation numbers and the enhancement of the decay to the third $2^+$ state compared to the first $2^+$.

The calculations show substantial differences in the average occupation numbers between the states in $^{66}$Ni. The $0^+$ and $2^+$ states which are not populated in the $\beta^-$ decay have a significantly larger occupation of the neutron $0g_{9/2}$ orbital and the proton orbitals above Z=28, compared to the states populated in the $\beta^-$ decay. The simultaneous increase of the proton and neutron excitations can be understood as a Type II shell evolution \cite{Tsunoda2014,Otsuka2016}. It was observed in other nickel isotopes that with the increase of the $\nu 0g_{9/2}$ occupation, the gap between $\pi 0f_{7/2}$ and $\pi 0f_{5/2}$ shells is reduced \cite{Tsunoda2014}. The differences in the configuration are also leading to shape coexistence, as can be deduced from the states T-plots (Fig. \ref{fig:tplots}). The $0^+$ and $2^+$ levels populated in the $\beta^-$ decay of the spherical $^{66}$Co ground state are also spherical, while the non-populated states are oblate- or prolate-deformed. 

\section{\label{sec:conclusions}Conclusions}

The excited states in the $^{65,66}$Fe, $^{66}$Co and $^{66}$Ni, populated in the $\beta^-$ decay of $^{66}$Mn, were studied be the means of $\gamma$ spectroscopy. The decay schemes were build using $\beta$-$\gamma$ and $\gamma$-$\gamma$ coincidence techniques. The half-life of two nuclei, $^{66}$Mn and $^{66}$Fe, and two isomeric states, $^{65m2}$Fe and $^{66m1}$Co, were determined in this analysis and compared with the previous experimental results. The spins and parities of the low-lying states were tentatively assigned based on the experimental data and theoretical calculations. The ground state $\beta$-branchings, which were obtained by analyzing the $\gamma$-ray intensities and by comparing the number of registered $\beta$ and $\gamma$ counts, are in contradiction with the previously reported values for the $^{66}$Mn and $^{66}$Fe decays while for the decay of $^{66}$Co it was determined for the first time.  

The Monte Carlo Shell Model calculations with the A3DA interactions were performed in order to obtain a better understanding of the structure of the analyzed nuclei. A strong $\beta$-feeding from a deformed $^{66}$Mn ground state to the deformed $1^+$ state at 2874 keV in $^{66}$Fe was well reproduced as well as the selective population of the $0^+$ and $2^+$ states in $^{66}$Ni in the $\beta^-$ decay of $^{66}$Co. The shell model calculations suggest an onset of deformation in the $A=66$ chain, from the spherical $^{66}$Ni and $^{66}$Co through transitional $^{66}$Fe towards prolate-deformed $^{66}$Mn, which is related to the occupation of the neutron $0g_{9/2}$ shell and the proton excitations across the magic number $Z=28$. 

\begin{acknowledgments}

M.S. would like to acknowledge E. Pompe and R. Stasi\'{n}ski from the University of Oxford and W. Gins from KU Leuven for their support with statistical analysis. 

We acknowledge the support of the ISOLDE Collaboration and technical teams. This project has received funding from the European Union’s Seventh Framework Programme for Research and Technological Development under grant agreement No 262010. This work has been funded by FWO-Vlaanderen (Belgium), by GOA/2010/010 (BOF KU Leuven), by the Interuniversity Attraction Poles Programme initiated by the Belgian Science Policy Office (BriX network P7/12). This material is based upon work supported by the U.S. Department of Energy, Office of Science, Office of Nuclear Physics under Award No. DE-FG02-94-ER40834, by Spanish MINECO via Project No. FPA2015-65035-P, by the Slovak grant agency VEGA (Contract No. 2/0129/17), and by the Slovak Research and Development Agency (Contract No. APVV-15-0225).

\end{acknowledgments}

\appendix
\section{\label{apx:511intensity}Intensity of the 511 keV transition in the $^{66}$Fe to $^{66}$Co decay}

To obtain the intensity of the 511~keV transition, its time behavior was analyzed. The utilized model (Eq. \ref{eq:model511}) describes the main sources of this $\gamma$-ray: $\beta^-$ decay of $^{66}$Fe ($N_{Fe}$), $^{66}$Mn decay high-energy $\gamma$-rays ($N_{Mn}$), Compton-scattered $\gamma$-rays ($N_{bkg}$), environmental background ($N_{off}$) and other ($N_{o}$).

\begin{equation}
N_{511}(t) = N_{Fe}(t) + N_{Mn}(t) + N_{bkg}(t) + N_{off}(t) + N_{o}(t)
\label{eq:model511}
\end{equation}

The 511~keV transitions from the $\beta^-$ decay of $^{66}$Fe have the same time behavior as the 471~keV transition as they are both originating from the same source. Hence, this part of the model is parameterized as

\begin{equation}
N_{Fe}(t) = \xi \times \gamma^{sig}_{471}(t) \mathrm{,}
\label{eq:model511_470}
\end{equation}

\noindent where $\gamma^{sig}_{471}(t)$ is the $\gamma$-decay curve of the 471 keV transition and $\xi$ is the scaling parameter which is equal to the ratio of the number of registered counts of the 511 keV and 471 keV transitions. By analogy, the high-energy $\gamma$-rays from the $^{66}$Mn decay have the same time behavior as the 574~keV transition, thus they are described as 

\begin{equation}
N_{Mn}(t) = \phi \times \gamma^{sig}_{574}(t) \mathrm{.}
\label{eq:model511_573}
\end{equation}

\noindent The Compton-scattered background ($N_{bkg}$) was described by an exponential decay model with a constant to include time-dependent and time-independent components. 

\begin{equation}
N_{bkg}(t) = N_{bkg0} e^{-\frac{ln(2)}{T_{bkg}}t} + C_{bkg}
\label{eq:comptonbackground}
\end{equation}

\noindent Both, the environmental background ($N_{off}$) and the other sources of the 511 keV transition ($N_{o}$) were parameterized using constants.

As described in Sec. \ref{sec:resultsgsf}, the $\gamma$-decay curves ($\gamma^{sig}_{471}$ and $\gamma^{sig}_{574}$ from Eqs. \ref{eq:model511_470} and \ref{eq:model511_573}) were constrained using for each, the 471 keV and 574 keV transitions, one dataset from the peak area and one from the background area. The Compton-scattered background part of the model was constrained by using the dataset from the 511 keV transition background area while the environmental background was constrained by the dataset from the 511 keV transition peak area collected in the \textit{laser-off} mode and scaled by the \textit{laser-on} to \textit{laser-off} acquisition time ratio. All the datasets were taken from the $\beta$-gated-$\gamma$ spectrum. 

The simultaneous fit of seven datasets with 18 free parameters was performed with SATLAS. The fitting range was set from 140 to 1000 ms after PP. Each parameter was set to be non-negative by using priors. The likelihood function was built assuming that the number of counts in each bin in all datasets are following the Poisson distribution. The random walk was performed with 60 walkers and 100000 steps, from which first 15\% were rejected as a burn-in. The fit results are presented in Fig. \ref{fig:intensity511}.

\begin{figure}
\includegraphics[width=\columnwidth]{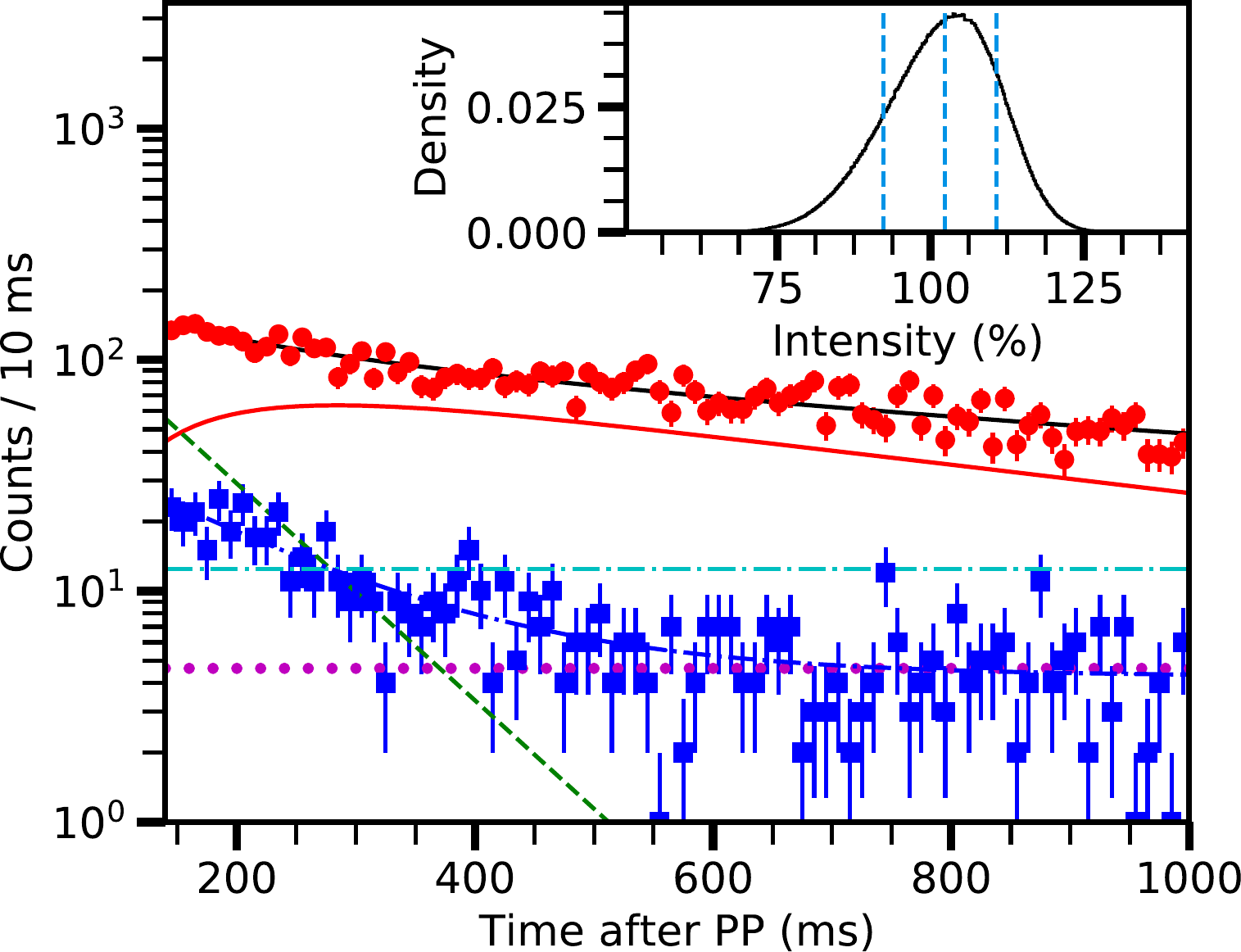}
\caption{\label{fig:intensity511} (Color online) The $\beta$-gated-$\gamma$ counts in the 511 keV transition peak area as a function of time after PP (red circles) with the fitted function (Eq. \ref{eq:model511}, black straight line) and te $\beta$-gated-$\gamma$ counts in the background area (blue squares) with the fitted function (Eq. \ref{eq:comptonbackground}, blue dash-dotted line). The contributions of the decay of $^{66}$Fe ($N_{Fe}$), $^{66}$Mn decay high-energy $\gamma$-rays ($N_{Mn}$), the environmental background ($N_{off}$) and other sources ($N_{o}$) are represented by the red straight line, the green dashed line, the cyan dash-dotted line and the purple dotted line, respectively. Insert: posterior probability density function of the 511 keV transition intensity. The 16, 50 and 84 percentiles are indicated with vertical, dotted lines.}
\end{figure}

The intensity of the 511 keV transition can be linked to the $\xi$ parameter using Eq. \ref{eq:ksiintensity}:

\begin{equation}
I_{511} =\frac{N_{511}}{N_{471}} = \frac{\sfrac{N^{R}_{511}}{eff_{511}}}{\sfrac{N^{R}_{471}}{eff_{471}}} = \xi \times \frac{eff_{471}}{eff_{511}} \mathrm{,}
\label{eq:ksiintensity}
\end{equation}

\noindent where $N_{511}$ and $N_{471}$ are the numbers of emitted $\gamma$-rays, $N^{R}_{511}$ and $N^{R}_{471}$ are the numbers of registered $\gamma$-rays and $eff_{511}$ and $eff_{471}$ are the detection efficiency for the 511 keV and 471 keV, respectively. 

After the marginalization, the $\xi$ parameter was corrected by the $\gamma$-detection detection efficiencies and the obtained intensity is equal $I_{511} = 102.3^{+8.4}_{-10.0}$. The posterior probability density function is presented as an insert in Fig. \ref{fig:intensity511}

\section{\label{apx:bateman}Bateman's equations}

\begin{equation}
\gamma_{Fe}^{sig}(t) = \alpha^{Fe}e^{-\frac{ln(2)}{T^{Mn66}_{1/2}}t}
\label{eq:gsfbatemanfe}
\end{equation}

\begin{equation}
\begin{aligned}
& \gamma_{Co66}^{sig}(t) = \alpha^{Co66}_{1}\frac{1/T^{Fe66}_{1/2}}{1/T^{Fe66}_{1/2} - 1/T^{Mn66}_{1/2}}  \\
& \times (e^{-\frac{ln(2)}{T^{Mn66}_{1/2}}t} - e^{-\frac{ln(2)}{T^{Fe66}_{1/2}}t}) + \alpha^{Co66}_{2} e^{-\frac{ln(2)}{T^{Fe66}_{1/2}}t}
\label{eq:gsfbatemanco}
\end{aligned}
\end{equation}

\begin{equation}
\begin{aligned}
& \gamma_{Ni66}^{sig}(t) = \alpha^{Ni66}_{1}\frac{1}{T^{Fe66}_{1/2}T^{Co66}_{1/2}} \\
& \times \Big(e^{-\frac{ln(2)}{T^{Mn66}_{1/2}} t} \big((\frac{1}{T^{Fe66}_{1/2}} - \frac{1}{T^{Mn66}_{1/2}})(\frac{1}{T^{Co66}_{1/2}} - \frac{1}{T^{Mn66}_{1/2}})\big)^{-1} \\
& + e^{-\frac{ln(2)}{T^{Fe66}_{1/2}} t} \big((\frac{1}{T^{Mn66}_{1/2}} - \frac{1}{T^{Fe66}_{1/2}})(\frac{1}{T^{Co66}_{1/2}} - \frac{1}{T^{Fe66}_{1/2}})\big)^{-1} \\
& + e^{-\frac{ln(2)}{T^{Co66}_{1/2}} t} \big((\frac{1}{T^{Mn66}_{1/2}} - \frac{1}{T^{Co66}_{1/2}})(\frac{1}{T^{Fe66}_{1/2}} - \frac{1}{T^{Co66}_{1/2}})\big)^{-1}\Big) \\
& + \alpha^{Ni66}_{2} \frac{1/T^{Co66}_{1/2}}{1/T^{Co66}_{1/2} - 1/T^{Fe66}_{1/2}} (e^{-\frac{ln(2)}{T^{Fe66}_{1/2}} t} - e^{-\frac{ln(2)}{T^{Co66}_{1/2}} t}) \\
& + \alpha^{Ni66}_{3} e^{-\frac{ln(2)}{T^{Co66}_{1/2}} t}
\label{eq:gsfbatemanni}
\end{aligned}
\end{equation}

\section{\label{apx:pneq}Derivation of $P_n$ equation}

The $P_n$ value is defined as a ratio of decays through delayed neutron channel to all the decays:

\begin{equation}
P_n = \frac{\beta_n}{\beta} \mathrm{.}
\label{eq:pnderived}
\end{equation}

\noindent Since both channels have the same time behavior and they are described by the Bateman's equation with the same half-life, it is enough to consider only amplitudes of these functions ($\alpha$ parameters). The total number of decays can be written as:

\begin{equation}
\beta = A_{Fe} \times \alpha^{Fe} \mathrm{,}
\label{eq:beta}
\end{equation}

\noindent where both parameters are determined from the fit. By analogy, the number of decays through delayed neutron channel can be written as: 

\begin{equation}
\beta_n = A_{Fe65} \times \alpha^{Fe65} \mathrm{.}
\label{eq:betan}
\end{equation}

\noindent The $\alpha^{Fe65}$ parameter is determined from the fit of the 364 keV transition while $A_{Fe65}$ is unknown, but it can be linked to the ground state feeding of $^{65}$Fe from $^{66}$Mn decay by using Eq. \ref{eq:gsfequation}:

\begin{equation}
gsf^{65} = 1 - \frac{1}{A_{Fe65}\times eff^{364}_\gamma \times f^{364}_I} \mathrm{,}
\end{equation}

\noindent which then can be transformed into:

\begin{equation}
A_{Fe65} = \frac{1}{1-gsf^{65}} \times \frac{1}{eff^{364}_\gamma \times f^{364}_I} \mathrm{.}
\label{eq:afe65}
\end{equation}

\noindent All the parameters in Eq. \ref{eq:afe65} are known from the analysis. By putting Eqs. \ref{eq:beta}, \ref{eq:betan} and \ref{eq:afe65} into Eq. \ref{eq:pnderived}, one can get:

\begin{eqnarray}
P_n = \frac{\beta_n}{\beta} = \frac{A_{Fe65} \times \alpha^{Fe65}}{A_{Fe} \times \alpha^{Fe}} = \frac{1}{A_{Fe}} \times \frac{\alpha^{Fe65}}{\alpha^{Fe}} \nonumber\\ \times \frac{1}{1-gsf^{65}} \times \frac{1}{eff^{364}_\gamma \times f^{364}_I} \mathrm{.}
\end{eqnarray}

\bibliography{2018_12_10_66Mn_Marek_accepted,other}

\begin{thebibliography}{100}%
\makeatletter
\providecommand \@ifxundefined [1]{%
 \@ifx{#1\undefined}
}%
\providecommand \@ifnum [1]{%
 \ifnum #1\expandafter \@firstoftwo
 \else \expandafter \@secondoftwo
 \fi
}%
\providecommand \@ifx [1]{%
 \ifx #1\expandafter \@firstoftwo
 \else \expandafter \@secondoftwo
 \fi
}%
\providecommand \natexlab [1]{#1}%
\providecommand \enquote  [1]{``#1''}%
\providecommand \bibnamefont  [1]{#1}%
\providecommand \bibfnamefont [1]{#1}%
\providecommand \citenamefont [1]{#1}%
\providecommand \href@noop [0]{\@secondoftwo}%
\providecommand \href [0]{\begingroup \@sanitize@url \@href}%
\providecommand \@href[1]{\@@startlink{#1}\@@href}%
\providecommand \@@href[1]{\endgroup#1\@@endlink}%
\providecommand \@sanitize@url [0]{\catcode `\\12\catcode `\$12\catcode
  `\&12\catcode `\#12\catcode `\^12\catcode `\_12\catcode `\%12\relax}%
\providecommand \@@startlink[1]{}%
\providecommand \@@endlink[0]{}%
\providecommand \url  [0]{\begingroup\@sanitize@url \@url }%
\providecommand \@url [1]{\endgroup\@href {#1}{\urlprefix }}%
\providecommand \urlprefix  [0]{URL }%
\providecommand \Eprint [0]{\href }%
\providecommand \doibase [0]{http://dx.doi.org/}%
\providecommand \selectlanguage [0]{\@gobble}%
\providecommand \bibinfo  [0]{\@secondoftwo}%
\providecommand \bibfield  [0]{\@secondoftwo}%
\providecommand \translation [1]{[#1]}%
\providecommand \BibitemOpen [0]{}%
\providecommand \bibitemStop [0]{}%
\providecommand \bibitemNoStop [0]{.\EOS\space}%
\providecommand \EOS [0]{\spacefactor3000\relax}%
\providecommand \BibitemShut  [1]{\csname bibitem#1\endcsname}%
\let\auto@bib@innerbib\@empty
\bibitem [{\citenamefont {Pf{\"{u}}tzner}\ \emph {et~al.}(2002)\citenamefont
  {Pf{\"{u}}tzner}, \citenamefont {Badura}, \citenamefont {Bingham},
  \citenamefont {Blank}, \citenamefont {Chartier}, \citenamefont {Geissel},
  \citenamefont {Giovinazzo}, \citenamefont {Grigorenko}, \citenamefont
  {Grzywacz}, \citenamefont {Hellstr{\"{o}}m}, \citenamefont {Janas},
  \citenamefont {Kurcewicz}, \citenamefont {Lalleman}, \citenamefont
  {Mazzocchi}, \citenamefont {Mukha}, \citenamefont {M{\"{u}}nzenberg},
  \citenamefont {Plettner}, \citenamefont {Roeckl}, \citenamefont
  {Rykaczewski}, \citenamefont {Schmidt}, \citenamefont {Simon}, \citenamefont
  {Stanoiu},\ and\ \citenamefont {Thomas}}]{Pfutzner2002}%
  \BibitemOpen
  \bibfield  {author} {\bibinfo {author} {\bibfnamefont {M.}~\bibnamefont
  {Pf{\"{u}}tzner}}, \bibinfo {author} {\bibfnamefont {E.}~\bibnamefont
  {Badura}}, \bibinfo {author} {\bibfnamefont {C.}~\bibnamefont {Bingham}},
  \bibinfo {author} {\bibfnamefont {B.}~\bibnamefont {Blank}}, \bibinfo
  {author} {\bibfnamefont {M.}~\bibnamefont {Chartier}}, \bibinfo {author}
  {\bibfnamefont {H.}~\bibnamefont {Geissel}}, \bibinfo {author} {\bibfnamefont
  {J.}~\bibnamefont {Giovinazzo}}, \bibinfo {author} {\bibfnamefont
  {L.}~\bibnamefont {Grigorenko}}, \bibinfo {author} {\bibfnamefont
  {R.}~\bibnamefont {Grzywacz}}, \bibinfo {author} {\bibfnamefont
  {M.}~\bibnamefont {Hellstr{\"{o}}m}}, \bibinfo {author} {\bibfnamefont
  {Z.}~\bibnamefont {Janas}}, \bibinfo {author} {\bibfnamefont
  {J.}~\bibnamefont {Kurcewicz}}, \bibinfo {author} {\bibfnamefont
  {A.}~\bibnamefont {Lalleman}}, \bibinfo {author} {\bibfnamefont
  {C.}~\bibnamefont {Mazzocchi}}, \bibinfo {author} {\bibfnamefont
  {I.}~\bibnamefont {Mukha}}, \bibinfo {author} {\bibfnamefont
  {G.}~\bibnamefont {M{\"{u}}nzenberg}}, \bibinfo {author} {\bibfnamefont
  {C.}~\bibnamefont {Plettner}}, \bibinfo {author} {\bibfnamefont
  {E.}~\bibnamefont {Roeckl}}, \bibinfo {author} {\bibfnamefont
  {K.}~\bibnamefont {Rykaczewski}}, \bibinfo {author} {\bibfnamefont
  {K.}~\bibnamefont {Schmidt}}, \bibinfo {author} {\bibfnamefont
  {R.}~\bibnamefont {Simon}}, \bibinfo {author} {\bibfnamefont
  {M.}~\bibnamefont {Stanoiu}}, \ and\ \bibinfo {author} {\bibfnamefont
  {J.-C.}\ \bibnamefont {Thomas}},\ }\href {\doibase
  10.1140/epja/i2002-10033-9} {\bibfield  {journal} {\bibinfo  {journal} {The
  European Physical Journal A}\ }\textbf {\bibinfo {volume} {14}},\ \bibinfo
  {pages} {279} (\bibinfo {year} {2002})}\BibitemShut {NoStop}%
\bibitem [{\citenamefont {Pomorski}\ \emph {et~al.}(2011)\citenamefont
  {Pomorski}, \citenamefont {Pf{\"{u}}tzner}, \citenamefont {Dominik},
  \citenamefont {Grzywacz}, \citenamefont {Baumann}, \citenamefont {Berryman},
  \citenamefont {Czyrkowski}, \citenamefont {D{\c{a}}browski}, \citenamefont
  {Ginter}, \citenamefont {Johnson}, \citenamefont {Kami{\'{n}}ski},
  \citenamefont {Ku{\'{z}}niak}, \citenamefont {Larson}, \citenamefont
  {Liddick}, \citenamefont {Madurga}, \citenamefont {Mazzocchi}, \citenamefont
  {Mianowski}, \citenamefont {Miernik}, \citenamefont {Miller}, \citenamefont
  {Paulauskas}, \citenamefont {Pereira}, \citenamefont {Rykaczewski},
  \citenamefont {Stolz},\ and\ \citenamefont {Suchyta}}]{Pomorski2011}%
  \BibitemOpen
  \bibfield  {author} {\bibinfo {author} {\bibfnamefont {M.}~\bibnamefont
  {Pomorski}}, \bibinfo {author} {\bibfnamefont {M.}~\bibnamefont
  {Pf{\"{u}}tzner}}, \bibinfo {author} {\bibfnamefont {W.}~\bibnamefont
  {Dominik}}, \bibinfo {author} {\bibfnamefont {R.}~\bibnamefont {Grzywacz}},
  \bibinfo {author} {\bibfnamefont {T.}~\bibnamefont {Baumann}}, \bibinfo
  {author} {\bibfnamefont {J.~S.}\ \bibnamefont {Berryman}}, \bibinfo {author}
  {\bibfnamefont {H.}~\bibnamefont {Czyrkowski}}, \bibinfo {author}
  {\bibfnamefont {R.}~\bibnamefont {D{\c{a}}browski}}, \bibinfo {author}
  {\bibfnamefont {T.}~\bibnamefont {Ginter}}, \bibinfo {author} {\bibfnamefont
  {J.}~\bibnamefont {Johnson}}, \bibinfo {author} {\bibfnamefont
  {G.}~\bibnamefont {Kami{\'{n}}ski}}, \bibinfo {author} {\bibfnamefont
  {A.}~\bibnamefont {Ku{\'{z}}niak}}, \bibinfo {author} {\bibfnamefont
  {N.}~\bibnamefont {Larson}}, \bibinfo {author} {\bibfnamefont {S.~N.}\
  \bibnamefont {Liddick}}, \bibinfo {author} {\bibfnamefont {M.}~\bibnamefont
  {Madurga}}, \bibinfo {author} {\bibfnamefont {C.}~\bibnamefont {Mazzocchi}},
  \bibinfo {author} {\bibfnamefont {S.}~\bibnamefont {Mianowski}}, \bibinfo
  {author} {\bibfnamefont {K.}~\bibnamefont {Miernik}}, \bibinfo {author}
  {\bibfnamefont {D.}~\bibnamefont {Miller}}, \bibinfo {author} {\bibfnamefont
  {S.}~\bibnamefont {Paulauskas}}, \bibinfo {author} {\bibfnamefont
  {J.}~\bibnamefont {Pereira}}, \bibinfo {author} {\bibfnamefont {K.~P.}\
  \bibnamefont {Rykaczewski}}, \bibinfo {author} {\bibfnamefont
  {A.}~\bibnamefont {Stolz}}, \ and\ \bibinfo {author} {\bibfnamefont
  {S.}~\bibnamefont {Suchyta}},\ }\href {\doibase 10.1103/PhysRevC.83.061303}
  {\bibfield  {journal} {\bibinfo  {journal} {Physical Review C}\ }\textbf
  {\bibinfo {volume} {83}},\ \bibinfo {pages} {061303} (\bibinfo {year}
  {2011})}\BibitemShut {NoStop}%
\bibitem [{\citenamefont {Pomorski}\ \emph {et~al.}(2014)\citenamefont
  {Pomorski}, \citenamefont {Pf{\"{u}}tzner}, \citenamefont {Dominik},
  \citenamefont {Grzywacz}, \citenamefont {Stolz}, \citenamefont {Baumann},
  \citenamefont {Berryman}, \citenamefont {Czyrkowski}, \citenamefont
  {D{\c{a}}browski}, \citenamefont {Fija{\l}kowska}, \citenamefont {Ginter},
  \citenamefont {Johnson}, \citenamefont {Kami{\'{n}}ski}, \citenamefont
  {Larson}, \citenamefont {Liddick}, \citenamefont {Madurga}, \citenamefont
  {Mazzocchi}, \citenamefont {Mianowski}, \citenamefont {Miernik},
  \citenamefont {Miller}, \citenamefont {Paulauskas}, \citenamefont {Pereira},
  \citenamefont {Rykaczewski},\ and\ \citenamefont {Suchyta}}]{Pomorski2014}%
  \BibitemOpen
  \bibfield  {author} {\bibinfo {author} {\bibfnamefont {M.}~\bibnamefont
  {Pomorski}}, \bibinfo {author} {\bibfnamefont {M.}~\bibnamefont
  {Pf{\"{u}}tzner}}, \bibinfo {author} {\bibfnamefont {W.}~\bibnamefont
  {Dominik}}, \bibinfo {author} {\bibfnamefont {R.}~\bibnamefont {Grzywacz}},
  \bibinfo {author} {\bibfnamefont {A.}~\bibnamefont {Stolz}}, \bibinfo
  {author} {\bibfnamefont {T.}~\bibnamefont {Baumann}}, \bibinfo {author}
  {\bibfnamefont {J.~S.}\ \bibnamefont {Berryman}}, \bibinfo {author}
  {\bibfnamefont {H.}~\bibnamefont {Czyrkowski}}, \bibinfo {author}
  {\bibfnamefont {R.}~\bibnamefont {D{\c{a}}browski}}, \bibinfo {author}
  {\bibfnamefont {A.}~\bibnamefont {Fija{\l}kowska}}, \bibinfo {author}
  {\bibfnamefont {T.}~\bibnamefont {Ginter}}, \bibinfo {author} {\bibfnamefont
  {J.}~\bibnamefont {Johnson}}, \bibinfo {author} {\bibfnamefont
  {G.}~\bibnamefont {Kami{\'{n}}ski}}, \bibinfo {author} {\bibfnamefont
  {N.}~\bibnamefont {Larson}}, \bibinfo {author} {\bibfnamefont {S.~N.}\
  \bibnamefont {Liddick}}, \bibinfo {author} {\bibfnamefont {M.}~\bibnamefont
  {Madurga}}, \bibinfo {author} {\bibfnamefont {C.}~\bibnamefont {Mazzocchi}},
  \bibinfo {author} {\bibfnamefont {S.}~\bibnamefont {Mianowski}}, \bibinfo
  {author} {\bibfnamefont {K.}~\bibnamefont {Miernik}}, \bibinfo {author}
  {\bibfnamefont {D.}~\bibnamefont {Miller}}, \bibinfo {author} {\bibfnamefont
  {S.}~\bibnamefont {Paulauskas}}, \bibinfo {author} {\bibfnamefont
  {J.}~\bibnamefont {Pereira}}, \bibinfo {author} {\bibfnamefont {K.~P.}\
  \bibnamefont {Rykaczewski}}, \ and\ \bibinfo {author} {\bibfnamefont
  {S.}~\bibnamefont {Suchyta}},\ }\href {\doibase 10.1103/PhysRevC.90.014311}
  {\bibfield  {journal} {\bibinfo  {journal} {Physical Review C}\ }\textbf
  {\bibinfo {volume} {90}},\ \bibinfo {pages} {014311} (\bibinfo {year}
  {2014})}\BibitemShut {NoStop}%
\bibitem [{\citenamefont {{Van de Walle}}\ \emph {et~al.}(2009)\citenamefont
  {{Van de Walle}}, \citenamefont {Aksouh}, \citenamefont {Behrens},
  \citenamefont {Bildstein}, \citenamefont {Blazhev}, \citenamefont
  {Cederk{\"{a}}ll}, \citenamefont {Cl{\'{e}}ment}, \citenamefont {Cocolios},
  \citenamefont {Davinson}, \citenamefont {Delahaye}, \citenamefont {Eberth},
  \citenamefont {Ekstr{\"{o}}m}, \citenamefont {Fedorov}, \citenamefont
  {Fedosseev}, \citenamefont {Fraile}, \citenamefont {Franchoo}, \citenamefont
  {Gernhauser}, \citenamefont {Georgiev}, \citenamefont {Habs}, \citenamefont
  {Heyde}, \citenamefont {Huber}, \citenamefont {Huyse}, \citenamefont
  {Ibrahim}, \citenamefont {Ivanov}, \citenamefont {Iwanicki}, \citenamefont
  {Jolie}, \citenamefont {Kester}, \citenamefont {K{\"{o}}ster}, \citenamefont
  {Kr{\"{o}}ll}, \citenamefont {Kr{\"{u}}cken}, \citenamefont {Lauer},
  \citenamefont {Lisetskiy}, \citenamefont {Lutter}, \citenamefont {Marsh},
  \citenamefont {Mayet}, \citenamefont {Niedermaier}, \citenamefont {Pantea},
  \citenamefont {Raabe}, \citenamefont {Reiter}, \citenamefont {Sawicka},
  \citenamefont {Scheit}, \citenamefont {Schrieder}, \citenamefont {Schwalm},
  \citenamefont {Seliverstov}, \citenamefont {Sieber}, \citenamefont {Sletten},
  \citenamefont {Smirnova}, \citenamefont {Stanoiu}, \citenamefont
  {Stefanescu}, \citenamefont {Thomas}, \citenamefont {Valiente-Dob{\'{o}}n},
  \citenamefont {{Van Duppen}}, \citenamefont {Verney}, \citenamefont {Voulot},
  \citenamefont {Warr}, \citenamefont {Weisshaar}, \citenamefont {Wenander},
  \citenamefont {Wolf},\ and\ \citenamefont
  {Zieli{\'{n}}ska}}]{VandeWalle2009}%
  \BibitemOpen
  \bibfield  {author} {\bibinfo {author} {\bibfnamefont {J.}~\bibnamefont {{Van
  de Walle}}}, \bibinfo {author} {\bibfnamefont {F.}~\bibnamefont {Aksouh}},
  \bibinfo {author} {\bibfnamefont {T.}~\bibnamefont {Behrens}}, \bibinfo
  {author} {\bibfnamefont {V.}~\bibnamefont {Bildstein}}, \bibinfo {author}
  {\bibfnamefont {A.}~\bibnamefont {Blazhev}}, \bibinfo {author} {\bibfnamefont
  {J.}~\bibnamefont {Cederk{\"{a}}ll}}, \bibinfo {author} {\bibfnamefont
  {E.}~\bibnamefont {Cl{\'{e}}ment}}, \bibinfo {author} {\bibfnamefont {T.~E.}\
  \bibnamefont {Cocolios}}, \bibinfo {author} {\bibfnamefont {T.}~\bibnamefont
  {Davinson}}, \bibinfo {author} {\bibfnamefont {P.}~\bibnamefont {Delahaye}},
  \bibinfo {author} {\bibfnamefont {J.}~\bibnamefont {Eberth}}, \bibinfo
  {author} {\bibfnamefont {A.}~\bibnamefont {Ekstr{\"{o}}m}}, \bibinfo {author}
  {\bibfnamefont {D.~V.}\ \bibnamefont {Fedorov}}, \bibinfo {author}
  {\bibfnamefont {V.~N.}\ \bibnamefont {Fedosseev}}, \bibinfo {author}
  {\bibfnamefont {L.~M.}\ \bibnamefont {Fraile}}, \bibinfo {author}
  {\bibfnamefont {S.}~\bibnamefont {Franchoo}}, \bibinfo {author}
  {\bibfnamefont {R.}~\bibnamefont {Gernhauser}}, \bibinfo {author}
  {\bibfnamefont {G.}~\bibnamefont {Georgiev}}, \bibinfo {author}
  {\bibfnamefont {D.}~\bibnamefont {Habs}}, \bibinfo {author} {\bibfnamefont
  {K.}~\bibnamefont {Heyde}}, \bibinfo {author} {\bibfnamefont
  {G.}~\bibnamefont {Huber}}, \bibinfo {author} {\bibfnamefont
  {M.}~\bibnamefont {Huyse}}, \bibinfo {author} {\bibfnamefont
  {F.}~\bibnamefont {Ibrahim}}, \bibinfo {author} {\bibfnamefont
  {O.}~\bibnamefont {Ivanov}}, \bibinfo {author} {\bibfnamefont
  {J.}~\bibnamefont {Iwanicki}}, \bibinfo {author} {\bibfnamefont
  {J.}~\bibnamefont {Jolie}}, \bibinfo {author} {\bibfnamefont
  {O.}~\bibnamefont {Kester}}, \bibinfo {author} {\bibfnamefont
  {U.}~\bibnamefont {K{\"{o}}ster}}, \bibinfo {author} {\bibfnamefont
  {T.}~\bibnamefont {Kr{\"{o}}ll}}, \bibinfo {author} {\bibfnamefont
  {R.}~\bibnamefont {Kr{\"{u}}cken}}, \bibinfo {author} {\bibfnamefont
  {M.}~\bibnamefont {Lauer}}, \bibinfo {author} {\bibfnamefont {A.~F.}\
  \bibnamefont {Lisetskiy}}, \bibinfo {author} {\bibfnamefont {R.}~\bibnamefont
  {Lutter}}, \bibinfo {author} {\bibfnamefont {B.~A.}\ \bibnamefont {Marsh}},
  \bibinfo {author} {\bibfnamefont {P.}~\bibnamefont {Mayet}}, \bibinfo
  {author} {\bibfnamefont {O.}~\bibnamefont {Niedermaier}}, \bibinfo {author}
  {\bibfnamefont {M.}~\bibnamefont {Pantea}}, \bibinfo {author} {\bibfnamefont
  {R.}~\bibnamefont {Raabe}}, \bibinfo {author} {\bibfnamefont
  {P.}~\bibnamefont {Reiter}}, \bibinfo {author} {\bibfnamefont
  {M.}~\bibnamefont {Sawicka}}, \bibinfo {author} {\bibfnamefont
  {H.}~\bibnamefont {Scheit}}, \bibinfo {author} {\bibfnamefont
  {G.}~\bibnamefont {Schrieder}}, \bibinfo {author} {\bibfnamefont
  {D.}~\bibnamefont {Schwalm}}, \bibinfo {author} {\bibfnamefont {M.~D.}\
  \bibnamefont {Seliverstov}}, \bibinfo {author} {\bibfnamefont
  {T.}~\bibnamefont {Sieber}}, \bibinfo {author} {\bibfnamefont
  {G.}~\bibnamefont {Sletten}}, \bibinfo {author} {\bibfnamefont
  {N.}~\bibnamefont {Smirnova}}, \bibinfo {author} {\bibfnamefont
  {M.}~\bibnamefont {Stanoiu}}, \bibinfo {author} {\bibfnamefont
  {I.}~\bibnamefont {Stefanescu}}, \bibinfo {author} {\bibfnamefont {J.-C.}\
  \bibnamefont {Thomas}}, \bibinfo {author} {\bibfnamefont {J.~J.}\
  \bibnamefont {Valiente-Dob{\'{o}}n}}, \bibinfo {author} {\bibfnamefont
  {P.}~\bibnamefont {{Van Duppen}}}, \bibinfo {author} {\bibfnamefont
  {D.}~\bibnamefont {Verney}}, \bibinfo {author} {\bibfnamefont
  {D.}~\bibnamefont {Voulot}}, \bibinfo {author} {\bibfnamefont
  {N.}~\bibnamefont {Warr}}, \bibinfo {author} {\bibfnamefont {D.}~\bibnamefont
  {Weisshaar}}, \bibinfo {author} {\bibfnamefont {F.}~\bibnamefont {Wenander}},
  \bibinfo {author} {\bibfnamefont {B.~H.}\ \bibnamefont {Wolf}}, \ and\
  \bibinfo {author} {\bibfnamefont {M.}~\bibnamefont {Zieli{\'{n}}ska}},\
  }\href {\doibase 10.1103/PhysRevC.79.014309} {\bibfield  {journal} {\bibinfo
  {journal} {Physical Review C}\ }\textbf {\bibinfo {volume} {79}},\ \bibinfo
  {pages} {014309} (\bibinfo {year} {2009})}\BibitemShut {NoStop}%
\bibitem [{\citenamefont {Padgett}\ \emph {et~al.}(2010)\citenamefont
  {Padgett}, \citenamefont {Madurga}, \citenamefont {Grzywacz}, \citenamefont
  {Darby}, \citenamefont {Liddick}, \citenamefont {Paulauskas}, \citenamefont
  {Cartegni}, \citenamefont {Bingham}, \citenamefont {Gross}, \citenamefont
  {Rykaczewski}, \citenamefont {Shapira}, \citenamefont {Stracener},
  \citenamefont {Mendez}, \citenamefont {Winger}, \citenamefont {Ilyushkin},
  \citenamefont {Korgul}, \citenamefont {Kr{\'{o}}las}, \citenamefont
  {Zganjar}, \citenamefont {Mazzocchi}, \citenamefont {Liu}, \citenamefont
  {Hamilton}, \citenamefont {Batchelder},\ and\ \citenamefont
  {Rajabali}}]{Padgett2010}%
  \BibitemOpen
  \bibfield  {author} {\bibinfo {author} {\bibfnamefont {S.}~\bibnamefont
  {Padgett}}, \bibinfo {author} {\bibfnamefont {M.}~\bibnamefont {Madurga}},
  \bibinfo {author} {\bibfnamefont {R.}~\bibnamefont {Grzywacz}}, \bibinfo
  {author} {\bibfnamefont {I.~G.}\ \bibnamefont {Darby}}, \bibinfo {author}
  {\bibfnamefont {S.~N.}\ \bibnamefont {Liddick}}, \bibinfo {author}
  {\bibfnamefont {S.~V.}\ \bibnamefont {Paulauskas}}, \bibinfo {author}
  {\bibfnamefont {L.}~\bibnamefont {Cartegni}}, \bibinfo {author}
  {\bibfnamefont {C.~R.}\ \bibnamefont {Bingham}}, \bibinfo {author}
  {\bibfnamefont {C.~J.}\ \bibnamefont {Gross}}, \bibinfo {author}
  {\bibfnamefont {K.}~\bibnamefont {Rykaczewski}}, \bibinfo {author}
  {\bibfnamefont {D.}~\bibnamefont {Shapira}}, \bibinfo {author} {\bibfnamefont
  {D.~W.}\ \bibnamefont {Stracener}}, \bibinfo {author} {\bibfnamefont {A.~J.}\
  \bibnamefont {Mendez}}, \bibinfo {author} {\bibfnamefont {J.~A.}\
  \bibnamefont {Winger}}, \bibinfo {author} {\bibfnamefont {S.~V.}\
  \bibnamefont {Ilyushkin}}, \bibinfo {author} {\bibfnamefont {A.}~\bibnamefont
  {Korgul}}, \bibinfo {author} {\bibfnamefont {W.}~\bibnamefont
  {Kr{\'{o}}las}}, \bibinfo {author} {\bibfnamefont {E.}~\bibnamefont
  {Zganjar}}, \bibinfo {author} {\bibfnamefont {C.}~\bibnamefont {Mazzocchi}},
  \bibinfo {author} {\bibfnamefont {S.}~\bibnamefont {Liu}}, \bibinfo {author}
  {\bibfnamefont {J.~H.}\ \bibnamefont {Hamilton}}, \bibinfo {author}
  {\bibfnamefont {J.~C.}\ \bibnamefont {Batchelder}}, \ and\ \bibinfo {author}
  {\bibfnamefont {M.~M.}\ \bibnamefont {Rajabali}},\ }\href {\doibase
  10.1103/PhysRevC.82.064314} {\bibfield  {journal} {\bibinfo  {journal}
  {Physical Review C}\ }\textbf {\bibinfo {volume} {82}},\ \bibinfo {pages}
  {064314} (\bibinfo {year} {2010})}\BibitemShut {NoStop}%
\bibitem [{\citenamefont {Xu}\ \emph {et~al.}(2014)\citenamefont {Xu},
  \citenamefont {Nishimura}, \citenamefont {Lorusso}, \citenamefont {Browne},
  \citenamefont {Doornenbal}, \citenamefont {Gey}, \citenamefont {Jung},
  \citenamefont {Li}, \citenamefont {Niikura}, \citenamefont
  {S{\"{o}}derstr{\"{o}}m}, \citenamefont {Sumikama}, \citenamefont {Taprogge},
  \citenamefont {Vajta}, \citenamefont {Watanabe}, \citenamefont {Wu},
  \citenamefont {Yagi}, \citenamefont {Yoshinaga}, \citenamefont {Baba},
  \citenamefont {Franchoo}, \citenamefont {Isobe}, \citenamefont {John},
  \citenamefont {Kojouharov}, \citenamefont {Kubono}, \citenamefont {Kurz},
  \citenamefont {Matea}, \citenamefont {Matsui}, \citenamefont {Mengoni},
  \citenamefont {Morfouace}, \citenamefont {Napoli}, \citenamefont {Naqvi},
  \citenamefont {Nishibata}, \citenamefont {Odahara}, \citenamefont {Şahin},
  \citenamefont {Sakurai}, \citenamefont {Schaffner}, \citenamefont {Stefan},
  \citenamefont {Suzuki}, \citenamefont {Taniuchi},\ and\ \citenamefont
  {Werner}}]{Xu2014}%
  \BibitemOpen
  \bibfield  {author} {\bibinfo {author} {\bibfnamefont {Z.~Y.}\ \bibnamefont
  {Xu}}, \bibinfo {author} {\bibfnamefont {S.}~\bibnamefont {Nishimura}},
  \bibinfo {author} {\bibfnamefont {G.}~\bibnamefont {Lorusso}}, \bibinfo
  {author} {\bibfnamefont {F.}~\bibnamefont {Browne}}, \bibinfo {author}
  {\bibfnamefont {P.}~\bibnamefont {Doornenbal}}, \bibinfo {author}
  {\bibfnamefont {G.}~\bibnamefont {Gey}}, \bibinfo {author} {\bibfnamefont
  {H.-S.}\ \bibnamefont {Jung}}, \bibinfo {author} {\bibfnamefont
  {Z.}~\bibnamefont {Li}}, \bibinfo {author} {\bibfnamefont {M.}~\bibnamefont
  {Niikura}}, \bibinfo {author} {\bibfnamefont {P.-A.}\ \bibnamefont
  {S{\"{o}}derstr{\"{o}}m}}, \bibinfo {author} {\bibfnamefont {T.}~\bibnamefont
  {Sumikama}}, \bibinfo {author} {\bibfnamefont {J.}~\bibnamefont {Taprogge}},
  \bibinfo {author} {\bibfnamefont {Z.}~\bibnamefont {Vajta}}, \bibinfo
  {author} {\bibfnamefont {H.}~\bibnamefont {Watanabe}}, \bibinfo {author}
  {\bibfnamefont {J.}~\bibnamefont {Wu}}, \bibinfo {author} {\bibfnamefont
  {A.}~\bibnamefont {Yagi}}, \bibinfo {author} {\bibfnamefont {K.}~\bibnamefont
  {Yoshinaga}}, \bibinfo {author} {\bibfnamefont {H.}~\bibnamefont {Baba}},
  \bibinfo {author} {\bibfnamefont {S.}~\bibnamefont {Franchoo}}, \bibinfo
  {author} {\bibfnamefont {T.}~\bibnamefont {Isobe}}, \bibinfo {author}
  {\bibfnamefont {P.~R.}\ \bibnamefont {John}}, \bibinfo {author}
  {\bibfnamefont {I.}~\bibnamefont {Kojouharov}}, \bibinfo {author}
  {\bibfnamefont {S.}~\bibnamefont {Kubono}}, \bibinfo {author} {\bibfnamefont
  {N.}~\bibnamefont {Kurz}}, \bibinfo {author} {\bibfnamefont {I.}~\bibnamefont
  {Matea}}, \bibinfo {author} {\bibfnamefont {K.}~\bibnamefont {Matsui}},
  \bibinfo {author} {\bibfnamefont {D.}~\bibnamefont {Mengoni}}, \bibinfo
  {author} {\bibfnamefont {P.}~\bibnamefont {Morfouace}}, \bibinfo {author}
  {\bibfnamefont {D.~R.}\ \bibnamefont {Napoli}}, \bibinfo {author}
  {\bibfnamefont {F.}~\bibnamefont {Naqvi}}, \bibinfo {author} {\bibfnamefont
  {H.}~\bibnamefont {Nishibata}}, \bibinfo {author} {\bibfnamefont
  {A.}~\bibnamefont {Odahara}}, \bibinfo {author} {\bibfnamefont
  {E.}~\bibnamefont {Şahin}}, \bibinfo {author} {\bibfnamefont
  {H.}~\bibnamefont {Sakurai}}, \bibinfo {author} {\bibfnamefont
  {H.}~\bibnamefont {Schaffner}}, \bibinfo {author} {\bibfnamefont {I.~G.}\
  \bibnamefont {Stefan}}, \bibinfo {author} {\bibfnamefont {D.}~\bibnamefont
  {Suzuki}}, \bibinfo {author} {\bibfnamefont {R.}~\bibnamefont {Taniuchi}}, \
  and\ \bibinfo {author} {\bibfnamefont {V.}~\bibnamefont {Werner}},\ }\href
  {\doibase 10.1103/PhysRevLett.113.032505} {\bibfield  {journal} {\bibinfo
  {journal} {Physical Review Letters}\ }\textbf {\bibinfo {volume} {113}},\
  \bibinfo {pages} {032505} (\bibinfo {year} {2014})}\BibitemShut {NoStop}%
\bibitem [{\citenamefont {Shiga}\ \emph {et~al.}(2016)\citenamefont {Shiga},
  \citenamefont {Yoneda}, \citenamefont {Steppenbeck}, \citenamefont {Aoi},
  \citenamefont {Doornenbal}, \citenamefont {Lee}, \citenamefont {Liu},
  \citenamefont {Matsushita}, \citenamefont {Takeuchi}, \citenamefont {Wang},
  \citenamefont {Baba}, \citenamefont {Bednarczyk}, \citenamefont {Dombradi},
  \citenamefont {Fulop}, \citenamefont {Go}, \citenamefont {Hashimoto},
  \citenamefont {Honma}, \citenamefont {Ideguchi}, \citenamefont {Ieki},
  \citenamefont {Kobayashi}, \citenamefont {Kondo}, \citenamefont {Minakata},
  \citenamefont {Motobayashi}, \citenamefont {Nishimura}, \citenamefont
  {Otsuka}, \citenamefont {Otsu}, \citenamefont {Sakurai}, \citenamefont
  {Shimizu}, \citenamefont {Sohler}, \citenamefont {Sun}, \citenamefont
  {Tamii}, \citenamefont {Tanaka}, \citenamefont {Tian}, \citenamefont
  {Tsunoda}, \citenamefont {Vajta}, \citenamefont {Yamamoto}, \citenamefont
  {Yang}, \citenamefont {Yang}, \citenamefont {Ye}, \citenamefont {Yokoyama},\
  and\ \citenamefont {Zenihiro}}]{Shiga2016}%
  \BibitemOpen
  \bibfield  {author} {\bibinfo {author} {\bibfnamefont {Y.}~\bibnamefont
  {Shiga}}, \bibinfo {author} {\bibfnamefont {K.}~\bibnamefont {Yoneda}},
  \bibinfo {author} {\bibfnamefont {D.}~\bibnamefont {Steppenbeck}}, \bibinfo
  {author} {\bibfnamefont {N.}~\bibnamefont {Aoi}}, \bibinfo {author}
  {\bibfnamefont {P.}~\bibnamefont {Doornenbal}}, \bibinfo {author}
  {\bibfnamefont {J.}~\bibnamefont {Lee}}, \bibinfo {author} {\bibfnamefont
  {H.}~\bibnamefont {Liu}}, \bibinfo {author} {\bibfnamefont {M.}~\bibnamefont
  {Matsushita}}, \bibinfo {author} {\bibfnamefont {S.}~\bibnamefont
  {Takeuchi}}, \bibinfo {author} {\bibfnamefont {H.}~\bibnamefont {Wang}},
  \bibinfo {author} {\bibfnamefont {H.}~\bibnamefont {Baba}}, \bibinfo {author}
  {\bibfnamefont {P.}~\bibnamefont {Bednarczyk}}, \bibinfo {author}
  {\bibfnamefont {Z.}~\bibnamefont {Dombradi}}, \bibinfo {author}
  {\bibfnamefont {Z.}~\bibnamefont {Fulop}}, \bibinfo {author} {\bibfnamefont
  {S.}~\bibnamefont {Go}}, \bibinfo {author} {\bibfnamefont {T.}~\bibnamefont
  {Hashimoto}}, \bibinfo {author} {\bibfnamefont {M.}~\bibnamefont {Honma}},
  \bibinfo {author} {\bibfnamefont {E.}~\bibnamefont {Ideguchi}}, \bibinfo
  {author} {\bibfnamefont {K.}~\bibnamefont {Ieki}}, \bibinfo {author}
  {\bibfnamefont {K.}~\bibnamefont {Kobayashi}}, \bibinfo {author}
  {\bibfnamefont {Y.}~\bibnamefont {Kondo}}, \bibinfo {author} {\bibfnamefont
  {R.}~\bibnamefont {Minakata}}, \bibinfo {author} {\bibfnamefont
  {T.}~\bibnamefont {Motobayashi}}, \bibinfo {author} {\bibfnamefont
  {D.}~\bibnamefont {Nishimura}}, \bibinfo {author} {\bibfnamefont
  {T.}~\bibnamefont {Otsuka}}, \bibinfo {author} {\bibfnamefont
  {H.}~\bibnamefont {Otsu}}, \bibinfo {author} {\bibfnamefont {H.}~\bibnamefont
  {Sakurai}}, \bibinfo {author} {\bibfnamefont {N.}~\bibnamefont {Shimizu}},
  \bibinfo {author} {\bibfnamefont {D.}~\bibnamefont {Sohler}}, \bibinfo
  {author} {\bibfnamefont {Y.}~\bibnamefont {Sun}}, \bibinfo {author}
  {\bibfnamefont {A.}~\bibnamefont {Tamii}}, \bibinfo {author} {\bibfnamefont
  {R.}~\bibnamefont {Tanaka}}, \bibinfo {author} {\bibfnamefont
  {Z.}~\bibnamefont {Tian}}, \bibinfo {author} {\bibfnamefont {Y.}~\bibnamefont
  {Tsunoda}}, \bibinfo {author} {\bibfnamefont {Z.}~\bibnamefont {Vajta}},
  \bibinfo {author} {\bibfnamefont {T.}~\bibnamefont {Yamamoto}}, \bibinfo
  {author} {\bibfnamefont {X.}~\bibnamefont {Yang}}, \bibinfo {author}
  {\bibfnamefont {Z.}~\bibnamefont {Yang}}, \bibinfo {author} {\bibfnamefont
  {Y.}~\bibnamefont {Ye}}, \bibinfo {author} {\bibfnamefont {R.}~\bibnamefont
  {Yokoyama}}, \ and\ \bibinfo {author} {\bibfnamefont {J.}~\bibnamefont
  {Zenihiro}},\ }\href {\doibase 10.1103/PhysRevC.93.024320} {\bibfield
  {journal} {\bibinfo  {journal} {Physical Review C}\ }\textbf {\bibinfo
  {volume} {93}},\ \bibinfo {pages} {024320} (\bibinfo {year}
  {2016})}\BibitemShut {NoStop}%
\bibitem [{\citenamefont {Alshudifat}\ \emph {et~al.}(2016)\citenamefont
  {Alshudifat}, \citenamefont {Grzywacz}, \citenamefont {Madurga},
  \citenamefont {Gross}, \citenamefont {Rykaczewski}, \citenamefont
  {Batchelder}, \citenamefont {Bingham}, \citenamefont {Borzov}, \citenamefont
  {Brewer}, \citenamefont {Cartegni}, \citenamefont {Fija{\l}kowska},
  \citenamefont {Hamilton}, \citenamefont {Hwang}, \citenamefont {Ilyushkin},
  \citenamefont {Jost}, \citenamefont {Karny}, \citenamefont {Korgul},
  \citenamefont {Kr{\'{o}}las}, \citenamefont {Liu}, \citenamefont {Mazzocchi},
  \citenamefont {Mendez}, \citenamefont {Miernik}, \citenamefont {Miller},
  \citenamefont {Padgett}, \citenamefont {Paulauskas}, \citenamefont {Ramayya},
  \citenamefont {Stracener}, \citenamefont {Surman}, \citenamefont {Winger},
  \citenamefont {Woli{\'{n}}ska-Cichocka},\ and\ \citenamefont
  {Zganjar}}]{Alshudifat2016}%
  \BibitemOpen
  \bibfield  {author} {\bibinfo {author} {\bibfnamefont {M.~F.}\ \bibnamefont
  {Alshudifat}}, \bibinfo {author} {\bibfnamefont {R.}~\bibnamefont
  {Grzywacz}}, \bibinfo {author} {\bibfnamefont {M.}~\bibnamefont {Madurga}},
  \bibinfo {author} {\bibfnamefont {C.~J.}\ \bibnamefont {Gross}}, \bibinfo
  {author} {\bibfnamefont {K.~P.}\ \bibnamefont {Rykaczewski}}, \bibinfo
  {author} {\bibfnamefont {J.~C.}\ \bibnamefont {Batchelder}}, \bibinfo
  {author} {\bibfnamefont {C.}~\bibnamefont {Bingham}}, \bibinfo {author}
  {\bibfnamefont {I.~N.}\ \bibnamefont {Borzov}}, \bibinfo {author}
  {\bibfnamefont {N.~T.}\ \bibnamefont {Brewer}}, \bibinfo {author}
  {\bibfnamefont {L.}~\bibnamefont {Cartegni}}, \bibinfo {author}
  {\bibfnamefont {A.}~\bibnamefont {Fija{\l}kowska}}, \bibinfo {author}
  {\bibfnamefont {J.~H.}\ \bibnamefont {Hamilton}}, \bibinfo {author}
  {\bibfnamefont {J.~K.}\ \bibnamefont {Hwang}}, \bibinfo {author}
  {\bibfnamefont {S.~V.}\ \bibnamefont {Ilyushkin}}, \bibinfo {author}
  {\bibfnamefont {C.}~\bibnamefont {Jost}}, \bibinfo {author} {\bibfnamefont
  {M.}~\bibnamefont {Karny}}, \bibinfo {author} {\bibfnamefont
  {A.}~\bibnamefont {Korgul}}, \bibinfo {author} {\bibfnamefont
  {W.}~\bibnamefont {Kr{\'{o}}las}}, \bibinfo {author} {\bibfnamefont {S.~H.}\
  \bibnamefont {Liu}}, \bibinfo {author} {\bibfnamefont {C.}~\bibnamefont
  {Mazzocchi}}, \bibinfo {author} {\bibfnamefont {A.~J.}\ \bibnamefont
  {Mendez}}, \bibinfo {author} {\bibfnamefont {K.}~\bibnamefont {Miernik}},
  \bibinfo {author} {\bibfnamefont {D.}~\bibnamefont {Miller}}, \bibinfo
  {author} {\bibfnamefont {S.~W.}\ \bibnamefont {Padgett}}, \bibinfo {author}
  {\bibfnamefont {S.~V.}\ \bibnamefont {Paulauskas}}, \bibinfo {author}
  {\bibfnamefont {A.~V.}\ \bibnamefont {Ramayya}}, \bibinfo {author}
  {\bibfnamefont {D.~W.}\ \bibnamefont {Stracener}}, \bibinfo {author}
  {\bibfnamefont {R.}~\bibnamefont {Surman}}, \bibinfo {author} {\bibfnamefont
  {J.~A.}\ \bibnamefont {Winger}}, \bibinfo {author} {\bibfnamefont
  {M.}~\bibnamefont {Woli{\'{n}}ska-Cichocka}}, \ and\ \bibinfo {author}
  {\bibfnamefont {E.~F.}\ \bibnamefont {Zganjar}},\ }\href {\doibase
  10.1103/PhysRevC.93.044325} {\bibfield  {journal} {\bibinfo  {journal}
  {Physical Review C}\ }\textbf {\bibinfo {volume} {93}},\ \bibinfo {pages}
  {044325} (\bibinfo {year} {2016})}\BibitemShut {NoStop}%
\bibitem [{\citenamefont {Olivier}\ \emph {et~al.}(2017)\citenamefont
  {Olivier}, \citenamefont {Franchoo}, \citenamefont {Niikura}, \citenamefont
  {Vajta}, \citenamefont {Sohler}, \citenamefont {Doornenbal}, \citenamefont
  {Obertelli}, \citenamefont {Tsunoda}, \citenamefont {Otsuka}, \citenamefont
  {Authelet}, \citenamefont {Baba}, \citenamefont {Calvet}, \citenamefont
  {Ch{\^{a}}teau}, \citenamefont {Corsi}, \citenamefont {Delbart},
  \citenamefont {Gheller}, \citenamefont {Gillibert}, \citenamefont {Isobe},
  \citenamefont {Lapoux}, \citenamefont {Matsushita}, \citenamefont {Momiyama},
  \citenamefont {Motobayashi}, \citenamefont {Otsu}, \citenamefont
  {P{\'{e}}ron}, \citenamefont {Peyaud}, \citenamefont {Pollacco},
  \citenamefont {Rouss{\'{e}}}, \citenamefont {Sakurai}, \citenamefont
  {Santamaria}, \citenamefont {Sasano}, \citenamefont {Shiga}, \citenamefont
  {Takeuchi}, \citenamefont {Taniuchi}, \citenamefont {Uesaka}, \citenamefont
  {Wang}, \citenamefont {Yoneda}, \citenamefont {Browne}, \citenamefont
  {Chung}, \citenamefont {Dombradi}, \citenamefont {Flavigny}, \citenamefont
  {Giacoppo}, \citenamefont {Gottardo}, \citenamefont
  {Hady{\'{n}}ska-Kl{\c{e}}k}, \citenamefont {Korkulu}, \citenamefont {Koyama},
  \citenamefont {Kubota}, \citenamefont {Lee}, \citenamefont {Lettmann},
  \citenamefont {Louchart}, \citenamefont {Lozeva}, \citenamefont {Matsui},
  \citenamefont {Miyazaki}, \citenamefont {Nishimura}, \citenamefont {Ogata},
  \citenamefont {Ota}, \citenamefont {Patel}, \citenamefont {Sahin},
  \citenamefont {Shand}, \citenamefont {S{\"{o}}derstr{\"{o}}m}, \citenamefont
  {Stefan}, \citenamefont {Steppenbeck}, \citenamefont {Sumikama},
  \citenamefont {Suzuki}, \citenamefont {Werner}, \citenamefont {Wu},\ and\
  \citenamefont {Xu}}]{Olivier2017}%
  \BibitemOpen
  \bibfield  {author} {\bibinfo {author} {\bibfnamefont {L.}~\bibnamefont
  {Olivier}}, \bibinfo {author} {\bibfnamefont {S.}~\bibnamefont {Franchoo}},
  \bibinfo {author} {\bibfnamefont {M.}~\bibnamefont {Niikura}}, \bibinfo
  {author} {\bibfnamefont {Z.}~\bibnamefont {Vajta}}, \bibinfo {author}
  {\bibfnamefont {D.}~\bibnamefont {Sohler}}, \bibinfo {author} {\bibfnamefont
  {P.}~\bibnamefont {Doornenbal}}, \bibinfo {author} {\bibfnamefont
  {A.}~\bibnamefont {Obertelli}}, \bibinfo {author} {\bibfnamefont
  {Y.}~\bibnamefont {Tsunoda}}, \bibinfo {author} {\bibfnamefont
  {T.}~\bibnamefont {Otsuka}}, \bibinfo {author} {\bibfnamefont
  {G.}~\bibnamefont {Authelet}}, \bibinfo {author} {\bibfnamefont
  {H.}~\bibnamefont {Baba}}, \bibinfo {author} {\bibfnamefont {D.}~\bibnamefont
  {Calvet}}, \bibinfo {author} {\bibfnamefont {F.}~\bibnamefont
  {Ch{\^{a}}teau}}, \bibinfo {author} {\bibfnamefont {A.}~\bibnamefont
  {Corsi}}, \bibinfo {author} {\bibfnamefont {A.}~\bibnamefont {Delbart}},
  \bibinfo {author} {\bibfnamefont {J.-M.}\ \bibnamefont {Gheller}}, \bibinfo
  {author} {\bibfnamefont {A.}~\bibnamefont {Gillibert}}, \bibinfo {author}
  {\bibfnamefont {T.}~\bibnamefont {Isobe}}, \bibinfo {author} {\bibfnamefont
  {V.}~\bibnamefont {Lapoux}}, \bibinfo {author} {\bibfnamefont
  {M.}~\bibnamefont {Matsushita}}, \bibinfo {author} {\bibfnamefont
  {S.}~\bibnamefont {Momiyama}}, \bibinfo {author} {\bibfnamefont
  {T.}~\bibnamefont {Motobayashi}}, \bibinfo {author} {\bibfnamefont
  {H.}~\bibnamefont {Otsu}}, \bibinfo {author} {\bibfnamefont {C.}~\bibnamefont
  {P{\'{e}}ron}}, \bibinfo {author} {\bibfnamefont {A.}~\bibnamefont {Peyaud}},
  \bibinfo {author} {\bibfnamefont {E.~C.}\ \bibnamefont {Pollacco}}, \bibinfo
  {author} {\bibfnamefont {J.-Y.}\ \bibnamefont {Rouss{\'{e}}}}, \bibinfo
  {author} {\bibfnamefont {H.}~\bibnamefont {Sakurai}}, \bibinfo {author}
  {\bibfnamefont {C.}~\bibnamefont {Santamaria}}, \bibinfo {author}
  {\bibfnamefont {M.}~\bibnamefont {Sasano}}, \bibinfo {author} {\bibfnamefont
  {Y.}~\bibnamefont {Shiga}}, \bibinfo {author} {\bibfnamefont
  {S.}~\bibnamefont {Takeuchi}}, \bibinfo {author} {\bibfnamefont
  {R.}~\bibnamefont {Taniuchi}}, \bibinfo {author} {\bibfnamefont
  {T.}~\bibnamefont {Uesaka}}, \bibinfo {author} {\bibfnamefont
  {H.}~\bibnamefont {Wang}}, \bibinfo {author} {\bibfnamefont {K.}~\bibnamefont
  {Yoneda}}, \bibinfo {author} {\bibfnamefont {F.}~\bibnamefont {Browne}},
  \bibinfo {author} {\bibfnamefont {L.~X.}\ \bibnamefont {Chung}}, \bibinfo
  {author} {\bibfnamefont {Z.}~\bibnamefont {Dombradi}}, \bibinfo {author}
  {\bibfnamefont {F.}~\bibnamefont {Flavigny}}, \bibinfo {author}
  {\bibfnamefont {F.}~\bibnamefont {Giacoppo}}, \bibinfo {author}
  {\bibfnamefont {A.}~\bibnamefont {Gottardo}}, \bibinfo {author}
  {\bibfnamefont {K.}~\bibnamefont {Hady{\'{n}}ska-Kl{\c{e}}k}}, \bibinfo
  {author} {\bibfnamefont {Z.}~\bibnamefont {Korkulu}}, \bibinfo {author}
  {\bibfnamefont {S.}~\bibnamefont {Koyama}}, \bibinfo {author} {\bibfnamefont
  {Y.}~\bibnamefont {Kubota}}, \bibinfo {author} {\bibfnamefont
  {J.}~\bibnamefont {Lee}}, \bibinfo {author} {\bibfnamefont {M.}~\bibnamefont
  {Lettmann}}, \bibinfo {author} {\bibfnamefont {C.}~\bibnamefont {Louchart}},
  \bibinfo {author} {\bibfnamefont {R.}~\bibnamefont {Lozeva}}, \bibinfo
  {author} {\bibfnamefont {K.}~\bibnamefont {Matsui}}, \bibinfo {author}
  {\bibfnamefont {T.}~\bibnamefont {Miyazaki}}, \bibinfo {author}
  {\bibfnamefont {S.}~\bibnamefont {Nishimura}}, \bibinfo {author}
  {\bibfnamefont {K.}~\bibnamefont {Ogata}}, \bibinfo {author} {\bibfnamefont
  {S.}~\bibnamefont {Ota}}, \bibinfo {author} {\bibfnamefont {Z.}~\bibnamefont
  {Patel}}, \bibinfo {author} {\bibfnamefont {E.}~\bibnamefont {Sahin}},
  \bibinfo {author} {\bibfnamefont {C.}~\bibnamefont {Shand}}, \bibinfo
  {author} {\bibfnamefont {P.-A.}\ \bibnamefont {S{\"{o}}derstr{\"{o}}m}},
  \bibinfo {author} {\bibfnamefont {I.}~\bibnamefont {Stefan}}, \bibinfo
  {author} {\bibfnamefont {D.}~\bibnamefont {Steppenbeck}}, \bibinfo {author}
  {\bibfnamefont {T.}~\bibnamefont {Sumikama}}, \bibinfo {author}
  {\bibfnamefont {D.}~\bibnamefont {Suzuki}}, \bibinfo {author} {\bibfnamefont
  {V.}~\bibnamefont {Werner}}, \bibinfo {author} {\bibfnamefont
  {J.}~\bibnamefont {Wu}}, \ and\ \bibinfo {author} {\bibfnamefont
  {Z.}~\bibnamefont {Xu}},\ }\href {\doibase 10.1103/PhysRevLett.119.192501}
  {\bibfield  {journal} {\bibinfo  {journal} {Physical Review Letters}\
  }\textbf {\bibinfo {volume} {119}},\ \bibinfo {pages} {192501} (\bibinfo
  {year} {2017})}\BibitemShut {NoStop}%
\bibitem [{\citenamefont {Welker}\ \emph {et~al.}(2017)\citenamefont {Welker},
  \citenamefont {Althubiti}, \citenamefont {Atanasov}, \citenamefont {Blaum},
  \citenamefont {Cocolios}, \citenamefont {Herfurth}, \citenamefont {Kreim},
  \citenamefont {Lunney}, \citenamefont {Manea}, \citenamefont {Mougeot},
  \citenamefont {Neidherr}, \citenamefont {Nowacki}, \citenamefont {Poves},
  \citenamefont {Rosenbusch}, \citenamefont {Schweikhard}, \citenamefont
  {Wienholtz}, \citenamefont {Wolf},\ and\ \citenamefont {Zuber}}]{Welker2017}%
  \BibitemOpen
  \bibfield  {author} {\bibinfo {author} {\bibfnamefont {A.}~\bibnamefont
  {Welker}}, \bibinfo {author} {\bibfnamefont {N.~A.~S.}\ \bibnamefont
  {Althubiti}}, \bibinfo {author} {\bibfnamefont {D.}~\bibnamefont {Atanasov}},
  \bibinfo {author} {\bibfnamefont {K.}~\bibnamefont {Blaum}}, \bibinfo
  {author} {\bibfnamefont {T.~E.}\ \bibnamefont {Cocolios}}, \bibinfo {author}
  {\bibfnamefont {F.}~\bibnamefont {Herfurth}}, \bibinfo {author}
  {\bibfnamefont {S.}~\bibnamefont {Kreim}}, \bibinfo {author} {\bibfnamefont
  {D.}~\bibnamefont {Lunney}}, \bibinfo {author} {\bibfnamefont
  {V.}~\bibnamefont {Manea}}, \bibinfo {author} {\bibfnamefont
  {M.}~\bibnamefont {Mougeot}}, \bibinfo {author} {\bibfnamefont
  {D.}~\bibnamefont {Neidherr}}, \bibinfo {author} {\bibfnamefont
  {F.}~\bibnamefont {Nowacki}}, \bibinfo {author} {\bibfnamefont
  {A.}~\bibnamefont {Poves}}, \bibinfo {author} {\bibfnamefont
  {M.}~\bibnamefont {Rosenbusch}}, \bibinfo {author} {\bibfnamefont
  {L.}~\bibnamefont {Schweikhard}}, \bibinfo {author} {\bibfnamefont
  {F.}~\bibnamefont {Wienholtz}}, \bibinfo {author} {\bibfnamefont {R.~N.}\
  \bibnamefont {Wolf}}, \ and\ \bibinfo {author} {\bibfnamefont
  {K.}~\bibnamefont {Zuber}},\ }\href {\doibase 10.1103/PhysRevLett.119.192502}
  {\bibfield  {journal} {\bibinfo  {journal} {Physical Review Letters}\
  }\textbf {\bibinfo {volume} {119}},\ \bibinfo {pages} {192502} (\bibinfo
  {year} {2017})}\BibitemShut {NoStop}%
\bibitem [{\citenamefont {Mueller}\ \emph {et~al.}(2000)\citenamefont
  {Mueller}, \citenamefont {Bruyneel}, \citenamefont {Franchoo}, \citenamefont
  {Huyse}, \citenamefont {Kurpeta}, \citenamefont {Kruglov}, \citenamefont
  {Kudryavtsev}, \citenamefont {Prasad}, \citenamefont {Raabe}, \citenamefont
  {Reusen}, \citenamefont {{Van Duppen}}, \citenamefont {{Van Roosbroeck}},
  \citenamefont {Vermeeren}, \citenamefont {Weissman}, \citenamefont {Janas},
  \citenamefont {Karny}, \citenamefont {Kszczot}, \citenamefont {P{\l}ochocki},
  \citenamefont {Kratz}, \citenamefont {Pfeiffer}, \citenamefont {Grawe},
  \citenamefont {K{\"{o}}ster}, \citenamefont {Thirolf},\ and\ \citenamefont
  {Walters}}]{Mueller2000}%
  \BibitemOpen
  \bibfield  {author} {\bibinfo {author} {\bibfnamefont {W.~F.}\ \bibnamefont
  {Mueller}}, \bibinfo {author} {\bibfnamefont {B.}~\bibnamefont {Bruyneel}},
  \bibinfo {author} {\bibfnamefont {S.}~\bibnamefont {Franchoo}}, \bibinfo
  {author} {\bibfnamefont {M.}~\bibnamefont {Huyse}}, \bibinfo {author}
  {\bibfnamefont {J.}~\bibnamefont {Kurpeta}}, \bibinfo {author} {\bibfnamefont
  {K.}~\bibnamefont {Kruglov}}, \bibinfo {author} {\bibfnamefont
  {Y.}~\bibnamefont {Kudryavtsev}}, \bibinfo {author} {\bibfnamefont {N.~V.
  S.~V.}\ \bibnamefont {Prasad}}, \bibinfo {author} {\bibfnamefont
  {R.}~\bibnamefont {Raabe}}, \bibinfo {author} {\bibfnamefont
  {I.}~\bibnamefont {Reusen}}, \bibinfo {author} {\bibfnamefont
  {P.}~\bibnamefont {{Van Duppen}}}, \bibinfo {author} {\bibfnamefont
  {J.}~\bibnamefont {{Van Roosbroeck}}}, \bibinfo {author} {\bibfnamefont
  {L.}~\bibnamefont {Vermeeren}}, \bibinfo {author} {\bibfnamefont
  {L.}~\bibnamefont {Weissman}}, \bibinfo {author} {\bibfnamefont
  {Z.}~\bibnamefont {Janas}}, \bibinfo {author} {\bibfnamefont
  {M.}~\bibnamefont {Karny}}, \bibinfo {author} {\bibfnamefont
  {T.}~\bibnamefont {Kszczot}}, \bibinfo {author} {\bibfnamefont
  {A.}~\bibnamefont {P{\l}ochocki}}, \bibinfo {author} {\bibfnamefont {K.-L.}\
  \bibnamefont {Kratz}}, \bibinfo {author} {\bibfnamefont {B.}~\bibnamefont
  {Pfeiffer}}, \bibinfo {author} {\bibfnamefont {H.}~\bibnamefont {Grawe}},
  \bibinfo {author} {\bibfnamefont {U.}~\bibnamefont {K{\"{o}}ster}}, \bibinfo
  {author} {\bibfnamefont {P.}~\bibnamefont {Thirolf}}, \ and\ \bibinfo
  {author} {\bibfnamefont {W.~B.}\ \bibnamefont {Walters}},\ }\href {\doibase
  10.1103/PhysRevC.61.054308} {\bibfield  {journal} {\bibinfo  {journal}
  {Physical Review C}\ }\textbf {\bibinfo {volume} {61}},\ \bibinfo {pages}
  {054308} (\bibinfo {year} {2000})}\BibitemShut {NoStop}%
\bibitem [{\citenamefont {Sorlin}\ \emph {et~al.}(2002)\citenamefont {Sorlin},
  \citenamefont {Leenhardt}, \citenamefont {Donzaud}, \citenamefont {Duprat},
  \citenamefont {Azaiez}, \citenamefont {Nowacki}, \citenamefont {Grawe},
  \citenamefont {Dombr{\'{a}}di}, \citenamefont {Amorini}, \citenamefont
  {Astier}, \citenamefont {Baiborodin}, \citenamefont {Belleguic},
  \citenamefont {Borcea}, \citenamefont {Bourgeois}, \citenamefont {Cullen},
  \citenamefont {Dlouhy}, \citenamefont {Dragulescu}, \citenamefont
  {G{\'{o}}rska}, \citenamefont {Gr{\'{e}}vy}, \citenamefont
  {Guillemaud-Mueller}, \citenamefont {Hagemann}, \citenamefont {Herskind},
  \citenamefont {Kiener}, \citenamefont {Lemmon}, \citenamefont {Lewitowicz},
  \citenamefont {Lukyanov}, \citenamefont {Mayet}, \citenamefont {{de Oliveira
  Santos}}, \citenamefont {Pantalica}, \citenamefont {Penionzhkevich},
  \citenamefont {Pougheon}, \citenamefont {Poves}, \citenamefont {Redon},
  \citenamefont {Saint-Laurent}, \citenamefont {Scarpaci}, \citenamefont
  {Sletten}, \citenamefont {Stanoiu}, \citenamefont {Tarasov},\ and\
  \citenamefont {Theisen}}]{Sorlin2002}%
  \BibitemOpen
  \bibfield  {author} {\bibinfo {author} {\bibfnamefont {O.}~\bibnamefont
  {Sorlin}}, \bibinfo {author} {\bibfnamefont {S.}~\bibnamefont {Leenhardt}},
  \bibinfo {author} {\bibfnamefont {C.}~\bibnamefont {Donzaud}}, \bibinfo
  {author} {\bibfnamefont {J.}~\bibnamefont {Duprat}}, \bibinfo {author}
  {\bibfnamefont {F.}~\bibnamefont {Azaiez}}, \bibinfo {author} {\bibfnamefont
  {F.}~\bibnamefont {Nowacki}}, \bibinfo {author} {\bibfnamefont
  {H.}~\bibnamefont {Grawe}}, \bibinfo {author} {\bibfnamefont
  {Z.}~\bibnamefont {Dombr{\'{a}}di}}, \bibinfo {author} {\bibfnamefont
  {F.}~\bibnamefont {Amorini}}, \bibinfo {author} {\bibfnamefont
  {A.}~\bibnamefont {Astier}}, \bibinfo {author} {\bibfnamefont
  {D.}~\bibnamefont {Baiborodin}}, \bibinfo {author} {\bibfnamefont
  {M.}~\bibnamefont {Belleguic}}, \bibinfo {author} {\bibfnamefont
  {C.}~\bibnamefont {Borcea}}, \bibinfo {author} {\bibfnamefont
  {C.}~\bibnamefont {Bourgeois}}, \bibinfo {author} {\bibfnamefont {D.~M.}\
  \bibnamefont {Cullen}}, \bibinfo {author} {\bibfnamefont {Z.}~\bibnamefont
  {Dlouhy}}, \bibinfo {author} {\bibfnamefont {E.}~\bibnamefont {Dragulescu}},
  \bibinfo {author} {\bibfnamefont {M.}~\bibnamefont {G{\'{o}}rska}}, \bibinfo
  {author} {\bibfnamefont {S.}~\bibnamefont {Gr{\'{e}}vy}}, \bibinfo {author}
  {\bibfnamefont {D.}~\bibnamefont {Guillemaud-Mueller}}, \bibinfo {author}
  {\bibfnamefont {G.}~\bibnamefont {Hagemann}}, \bibinfo {author}
  {\bibfnamefont {B.}~\bibnamefont {Herskind}}, \bibinfo {author}
  {\bibfnamefont {J.}~\bibnamefont {Kiener}}, \bibinfo {author} {\bibfnamefont
  {R.}~\bibnamefont {Lemmon}}, \bibinfo {author} {\bibfnamefont
  {M.}~\bibnamefont {Lewitowicz}}, \bibinfo {author} {\bibfnamefont {S.~M.}\
  \bibnamefont {Lukyanov}}, \bibinfo {author} {\bibfnamefont {P.}~\bibnamefont
  {Mayet}}, \bibinfo {author} {\bibfnamefont {F.}~\bibnamefont {{de Oliveira
  Santos}}}, \bibinfo {author} {\bibfnamefont {D.}~\bibnamefont {Pantalica}},
  \bibinfo {author} {\bibfnamefont {Y.-E.}\ \bibnamefont {Penionzhkevich}},
  \bibinfo {author} {\bibfnamefont {F.}~\bibnamefont {Pougheon}}, \bibinfo
  {author} {\bibfnamefont {A.}~\bibnamefont {Poves}}, \bibinfo {author}
  {\bibfnamefont {N.}~\bibnamefont {Redon}}, \bibinfo {author} {\bibfnamefont
  {M.~G.}\ \bibnamefont {Saint-Laurent}}, \bibinfo {author} {\bibfnamefont
  {J.~A.}\ \bibnamefont {Scarpaci}}, \bibinfo {author} {\bibfnamefont
  {G.}~\bibnamefont {Sletten}}, \bibinfo {author} {\bibfnamefont
  {M.}~\bibnamefont {Stanoiu}}, \bibinfo {author} {\bibfnamefont
  {O.}~\bibnamefont {Tarasov}}, \ and\ \bibinfo {author} {\bibfnamefont
  {C.}~\bibnamefont {Theisen}},\ }\href {\doibase
  10.1103/PhysRevLett.88.092501} {\bibfield  {journal} {\bibinfo  {journal}
  {Physical Review Letters}\ }\textbf {\bibinfo {volume} {88}},\ \bibinfo
  {pages} {092501} (\bibinfo {year} {2002})}\BibitemShut {NoStop}%
\bibitem [{\citenamefont {Bree}\ \emph {et~al.}(2008)\citenamefont {Bree},
  \citenamefont {Stefanescu}, \citenamefont {Butler}, \citenamefont
  {Cederk{\"{a}}ll}, \citenamefont {Davinson}, \citenamefont {Delahaye},
  \citenamefont {Eberth}, \citenamefont {Fedorov}, \citenamefont {Fedosseev},
  \citenamefont {Fraile}, \citenamefont {Franchoo}, \citenamefont {Georgiev},
  \citenamefont {Gladnishki}, \citenamefont {Huyse}, \citenamefont {Ivanov},
  \citenamefont {Iwanicki}, \citenamefont {Jolie}, \citenamefont
  {K{\"{o}}ster}, \citenamefont {Kr{\"{o}}ll}, \citenamefont {Kr{\"{u}}cken},
  \citenamefont {Marsh}, \citenamefont {Niedermaier}, \citenamefont {Reiter},
  \citenamefont {Scheit}, \citenamefont {Schwalm}, \citenamefont {Sieber},
  \citenamefont {{Van de Walle}}, \citenamefont {{Van Duppen}}, \citenamefont
  {Warr}, \citenamefont {Weisshaar}, \citenamefont {Wenander},\ and\
  \citenamefont {Zemlyanoy}}]{Bree2008}%
  \BibitemOpen
  \bibfield  {author} {\bibinfo {author} {\bibfnamefont {N.}~\bibnamefont
  {Bree}}, \bibinfo {author} {\bibfnamefont {I.}~\bibnamefont {Stefanescu}},
  \bibinfo {author} {\bibfnamefont {P.~A.}\ \bibnamefont {Butler}}, \bibinfo
  {author} {\bibfnamefont {J.}~\bibnamefont {Cederk{\"{a}}ll}}, \bibinfo
  {author} {\bibfnamefont {T.}~\bibnamefont {Davinson}}, \bibinfo {author}
  {\bibfnamefont {P.}~\bibnamefont {Delahaye}}, \bibinfo {author}
  {\bibfnamefont {J.}~\bibnamefont {Eberth}}, \bibinfo {author} {\bibfnamefont
  {D.}~\bibnamefont {Fedorov}}, \bibinfo {author} {\bibfnamefont {V.~N.}\
  \bibnamefont {Fedosseev}}, \bibinfo {author} {\bibfnamefont {L.~M.}\
  \bibnamefont {Fraile}}, \bibinfo {author} {\bibfnamefont {S.}~\bibnamefont
  {Franchoo}}, \bibinfo {author} {\bibfnamefont {G.}~\bibnamefont {Georgiev}},
  \bibinfo {author} {\bibfnamefont {K.}~\bibnamefont {Gladnishki}}, \bibinfo
  {author} {\bibfnamefont {M.}~\bibnamefont {Huyse}}, \bibinfo {author}
  {\bibfnamefont {O.}~\bibnamefont {Ivanov}}, \bibinfo {author} {\bibfnamefont
  {J.}~\bibnamefont {Iwanicki}}, \bibinfo {author} {\bibfnamefont
  {J.}~\bibnamefont {Jolie}}, \bibinfo {author} {\bibfnamefont
  {U.}~\bibnamefont {K{\"{o}}ster}}, \bibinfo {author} {\bibfnamefont
  {T.}~\bibnamefont {Kr{\"{o}}ll}}, \bibinfo {author} {\bibfnamefont
  {R.}~\bibnamefont {Kr{\"{u}}cken}}, \bibinfo {author} {\bibfnamefont {B.~A.}\
  \bibnamefont {Marsh}}, \bibinfo {author} {\bibfnamefont {O.}~\bibnamefont
  {Niedermaier}}, \bibinfo {author} {\bibfnamefont {P.}~\bibnamefont {Reiter}},
  \bibinfo {author} {\bibfnamefont {H.}~\bibnamefont {Scheit}}, \bibinfo
  {author} {\bibfnamefont {D.}~\bibnamefont {Schwalm}}, \bibinfo {author}
  {\bibfnamefont {T.}~\bibnamefont {Sieber}}, \bibinfo {author} {\bibfnamefont
  {J.}~\bibnamefont {{Van de Walle}}}, \bibinfo {author} {\bibfnamefont
  {P.}~\bibnamefont {{Van Duppen}}}, \bibinfo {author} {\bibfnamefont
  {N.}~\bibnamefont {Warr}}, \bibinfo {author} {\bibfnamefont {D.}~\bibnamefont
  {Weisshaar}}, \bibinfo {author} {\bibfnamefont {F.}~\bibnamefont {Wenander}},
  \ and\ \bibinfo {author} {\bibfnamefont {S.}~\bibnamefont {Zemlyanoy}},\
  }\href {\doibase 10.1103/PhysRevC.78.047301} {\bibfield  {journal} {\bibinfo
  {journal} {Physical Review C}\ }\textbf {\bibinfo {volume} {78}},\ \bibinfo
  {pages} {047301} (\bibinfo {year} {2008})}\BibitemShut {NoStop}%
\bibitem [{\citenamefont {Hannawald}\ \emph {et~al.}(1999)\citenamefont
  {Hannawald}, \citenamefont {Kautzsch}, \citenamefont {W{\"{o}}hr},
  \citenamefont {Walters}, \citenamefont {Kratz}, \citenamefont {Fedoseyev},
  \citenamefont {Mishin}, \citenamefont {B{\"{o}}hmer}, \citenamefont
  {Pfeiffer}, \citenamefont {Sebastian}, \citenamefont {Jading}, \citenamefont
  {K{\"{o}}ster}, \citenamefont {Lettry}, \citenamefont {Ravn},\ and\
  \citenamefont {{the ISOLDE Collaboration}}}]{Hannawald1999}%
  \BibitemOpen
  \bibfield  {author} {\bibinfo {author} {\bibfnamefont {M.}~\bibnamefont
  {Hannawald}}, \bibinfo {author} {\bibfnamefont {T.}~\bibnamefont {Kautzsch}},
  \bibinfo {author} {\bibfnamefont {A.}~\bibnamefont {W{\"{o}}hr}}, \bibinfo
  {author} {\bibfnamefont {W.~B.}\ \bibnamefont {Walters}}, \bibinfo {author}
  {\bibfnamefont {K.-L.}\ \bibnamefont {Kratz}}, \bibinfo {author}
  {\bibfnamefont {V.~N.}\ \bibnamefont {Fedoseyev}}, \bibinfo {author}
  {\bibfnamefont {V.~I.}\ \bibnamefont {Mishin}}, \bibinfo {author}
  {\bibfnamefont {W.}~\bibnamefont {B{\"{o}}hmer}}, \bibinfo {author}
  {\bibfnamefont {B.}~\bibnamefont {Pfeiffer}}, \bibinfo {author}
  {\bibfnamefont {V.}~\bibnamefont {Sebastian}}, \bibinfo {author}
  {\bibfnamefont {Y.}~\bibnamefont {Jading}}, \bibinfo {author} {\bibfnamefont
  {U.}~\bibnamefont {K{\"{o}}ster}}, \bibinfo {author} {\bibfnamefont
  {J.}~\bibnamefont {Lettry}}, \bibinfo {author} {\bibfnamefont {H.~L.}\
  \bibnamefont {Ravn}}, \ and\ \bibinfo {author} {\bibnamefont {{the ISOLDE
  Collaboration}}},\ }\href {\doibase 10.1103/PhysRevLett.82.1391} {\bibfield
  {journal} {\bibinfo  {journal} {Physical Review Letters}\ }\textbf {\bibinfo
  {volume} {82}},\ \bibinfo {pages} {1391} (\bibinfo {year}
  {1999})}\BibitemShut {NoStop}%
\bibitem [{\citenamefont {Gade}\ \emph {et~al.}(2010)\citenamefont {Gade},
  \citenamefont {Janssens}, \citenamefont {Baugher}, \citenamefont {Bazin},
  \citenamefont {Brown}, \citenamefont {Carpenter}, \citenamefont {Chiara},
  \citenamefont {Deacon}, \citenamefont {Freeman}, \citenamefont {Grinyer},
  \citenamefont {Hoffman}, \citenamefont {Kay}, \citenamefont {Kondev},
  \citenamefont {Lauritsen}, \citenamefont {McDaniel}, \citenamefont
  {Meierbachtol}, \citenamefont {Ratkiewicz}, \citenamefont {Stroberg},
  \citenamefont {Walsh}, \citenamefont {Weisshaar}, \citenamefont {Winkler},\
  and\ \citenamefont {Zhu}}]{Gade2010}%
  \BibitemOpen
  \bibfield  {author} {\bibinfo {author} {\bibfnamefont {A.}~\bibnamefont
  {Gade}}, \bibinfo {author} {\bibfnamefont {R.~V.~F.}\ \bibnamefont
  {Janssens}}, \bibinfo {author} {\bibfnamefont {T.}~\bibnamefont {Baugher}},
  \bibinfo {author} {\bibfnamefont {D.}~\bibnamefont {Bazin}}, \bibinfo
  {author} {\bibfnamefont {B.~A.}\ \bibnamefont {Brown}}, \bibinfo {author}
  {\bibfnamefont {M.~P.}\ \bibnamefont {Carpenter}}, \bibinfo {author}
  {\bibfnamefont {C.~J.}\ \bibnamefont {Chiara}}, \bibinfo {author}
  {\bibfnamefont {A.~N.}\ \bibnamefont {Deacon}}, \bibinfo {author}
  {\bibfnamefont {S.~J.}\ \bibnamefont {Freeman}}, \bibinfo {author}
  {\bibfnamefont {G.~F.}\ \bibnamefont {Grinyer}}, \bibinfo {author}
  {\bibfnamefont {C.~R.}\ \bibnamefont {Hoffman}}, \bibinfo {author}
  {\bibfnamefont {B.~P.}\ \bibnamefont {Kay}}, \bibinfo {author} {\bibfnamefont
  {F.~G.}\ \bibnamefont {Kondev}}, \bibinfo {author} {\bibfnamefont
  {T.}~\bibnamefont {Lauritsen}}, \bibinfo {author} {\bibfnamefont
  {S.}~\bibnamefont {McDaniel}}, \bibinfo {author} {\bibfnamefont
  {K.}~\bibnamefont {Meierbachtol}}, \bibinfo {author} {\bibfnamefont
  {A.}~\bibnamefont {Ratkiewicz}}, \bibinfo {author} {\bibfnamefont {S.~R.}\
  \bibnamefont {Stroberg}}, \bibinfo {author} {\bibfnamefont {K.~A.}\
  \bibnamefont {Walsh}}, \bibinfo {author} {\bibfnamefont {D.}~\bibnamefont
  {Weisshaar}}, \bibinfo {author} {\bibfnamefont {R.}~\bibnamefont {Winkler}},
  \ and\ \bibinfo {author} {\bibfnamefont {S.}~\bibnamefont {Zhu}},\ }\href
  {\doibase 10.1103/PhysRevC.81.051304} {\bibfield  {journal} {\bibinfo
  {journal} {Physical Review C}\ }\textbf {\bibinfo {volume} {81}},\ \bibinfo
  {pages} {051304} (\bibinfo {year} {2010})}\BibitemShut {NoStop}%
\bibitem [{\citenamefont {Crawford}\ \emph {et~al.}(2013)\citenamefont
  {Crawford}, \citenamefont {Clark}, \citenamefont {Fallon}, \citenamefont
  {Macchiavelli}, \citenamefont {Baugher}, \citenamefont {Bazin}, \citenamefont
  {Beausang}, \citenamefont {Berryman}, \citenamefont {Bleuel}, \citenamefont
  {Campbell}, \citenamefont {Cromaz}, \citenamefont {de~Angelis}, \citenamefont
  {Gade}, \citenamefont {Hughes}, \citenamefont {Lee}, \citenamefont {Lenzi},
  \citenamefont {Nowacki}, \citenamefont {Paschalis}, \citenamefont {Petri},
  \citenamefont {Poves}, \citenamefont {Ratkiewicz}, \citenamefont {Ross},
  \citenamefont {Sahin}, \citenamefont {Weisshaar}, \citenamefont {Wimmer},\
  and\ \citenamefont {Winkler}}]{Crawford2013}%
  \BibitemOpen
  \bibfield  {author} {\bibinfo {author} {\bibfnamefont {H.~L.}\ \bibnamefont
  {Crawford}}, \bibinfo {author} {\bibfnamefont {R.~M.}\ \bibnamefont {Clark}},
  \bibinfo {author} {\bibfnamefont {P.}~\bibnamefont {Fallon}}, \bibinfo
  {author} {\bibfnamefont {A.~O.}\ \bibnamefont {Macchiavelli}}, \bibinfo
  {author} {\bibfnamefont {T.}~\bibnamefont {Baugher}}, \bibinfo {author}
  {\bibfnamefont {D.}~\bibnamefont {Bazin}}, \bibinfo {author} {\bibfnamefont
  {C.~W.}\ \bibnamefont {Beausang}}, \bibinfo {author} {\bibfnamefont {J.~S.}\
  \bibnamefont {Berryman}}, \bibinfo {author} {\bibfnamefont {D.~L.}\
  \bibnamefont {Bleuel}}, \bibinfo {author} {\bibfnamefont {C.~M.}\
  \bibnamefont {Campbell}}, \bibinfo {author} {\bibfnamefont {M.}~\bibnamefont
  {Cromaz}}, \bibinfo {author} {\bibfnamefont {G.}~\bibnamefont {de~Angelis}},
  \bibinfo {author} {\bibfnamefont {A.}~\bibnamefont {Gade}}, \bibinfo {author}
  {\bibfnamefont {R.~O.}\ \bibnamefont {Hughes}}, \bibinfo {author}
  {\bibfnamefont {I.~Y.}\ \bibnamefont {Lee}}, \bibinfo {author} {\bibfnamefont
  {S.~M.}\ \bibnamefont {Lenzi}}, \bibinfo {author} {\bibfnamefont
  {F.}~\bibnamefont {Nowacki}}, \bibinfo {author} {\bibfnamefont
  {S.}~\bibnamefont {Paschalis}}, \bibinfo {author} {\bibfnamefont
  {M.}~\bibnamefont {Petri}}, \bibinfo {author} {\bibfnamefont
  {A.}~\bibnamefont {Poves}}, \bibinfo {author} {\bibfnamefont
  {A.}~\bibnamefont {Ratkiewicz}}, \bibinfo {author} {\bibfnamefont {T.~J.}\
  \bibnamefont {Ross}}, \bibinfo {author} {\bibfnamefont {E.}~\bibnamefont
  {Sahin}}, \bibinfo {author} {\bibfnamefont {D.}~\bibnamefont {Weisshaar}},
  \bibinfo {author} {\bibfnamefont {K.}~\bibnamefont {Wimmer}}, \ and\ \bibinfo
  {author} {\bibfnamefont {R.}~\bibnamefont {Winkler}},\ }\href {\doibase
  10.1103/PhysRevLett.110.242701} {\bibfield  {journal} {\bibinfo  {journal}
  {Physical Review Letters}\ }\textbf {\bibinfo {volume} {110}},\ \bibinfo
  {pages} {242701} (\bibinfo {year} {2013})}\BibitemShut {NoStop}%
\bibitem [{\citenamefont {{\v{C}}elikovi{\'{c}}}\ \emph
  {et~al.}(2013)\citenamefont {{\v{C}}elikovi{\'{c}}}, \citenamefont {Dijon},
  \citenamefont {Cl{\'{e}}ment}, \citenamefont {de~France}, \citenamefont {{Van
  Isacker}}, \citenamefont {Ljungvall}, \citenamefont {Fransen}, \citenamefont
  {Georgiev}, \citenamefont {G{\"{o}}rgen}, \citenamefont {Gottardo},
  \citenamefont {Hackstein}, \citenamefont {Hagen}, \citenamefont {Louchart},
  \citenamefont {Napiorkowski}, \citenamefont {Obertelli}, \citenamefont
  {Recchia}, \citenamefont {Rother}, \citenamefont {Siem}, \citenamefont
  {Sulignano}, \citenamefont {Uji{\'{c}}}, \citenamefont
  {Valiente-Dob{\'{o}}n},\ and\ \citenamefont
  {Zieli{\'{n}}ska}}]{Celikovic2013}%
  \BibitemOpen
  \bibfield  {author} {\bibinfo {author} {\bibfnamefont {I.}~\bibnamefont
  {{\v{C}}elikovi{\'{c}}}}, \bibinfo {author} {\bibfnamefont {A.}~\bibnamefont
  {Dijon}}, \bibinfo {author} {\bibfnamefont {E.}~\bibnamefont
  {Cl{\'{e}}ment}}, \bibinfo {author} {\bibfnamefont {G.}~\bibnamefont
  {de~France}}, \bibinfo {author} {\bibfnamefont {P.}~\bibnamefont {{Van
  Isacker}}}, \bibinfo {author} {\bibfnamefont {J.}~\bibnamefont {Ljungvall}},
  \bibinfo {author} {\bibfnamefont {C.}~\bibnamefont {Fransen}}, \bibinfo
  {author} {\bibfnamefont {G.}~\bibnamefont {Georgiev}}, \bibinfo {author}
  {\bibfnamefont {A.}~\bibnamefont {G{\"{o}}rgen}}, \bibinfo {author}
  {\bibfnamefont {A.}~\bibnamefont {Gottardo}}, \bibinfo {author}
  {\bibfnamefont {M.}~\bibnamefont {Hackstein}}, \bibinfo {author}
  {\bibfnamefont {T.}~\bibnamefont {Hagen}}, \bibinfo {author} {\bibfnamefont
  {C.}~\bibnamefont {Louchart}}, \bibinfo {author} {\bibfnamefont
  {P.}~\bibnamefont {Napiorkowski}}, \bibinfo {author} {\bibfnamefont
  {A.}~\bibnamefont {Obertelli}}, \bibinfo {author} {\bibfnamefont
  {F.}~\bibnamefont {Recchia}}, \bibinfo {author} {\bibfnamefont
  {W.}~\bibnamefont {Rother}}, \bibinfo {author} {\bibfnamefont
  {S.}~\bibnamefont {Siem}}, \bibinfo {author} {\bibfnamefont {B.}~\bibnamefont
  {Sulignano}}, \bibinfo {author} {\bibfnamefont {P.}~\bibnamefont
  {Uji{\'{c}}}}, \bibinfo {author} {\bibfnamefont {J.}~\bibnamefont
  {Valiente-Dob{\'{o}}n}}, \ and\ \bibinfo {author} {\bibfnamefont
  {M.}~\bibnamefont {Zieli{\'{n}}ska}},\ }\href {\doibase
  10.5506/APhysPolB.44.375} {\bibfield  {journal} {\bibinfo  {journal} {Acta
  Physica Polonica B}\ }\textbf {\bibinfo {volume} {44}},\ \bibinfo {pages}
  {375} (\bibinfo {year} {2013})}\BibitemShut {NoStop}%
\bibitem [{\citenamefont {Louchart}\ \emph {et~al.}(2013)\citenamefont
  {Louchart}, \citenamefont {Obertelli}, \citenamefont {G{\"{o}}rgen},
  \citenamefont {Korten}, \citenamefont {Bazzacco}, \citenamefont {Birkenbach},
  \citenamefont {Bruyneel}, \citenamefont {Cl{\'{e}}ment}, \citenamefont
  {Coleman-Smith}, \citenamefont {Corradi}, \citenamefont {Curien},
  \citenamefont {de~Angelis}, \citenamefont {de~France}, \citenamefont
  {Delaroche}, \citenamefont {Dewald}, \citenamefont {Didierjean},
  \citenamefont {Doncel}, \citenamefont {Duch{\^{e}}ne}, \citenamefont
  {Eberth}, \citenamefont {Erduran}, \citenamefont {Farnea}, \citenamefont
  {Finck}, \citenamefont {Fioretto}, \citenamefont {Fransen}, \citenamefont
  {Gadea}, \citenamefont {Girod}, \citenamefont {Gottardo}, \citenamefont
  {Grebosz}, \citenamefont {Habermann}, \citenamefont {Hackstein},
  \citenamefont {Huyuk}, \citenamefont {Jolie}, \citenamefont {Judson},
  \citenamefont {Jungclaus}, \citenamefont {Karkour}, \citenamefont {Klupp},
  \citenamefont {Kr{\"{u}}cken}, \citenamefont {Kusoglu}, \citenamefont
  {Lenzi}, \citenamefont {Libert}, \citenamefont {Ljungvall}, \citenamefont
  {Lunardi}, \citenamefont {Maron}, \citenamefont {Menegazzo}, \citenamefont
  {Mengoni}, \citenamefont {Michelagnoli}, \citenamefont {Million},
  \citenamefont {Molini}, \citenamefont {M{\"{o}}ller}, \citenamefont
  {Montagnoli}, \citenamefont {Montanari}, \citenamefont {Napoli},
  \citenamefont {Orlandi}, \citenamefont {Pollarolo}, \citenamefont {Prieto},
  \citenamefont {Pullia}, \citenamefont {Quintana}, \citenamefont {Recchia},
  \citenamefont {Reiter}, \citenamefont {Rosso}, \citenamefont {Rother},
  \citenamefont {Sahin}, \citenamefont {Salsac}, \citenamefont {Scarlassara},
  \citenamefont {Schlarb}, \citenamefont {Siem}, \citenamefont {Singh},
  \citenamefont {S{\"{o}}derstr{\"{o}}m}, \citenamefont {Stefanini},
  \citenamefont {St{\'{e}}zowski}, \citenamefont {Sulignano}, \citenamefont
  {Szilner}, \citenamefont {Theisen}, \citenamefont {Ur}, \citenamefont
  {Valiente-Dob{\'{o}}n},\ and\ \citenamefont {Zielinska}}]{Louchart2013}%
  \BibitemOpen
  \bibfield  {author} {\bibinfo {author} {\bibfnamefont {C.}~\bibnamefont
  {Louchart}}, \bibinfo {author} {\bibfnamefont {A.}~\bibnamefont {Obertelli}},
  \bibinfo {author} {\bibfnamefont {A.}~\bibnamefont {G{\"{o}}rgen}}, \bibinfo
  {author} {\bibfnamefont {W.}~\bibnamefont {Korten}}, \bibinfo {author}
  {\bibfnamefont {D.}~\bibnamefont {Bazzacco}}, \bibinfo {author}
  {\bibfnamefont {B.}~\bibnamefont {Birkenbach}}, \bibinfo {author}
  {\bibfnamefont {B.}~\bibnamefont {Bruyneel}}, \bibinfo {author}
  {\bibfnamefont {E.}~\bibnamefont {Cl{\'{e}}ment}}, \bibinfo {author}
  {\bibfnamefont {P.~J.}\ \bibnamefont {Coleman-Smith}}, \bibinfo {author}
  {\bibfnamefont {L.}~\bibnamefont {Corradi}}, \bibinfo {author} {\bibfnamefont
  {D.}~\bibnamefont {Curien}}, \bibinfo {author} {\bibfnamefont
  {G.}~\bibnamefont {de~Angelis}}, \bibinfo {author} {\bibfnamefont
  {G.}~\bibnamefont {de~France}}, \bibinfo {author} {\bibfnamefont {J.-P.}\
  \bibnamefont {Delaroche}}, \bibinfo {author} {\bibfnamefont {A.}~\bibnamefont
  {Dewald}}, \bibinfo {author} {\bibfnamefont {F.}~\bibnamefont {Didierjean}},
  \bibinfo {author} {\bibfnamefont {M.}~\bibnamefont {Doncel}}, \bibinfo
  {author} {\bibfnamefont {G.}~\bibnamefont {Duch{\^{e}}ne}}, \bibinfo {author}
  {\bibfnamefont {J.}~\bibnamefont {Eberth}}, \bibinfo {author} {\bibfnamefont
  {M.~N.}\ \bibnamefont {Erduran}}, \bibinfo {author} {\bibfnamefont
  {E.}~\bibnamefont {Farnea}}, \bibinfo {author} {\bibfnamefont
  {C.}~\bibnamefont {Finck}}, \bibinfo {author} {\bibfnamefont
  {E.}~\bibnamefont {Fioretto}}, \bibinfo {author} {\bibfnamefont
  {C.}~\bibnamefont {Fransen}}, \bibinfo {author} {\bibfnamefont
  {A.}~\bibnamefont {Gadea}}, \bibinfo {author} {\bibfnamefont
  {M.}~\bibnamefont {Girod}}, \bibinfo {author} {\bibfnamefont
  {A.}~\bibnamefont {Gottardo}}, \bibinfo {author} {\bibfnamefont
  {J.}~\bibnamefont {Grebosz}}, \bibinfo {author} {\bibfnamefont
  {T.}~\bibnamefont {Habermann}}, \bibinfo {author} {\bibfnamefont
  {M.}~\bibnamefont {Hackstein}}, \bibinfo {author} {\bibfnamefont
  {T.}~\bibnamefont {Huyuk}}, \bibinfo {author} {\bibfnamefont
  {J.}~\bibnamefont {Jolie}}, \bibinfo {author} {\bibfnamefont
  {D.}~\bibnamefont {Judson}}, \bibinfo {author} {\bibfnamefont
  {A.}~\bibnamefont {Jungclaus}}, \bibinfo {author} {\bibfnamefont
  {N.}~\bibnamefont {Karkour}}, \bibinfo {author} {\bibfnamefont
  {S.}~\bibnamefont {Klupp}}, \bibinfo {author} {\bibfnamefont
  {R.}~\bibnamefont {Kr{\"{u}}cken}}, \bibinfo {author} {\bibfnamefont
  {A.}~\bibnamefont {Kusoglu}}, \bibinfo {author} {\bibfnamefont {S.~M.}\
  \bibnamefont {Lenzi}}, \bibinfo {author} {\bibfnamefont {J.}~\bibnamefont
  {Libert}}, \bibinfo {author} {\bibfnamefont {J.}~\bibnamefont {Ljungvall}},
  \bibinfo {author} {\bibfnamefont {S.}~\bibnamefont {Lunardi}}, \bibinfo
  {author} {\bibfnamefont {G.}~\bibnamefont {Maron}}, \bibinfo {author}
  {\bibfnamefont {R.}~\bibnamefont {Menegazzo}}, \bibinfo {author}
  {\bibfnamefont {D.}~\bibnamefont {Mengoni}}, \bibinfo {author} {\bibfnamefont
  {C.}~\bibnamefont {Michelagnoli}}, \bibinfo {author} {\bibfnamefont
  {B.}~\bibnamefont {Million}}, \bibinfo {author} {\bibfnamefont
  {P.}~\bibnamefont {Molini}}, \bibinfo {author} {\bibfnamefont
  {O.}~\bibnamefont {M{\"{o}}ller}}, \bibinfo {author} {\bibfnamefont
  {G.}~\bibnamefont {Montagnoli}}, \bibinfo {author} {\bibfnamefont
  {D.}~\bibnamefont {Montanari}}, \bibinfo {author} {\bibfnamefont {D.~R.}\
  \bibnamefont {Napoli}}, \bibinfo {author} {\bibfnamefont {R.}~\bibnamefont
  {Orlandi}}, \bibinfo {author} {\bibfnamefont {G.}~\bibnamefont {Pollarolo}},
  \bibinfo {author} {\bibfnamefont {A.}~\bibnamefont {Prieto}}, \bibinfo
  {author} {\bibfnamefont {A.}~\bibnamefont {Pullia}}, \bibinfo {author}
  {\bibfnamefont {B.}~\bibnamefont {Quintana}}, \bibinfo {author}
  {\bibfnamefont {F.}~\bibnamefont {Recchia}}, \bibinfo {author} {\bibfnamefont
  {P.}~\bibnamefont {Reiter}}, \bibinfo {author} {\bibfnamefont
  {D.}~\bibnamefont {Rosso}}, \bibinfo {author} {\bibfnamefont
  {W.}~\bibnamefont {Rother}}, \bibinfo {author} {\bibfnamefont
  {E.}~\bibnamefont {Sahin}}, \bibinfo {author} {\bibfnamefont {M.-D.}\
  \bibnamefont {Salsac}}, \bibinfo {author} {\bibfnamefont {F.}~\bibnamefont
  {Scarlassara}}, \bibinfo {author} {\bibfnamefont {M.}~\bibnamefont
  {Schlarb}}, \bibinfo {author} {\bibfnamefont {S.}~\bibnamefont {Siem}},
  \bibinfo {author} {\bibfnamefont {P.~P.}\ \bibnamefont {Singh}}, \bibinfo
  {author} {\bibfnamefont {P.-A.}\ \bibnamefont {S{\"{o}}derstr{\"{o}}m}},
  \bibinfo {author} {\bibfnamefont {A.~M.}\ \bibnamefont {Stefanini}}, \bibinfo
  {author} {\bibfnamefont {O.}~\bibnamefont {St{\'{e}}zowski}}, \bibinfo
  {author} {\bibfnamefont {B.}~\bibnamefont {Sulignano}}, \bibinfo {author}
  {\bibfnamefont {S.}~\bibnamefont {Szilner}}, \bibinfo {author} {\bibfnamefont
  {C.}~\bibnamefont {Theisen}}, \bibinfo {author} {\bibfnamefont {C.~A.}\
  \bibnamefont {Ur}}, \bibinfo {author} {\bibfnamefont {J.~J.}\ \bibnamefont
  {Valiente-Dob{\'{o}}n}}, \ and\ \bibinfo {author} {\bibfnamefont
  {M.}~\bibnamefont {Zielinska}},\ }\href {\doibase 10.1103/PhysRevC.87.054302}
  {\bibfield  {journal} {\bibinfo  {journal} {Physical Review C}\ }\textbf
  {\bibinfo {volume} {87}},\ \bibinfo {pages} {054302} (\bibinfo {year}
  {2013})}\BibitemShut {NoStop}%
\bibitem [{\citenamefont {Suchyta}\ \emph {et~al.}(2014)\citenamefont
  {Suchyta}, \citenamefont {Liddick}, \citenamefont {Chiara}, \citenamefont
  {Walters}, \citenamefont {Carpenter}, \citenamefont {Crawford}, \citenamefont
  {Grinyer}, \citenamefont {G{\"{u}}rdal}, \citenamefont {Klose}, \citenamefont
  {McCutchan}, \citenamefont {Pereira},\ and\ \citenamefont
  {Zhu}}]{Suchyta2014}%
  \BibitemOpen
  \bibfield  {author} {\bibinfo {author} {\bibfnamefont {S.}~\bibnamefont
  {Suchyta}}, \bibinfo {author} {\bibfnamefont {S.~N.}\ \bibnamefont
  {Liddick}}, \bibinfo {author} {\bibfnamefont {C.~J.}\ \bibnamefont {Chiara}},
  \bibinfo {author} {\bibfnamefont {W.~B.}\ \bibnamefont {Walters}}, \bibinfo
  {author} {\bibfnamefont {M.~P.}\ \bibnamefont {Carpenter}}, \bibinfo {author}
  {\bibfnamefont {H.~L.}\ \bibnamefont {Crawford}}, \bibinfo {author}
  {\bibfnamefont {G.~F.}\ \bibnamefont {Grinyer}}, \bibinfo {author}
  {\bibfnamefont {G.}~\bibnamefont {G{\"{u}}rdal}}, \bibinfo {author}
  {\bibfnamefont {A.}~\bibnamefont {Klose}}, \bibinfo {author} {\bibfnamefont
  {E.~A.}\ \bibnamefont {McCutchan}}, \bibinfo {author} {\bibfnamefont
  {J.}~\bibnamefont {Pereira}}, \ and\ \bibinfo {author} {\bibfnamefont
  {S.}~\bibnamefont {Zhu}},\ }\href {\doibase 10.1103/PhysRevC.89.067303}
  {\bibfield  {journal} {\bibinfo  {journal} {Physical Review C}\ }\textbf
  {\bibinfo {volume} {89}},\ \bibinfo {pages} {067303} (\bibinfo {year}
  {2014})}\BibitemShut {NoStop}%
\bibitem [{\citenamefont {Lenzi}\ \emph {et~al.}(2010)\citenamefont {Lenzi},
  \citenamefont {Nowacki}, \citenamefont {Poves},\ and\ \citenamefont
  {Sieja}}]{Lenzi2010}%
  \BibitemOpen
  \bibfield  {author} {\bibinfo {author} {\bibfnamefont {S.~M.}\ \bibnamefont
  {Lenzi}}, \bibinfo {author} {\bibfnamefont {F.}~\bibnamefont {Nowacki}},
  \bibinfo {author} {\bibfnamefont {A.}~\bibnamefont {Poves}}, \ and\ \bibinfo
  {author} {\bibfnamefont {K.}~\bibnamefont {Sieja}},\ }\href {\doibase
  10.1103/PhysRevC.82.054301} {\bibfield  {journal} {\bibinfo  {journal}
  {Physical Review C}\ }\textbf {\bibinfo {volume} {82}},\ \bibinfo {pages}
  {054301} (\bibinfo {year} {2010})}\BibitemShut {NoStop}%
\bibitem [{\citenamefont {Tsunoda}\ \emph {et~al.}(2014)\citenamefont
  {Tsunoda}, \citenamefont {Otsuka}, \citenamefont {Shimizu}, \citenamefont
  {Honma},\ and\ \citenamefont {Utsuno}}]{Tsunoda2014}%
  \BibitemOpen
  \bibfield  {author} {\bibinfo {author} {\bibfnamefont {Y.}~\bibnamefont
  {Tsunoda}}, \bibinfo {author} {\bibfnamefont {T.}~\bibnamefont {Otsuka}},
  \bibinfo {author} {\bibfnamefont {N.}~\bibnamefont {Shimizu}}, \bibinfo
  {author} {\bibfnamefont {M.}~\bibnamefont {Honma}}, \ and\ \bibinfo {author}
  {\bibfnamefont {Y.}~\bibnamefont {Utsuno}},\ }\href {\doibase
  10.1103/PhysRevC.89.031301} {\bibfield  {journal} {\bibinfo  {journal}
  {Physical Review C}\ }\textbf {\bibinfo {volume} {89}},\ \bibinfo {pages}
  {031301} (\bibinfo {year} {2014})}\BibitemShut {NoStop}%
\bibitem [{\citenamefont {Santamaria}\ \emph {et~al.}(2015)\citenamefont
  {Santamaria}, \citenamefont {Louchart}, \citenamefont {Obertelli},
  \citenamefont {Werner}, \citenamefont {Doornenbal}, \citenamefont {Nowacki},
  \citenamefont {Authelet}, \citenamefont {Baba}, \citenamefont {Calvet},
  \citenamefont {Ch{\^{a}}teau}, \citenamefont {Corsi}, \citenamefont
  {Delbart}, \citenamefont {Gheller}, \citenamefont {Gillibert}, \citenamefont
  {Isobe}, \citenamefont {Lapoux}, \citenamefont {Matsushita}, \citenamefont
  {Momiyama}, \citenamefont {Motobayashi}, \citenamefont {Niikura},
  \citenamefont {Otsu}, \citenamefont {P{\'{e}}ron}, \citenamefont {Peyaud},
  \citenamefont {Pollacco}, \citenamefont {Rouss{\'{e}}}, \citenamefont
  {Sakurai}, \citenamefont {Sasano}, \citenamefont {Shiga}, \citenamefont
  {Takeuchi}, \citenamefont {Taniuchi}, \citenamefont {Uesaka}, \citenamefont
  {Wang}, \citenamefont {Yoneda}, \citenamefont {Browne}, \citenamefont
  {Chung}, \citenamefont {Dombradi}, \citenamefont {Franchoo}, \citenamefont
  {Giacoppo}, \citenamefont {Gottardo}, \citenamefont {Hadynska-Klek},
  \citenamefont {Korkulu}, \citenamefont {Koyama}, \citenamefont {Kubota},
  \citenamefont {Lee}, \citenamefont {Lettmann}, \citenamefont {Lozeva},
  \citenamefont {Matsui}, \citenamefont {Miyazaki}, \citenamefont {Nishimura},
  \citenamefont {Olivier}, \citenamefont {Ota}, \citenamefont {Patel},
  \citenamefont {Pietralla}, \citenamefont {Sahin}, \citenamefont {Shand},
  \citenamefont {S{\"{o}}derstr{\"{o}}m}, \citenamefont {Stefan}, \citenamefont
  {Steppenbeck}, \citenamefont {Sumikama}, \citenamefont {Suzuki},
  \citenamefont {Vajta}, \citenamefont {Wu},\ and\ \citenamefont
  {Xu}}]{Santamaria2015}%
  \BibitemOpen
  \bibfield  {author} {\bibinfo {author} {\bibfnamefont {C.}~\bibnamefont
  {Santamaria}}, \bibinfo {author} {\bibfnamefont {C.}~\bibnamefont
  {Louchart}}, \bibinfo {author} {\bibfnamefont {A.}~\bibnamefont {Obertelli}},
  \bibinfo {author} {\bibfnamefont {V.}~\bibnamefont {Werner}}, \bibinfo
  {author} {\bibfnamefont {P.}~\bibnamefont {Doornenbal}}, \bibinfo {author}
  {\bibfnamefont {F.}~\bibnamefont {Nowacki}}, \bibinfo {author} {\bibfnamefont
  {G.}~\bibnamefont {Authelet}}, \bibinfo {author} {\bibfnamefont
  {H.}~\bibnamefont {Baba}}, \bibinfo {author} {\bibfnamefont {D.}~\bibnamefont
  {Calvet}}, \bibinfo {author} {\bibfnamefont {F.}~\bibnamefont
  {Ch{\^{a}}teau}}, \bibinfo {author} {\bibfnamefont {A.}~\bibnamefont
  {Corsi}}, \bibinfo {author} {\bibfnamefont {A.}~\bibnamefont {Delbart}},
  \bibinfo {author} {\bibfnamefont {J.-M.}\ \bibnamefont {Gheller}}, \bibinfo
  {author} {\bibfnamefont {A.}~\bibnamefont {Gillibert}}, \bibinfo {author}
  {\bibfnamefont {T.}~\bibnamefont {Isobe}}, \bibinfo {author} {\bibfnamefont
  {V.}~\bibnamefont {Lapoux}}, \bibinfo {author} {\bibfnamefont
  {M.}~\bibnamefont {Matsushita}}, \bibinfo {author} {\bibfnamefont
  {S.}~\bibnamefont {Momiyama}}, \bibinfo {author} {\bibfnamefont
  {T.}~\bibnamefont {Motobayashi}}, \bibinfo {author} {\bibfnamefont
  {M.}~\bibnamefont {Niikura}}, \bibinfo {author} {\bibfnamefont
  {H.}~\bibnamefont {Otsu}}, \bibinfo {author} {\bibfnamefont {C.}~\bibnamefont
  {P{\'{e}}ron}}, \bibinfo {author} {\bibfnamefont {A.}~\bibnamefont {Peyaud}},
  \bibinfo {author} {\bibfnamefont {E.~C.}\ \bibnamefont {Pollacco}}, \bibinfo
  {author} {\bibfnamefont {J.-Y.}\ \bibnamefont {Rouss{\'{e}}}}, \bibinfo
  {author} {\bibfnamefont {H.}~\bibnamefont {Sakurai}}, \bibinfo {author}
  {\bibfnamefont {M.}~\bibnamefont {Sasano}}, \bibinfo {author} {\bibfnamefont
  {Y.}~\bibnamefont {Shiga}}, \bibinfo {author} {\bibfnamefont
  {S.}~\bibnamefont {Takeuchi}}, \bibinfo {author} {\bibfnamefont
  {R.}~\bibnamefont {Taniuchi}}, \bibinfo {author} {\bibfnamefont
  {T.}~\bibnamefont {Uesaka}}, \bibinfo {author} {\bibfnamefont
  {H.}~\bibnamefont {Wang}}, \bibinfo {author} {\bibfnamefont {K.}~\bibnamefont
  {Yoneda}}, \bibinfo {author} {\bibfnamefont {F.}~\bibnamefont {Browne}},
  \bibinfo {author} {\bibfnamefont {L.~X.}\ \bibnamefont {Chung}}, \bibinfo
  {author} {\bibfnamefont {Z.}~\bibnamefont {Dombradi}}, \bibinfo {author}
  {\bibfnamefont {S.}~\bibnamefont {Franchoo}}, \bibinfo {author}
  {\bibfnamefont {F.}~\bibnamefont {Giacoppo}}, \bibinfo {author}
  {\bibfnamefont {A.}~\bibnamefont {Gottardo}}, \bibinfo {author}
  {\bibfnamefont {K.}~\bibnamefont {Hadynska-Klek}}, \bibinfo {author}
  {\bibfnamefont {Z.}~\bibnamefont {Korkulu}}, \bibinfo {author} {\bibfnamefont
  {S.}~\bibnamefont {Koyama}}, \bibinfo {author} {\bibfnamefont
  {Y.}~\bibnamefont {Kubota}}, \bibinfo {author} {\bibfnamefont
  {J.}~\bibnamefont {Lee}}, \bibinfo {author} {\bibfnamefont {M.}~\bibnamefont
  {Lettmann}}, \bibinfo {author} {\bibfnamefont {R.}~\bibnamefont {Lozeva}},
  \bibinfo {author} {\bibfnamefont {K.}~\bibnamefont {Matsui}}, \bibinfo
  {author} {\bibfnamefont {T.}~\bibnamefont {Miyazaki}}, \bibinfo {author}
  {\bibfnamefont {S.}~\bibnamefont {Nishimura}}, \bibinfo {author}
  {\bibfnamefont {L.}~\bibnamefont {Olivier}}, \bibinfo {author} {\bibfnamefont
  {S.}~\bibnamefont {Ota}}, \bibinfo {author} {\bibfnamefont {Z.}~\bibnamefont
  {Patel}}, \bibinfo {author} {\bibfnamefont {N.}~\bibnamefont {Pietralla}},
  \bibinfo {author} {\bibfnamefont {E.}~\bibnamefont {Sahin}}, \bibinfo
  {author} {\bibfnamefont {C.}~\bibnamefont {Shand}}, \bibinfo {author}
  {\bibfnamefont {P.-A.}\ \bibnamefont {S{\"{o}}derstr{\"{o}}m}}, \bibinfo
  {author} {\bibfnamefont {I.}~\bibnamefont {Stefan}}, \bibinfo {author}
  {\bibfnamefont {D.}~\bibnamefont {Steppenbeck}}, \bibinfo {author}
  {\bibfnamefont {T.}~\bibnamefont {Sumikama}}, \bibinfo {author}
  {\bibfnamefont {D.}~\bibnamefont {Suzuki}}, \bibinfo {author} {\bibfnamefont
  {Z.}~\bibnamefont {Vajta}}, \bibinfo {author} {\bibfnamefont
  {J.}~\bibnamefont {Wu}}, \ and\ \bibinfo {author} {\bibfnamefont
  {Z.}~\bibnamefont {Xu}},\ }\href {\doibase 10.1103/PhysRevLett.115.192501}
  {\bibfield  {journal} {\bibinfo  {journal} {Physical Review Letters}\
  }\textbf {\bibinfo {volume} {115}},\ \bibinfo {pages} {192501} (\bibinfo
  {year} {2015})}\BibitemShut {NoStop}%
\bibitem [{\citenamefont {Togashi}\ \emph {et~al.}(2015)\citenamefont
  {Togashi}, \citenamefont {Shimizu}, \citenamefont {Utsuno}, \citenamefont
  {Otsuka},\ and\ \citenamefont {Honma}}]{Togashi2015}%
  \BibitemOpen
  \bibfield  {author} {\bibinfo {author} {\bibfnamefont {T.}~\bibnamefont
  {Togashi}}, \bibinfo {author} {\bibfnamefont {N.}~\bibnamefont {Shimizu}},
  \bibinfo {author} {\bibfnamefont {Y.}~\bibnamefont {Utsuno}}, \bibinfo
  {author} {\bibfnamefont {T.}~\bibnamefont {Otsuka}}, \ and\ \bibinfo {author}
  {\bibfnamefont {M.}~\bibnamefont {Honma}},\ }\href {\doibase
  10.1103/PhysRevC.91.024320} {\bibfield  {journal} {\bibinfo  {journal}
  {Physical Review C}\ }\textbf {\bibinfo {volume} {91}},\ \bibinfo {pages}
  {024320} (\bibinfo {year} {2015})}\BibitemShut {NoStop}%
\bibitem [{\citenamefont {Mougeot}\ \emph {et~al.}(2018)\citenamefont
  {Mougeot}, \citenamefont {Atanasov}, \citenamefont {Blaum}, \citenamefont
  {Chrysalidis}, \citenamefont {Goodacre}, \citenamefont {Fedorov},
  \citenamefont {Fedosseev}, \citenamefont {George}, \citenamefont {Herfurth},
  \citenamefont {Holt}, \citenamefont {Lunney}, \citenamefont {Manea},
  \citenamefont {Marsh}, \citenamefont {Neidherr}, \citenamefont {Rosenbusch},
  \citenamefont {Rothe}, \citenamefont {Schweikhard}, \citenamefont {Schwenk},
  \citenamefont {Seiffert}, \citenamefont {Simonis}, \citenamefont {Stroberg},
  \citenamefont {Welker}, \citenamefont {Wienholtz}, \citenamefont {Wolf},\
  and\ \citenamefont {Zuber}}]{Mougeot2018}%
  \BibitemOpen
  \bibfield  {author} {\bibinfo {author} {\bibfnamefont {M.}~\bibnamefont
  {Mougeot}}, \bibinfo {author} {\bibfnamefont {D.}~\bibnamefont {Atanasov}},
  \bibinfo {author} {\bibfnamefont {K.}~\bibnamefont {Blaum}}, \bibinfo
  {author} {\bibfnamefont {K.}~\bibnamefont {Chrysalidis}}, \bibinfo {author}
  {\bibfnamefont {T.~D.}\ \bibnamefont {Goodacre}}, \bibinfo {author}
  {\bibfnamefont {D.}~\bibnamefont {Fedorov}}, \bibinfo {author} {\bibfnamefont
  {V.}~\bibnamefont {Fedosseev}}, \bibinfo {author} {\bibfnamefont
  {S.}~\bibnamefont {George}}, \bibinfo {author} {\bibfnamefont
  {F.}~\bibnamefont {Herfurth}}, \bibinfo {author} {\bibfnamefont {J.~D.}\
  \bibnamefont {Holt}}, \bibinfo {author} {\bibfnamefont {D.}~\bibnamefont
  {Lunney}}, \bibinfo {author} {\bibfnamefont {V.}~\bibnamefont {Manea}},
  \bibinfo {author} {\bibfnamefont {B.}~\bibnamefont {Marsh}}, \bibinfo
  {author} {\bibfnamefont {D.}~\bibnamefont {Neidherr}}, \bibinfo {author}
  {\bibfnamefont {M.}~\bibnamefont {Rosenbusch}}, \bibinfo {author}
  {\bibfnamefont {S.}~\bibnamefont {Rothe}}, \bibinfo {author} {\bibfnamefont
  {L.}~\bibnamefont {Schweikhard}}, \bibinfo {author} {\bibfnamefont
  {A.}~\bibnamefont {Schwenk}}, \bibinfo {author} {\bibfnamefont
  {C.}~\bibnamefont {Seiffert}}, \bibinfo {author} {\bibfnamefont
  {J.}~\bibnamefont {Simonis}}, \bibinfo {author} {\bibfnamefont {S.~R.}\
  \bibnamefont {Stroberg}}, \bibinfo {author} {\bibfnamefont {A.}~\bibnamefont
  {Welker}}, \bibinfo {author} {\bibfnamefont {F.}~\bibnamefont {Wienholtz}},
  \bibinfo {author} {\bibfnamefont {R.~N.}\ \bibnamefont {Wolf}}, \ and\
  \bibinfo {author} {\bibfnamefont {K.}~\bibnamefont {Zuber}},\ }\href
  {\doibase 10.1103/PhysRevLett.120.232501} {\bibfield  {journal} {\bibinfo
  {journal} {Physical Review Letters}\ }\textbf {\bibinfo {volume} {120}},\
  \bibinfo {pages} {232501} (\bibinfo {year} {2018})}\BibitemShut {NoStop}%
\bibitem [{\citenamefont {Otsuka}\ \emph {et~al.}(2006)\citenamefont {Otsuka},
  \citenamefont {Matsuo},\ and\ \citenamefont {Abe}}]{Otsuka2006}%
  \BibitemOpen
  \bibfield  {author} {\bibinfo {author} {\bibfnamefont {T.}~\bibnamefont
  {Otsuka}}, \bibinfo {author} {\bibfnamefont {T.}~\bibnamefont {Matsuo}}, \
  and\ \bibinfo {author} {\bibfnamefont {D.}~\bibnamefont {Abe}},\ }\href
  {\doibase 10.1103/PhysRevLett.97.162501} {\bibfield  {journal} {\bibinfo
  {journal} {Physical Review Letters}\ }\textbf {\bibinfo {volume} {97}},\
  \bibinfo {pages} {162501} (\bibinfo {year} {2006})}\BibitemShut {NoStop}%
\bibitem [{\citenamefont {Otsuka}(2013)}]{Otsuka2013}%
  \BibitemOpen
  \bibfield  {author} {\bibinfo {author} {\bibfnamefont {T.}~\bibnamefont
  {Otsuka}},\ }\href {\doibase 10.1088/0031-8949/2013/T152/014007} {\bibfield
  {journal} {\bibinfo  {journal} {Physica Scripta}\ }\textbf {\bibinfo {volume}
  {T152}},\ \bibinfo {pages} {014007} (\bibinfo {year} {2013})}\BibitemShut
  {NoStop}%
\bibitem [{\citenamefont {Bastin}\ \emph {et~al.}(2007)\citenamefont {Bastin},
  \citenamefont {Gr{\'{e}}vy}, \citenamefont {Sohler}, \citenamefont {Sorlin},
  \citenamefont {Dombr{\'{a}}di}, \citenamefont {Achouri}, \citenamefont
  {Ang{\'{e}}lique}, \citenamefont {Azaiez}, \citenamefont {Baiborodin},
  \citenamefont {Borcea}, \citenamefont {Bourgeois}, \citenamefont {Buta},
  \citenamefont {B{\"{u}}rger}, \citenamefont {Chapman}, \citenamefont
  {Dalouzy}, \citenamefont {Dlouhy}, \citenamefont {Drouard}, \citenamefont
  {Elekes}, \citenamefont {Franchoo}, \citenamefont {Iacob}, \citenamefont
  {Laurent}, \citenamefont {Lazar}, \citenamefont {Liang}, \citenamefont
  {Li{\'{e}}nard}, \citenamefont {Mrazek}, \citenamefont {Nalpas},
  \citenamefont {Negoita}, \citenamefont {Orr}, \citenamefont {Penionzhkevich},
  \citenamefont {Podoly{\'{a}}k}, \citenamefont {Pougheon}, \citenamefont
  {Roussel-Chomaz}, \citenamefont {Saint-Laurent}, \citenamefont {Stanoiu},
  \citenamefont {Stefan}, \citenamefont {Nowacki},\ and\ \citenamefont
  {Poves}}]{Bastin2007}%
  \BibitemOpen
  \bibfield  {author} {\bibinfo {author} {\bibfnamefont {B.}~\bibnamefont
  {Bastin}}, \bibinfo {author} {\bibfnamefont {S.}~\bibnamefont {Gr{\'{e}}vy}},
  \bibinfo {author} {\bibfnamefont {D.}~\bibnamefont {Sohler}}, \bibinfo
  {author} {\bibfnamefont {O.}~\bibnamefont {Sorlin}}, \bibinfo {author}
  {\bibfnamefont {Z.}~\bibnamefont {Dombr{\'{a}}di}}, \bibinfo {author}
  {\bibfnamefont {N.~L.}\ \bibnamefont {Achouri}}, \bibinfo {author}
  {\bibfnamefont {J.~C.}\ \bibnamefont {Ang{\'{e}}lique}}, \bibinfo {author}
  {\bibfnamefont {F.}~\bibnamefont {Azaiez}}, \bibinfo {author} {\bibfnamefont
  {D.}~\bibnamefont {Baiborodin}}, \bibinfo {author} {\bibfnamefont
  {R.}~\bibnamefont {Borcea}}, \bibinfo {author} {\bibfnamefont
  {C.}~\bibnamefont {Bourgeois}}, \bibinfo {author} {\bibfnamefont
  {A.}~\bibnamefont {Buta}}, \bibinfo {author} {\bibfnamefont {A.}~\bibnamefont
  {B{\"{u}}rger}}, \bibinfo {author} {\bibfnamefont {R.}~\bibnamefont
  {Chapman}}, \bibinfo {author} {\bibfnamefont {J.~C.}\ \bibnamefont
  {Dalouzy}}, \bibinfo {author} {\bibfnamefont {Z.}~\bibnamefont {Dlouhy}},
  \bibinfo {author} {\bibfnamefont {A.}~\bibnamefont {Drouard}}, \bibinfo
  {author} {\bibfnamefont {Z.}~\bibnamefont {Elekes}}, \bibinfo {author}
  {\bibfnamefont {S.}~\bibnamefont {Franchoo}}, \bibinfo {author}
  {\bibfnamefont {S.}~\bibnamefont {Iacob}}, \bibinfo {author} {\bibfnamefont
  {B.}~\bibnamefont {Laurent}}, \bibinfo {author} {\bibfnamefont
  {M.}~\bibnamefont {Lazar}}, \bibinfo {author} {\bibfnamefont
  {X.}~\bibnamefont {Liang}}, \bibinfo {author} {\bibfnamefont
  {E.}~\bibnamefont {Li{\'{e}}nard}}, \bibinfo {author} {\bibfnamefont
  {J.}~\bibnamefont {Mrazek}}, \bibinfo {author} {\bibfnamefont
  {L.}~\bibnamefont {Nalpas}}, \bibinfo {author} {\bibfnamefont
  {F.}~\bibnamefont {Negoita}}, \bibinfo {author} {\bibfnamefont {N.~A.}\
  \bibnamefont {Orr}}, \bibinfo {author} {\bibfnamefont {Y.}~\bibnamefont
  {Penionzhkevich}}, \bibinfo {author} {\bibfnamefont {Z.}~\bibnamefont
  {Podoly{\'{a}}k}}, \bibinfo {author} {\bibfnamefont {F.}~\bibnamefont
  {Pougheon}}, \bibinfo {author} {\bibfnamefont {P.}~\bibnamefont
  {Roussel-Chomaz}}, \bibinfo {author} {\bibfnamefont {M.~G.}\ \bibnamefont
  {Saint-Laurent}}, \bibinfo {author} {\bibfnamefont {M.}~\bibnamefont
  {Stanoiu}}, \bibinfo {author} {\bibfnamefont {I.}~\bibnamefont {Stefan}},
  \bibinfo {author} {\bibfnamefont {F.}~\bibnamefont {Nowacki}}, \ and\
  \bibinfo {author} {\bibfnamefont {A.}~\bibnamefont {Poves}},\ }\href
  {\doibase 10.1103/PhysRevLett.99.022503} {\bibfield  {journal} {\bibinfo
  {journal} {Physical Review Letters}\ }\textbf {\bibinfo {volume} {99}},\
  \bibinfo {pages} {022503} (\bibinfo {year} {2007})}\BibitemShut {NoStop}%
\bibitem [{\citenamefont {Steppenbeck}\ \emph {et~al.}(2013)\citenamefont
  {Steppenbeck}, \citenamefont {Takeuchi}, \citenamefont {Aoi}, \citenamefont
  {Doornenbal}, \citenamefont {Matsushita}, \citenamefont {Wang}, \citenamefont
  {Baba}, \citenamefont {Fukuda}, \citenamefont {Go}, \citenamefont {Honma},
  \citenamefont {Lee}, \citenamefont {Matsui}, \citenamefont {Michimasa},
  \citenamefont {Motobayashi}, \citenamefont {Nishimura}, \citenamefont
  {Otsuka}, \citenamefont {Sakurai}, \citenamefont {Shiga}, \citenamefont
  {S{\"{o}}derstr{\"{o}}m}, \citenamefont {Sumikama}, \citenamefont {Suzuki},
  \citenamefont {Taniuchi}, \citenamefont {Utsuno}, \citenamefont
  {Valiente-Dob{\'{o}}n},\ and\ \citenamefont {Yoneda}}]{Steppenbeck2013}%
  \BibitemOpen
  \bibfield  {author} {\bibinfo {author} {\bibfnamefont {D.}~\bibnamefont
  {Steppenbeck}}, \bibinfo {author} {\bibfnamefont {S.}~\bibnamefont
  {Takeuchi}}, \bibinfo {author} {\bibfnamefont {N.}~\bibnamefont {Aoi}},
  \bibinfo {author} {\bibfnamefont {P.}~\bibnamefont {Doornenbal}}, \bibinfo
  {author} {\bibfnamefont {M.}~\bibnamefont {Matsushita}}, \bibinfo {author}
  {\bibfnamefont {H.}~\bibnamefont {Wang}}, \bibinfo {author} {\bibfnamefont
  {H.}~\bibnamefont {Baba}}, \bibinfo {author} {\bibfnamefont {N.}~\bibnamefont
  {Fukuda}}, \bibinfo {author} {\bibfnamefont {S.}~\bibnamefont {Go}}, \bibinfo
  {author} {\bibfnamefont {M.}~\bibnamefont {Honma}}, \bibinfo {author}
  {\bibfnamefont {J.}~\bibnamefont {Lee}}, \bibinfo {author} {\bibfnamefont
  {K.}~\bibnamefont {Matsui}}, \bibinfo {author} {\bibfnamefont
  {S.}~\bibnamefont {Michimasa}}, \bibinfo {author} {\bibfnamefont
  {T.}~\bibnamefont {Motobayashi}}, \bibinfo {author} {\bibfnamefont
  {D.}~\bibnamefont {Nishimura}}, \bibinfo {author} {\bibfnamefont
  {T.}~\bibnamefont {Otsuka}}, \bibinfo {author} {\bibfnamefont
  {H.}~\bibnamefont {Sakurai}}, \bibinfo {author} {\bibfnamefont
  {Y.}~\bibnamefont {Shiga}}, \bibinfo {author} {\bibfnamefont {P.-A.}\
  \bibnamefont {S{\"{o}}derstr{\"{o}}m}}, \bibinfo {author} {\bibfnamefont
  {T.}~\bibnamefont {Sumikama}}, \bibinfo {author} {\bibfnamefont
  {H.}~\bibnamefont {Suzuki}}, \bibinfo {author} {\bibfnamefont
  {R.}~\bibnamefont {Taniuchi}}, \bibinfo {author} {\bibfnamefont
  {Y.}~\bibnamefont {Utsuno}}, \bibinfo {author} {\bibfnamefont {J.~J.}\
  \bibnamefont {Valiente-Dob{\'{o}}n}}, \ and\ \bibinfo {author} {\bibfnamefont
  {K.}~\bibnamefont {Yoneda}},\ }\href {\doibase 10.1038/nature12522}
  {\bibfield  {journal} {\bibinfo  {journal} {Nature}\ }\textbf {\bibinfo
  {volume} {502}},\ \bibinfo {pages} {207} (\bibinfo {year}
  {2013})}\BibitemShut {NoStop}%
\bibitem [{\citenamefont {Utsuno}\ \emph {et~al.}(2014)\citenamefont {Utsuno},
  \citenamefont {Otsuka}, \citenamefont {Shimizu}, \citenamefont {Honma},
  \citenamefont {Mizusaki}, \citenamefont {Tsunoda},\ and\ \citenamefont
  {Abe}}]{Utsuno2014}%
  \BibitemOpen
  \bibfield  {author} {\bibinfo {author} {\bibfnamefont {Y.}~\bibnamefont
  {Utsuno}}, \bibinfo {author} {\bibfnamefont {T.}~\bibnamefont {Otsuka}},
  \bibinfo {author} {\bibfnamefont {N.}~\bibnamefont {Shimizu}}, \bibinfo
  {author} {\bibfnamefont {M.}~\bibnamefont {Honma}}, \bibinfo {author}
  {\bibfnamefont {T.}~\bibnamefont {Mizusaki}}, \bibinfo {author}
  {\bibfnamefont {Y.}~\bibnamefont {Tsunoda}}, \ and\ \bibinfo {author}
  {\bibfnamefont {T.}~\bibnamefont {Abe}},\ }\href {\doibase
  10.1051/epjconf/20146602106} {\bibfield  {journal} {\bibinfo  {journal} {EPJ
  Web of Conferences}\ }\textbf {\bibinfo {volume} {66}},\ \bibinfo {pages}
  {02106} (\bibinfo {year} {2014})}\BibitemShut {NoStop}%
\bibitem [{\citenamefont {Franchoo}\ \emph {et~al.}(1998)\citenamefont
  {Franchoo}, \citenamefont {Huyse}, \citenamefont {Kruglov}, \citenamefont
  {Kudryavtsev}, \citenamefont {Mueller}, \citenamefont {Raabe}, \citenamefont
  {Reusen}, \citenamefont {{Van Duppen}}, \citenamefont {{Van Roosbroeck}},
  \citenamefont {Vermeeren}, \citenamefont {W{\"{o}}hr}, \citenamefont {Kratz},
  \citenamefont {Pfeiffer},\ and\ \citenamefont {Walters}}]{Franchoo1998}%
  \BibitemOpen
  \bibfield  {author} {\bibinfo {author} {\bibfnamefont {S.}~\bibnamefont
  {Franchoo}}, \bibinfo {author} {\bibfnamefont {M.}~\bibnamefont {Huyse}},
  \bibinfo {author} {\bibfnamefont {K.}~\bibnamefont {Kruglov}}, \bibinfo
  {author} {\bibfnamefont {Y.}~\bibnamefont {Kudryavtsev}}, \bibinfo {author}
  {\bibfnamefont {W.~F.}\ \bibnamefont {Mueller}}, \bibinfo {author}
  {\bibfnamefont {R.}~\bibnamefont {Raabe}}, \bibinfo {author} {\bibfnamefont
  {I.}~\bibnamefont {Reusen}}, \bibinfo {author} {\bibfnamefont
  {P.}~\bibnamefont {{Van Duppen}}}, \bibinfo {author} {\bibfnamefont
  {J.}~\bibnamefont {{Van Roosbroeck}}}, \bibinfo {author} {\bibfnamefont
  {L.}~\bibnamefont {Vermeeren}}, \bibinfo {author} {\bibfnamefont
  {A.}~\bibnamefont {W{\"{o}}hr}}, \bibinfo {author} {\bibfnamefont {K.-L.}\
  \bibnamefont {Kratz}}, \bibinfo {author} {\bibfnamefont {B.}~\bibnamefont
  {Pfeiffer}}, \ and\ \bibinfo {author} {\bibfnamefont {W.~B.}\ \bibnamefont
  {Walters}},\ }\href {\doibase 10.1103/PhysRevLett.81.3100} {\bibfield
  {journal} {\bibinfo  {journal} {Physical Review Letters}\ }\textbf {\bibinfo
  {volume} {81}},\ \bibinfo {pages} {3100} (\bibinfo {year}
  {1998})}\BibitemShut {NoStop}%
\bibitem [{\citenamefont {Flanagan}\ \emph {et~al.}(2009)\citenamefont
  {Flanagan}, \citenamefont {Vingerhoets}, \citenamefont {Avgoulea},
  \citenamefont {Billowes}, \citenamefont {Bissell}, \citenamefont {Blaum},
  \citenamefont {Cheal}, \citenamefont {{De Rydt}}, \citenamefont {Fedosseev},
  \citenamefont {Forest}, \citenamefont {Geppert}, \citenamefont
  {K{\"{o}}ster}, \citenamefont {Kowalska}, \citenamefont {Kr{\"{a}}mer},
  \citenamefont {Kratz}, \citenamefont {Krieger}, \citenamefont {Man{\'{e}}},
  \citenamefont {Marsh}, \citenamefont {Materna}, \citenamefont {Mathieu},
  \citenamefont {Molkanov}, \citenamefont {Neugart}, \citenamefont {Neyens},
  \citenamefont {N{\"{o}}rtersh{\"{a}}user}, \citenamefont {Seliverstov},
  \citenamefont {Serot}, \citenamefont {Schug}, \citenamefont {Sjoedin},
  \citenamefont {Stone}, \citenamefont {Stone}, \citenamefont {Stroke},
  \citenamefont {Tungate}, \citenamefont {Yordanov},\ and\ \citenamefont
  {Volkov}}]{Flanagan2009}%
  \BibitemOpen
  \bibfield  {author} {\bibinfo {author} {\bibfnamefont {K.~T.}\ \bibnamefont
  {Flanagan}}, \bibinfo {author} {\bibfnamefont {P.}~\bibnamefont
  {Vingerhoets}}, \bibinfo {author} {\bibfnamefont {M.}~\bibnamefont
  {Avgoulea}}, \bibinfo {author} {\bibfnamefont {J.}~\bibnamefont {Billowes}},
  \bibinfo {author} {\bibfnamefont {M.~L.}\ \bibnamefont {Bissell}}, \bibinfo
  {author} {\bibfnamefont {K.}~\bibnamefont {Blaum}}, \bibinfo {author}
  {\bibfnamefont {B.}~\bibnamefont {Cheal}}, \bibinfo {author} {\bibfnamefont
  {M.}~\bibnamefont {{De Rydt}}}, \bibinfo {author} {\bibfnamefont {V.~N.}\
  \bibnamefont {Fedosseev}}, \bibinfo {author} {\bibfnamefont {D.~H.}\
  \bibnamefont {Forest}}, \bibinfo {author} {\bibfnamefont {C.}~\bibnamefont
  {Geppert}}, \bibinfo {author} {\bibfnamefont {U.}~\bibnamefont
  {K{\"{o}}ster}}, \bibinfo {author} {\bibfnamefont {M.}~\bibnamefont
  {Kowalska}}, \bibinfo {author} {\bibfnamefont {J.}~\bibnamefont
  {Kr{\"{a}}mer}}, \bibinfo {author} {\bibfnamefont {K.~L.}\ \bibnamefont
  {Kratz}}, \bibinfo {author} {\bibfnamefont {A.}~\bibnamefont {Krieger}},
  \bibinfo {author} {\bibfnamefont {E.}~\bibnamefont {Man{\'{e}}}}, \bibinfo
  {author} {\bibfnamefont {B.~A.}\ \bibnamefont {Marsh}}, \bibinfo {author}
  {\bibfnamefont {T.}~\bibnamefont {Materna}}, \bibinfo {author} {\bibfnamefont
  {L.}~\bibnamefont {Mathieu}}, \bibinfo {author} {\bibfnamefont {P.~L.}\
  \bibnamefont {Molkanov}}, \bibinfo {author} {\bibfnamefont {R.}~\bibnamefont
  {Neugart}}, \bibinfo {author} {\bibfnamefont {G.}~\bibnamefont {Neyens}},
  \bibinfo {author} {\bibfnamefont {W.}~\bibnamefont
  {N{\"{o}}rtersh{\"{a}}user}}, \bibinfo {author} {\bibfnamefont {M.~D.}\
  \bibnamefont {Seliverstov}}, \bibinfo {author} {\bibfnamefont
  {O.}~\bibnamefont {Serot}}, \bibinfo {author} {\bibfnamefont
  {M.}~\bibnamefont {Schug}}, \bibinfo {author} {\bibfnamefont {M.~A.}\
  \bibnamefont {Sjoedin}}, \bibinfo {author} {\bibfnamefont {J.~R.}\
  \bibnamefont {Stone}}, \bibinfo {author} {\bibfnamefont {N.~J.}\ \bibnamefont
  {Stone}}, \bibinfo {author} {\bibfnamefont {H.~H.}\ \bibnamefont {Stroke}},
  \bibinfo {author} {\bibfnamefont {G.}~\bibnamefont {Tungate}}, \bibinfo
  {author} {\bibfnamefont {D.~T.}\ \bibnamefont {Yordanov}}, \ and\ \bibinfo
  {author} {\bibfnamefont {Y.~M.}\ \bibnamefont {Volkov}},\ }\href {\doibase
  10.1103/PhysRevLett.103.142501} {\bibfield  {journal} {\bibinfo  {journal}
  {Physical Review Letters}\ }\textbf {\bibinfo {volume} {103}},\ \bibinfo
  {pages} {142501} (\bibinfo {year} {2009})}\BibitemShut {NoStop}%
\bibitem [{\citenamefont {de~Groote}\ \emph {et~al.}(2017)\citenamefont
  {de~Groote}, \citenamefont {Billowes}, \citenamefont {Binnersley},
  \citenamefont {Bissell}, \citenamefont {Cocolios}, \citenamefont {{Day
  Goodacre}}, \citenamefont {Farooq-Smith}, \citenamefont {Fedorov},
  \citenamefont {Flanagan}, \citenamefont {Franchoo}, \citenamefont {{Garcia
  Ruiz}}, \citenamefont {Koszor{\'{u}}s}, \citenamefont {Lynch}, \citenamefont
  {Neyens}, \citenamefont {Nowacki}, \citenamefont {Otsuka}, \citenamefont
  {Rothe}, \citenamefont {Stroke}, \citenamefont {Tsunoda}, \citenamefont
  {Vernon}, \citenamefont {Wendt}, \citenamefont {Wilkins}, \citenamefont
  {Xu},\ and\ \citenamefont {Yang}}]{DeGroote2017}%
  \BibitemOpen
  \bibfield  {author} {\bibinfo {author} {\bibfnamefont {R.~P.}\ \bibnamefont
  {de~Groote}}, \bibinfo {author} {\bibfnamefont {J.}~\bibnamefont {Billowes}},
  \bibinfo {author} {\bibfnamefont {C.~L.}\ \bibnamefont {Binnersley}},
  \bibinfo {author} {\bibfnamefont {M.~L.}\ \bibnamefont {Bissell}}, \bibinfo
  {author} {\bibfnamefont {T.~E.}\ \bibnamefont {Cocolios}}, \bibinfo {author}
  {\bibfnamefont {T.}~\bibnamefont {{Day Goodacre}}}, \bibinfo {author}
  {\bibfnamefont {G.~J.}\ \bibnamefont {Farooq-Smith}}, \bibinfo {author}
  {\bibfnamefont {D.~V.}\ \bibnamefont {Fedorov}}, \bibinfo {author}
  {\bibfnamefont {K.~T.}\ \bibnamefont {Flanagan}}, \bibinfo {author}
  {\bibfnamefont {S.}~\bibnamefont {Franchoo}}, \bibinfo {author}
  {\bibfnamefont {R.~F.}\ \bibnamefont {{Garcia Ruiz}}}, \bibinfo {author}
  {\bibfnamefont {{\'{A}}.}~\bibnamefont {Koszor{\'{u}}s}}, \bibinfo {author}
  {\bibfnamefont {K.~M.}\ \bibnamefont {Lynch}}, \bibinfo {author}
  {\bibfnamefont {G.}~\bibnamefont {Neyens}}, \bibinfo {author} {\bibfnamefont
  {F.}~\bibnamefont {Nowacki}}, \bibinfo {author} {\bibfnamefont
  {T.}~\bibnamefont {Otsuka}}, \bibinfo {author} {\bibfnamefont
  {S.}~\bibnamefont {Rothe}}, \bibinfo {author} {\bibfnamefont {H.~H.}\
  \bibnamefont {Stroke}}, \bibinfo {author} {\bibfnamefont {Y.}~\bibnamefont
  {Tsunoda}}, \bibinfo {author} {\bibfnamefont {A.~R.}\ \bibnamefont {Vernon}},
  \bibinfo {author} {\bibfnamefont {K.~D.~A.}\ \bibnamefont {Wendt}}, \bibinfo
  {author} {\bibfnamefont {S.~G.}\ \bibnamefont {Wilkins}}, \bibinfo {author}
  {\bibfnamefont {Z.~Y.}\ \bibnamefont {Xu}}, \ and\ \bibinfo {author}
  {\bibfnamefont {X.~F.}\ \bibnamefont {Yang}},\ }\href {\doibase
  10.1103/PhysRevC.96.041302} {\bibfield  {journal} {\bibinfo  {journal}
  {Physical Review C}\ }\textbf {\bibinfo {volume} {96}},\ \bibinfo {pages}
  {041302} (\bibinfo {year} {2017})}\BibitemShut {NoStop}%
\bibitem [{\citenamefont {Walters}\ \emph {et~al.}(2015)\citenamefont
  {Walters}, \citenamefont {Chiara}, \citenamefont {Janssens}, \citenamefont
  {Weisshaar}, \citenamefont {Otsuka}, \citenamefont {Tsunoda}, \citenamefont
  {Recchia}, \citenamefont {Gade}, \citenamefont {Harker}, \citenamefont
  {Albers}, \citenamefont {Alcorta}, \citenamefont {Bader}, \citenamefont
  {Baugher}, \citenamefont {Bazin}, \citenamefont {Berryman}, \citenamefont
  {Bertone}, \citenamefont {Campbell}, \citenamefont {Carpenter}, \citenamefont
  {Chen}, \citenamefont {Crawford}, \citenamefont {David}, \citenamefont
  {Doherty}, \citenamefont {Hoffman}, \citenamefont {Honma}, \citenamefont
  {Kondev}, \citenamefont {Korichi}, \citenamefont {Langer}, \citenamefont
  {Larson}, \citenamefont {Lauritsen}, \citenamefont {Liddick}, \citenamefont
  {Lunderberg}, \citenamefont {Macchiavelli}, \citenamefont {Noji},
  \citenamefont {Prokop}, \citenamefont {Rogers}, \citenamefont {Seweryniak},
  \citenamefont {Shimizu}, \citenamefont {Stroberg}, \citenamefont {Suchyta},
  \citenamefont {Utsuno}, \citenamefont {Williams}, \citenamefont {Wimmer},\
  and\ \citenamefont {Zhu}}]{Walters2015}%
  \BibitemOpen
  \bibfield  {author} {\bibinfo {author} {\bibfnamefont {W.~B.}\ \bibnamefont
  {Walters}}, \bibinfo {author} {\bibfnamefont {C.~J.}\ \bibnamefont {Chiara}},
  \bibinfo {author} {\bibfnamefont {R.~V.~F.}\ \bibnamefont {Janssens}},
  \bibinfo {author} {\bibfnamefont {D.}~\bibnamefont {Weisshaar}}, \bibinfo
  {author} {\bibfnamefont {T.}~\bibnamefont {Otsuka}}, \bibinfo {author}
  {\bibfnamefont {Y.}~\bibnamefont {Tsunoda}}, \bibinfo {author} {\bibfnamefont
  {F.}~\bibnamefont {Recchia}}, \bibinfo {author} {\bibfnamefont
  {A.}~\bibnamefont {Gade}}, \bibinfo {author} {\bibfnamefont {J.~L.}\
  \bibnamefont {Harker}}, \bibinfo {author} {\bibfnamefont {M.}~\bibnamefont
  {Albers}}, \bibinfo {author} {\bibfnamefont {M.}~\bibnamefont {Alcorta}},
  \bibinfo {author} {\bibfnamefont {V.~M.}\ \bibnamefont {Bader}}, \bibinfo
  {author} {\bibfnamefont {T.}~\bibnamefont {Baugher}}, \bibinfo {author}
  {\bibfnamefont {D.}~\bibnamefont {Bazin}}, \bibinfo {author} {\bibfnamefont
  {J.~S.}\ \bibnamefont {Berryman}}, \bibinfo {author} {\bibfnamefont {P.~F.}\
  \bibnamefont {Bertone}}, \bibinfo {author} {\bibfnamefont {C.~M.}\
  \bibnamefont {Campbell}}, \bibinfo {author} {\bibfnamefont {M.~P.}\
  \bibnamefont {Carpenter}}, \bibinfo {author} {\bibfnamefont {J.}~\bibnamefont
  {Chen}}, \bibinfo {author} {\bibfnamefont {H.~L.}\ \bibnamefont {Crawford}},
  \bibinfo {author} {\bibfnamefont {H.~M.}\ \bibnamefont {David}}, \bibinfo
  {author} {\bibfnamefont {D.~T.}\ \bibnamefont {Doherty}}, \bibinfo {author}
  {\bibfnamefont {C.~R.}\ \bibnamefont {Hoffman}}, \bibinfo {author}
  {\bibfnamefont {M.}~\bibnamefont {Honma}}, \bibinfo {author} {\bibfnamefont
  {F.~G.}\ \bibnamefont {Kondev}}, \bibinfo {author} {\bibfnamefont
  {A.}~\bibnamefont {Korichi}}, \bibinfo {author} {\bibfnamefont
  {C.}~\bibnamefont {Langer}}, \bibinfo {author} {\bibfnamefont
  {N.}~\bibnamefont {Larson}}, \bibinfo {author} {\bibfnamefont
  {T.}~\bibnamefont {Lauritsen}}, \bibinfo {author} {\bibfnamefont {S.~N.}\
  \bibnamefont {Liddick}}, \bibinfo {author} {\bibfnamefont {E.}~\bibnamefont
  {Lunderberg}}, \bibinfo {author} {\bibfnamefont {A.~O.}\ \bibnamefont
  {Macchiavelli}}, \bibinfo {author} {\bibfnamefont {S.}~\bibnamefont {Noji}},
  \bibinfo {author} {\bibfnamefont {C.}~\bibnamefont {Prokop}}, \bibinfo
  {author} {\bibfnamefont {A.~M.}\ \bibnamefont {Rogers}}, \bibinfo {author}
  {\bibfnamefont {D.}~\bibnamefont {Seweryniak}}, \bibinfo {author}
  {\bibfnamefont {N.}~\bibnamefont {Shimizu}}, \bibinfo {author} {\bibfnamefont
  {S.~R.}\ \bibnamefont {Stroberg}}, \bibinfo {author} {\bibfnamefont
  {S.}~\bibnamefont {Suchyta}}, \bibinfo {author} {\bibfnamefont
  {Y.}~\bibnamefont {Utsuno}}, \bibinfo {author} {\bibfnamefont {S.~J.}\
  \bibnamefont {Williams}}, \bibinfo {author} {\bibfnamefont {K.}~\bibnamefont
  {Wimmer}}, \ and\ \bibinfo {author} {\bibfnamefont {S.}~\bibnamefont {Zhu}},\
  }in\ \href {\doibase 10.1063/1.4932251} {\emph {\bibinfo {booktitle} {AIP
  Conference Proceedings}}},\ Vol.\ \bibinfo {volume} {1681}\ (\bibinfo {year}
  {2015})\ p.\ \bibinfo {pages} {030007}\BibitemShut {NoStop}%
\bibitem [{\citenamefont {Otsuka}\ and\ \citenamefont
  {Tsunoda}(2016)}]{Otsuka2016}%
  \BibitemOpen
  \bibfield  {author} {\bibinfo {author} {\bibfnamefont {T.}~\bibnamefont
  {Otsuka}}\ and\ \bibinfo {author} {\bibfnamefont {Y.}~\bibnamefont
  {Tsunoda}},\ }\href {\doibase 10.1088/0954-3899/43/2/024009} {\bibfield
  {journal} {\bibinfo  {journal} {Journal of Physics G: Nuclear and Particle
  Physics}\ }\textbf {\bibinfo {volume} {43}},\ \bibinfo {pages} {024009}
  (\bibinfo {year} {2016})}\BibitemShut {NoStop}%
\bibitem [{\citenamefont {Leoni}\ \emph {et~al.}(2017)\citenamefont {Leoni},
  \citenamefont {Fornal}, \citenamefont {Mărginean}, \citenamefont
  {Sferrazza}, \citenamefont {Tsunoda}, \citenamefont {Otsuka}, \citenamefont
  {Bocchi}, \citenamefont {Crespi}, \citenamefont {Bracco}, \citenamefont
  {Aydin}, \citenamefont {Boromiza}, \citenamefont {Bucurescu}, \citenamefont
  {Cieplicka-Oryǹczak}, \citenamefont {Costache}, \citenamefont {Călinescu},
  \citenamefont {Florea}, \citenamefont {Ghiţă}, \citenamefont {Glodariu},
  \citenamefont {Ionescu}, \citenamefont {Iskra}, \citenamefont {Krzysiek},
  \citenamefont {Mărginean}, \citenamefont {Mihai}, \citenamefont {Mihai},
  \citenamefont {Mitu}, \citenamefont {Negreţ}, \citenamefont {Niţă},
  \citenamefont {Olăcel}, \citenamefont {Oprea}, \citenamefont {Pascu},
  \citenamefont {Petkov}, \citenamefont {Petrone}, \citenamefont {Porzio},
  \citenamefont {Şerban}, \citenamefont {Sotty}, \citenamefont {Stan},
  \citenamefont {Ştiru}, \citenamefont {Stroe}, \citenamefont {Şuvăilă},
  \citenamefont {Toma}, \citenamefont {Turturică}, \citenamefont {Ujeniuc},\
  and\ \citenamefont {Ur}}]{Leoni2017}%
  \BibitemOpen
  \bibfield  {author} {\bibinfo {author} {\bibfnamefont {S.}~\bibnamefont
  {Leoni}}, \bibinfo {author} {\bibfnamefont {B.}~\bibnamefont {Fornal}},
  \bibinfo {author} {\bibfnamefont {N.}~\bibnamefont {Mărginean}}, \bibinfo
  {author} {\bibfnamefont {M.}~\bibnamefont {Sferrazza}}, \bibinfo {author}
  {\bibfnamefont {Y.}~\bibnamefont {Tsunoda}}, \bibinfo {author} {\bibfnamefont
  {T.}~\bibnamefont {Otsuka}}, \bibinfo {author} {\bibfnamefont
  {G.}~\bibnamefont {Bocchi}}, \bibinfo {author} {\bibfnamefont {F.~C.~L.}\
  \bibnamefont {Crespi}}, \bibinfo {author} {\bibfnamefont {A.}~\bibnamefont
  {Bracco}}, \bibinfo {author} {\bibfnamefont {S.}~\bibnamefont {Aydin}},
  \bibinfo {author} {\bibfnamefont {M.}~\bibnamefont {Boromiza}}, \bibinfo
  {author} {\bibfnamefont {D.}~\bibnamefont {Bucurescu}}, \bibinfo {author}
  {\bibfnamefont {N.}~\bibnamefont {Cieplicka-Oryǹczak}}, \bibinfo {author}
  {\bibfnamefont {C.}~\bibnamefont {Costache}}, \bibinfo {author}
  {\bibfnamefont {S.}~\bibnamefont {Călinescu}}, \bibinfo {author}
  {\bibfnamefont {N.}~\bibnamefont {Florea}}, \bibinfo {author} {\bibfnamefont
  {D.~G.}\ \bibnamefont {Ghiţă}}, \bibinfo {author} {\bibfnamefont
  {T.}~\bibnamefont {Glodariu}}, \bibinfo {author} {\bibfnamefont
  {A.}~\bibnamefont {Ionescu}}, \bibinfo {author} {\bibfnamefont
  {{\L}.}~\bibnamefont {Iskra}}, \bibinfo {author} {\bibfnamefont
  {M.}~\bibnamefont {Krzysiek}}, \bibinfo {author} {\bibfnamefont
  {R.}~\bibnamefont {Mărginean}}, \bibinfo {author} {\bibfnamefont
  {C.}~\bibnamefont {Mihai}}, \bibinfo {author} {\bibfnamefont {R.~E.}\
  \bibnamefont {Mihai}}, \bibinfo {author} {\bibfnamefont {A.}~\bibnamefont
  {Mitu}}, \bibinfo {author} {\bibfnamefont {A.}~\bibnamefont {Negreţ}},
  \bibinfo {author} {\bibfnamefont {C.~R.}\ \bibnamefont {Niţă}}, \bibinfo
  {author} {\bibfnamefont {A.}~\bibnamefont {Olăcel}}, \bibinfo {author}
  {\bibfnamefont {A.}~\bibnamefont {Oprea}}, \bibinfo {author} {\bibfnamefont
  {S.}~\bibnamefont {Pascu}}, \bibinfo {author} {\bibfnamefont
  {P.}~\bibnamefont {Petkov}}, \bibinfo {author} {\bibfnamefont
  {C.}~\bibnamefont {Petrone}}, \bibinfo {author} {\bibfnamefont
  {G.}~\bibnamefont {Porzio}}, \bibinfo {author} {\bibfnamefont
  {A.}~\bibnamefont {Şerban}}, \bibinfo {author} {\bibfnamefont
  {C.}~\bibnamefont {Sotty}}, \bibinfo {author} {\bibfnamefont
  {L.}~\bibnamefont {Stan}}, \bibinfo {author} {\bibfnamefont {I.}~\bibnamefont
  {Ştiru}}, \bibinfo {author} {\bibfnamefont {L.}~\bibnamefont {Stroe}},
  \bibinfo {author} {\bibfnamefont {R.}~\bibnamefont {Şuvăilă}}, \bibinfo
  {author} {\bibfnamefont {S.}~\bibnamefont {Toma}}, \bibinfo {author}
  {\bibfnamefont {A.}~\bibnamefont {Turturică}}, \bibinfo {author}
  {\bibfnamefont {S.}~\bibnamefont {Ujeniuc}}, \ and\ \bibinfo {author}
  {\bibfnamefont {C.~A.}\ \bibnamefont {Ur}},\ }\href {\doibase
  10.1103/PhysRevLett.118.162502} {\bibfield  {journal} {\bibinfo  {journal}
  {Physical Review Letters}\ }\textbf {\bibinfo {volume} {118}},\ \bibinfo
  {pages} {162502} (\bibinfo {year} {2017})}\BibitemShut {NoStop}%
\bibitem [{\citenamefont {Girod}\ \emph {et~al.}(1988)\citenamefont {Girod},
  \citenamefont {Delaroche},\ and\ \citenamefont {Berger}}]{Girod1988}%
  \BibitemOpen
  \bibfield  {author} {\bibinfo {author} {\bibfnamefont {M.}~\bibnamefont
  {Girod}}, \bibinfo {author} {\bibfnamefont {J.~P.}\ \bibnamefont
  {Delaroche}}, \ and\ \bibinfo {author} {\bibfnamefont {J.~F.}\ \bibnamefont
  {Berger}},\ }\href {\doibase 10.1103/PhysRevC.38.1519} {\bibfield  {journal}
  {\bibinfo  {journal} {Physical Review C}\ }\textbf {\bibinfo {volume} {38}},\
  \bibinfo {pages} {1519} (\bibinfo {year} {1988})}\BibitemShut {NoStop}%
\bibitem [{\citenamefont {Girod}\ \emph {et~al.}(1989)\citenamefont {Girod},
  \citenamefont {Delaroche}, \citenamefont {Gogny},\ and\ \citenamefont
  {Berger}}]{Girod1989}%
  \BibitemOpen
  \bibfield  {author} {\bibinfo {author} {\bibfnamefont {M.}~\bibnamefont
  {Girod}}, \bibinfo {author} {\bibfnamefont {J.~P.}\ \bibnamefont
  {Delaroche}}, \bibinfo {author} {\bibfnamefont {D.}~\bibnamefont {Gogny}}, \
  and\ \bibinfo {author} {\bibfnamefont {J.~F.}\ \bibnamefont {Berger}},\
  }\href {\doibase 10.1103/PhysRevLett.62.2452} {\bibfield  {journal} {\bibinfo
   {journal} {Physical Review Letters}\ }\textbf {\bibinfo {volume} {62}},\
  \bibinfo {pages} {2452} (\bibinfo {year} {1989})}\BibitemShut {NoStop}%
\bibitem [{\citenamefont {Bonche}\ \emph {et~al.}(1989)\citenamefont {Bonche},
  \citenamefont {Krieger}, \citenamefont {Quentin}, \citenamefont {Weiss},
  \citenamefont {Meyer}, \citenamefont {Meyer}, \citenamefont {Redon},
  \citenamefont {Flocard},\ and\ \citenamefont {Heenen}}]{Bonche1989}%
  \BibitemOpen
  \bibfield  {author} {\bibinfo {author} {\bibfnamefont {P.}~\bibnamefont
  {Bonche}}, \bibinfo {author} {\bibfnamefont {S.}~\bibnamefont {Krieger}},
  \bibinfo {author} {\bibfnamefont {P.}~\bibnamefont {Quentin}}, \bibinfo
  {author} {\bibfnamefont {M.}~\bibnamefont {Weiss}}, \bibinfo {author}
  {\bibfnamefont {J.}~\bibnamefont {Meyer}}, \bibinfo {author} {\bibfnamefont
  {M.}~\bibnamefont {Meyer}}, \bibinfo {author} {\bibfnamefont
  {N.}~\bibnamefont {Redon}}, \bibinfo {author} {\bibfnamefont
  {H.}~\bibnamefont {Flocard}}, \ and\ \bibinfo {author} {\bibfnamefont
  {P.-H.}\ \bibnamefont {Heenen}},\ }\href {\doibase
  10.1016/0375-9474(89)90426-0} {\bibfield  {journal} {\bibinfo  {journal}
  {Nuclear Physics A}\ }\textbf {\bibinfo {volume} {500}},\ \bibinfo {pages}
  {308} (\bibinfo {year} {1989})}\BibitemShut {NoStop}%
\bibitem [{\citenamefont {M{\"{o}}ller}\ \emph {et~al.}(2009)\citenamefont
  {M{\"{o}}ller}, \citenamefont {Sierk}, \citenamefont {Bengtsson},
  \citenamefont {Sagawa},\ and\ \citenamefont {Ichikawa}}]{Moller2009}%
  \BibitemOpen
  \bibfield  {author} {\bibinfo {author} {\bibfnamefont {P.}~\bibnamefont
  {M{\"{o}}ller}}, \bibinfo {author} {\bibfnamefont {A.~J.}\ \bibnamefont
  {Sierk}}, \bibinfo {author} {\bibfnamefont {R.}~\bibnamefont {Bengtsson}},
  \bibinfo {author} {\bibfnamefont {H.}~\bibnamefont {Sagawa}}, \ and\ \bibinfo
  {author} {\bibfnamefont {T.}~\bibnamefont {Ichikawa}},\ }\href {\doibase
  10.1103/PhysRevLett.103.212501} {\bibfield  {journal} {\bibinfo  {journal}
  {Physical Review Letters}\ }\textbf {\bibinfo {volume} {103}},\ \bibinfo
  {pages} {212501} (\bibinfo {year} {2009})}\BibitemShut {NoStop}%
\bibitem [{\citenamefont {Grzywacz}\ \emph {et~al.}(1998)\citenamefont
  {Grzywacz}, \citenamefont {B{\'{e}}raud}, \citenamefont {Borcea},
  \citenamefont {Emsallem}, \citenamefont {Glogowski}, \citenamefont {Grawe},
  \citenamefont {Guillemaud-Mueller}, \citenamefont {Hjorth-Jensen},
  \citenamefont {Houry}, \citenamefont {Lewitowicz}, \citenamefont {Mueller},
  \citenamefont {Nowak}, \citenamefont {P{\l}ochocki}, \citenamefont
  {Pf{\"{u}}tzner}, \citenamefont {Rykaczewski}, \citenamefont {Saint-Laurent},
  \citenamefont {Sauvestre}, \citenamefont {Schaefer}, \citenamefont {Sorlin},
  \citenamefont {Szerypo}, \citenamefont {Trinder}, \citenamefont {Viteritti},\
  and\ \citenamefont {Winfield}}]{Grzywacz1998}%
  \BibitemOpen
  \bibfield  {author} {\bibinfo {author} {\bibfnamefont {R.}~\bibnamefont
  {Grzywacz}}, \bibinfo {author} {\bibfnamefont {R.}~\bibnamefont
  {B{\'{e}}raud}}, \bibinfo {author} {\bibfnamefont {C.}~\bibnamefont
  {Borcea}}, \bibinfo {author} {\bibfnamefont {A.}~\bibnamefont {Emsallem}},
  \bibinfo {author} {\bibfnamefont {M.}~\bibnamefont {Glogowski}}, \bibinfo
  {author} {\bibfnamefont {H.}~\bibnamefont {Grawe}}, \bibinfo {author}
  {\bibfnamefont {D.}~\bibnamefont {Guillemaud-Mueller}}, \bibinfo {author}
  {\bibfnamefont {M.}~\bibnamefont {Hjorth-Jensen}}, \bibinfo {author}
  {\bibfnamefont {M.}~\bibnamefont {Houry}}, \bibinfo {author} {\bibfnamefont
  {M.}~\bibnamefont {Lewitowicz}}, \bibinfo {author} {\bibfnamefont {A.~C.}\
  \bibnamefont {Mueller}}, \bibinfo {author} {\bibfnamefont {A.}~\bibnamefont
  {Nowak}}, \bibinfo {author} {\bibfnamefont {A.}~\bibnamefont {P{\l}ochocki}},
  \bibinfo {author} {\bibfnamefont {M.}~\bibnamefont {Pf{\"{u}}tzner}},
  \bibinfo {author} {\bibfnamefont {K.}~\bibnamefont {Rykaczewski}}, \bibinfo
  {author} {\bibfnamefont {M.~G.}\ \bibnamefont {Saint-Laurent}}, \bibinfo
  {author} {\bibfnamefont {J.~E.}\ \bibnamefont {Sauvestre}}, \bibinfo {author}
  {\bibfnamefont {M.}~\bibnamefont {Schaefer}}, \bibinfo {author}
  {\bibfnamefont {O.}~\bibnamefont {Sorlin}}, \bibinfo {author} {\bibfnamefont
  {J.}~\bibnamefont {Szerypo}}, \bibinfo {author} {\bibfnamefont
  {W.}~\bibnamefont {Trinder}}, \bibinfo {author} {\bibfnamefont
  {S.}~\bibnamefont {Viteritti}}, \ and\ \bibinfo {author} {\bibfnamefont
  {J.}~\bibnamefont {Winfield}},\ }\href {\doibase 10.1103/PhysRevLett.81.766}
  {\bibfield  {journal} {\bibinfo  {journal} {Physical Review Letters}\
  }\textbf {\bibinfo {volume} {81}},\ \bibinfo {pages} {766} (\bibinfo {year}
  {1998})}\BibitemShut {NoStop}%
\bibitem [{\citenamefont {Recchia}\ \emph {et~al.}(2012)\citenamefont
  {Recchia}, \citenamefont {Lenzi}, \citenamefont {Lunardi}, \citenamefont
  {Farnea}, \citenamefont {Gadea}, \citenamefont {Mărginean}, \citenamefont
  {Napoli}, \citenamefont {Nowacki}, \citenamefont {Poves}, \citenamefont
  {Valiente-Dob{\'{o}}n}, \citenamefont {Axiotis}, \citenamefont {Aydin},
  \citenamefont {Bazzacco}, \citenamefont {Benzoni}, \citenamefont {Bizzeti},
  \citenamefont {Bizzeti-Sona}, \citenamefont {Bracco}, \citenamefont
  {Bucurescu}, \citenamefont {Caurier}, \citenamefont {Corradi}, \citenamefont
  {de~Angelis}, \citenamefont {{Della Vedova}}, \citenamefont {Fioretto},
  \citenamefont {Gottardo}, \citenamefont {Ionescu-Bujor}, \citenamefont
  {Iordachescu}, \citenamefont {Leoni}, \citenamefont {Mărginean},
  \citenamefont {Mason}, \citenamefont {Menegazzo}, \citenamefont {Mengoni},
  \citenamefont {Million}, \citenamefont {Montagnoli}, \citenamefont {Orlandi},
  \citenamefont {Pollarolo}, \citenamefont {Sahin}, \citenamefont
  {Scarlassara}, \citenamefont {Singh}, \citenamefont {Stefanini},
  \citenamefont {Szilner}, \citenamefont {Ur},\ and\ \citenamefont
  {Wieland}}]{Recchia2012}%
  \BibitemOpen
  \bibfield  {author} {\bibinfo {author} {\bibfnamefont {F.}~\bibnamefont
  {Recchia}}, \bibinfo {author} {\bibfnamefont {S.~M.}\ \bibnamefont {Lenzi}},
  \bibinfo {author} {\bibfnamefont {S.}~\bibnamefont {Lunardi}}, \bibinfo
  {author} {\bibfnamefont {E.}~\bibnamefont {Farnea}}, \bibinfo {author}
  {\bibfnamefont {A.}~\bibnamefont {Gadea}}, \bibinfo {author} {\bibfnamefont
  {N.}~\bibnamefont {Mărginean}}, \bibinfo {author} {\bibfnamefont {D.~R.}\
  \bibnamefont {Napoli}}, \bibinfo {author} {\bibfnamefont {F.}~\bibnamefont
  {Nowacki}}, \bibinfo {author} {\bibfnamefont {A.}~\bibnamefont {Poves}},
  \bibinfo {author} {\bibfnamefont {J.~J.}\ \bibnamefont
  {Valiente-Dob{\'{o}}n}}, \bibinfo {author} {\bibfnamefont {M.}~\bibnamefont
  {Axiotis}}, \bibinfo {author} {\bibfnamefont {S.}~\bibnamefont {Aydin}},
  \bibinfo {author} {\bibfnamefont {D.}~\bibnamefont {Bazzacco}}, \bibinfo
  {author} {\bibfnamefont {G.}~\bibnamefont {Benzoni}}, \bibinfo {author}
  {\bibfnamefont {P.~G.}\ \bibnamefont {Bizzeti}}, \bibinfo {author}
  {\bibfnamefont {A.~M.}\ \bibnamefont {Bizzeti-Sona}}, \bibinfo {author}
  {\bibfnamefont {A.}~\bibnamefont {Bracco}}, \bibinfo {author} {\bibfnamefont
  {D.}~\bibnamefont {Bucurescu}}, \bibinfo {author} {\bibfnamefont
  {E.}~\bibnamefont {Caurier}}, \bibinfo {author} {\bibfnamefont
  {L.}~\bibnamefont {Corradi}}, \bibinfo {author} {\bibfnamefont
  {G.}~\bibnamefont {de~Angelis}}, \bibinfo {author} {\bibfnamefont
  {F.}~\bibnamefont {{Della Vedova}}}, \bibinfo {author} {\bibfnamefont
  {E.}~\bibnamefont {Fioretto}}, \bibinfo {author} {\bibfnamefont
  {A.}~\bibnamefont {Gottardo}}, \bibinfo {author} {\bibfnamefont
  {M.}~\bibnamefont {Ionescu-Bujor}}, \bibinfo {author} {\bibfnamefont
  {A.}~\bibnamefont {Iordachescu}}, \bibinfo {author} {\bibfnamefont
  {S.}~\bibnamefont {Leoni}}, \bibinfo {author} {\bibfnamefont
  {R.}~\bibnamefont {Mărginean}}, \bibinfo {author} {\bibfnamefont
  {P.}~\bibnamefont {Mason}}, \bibinfo {author} {\bibfnamefont
  {R.}~\bibnamefont {Menegazzo}}, \bibinfo {author} {\bibfnamefont
  {D.}~\bibnamefont {Mengoni}}, \bibinfo {author} {\bibfnamefont
  {B.}~\bibnamefont {Million}}, \bibinfo {author} {\bibfnamefont
  {G.}~\bibnamefont {Montagnoli}}, \bibinfo {author} {\bibfnamefont
  {R.}~\bibnamefont {Orlandi}}, \bibinfo {author} {\bibfnamefont
  {G.}~\bibnamefont {Pollarolo}}, \bibinfo {author} {\bibfnamefont
  {E.}~\bibnamefont {Sahin}}, \bibinfo {author} {\bibfnamefont
  {F.}~\bibnamefont {Scarlassara}}, \bibinfo {author} {\bibfnamefont {R.~P.}\
  \bibnamefont {Singh}}, \bibinfo {author} {\bibfnamefont {A.~M.}\ \bibnamefont
  {Stefanini}}, \bibinfo {author} {\bibfnamefont {S.}~\bibnamefont {Szilner}},
  \bibinfo {author} {\bibfnamefont {C.~A.}\ \bibnamefont {Ur}}, \ and\ \bibinfo
  {author} {\bibfnamefont {O.}~\bibnamefont {Wieland}},\ }\href {\doibase
  10.1103/PhysRevC.85.064305} {\bibfield  {journal} {\bibinfo  {journal}
  {Physical Review C}\ }\textbf {\bibinfo {volume} {85}},\ \bibinfo {pages}
  {064305} (\bibinfo {year} {2012})}\BibitemShut {NoStop}%
\bibitem [{\citenamefont {Liddick}\ \emph {et~al.}(2012)\citenamefont
  {Liddick}, \citenamefont {Abromeit}, \citenamefont {Ayres}, \citenamefont
  {Bey}, \citenamefont {Bingham}, \citenamefont {Bolla}, \citenamefont
  {Cartegni}, \citenamefont {Crawford}, \citenamefont {Darby}, \citenamefont
  {Grzywacz}, \citenamefont {Ilyushkin}, \citenamefont {Larson}, \citenamefont
  {Madurga}, \citenamefont {Miller}, \citenamefont {Padgett}, \citenamefont
  {Paulauskas}, \citenamefont {Rajabali}, \citenamefont {Rykaczewski},\ and\
  \citenamefont {Suchyta}}]{Liddick2012}%
  \BibitemOpen
  \bibfield  {author} {\bibinfo {author} {\bibfnamefont {S.~N.}\ \bibnamefont
  {Liddick}}, \bibinfo {author} {\bibfnamefont {B.}~\bibnamefont {Abromeit}},
  \bibinfo {author} {\bibfnamefont {A.}~\bibnamefont {Ayres}}, \bibinfo
  {author} {\bibfnamefont {A.}~\bibnamefont {Bey}}, \bibinfo {author}
  {\bibfnamefont {C.~R.}\ \bibnamefont {Bingham}}, \bibinfo {author}
  {\bibfnamefont {M.}~\bibnamefont {Bolla}}, \bibinfo {author} {\bibfnamefont
  {L.}~\bibnamefont {Cartegni}}, \bibinfo {author} {\bibfnamefont {H.~L.}\
  \bibnamefont {Crawford}}, \bibinfo {author} {\bibfnamefont {I.~G.}\
  \bibnamefont {Darby}}, \bibinfo {author} {\bibfnamefont {R.}~\bibnamefont
  {Grzywacz}}, \bibinfo {author} {\bibfnamefont {S.}~\bibnamefont {Ilyushkin}},
  \bibinfo {author} {\bibfnamefont {N.}~\bibnamefont {Larson}}, \bibinfo
  {author} {\bibfnamefont {M.}~\bibnamefont {Madurga}}, \bibinfo {author}
  {\bibfnamefont {D.}~\bibnamefont {Miller}}, \bibinfo {author} {\bibfnamefont
  {S.}~\bibnamefont {Padgett}}, \bibinfo {author} {\bibfnamefont
  {S.}~\bibnamefont {Paulauskas}}, \bibinfo {author} {\bibfnamefont {M.~M.}\
  \bibnamefont {Rajabali}}, \bibinfo {author} {\bibfnamefont {K.}~\bibnamefont
  {Rykaczewski}}, \ and\ \bibinfo {author} {\bibfnamefont {S.}~\bibnamefont
  {Suchyta}},\ }\href {\doibase 10.1103/PhysRevC.85.014328} {\bibfield
  {journal} {\bibinfo  {journal} {Physical Review C}\ }\textbf {\bibinfo
  {volume} {85}},\ \bibinfo {pages} {014328} (\bibinfo {year}
  {2012})}\BibitemShut {NoStop}%
\bibitem [{\citenamefont {Liddick}\ \emph {et~al.}(2013)\citenamefont
  {Liddick}, \citenamefont {Abromeit}, \citenamefont {Ayres}, \citenamefont
  {Bey}, \citenamefont {Bingham}, \citenamefont {Brown}, \citenamefont
  {Cartegni}, \citenamefont {Crawford}, \citenamefont {Darby}, \citenamefont
  {Grzywacz}, \citenamefont {Ilyushkin}, \citenamefont {Hjorth-Jensen},
  \citenamefont {Larson}, \citenamefont {Madurga}, \citenamefont {Miller},
  \citenamefont {Padgett}, \citenamefont {Paulauskas}, \citenamefont
  {Rajabali}, \citenamefont {Rykaczewski},\ and\ \citenamefont
  {Suchyta}}]{Liddick2013}%
  \BibitemOpen
  \bibfield  {author} {\bibinfo {author} {\bibfnamefont {S.~N.}\ \bibnamefont
  {Liddick}}, \bibinfo {author} {\bibfnamefont {B.}~\bibnamefont {Abromeit}},
  \bibinfo {author} {\bibfnamefont {A.}~\bibnamefont {Ayres}}, \bibinfo
  {author} {\bibfnamefont {A.}~\bibnamefont {Bey}}, \bibinfo {author}
  {\bibfnamefont {C.~R.}\ \bibnamefont {Bingham}}, \bibinfo {author}
  {\bibfnamefont {B.~A.}\ \bibnamefont {Brown}}, \bibinfo {author}
  {\bibfnamefont {L.}~\bibnamefont {Cartegni}}, \bibinfo {author}
  {\bibfnamefont {H.~L.}\ \bibnamefont {Crawford}}, \bibinfo {author}
  {\bibfnamefont {I.~G.}\ \bibnamefont {Darby}}, \bibinfo {author}
  {\bibfnamefont {R.}~\bibnamefont {Grzywacz}}, \bibinfo {author}
  {\bibfnamefont {S.}~\bibnamefont {Ilyushkin}}, \bibinfo {author}
  {\bibfnamefont {M.}~\bibnamefont {Hjorth-Jensen}}, \bibinfo {author}
  {\bibfnamefont {N.}~\bibnamefont {Larson}}, \bibinfo {author} {\bibfnamefont
  {M.}~\bibnamefont {Madurga}}, \bibinfo {author} {\bibfnamefont
  {D.}~\bibnamefont {Miller}}, \bibinfo {author} {\bibfnamefont
  {S.}~\bibnamefont {Padgett}}, \bibinfo {author} {\bibfnamefont {S.~V.}\
  \bibnamefont {Paulauskas}}, \bibinfo {author} {\bibfnamefont {M.~M.}\
  \bibnamefont {Rajabali}}, \bibinfo {author} {\bibfnamefont {K.}~\bibnamefont
  {Rykaczewski}}, \ and\ \bibinfo {author} {\bibfnamefont {S.}~\bibnamefont
  {Suchyta}},\ }\href {\doibase 10.1103/PhysRevC.87.014325} {\bibfield
  {journal} {\bibinfo  {journal} {Physical Review C}\ }\textbf {\bibinfo
  {volume} {87}},\ \bibinfo {pages} {014325} (\bibinfo {year}
  {2013})}\BibitemShut {NoStop}%
\bibitem [{\citenamefont {Olaizola}\ \emph
  {et~al.}(2017{\natexlab{a}})\citenamefont {Olaizola}, \citenamefont {Fraile},
  \citenamefont {Mach}, \citenamefont {Poves}, \citenamefont {Nowacki},
  \citenamefont {Aprahamian}, \citenamefont {Briz}, \citenamefont
  {Cal-Gonz{\'{a}}lez}, \citenamefont {Ghiţa}, \citenamefont {K{\"{o}}ster},
  \citenamefont {Kurcewicz}, \citenamefont {Lesher}, \citenamefont {Pauwels},
  \citenamefont {Picado}, \citenamefont {Radulov}, \citenamefont {Simpson},\
  and\ \citenamefont {Ud{\'{i}}as}}]{Olaizola2017}%
  \BibitemOpen
  \bibfield  {author} {\bibinfo {author} {\bibfnamefont {B.}~\bibnamefont
  {Olaizola}}, \bibinfo {author} {\bibfnamefont {L.~M.}\ \bibnamefont
  {Fraile}}, \bibinfo {author} {\bibfnamefont {H.}~\bibnamefont {Mach}},
  \bibinfo {author} {\bibfnamefont {A.}~\bibnamefont {Poves}}, \bibinfo
  {author} {\bibfnamefont {F.}~\bibnamefont {Nowacki}}, \bibinfo {author}
  {\bibfnamefont {A.}~\bibnamefont {Aprahamian}}, \bibinfo {author}
  {\bibfnamefont {J.~A.}\ \bibnamefont {Briz}}, \bibinfo {author}
  {\bibfnamefont {J.}~\bibnamefont {Cal-Gonz{\'{a}}lez}}, \bibinfo {author}
  {\bibfnamefont {D.}~\bibnamefont {Ghiţa}}, \bibinfo {author} {\bibfnamefont
  {U.}~\bibnamefont {K{\"{o}}ster}}, \bibinfo {author} {\bibfnamefont
  {W.}~\bibnamefont {Kurcewicz}}, \bibinfo {author} {\bibfnamefont {S.~R.}\
  \bibnamefont {Lesher}}, \bibinfo {author} {\bibfnamefont {D.}~\bibnamefont
  {Pauwels}}, \bibinfo {author} {\bibfnamefont {E.}~\bibnamefont {Picado}},
  \bibinfo {author} {\bibfnamefont {D.}~\bibnamefont {Radulov}}, \bibinfo
  {author} {\bibfnamefont {G.~S.}\ \bibnamefont {Simpson}}, \ and\ \bibinfo
  {author} {\bibfnamefont {J.~M.}\ \bibnamefont {Ud{\'{i}}as}},\ }\href
  {\doibase 10.1103/PhysRevC.95.061303} {\bibfield  {journal} {\bibinfo
  {journal} {Physical Review C}\ }\textbf {\bibinfo {volume} {95}},\ \bibinfo
  {pages} {061303} (\bibinfo {year} {2017}{\natexlab{a}})}\BibitemShut
  {NoStop}%
\bibitem [{\citenamefont {Olaizola}\ \emph
  {et~al.}(2017{\natexlab{b}})\citenamefont {Olaizola}, \citenamefont {Fraile},
  \citenamefont {Mach}, \citenamefont {Poves}, \citenamefont {Aprahamian},
  \citenamefont {Briz}, \citenamefont {Cal-Gonz{\'{a}}lez}, \citenamefont
  {Ghiţa}, \citenamefont {K{\"{o}}ster}, \citenamefont {Kurcewicz},
  \citenamefont {Lesher}, \citenamefont {Pauwels}, \citenamefont {Picado},
  \citenamefont {Radulov}, \citenamefont {Simpson},\ and\ \citenamefont
  {Ud{\'{i}}as}}]{Olaizola2017a}%
  \BibitemOpen
  \bibfield  {author} {\bibinfo {author} {\bibfnamefont {B.}~\bibnamefont
  {Olaizola}}, \bibinfo {author} {\bibfnamefont {L.~M.}\ \bibnamefont
  {Fraile}}, \bibinfo {author} {\bibfnamefont {H.}~\bibnamefont {Mach}},
  \bibinfo {author} {\bibfnamefont {A.}~\bibnamefont {Poves}}, \bibinfo
  {author} {\bibfnamefont {A.}~\bibnamefont {Aprahamian}}, \bibinfo {author}
  {\bibfnamefont {J.~A.}\ \bibnamefont {Briz}}, \bibinfo {author}
  {\bibfnamefont {J.}~\bibnamefont {Cal-Gonz{\'{a}}lez}}, \bibinfo {author}
  {\bibfnamefont {D.}~\bibnamefont {Ghiţa}}, \bibinfo {author} {\bibfnamefont
  {U.}~\bibnamefont {K{\"{o}}ster}}, \bibinfo {author} {\bibfnamefont
  {W.}~\bibnamefont {Kurcewicz}}, \bibinfo {author} {\bibfnamefont {S.~R.}\
  \bibnamefont {Lesher}}, \bibinfo {author} {\bibfnamefont {D.}~\bibnamefont
  {Pauwels}}, \bibinfo {author} {\bibfnamefont {E.}~\bibnamefont {Picado}},
  \bibinfo {author} {\bibfnamefont {D.}~\bibnamefont {Radulov}}, \bibinfo
  {author} {\bibfnamefont {G.~S.}\ \bibnamefont {Simpson}}, \ and\ \bibinfo
  {author} {\bibfnamefont {J.~M.}\ \bibnamefont {Ud{\'{i}}as}},\ }\href
  {\doibase 10.1088/1361-6471/aa915e} {\bibfield  {journal} {\bibinfo
  {journal} {Journal of Physics G: Nuclear and Particle Physics}\ }\textbf
  {\bibinfo {volume} {44}},\ \bibinfo {pages} {125103} (\bibinfo {year}
  {2017}{\natexlab{b}})}\BibitemShut {NoStop}%
\bibitem [{\citenamefont {Pauwels}\ \emph {et~al.}(2012)\citenamefont
  {Pauwels}, \citenamefont {Radulov}, \citenamefont {Walters}, \citenamefont
  {Darby}, \citenamefont {{De Witte}}, \citenamefont {Diriken}, \citenamefont
  {Fedorov}, \citenamefont {Fedosseev}, \citenamefont {Fraile}, \citenamefont
  {Huyse}, \citenamefont {K{\"{o}}ster}, \citenamefont {Marsh}, \citenamefont
  {Popescu}, \citenamefont {Seliverstov}, \citenamefont {Sj{\"{o}}din},
  \citenamefont {{Van den Bergh}}, \citenamefont {{Van de Walle}},
  \citenamefont {{Van Duppen}}, \citenamefont {Venhart},\ and\ \citenamefont
  {Wimmer}}]{Pauwels2012}%
  \BibitemOpen
  \bibfield  {author} {\bibinfo {author} {\bibfnamefont {D.}~\bibnamefont
  {Pauwels}}, \bibinfo {author} {\bibfnamefont {D.}~\bibnamefont {Radulov}},
  \bibinfo {author} {\bibfnamefont {W.~B.}\ \bibnamefont {Walters}}, \bibinfo
  {author} {\bibfnamefont {I.~G.}\ \bibnamefont {Darby}}, \bibinfo {author}
  {\bibfnamefont {H.}~\bibnamefont {{De Witte}}}, \bibinfo {author}
  {\bibfnamefont {J.}~\bibnamefont {Diriken}}, \bibinfo {author} {\bibfnamefont
  {D.~V.}\ \bibnamefont {Fedorov}}, \bibinfo {author} {\bibfnamefont {V.~N.}\
  \bibnamefont {Fedosseev}}, \bibinfo {author} {\bibfnamefont {L.~M.}\
  \bibnamefont {Fraile}}, \bibinfo {author} {\bibfnamefont {M.}~\bibnamefont
  {Huyse}}, \bibinfo {author} {\bibfnamefont {U.}~\bibnamefont {K{\"{o}}ster}},
  \bibinfo {author} {\bibfnamefont {B.~A.}\ \bibnamefont {Marsh}}, \bibinfo
  {author} {\bibfnamefont {L.}~\bibnamefont {Popescu}}, \bibinfo {author}
  {\bibfnamefont {M.~D.}\ \bibnamefont {Seliverstov}}, \bibinfo {author}
  {\bibfnamefont {A.~M.}\ \bibnamefont {Sj{\"{o}}din}}, \bibinfo {author}
  {\bibfnamefont {P.}~\bibnamefont {{Van den Bergh}}}, \bibinfo {author}
  {\bibfnamefont {J.}~\bibnamefont {{Van de Walle}}}, \bibinfo {author}
  {\bibfnamefont {P.}~\bibnamefont {{Van Duppen}}}, \bibinfo {author}
  {\bibfnamefont {M.}~\bibnamefont {Venhart}}, \ and\ \bibinfo {author}
  {\bibfnamefont {K.}~\bibnamefont {Wimmer}},\ }\href {\doibase
  10.1103/PhysRevC.86.064318} {\bibfield  {journal} {\bibinfo  {journal}
  {Physical Review C}\ }\textbf {\bibinfo {volume} {86}},\ \bibinfo {pages}
  {064318} (\bibinfo {year} {2012})}\BibitemShut {NoStop}%
\bibitem [{\citenamefont {Radulov}\ \emph {et~al.}(2013)\citenamefont
  {Radulov}, \citenamefont {Chiara}, \citenamefont {Darby}, \citenamefont {{De
  Witte}}, \citenamefont {Diriken}, \citenamefont {Fedorov}, \citenamefont
  {Fedosseev}, \citenamefont {Fraile}, \citenamefont {Huyse}, \citenamefont
  {K{\"{o}}ster}, \citenamefont {Marsh}, \citenamefont {Pauwels}, \citenamefont
  {Popescu}, \citenamefont {Seliverstov}, \citenamefont {Sj{\"{o}}din},
  \citenamefont {{Van den Bergh}}, \citenamefont {{Van Duppen}}, \citenamefont
  {Venhart}, \citenamefont {Walters},\ and\ \citenamefont
  {Wimmer}}]{Radulov2013}%
  \BibitemOpen
  \bibfield  {author} {\bibinfo {author} {\bibfnamefont {D.}~\bibnamefont
  {Radulov}}, \bibinfo {author} {\bibfnamefont {C.~J.}\ \bibnamefont {Chiara}},
  \bibinfo {author} {\bibfnamefont {I.~G.}\ \bibnamefont {Darby}}, \bibinfo
  {author} {\bibfnamefont {H.}~\bibnamefont {{De Witte}}}, \bibinfo {author}
  {\bibfnamefont {J.}~\bibnamefont {Diriken}}, \bibinfo {author} {\bibfnamefont
  {D.~V.}\ \bibnamefont {Fedorov}}, \bibinfo {author} {\bibfnamefont {V.~N.}\
  \bibnamefont {Fedosseev}}, \bibinfo {author} {\bibfnamefont {L.~M.}\
  \bibnamefont {Fraile}}, \bibinfo {author} {\bibfnamefont {M.}~\bibnamefont
  {Huyse}}, \bibinfo {author} {\bibfnamefont {U.}~\bibnamefont {K{\"{o}}ster}},
  \bibinfo {author} {\bibfnamefont {B.~A.}\ \bibnamefont {Marsh}}, \bibinfo
  {author} {\bibfnamefont {D.}~\bibnamefont {Pauwels}}, \bibinfo {author}
  {\bibfnamefont {L.}~\bibnamefont {Popescu}}, \bibinfo {author} {\bibfnamefont
  {M.~D.}\ \bibnamefont {Seliverstov}}, \bibinfo {author} {\bibfnamefont
  {A.~M.}\ \bibnamefont {Sj{\"{o}}din}}, \bibinfo {author} {\bibfnamefont
  {P.}~\bibnamefont {{Van den Bergh}}}, \bibinfo {author} {\bibfnamefont
  {P.}~\bibnamefont {{Van Duppen}}}, \bibinfo {author} {\bibfnamefont
  {M.}~\bibnamefont {Venhart}}, \bibinfo {author} {\bibfnamefont {W.~B.}\
  \bibnamefont {Walters}}, \ and\ \bibinfo {author} {\bibfnamefont
  {K.}~\bibnamefont {Wimmer}},\ }\href {\doibase 10.1103/PhysRevC.88.014307}
  {\bibfield  {journal} {\bibinfo  {journal} {Physical Review C}\ }\textbf
  {\bibinfo {volume} {88}},\ \bibinfo {pages} {014307} (\bibinfo {year}
  {2013})}\BibitemShut {NoStop}%
\bibitem [{\citenamefont {Radulov}(2014)}]{Radulov2014}%
  \BibitemOpen
  \bibfield  {author} {\bibinfo {author} {\bibfnamefont {D.}~\bibnamefont
  {Radulov}},\ }\emph {\bibinfo {title} {{Investigating the nuclear structure
  of the neutron-rich odd-mass Fe isotopes, in the beta-decay of their parent -
  Mn}}},\ \href
  {https://fys.kuleuven.be/iks/ns/files/thesis/final-thesis-radulov.pdf} {Ph.D.
  thesis},\ \bibinfo  {school} {KU Leuven} (\bibinfo {year} {2014})\BibitemShut
  {NoStop}%
\bibitem [{\citenamefont {Flavigny}\ \emph {et~al.}(2015)\citenamefont
  {Flavigny}, \citenamefont {Pauwels}, \citenamefont {Radulov}, \citenamefont
  {Darby}, \citenamefont {{De Witte}}, \citenamefont {Diriken}, \citenamefont
  {Fedorov}, \citenamefont {Fedosseev}, \citenamefont {Fraile}, \citenamefont
  {Huyse}, \citenamefont {Ivanov}, \citenamefont {K{\"{o}}ster}, \citenamefont
  {Marsh}, \citenamefont {Otsuka}, \citenamefont {Popescu}, \citenamefont
  {Raabe}, \citenamefont {Seliverstov}, \citenamefont {Shimizu}, \citenamefont
  {Sj{\"{o}}din}, \citenamefont {Tsunoda}, \citenamefont {{Van den Bergh}},
  \citenamefont {{Van Duppen}}, \citenamefont {{Van de Walle}}, \citenamefont
  {Venhart}, \citenamefont {Walters},\ and\ \citenamefont
  {Wimmer}}]{Flavigny2015}%
  \BibitemOpen
  \bibfield  {author} {\bibinfo {author} {\bibfnamefont {F.}~\bibnamefont
  {Flavigny}}, \bibinfo {author} {\bibfnamefont {D.}~\bibnamefont {Pauwels}},
  \bibinfo {author} {\bibfnamefont {D.}~\bibnamefont {Radulov}}, \bibinfo
  {author} {\bibfnamefont {I.~J.}\ \bibnamefont {Darby}}, \bibinfo {author}
  {\bibfnamefont {H.}~\bibnamefont {{De Witte}}}, \bibinfo {author}
  {\bibfnamefont {J.}~\bibnamefont {Diriken}}, \bibinfo {author} {\bibfnamefont
  {D.~V.}\ \bibnamefont {Fedorov}}, \bibinfo {author} {\bibfnamefont {V.~N.}\
  \bibnamefont {Fedosseev}}, \bibinfo {author} {\bibfnamefont {L.~M.}\
  \bibnamefont {Fraile}}, \bibinfo {author} {\bibfnamefont {M.}~\bibnamefont
  {Huyse}}, \bibinfo {author} {\bibfnamefont {V.~S.}\ \bibnamefont {Ivanov}},
  \bibinfo {author} {\bibfnamefont {U.}~\bibnamefont {K{\"{o}}ster}}, \bibinfo
  {author} {\bibfnamefont {B.~A.}\ \bibnamefont {Marsh}}, \bibinfo {author}
  {\bibfnamefont {T.}~\bibnamefont {Otsuka}}, \bibinfo {author} {\bibfnamefont
  {L.}~\bibnamefont {Popescu}}, \bibinfo {author} {\bibfnamefont
  {R.}~\bibnamefont {Raabe}}, \bibinfo {author} {\bibfnamefont {M.~D.}\
  \bibnamefont {Seliverstov}}, \bibinfo {author} {\bibfnamefont
  {N.}~\bibnamefont {Shimizu}}, \bibinfo {author} {\bibfnamefont {A.~M.}\
  \bibnamefont {Sj{\"{o}}din}}, \bibinfo {author} {\bibfnamefont
  {Y.}~\bibnamefont {Tsunoda}}, \bibinfo {author} {\bibfnamefont
  {P.}~\bibnamefont {{Van den Bergh}}}, \bibinfo {author} {\bibfnamefont
  {P.}~\bibnamefont {{Van Duppen}}}, \bibinfo {author} {\bibfnamefont
  {J.}~\bibnamefont {{Van de Walle}}}, \bibinfo {author} {\bibfnamefont
  {M.}~\bibnamefont {Venhart}}, \bibinfo {author} {\bibfnamefont {W.~B.}\
  \bibnamefont {Walters}}, \ and\ \bibinfo {author} {\bibfnamefont
  {K.}~\bibnamefont {Wimmer}},\ }\href {\doibase 10.1103/PhysRevC.91.034310}
  {\bibfield  {journal} {\bibinfo  {journal} {Physical Review C}\ }\textbf
  {\bibinfo {volume} {91}},\ \bibinfo {pages} {034310} (\bibinfo {year}
  {2015})}\BibitemShut {NoStop}%
\bibitem [{\citenamefont {Monta{\~{n}}o}\ \emph {et~al.}(2013)\citenamefont
  {Monta{\~{n}}o}, \citenamefont {Giles},\ and\ \citenamefont
  {Gottberg}}]{Montano2013}%
  \BibitemOpen
  \bibfield  {author} {\bibinfo {author} {\bibfnamefont {J.}~\bibnamefont
  {Monta{\~{n}}o}}, \bibinfo {author} {\bibfnamefont {T.}~\bibnamefont
  {Giles}}, \ and\ \bibinfo {author} {\bibfnamefont {A.}~\bibnamefont
  {Gottberg}},\ }\href {\doibase 10.1016/j.nimb.2013.08.019} {\bibfield
  {journal} {\bibinfo  {journal} {Nuclear Instruments and Methods in Physics
  Research Section B: Beam Interactions with Materials and Atoms}\ }\textbf
  {\bibinfo {volume} {317}},\ \bibinfo {pages} {430} (\bibinfo {year}
  {2013})}\BibitemShut {NoStop}%
\bibitem [{\citenamefont {Fedosseev}\ \emph {et~al.}(2012)\citenamefont
  {Fedosseev}, \citenamefont {Berg}, \citenamefont {Fedorov}, \citenamefont
  {Fink}, \citenamefont {Launila}, \citenamefont {Losito}, \citenamefont
  {Marsh}, \citenamefont {Rossel}, \citenamefont {Rothe}, \citenamefont
  {Seliverstov}, \citenamefont {Sj{\"{o}}din},\ and\ \citenamefont
  {Wendt}}]{Fedosseev2012}%
  \BibitemOpen
  \bibfield  {author} {\bibinfo {author} {\bibfnamefont {V.~N.}\ \bibnamefont
  {Fedosseev}}, \bibinfo {author} {\bibfnamefont {L.-E.}\ \bibnamefont {Berg}},
  \bibinfo {author} {\bibfnamefont {D.~V.}\ \bibnamefont {Fedorov}}, \bibinfo
  {author} {\bibfnamefont {D.}~\bibnamefont {Fink}}, \bibinfo {author}
  {\bibfnamefont {O.~J.}\ \bibnamefont {Launila}}, \bibinfo {author}
  {\bibfnamefont {R.}~\bibnamefont {Losito}}, \bibinfo {author} {\bibfnamefont
  {B.~A.}\ \bibnamefont {Marsh}}, \bibinfo {author} {\bibfnamefont {R.~E.}\
  \bibnamefont {Rossel}}, \bibinfo {author} {\bibfnamefont {S.}~\bibnamefont
  {Rothe}}, \bibinfo {author} {\bibfnamefont {M.~D.}\ \bibnamefont
  {Seliverstov}}, \bibinfo {author} {\bibfnamefont {A.~M.}\ \bibnamefont
  {Sj{\"{o}}din}}, \ and\ \bibinfo {author} {\bibfnamefont {K.~D.~A.}\
  \bibnamefont {Wendt}},\ }\href {\doibase 10.1063/1.3662206} {\bibfield
  {journal} {\bibinfo  {journal} {Review of Scientific Instruments}\ }\textbf
  {\bibinfo {volume} {83}},\ \bibinfo {pages} {02A903} (\bibinfo {year}
  {2012})}\BibitemShut {NoStop}%
\bibitem [{\citenamefont {Pauwels}\ \emph {et~al.}(2008)\citenamefont
  {Pauwels}, \citenamefont {Ivanov}, \citenamefont {B{\"{u}}scher},
  \citenamefont {Cocolios}, \citenamefont {Gentens}, \citenamefont {Huyse},
  \citenamefont {Korgul}, \citenamefont {Kudryavtsev}, \citenamefont {Raabe},
  \citenamefont {Sawicka}, \citenamefont {Stefanescu}, \citenamefont {{Van de
  Walle}}, \citenamefont {{Van den Bergh}},\ and\ \citenamefont {{Van
  Duppen}}}]{Pauwels2008}%
  \BibitemOpen
  \bibfield  {author} {\bibinfo {author} {\bibfnamefont {D.}~\bibnamefont
  {Pauwels}}, \bibinfo {author} {\bibfnamefont {O.}~\bibnamefont {Ivanov}},
  \bibinfo {author} {\bibfnamefont {J.}~\bibnamefont {B{\"{u}}scher}}, \bibinfo
  {author} {\bibfnamefont {T.}~\bibnamefont {Cocolios}}, \bibinfo {author}
  {\bibfnamefont {J.}~\bibnamefont {Gentens}}, \bibinfo {author} {\bibfnamefont
  {M.}~\bibnamefont {Huyse}}, \bibinfo {author} {\bibfnamefont
  {A.}~\bibnamefont {Korgul}}, \bibinfo {author} {\bibfnamefont
  {Y.}~\bibnamefont {Kudryavtsev}}, \bibinfo {author} {\bibfnamefont
  {R.}~\bibnamefont {Raabe}}, \bibinfo {author} {\bibfnamefont
  {M.}~\bibnamefont {Sawicka}}, \bibinfo {author} {\bibfnamefont
  {I.}~\bibnamefont {Stefanescu}}, \bibinfo {author} {\bibfnamefont
  {J.}~\bibnamefont {{Van de Walle}}}, \bibinfo {author} {\bibfnamefont
  {P.}~\bibnamefont {{Van den Bergh}}}, \ and\ \bibinfo {author} {\bibfnamefont
  {P.}~\bibnamefont {{Van Duppen}}},\ }\href {\doibase
  10.1016/j.nimb.2008.05.083} {\bibfield  {journal} {\bibinfo  {journal}
  {Nuclear Instruments and Methods in Physics Research Section B: Beam
  Interactions with Materials and Atoms}\ }\textbf {\bibinfo {volume} {266}},\
  \bibinfo {pages} {4600} (\bibinfo {year} {2008})}\BibitemShut {NoStop}%
\bibitem [{\citenamefont {Eberth}\ \emph {et~al.}(2001)\citenamefont {Eberth},
  \citenamefont {Pascovici}, \citenamefont {Thomas}, \citenamefont {Warr},
  \citenamefont {Weisshaar}, \citenamefont {Habs}, \citenamefont {Reiter},
  \citenamefont {Thirolf}, \citenamefont {Schwalm}, \citenamefont {Gund},
  \citenamefont {Scheit}, \citenamefont {Lauer}, \citenamefont {{Van Duppen}},
  \citenamefont {Franchoo}, \citenamefont {Huyse}, \citenamefont {Lieder},
  \citenamefont {Gast}, \citenamefont {Gerl},\ and\ \citenamefont
  {Lieb}}]{Eberth2001}%
  \BibitemOpen
  \bibfield  {author} {\bibinfo {author} {\bibfnamefont {J.}~\bibnamefont
  {Eberth}}, \bibinfo {author} {\bibfnamefont {G.}~\bibnamefont {Pascovici}},
  \bibinfo {author} {\bibfnamefont {H.}~\bibnamefont {Thomas}}, \bibinfo
  {author} {\bibfnamefont {N.}~\bibnamefont {Warr}}, \bibinfo {author}
  {\bibfnamefont {D.}~\bibnamefont {Weisshaar}}, \bibinfo {author}
  {\bibfnamefont {D.}~\bibnamefont {Habs}}, \bibinfo {author} {\bibfnamefont
  {P.}~\bibnamefont {Reiter}}, \bibinfo {author} {\bibfnamefont
  {P.}~\bibnamefont {Thirolf}}, \bibinfo {author} {\bibfnamefont
  {D.}~\bibnamefont {Schwalm}}, \bibinfo {author} {\bibfnamefont
  {C.}~\bibnamefont {Gund}}, \bibinfo {author} {\bibfnamefont {H.}~\bibnamefont
  {Scheit}}, \bibinfo {author} {\bibfnamefont {M.}~\bibnamefont {Lauer}},
  \bibinfo {author} {\bibfnamefont {P.}~\bibnamefont {{Van Duppen}}}, \bibinfo
  {author} {\bibfnamefont {S.}~\bibnamefont {Franchoo}}, \bibinfo {author}
  {\bibfnamefont {M.}~\bibnamefont {Huyse}}, \bibinfo {author} {\bibfnamefont
  {R.}~\bibnamefont {Lieder}}, \bibinfo {author} {\bibfnamefont
  {W.}~\bibnamefont {Gast}}, \bibinfo {author} {\bibfnamefont {J.}~\bibnamefont
  {Gerl}}, \ and\ \bibinfo {author} {\bibfnamefont {K.}~\bibnamefont {Lieb}},\
  }\href {\doibase 10.1016/S0146-6410(01)00145-4} {\bibfield  {journal}
  {\bibinfo  {journal} {Progress in Particle and Nuclear Physics}\ }\textbf
  {\bibinfo {volume} {46}},\ \bibinfo {pages} {389} (\bibinfo {year}
  {2001})}\BibitemShut {NoStop}%
\bibitem [{DGF(2009)}]{DGFmanual}%
  \BibitemOpen
  \href@noop {} {\emph {\bibinfo {title} {User's Manual, Digital Gamma Finder
  (DGF)}}} (\bibinfo {year} {2009}),\ \bibinfo {note}
  {\url{http://www.xia.com/Manuals/DGF_UserManual.pdf}}\BibitemShut {NoStop}%
\bibitem [{\citenamefont {Gins}\ \emph {et~al.}(2018)\citenamefont {Gins},
  \citenamefont {de~Groote}, \citenamefont {Bissell}, \citenamefont {{Granados
  Buitrago}}, \citenamefont {Ferrer}, \citenamefont {Lynch}, \citenamefont
  {Neyens},\ and\ \citenamefont {Sels}}]{Gins2017}%
  \BibitemOpen
  \bibfield  {author} {\bibinfo {author} {\bibfnamefont {W.}~\bibnamefont
  {Gins}}, \bibinfo {author} {\bibfnamefont {R.}~\bibnamefont {de~Groote}},
  \bibinfo {author} {\bibfnamefont {M.}~\bibnamefont {Bissell}}, \bibinfo
  {author} {\bibfnamefont {C.}~\bibnamefont {{Granados Buitrago}}}, \bibinfo
  {author} {\bibfnamefont {R.}~\bibnamefont {Ferrer}}, \bibinfo {author}
  {\bibfnamefont {K.}~\bibnamefont {Lynch}}, \bibinfo {author} {\bibfnamefont
  {G.}~\bibnamefont {Neyens}}, \ and\ \bibinfo {author} {\bibfnamefont
  {S.}~\bibnamefont {Sels}},\ }\href {\doibase 10.1016/j.cpc.2017.09.012}
  {\bibfield  {journal} {\bibinfo  {journal} {Computer Physics Communications}\
  }\textbf {\bibinfo {volume} {222}},\ \bibinfo {pages} {286} (\bibinfo {year}
  {2018})}\BibitemShut {NoStop}%
\bibitem [{\citenamefont {Foreman-Mackey}\ \emph {et~al.}(2013)\citenamefont
  {Foreman-Mackey}, \citenamefont {Hogg}, \citenamefont {Lang},\ and\
  \citenamefont {Goodman}}]{Foreman-Mackey2013}%
  \BibitemOpen
  \bibfield  {author} {\bibinfo {author} {\bibfnamefont {D.}~\bibnamefont
  {Foreman-Mackey}}, \bibinfo {author} {\bibfnamefont {D.~W.}\ \bibnamefont
  {Hogg}}, \bibinfo {author} {\bibfnamefont {D.}~\bibnamefont {Lang}}, \ and\
  \bibinfo {author} {\bibfnamefont {J.}~\bibnamefont {Goodman}},\ }\href
  {\doibase 10.1086/670067} {\bibfield  {journal} {\bibinfo  {journal}
  {Publications of the Astronomical Society of the Pacific}\ }\textbf {\bibinfo
  {volume} {125}},\ \bibinfo {pages} {306} (\bibinfo {year}
  {2013})}\BibitemShut {NoStop}%
\bibitem [{\citenamefont {Lunardi}\ \emph {et~al.}(2007)\citenamefont
  {Lunardi}, \citenamefont {Lenzi}, \citenamefont {Vedova}, \citenamefont
  {Farnea}, \citenamefont {Gadea}, \citenamefont {Mărginean}, \citenamefont
  {Bazzacco}, \citenamefont {Beghini}, \citenamefont {Bizzeti}, \citenamefont
  {Bizzeti-Sona}, \citenamefont {Bucurescu}, \citenamefont {Corradi},
  \citenamefont {Deacon}, \citenamefont {de~Angelis}, \citenamefont {Fioretto},
  \citenamefont {Freeman}, \citenamefont {Ionescu-Bujor}, \citenamefont
  {Iordachescu}, \citenamefont {Mason}, \citenamefont {Mengoni}, \citenamefont
  {Montagnoli}, \citenamefont {Napoli}, \citenamefont {Nowacki}, \citenamefont
  {Orlandi}, \citenamefont {Pollarolo}, \citenamefont {Recchia}, \citenamefont
  {Scarlassara}, \citenamefont {Smith}, \citenamefont {Stefanini},
  \citenamefont {Szilner}, \citenamefont {Ur}, \citenamefont
  {Valiente-Dob{\'{o}}n},\ and\ \citenamefont {Varley}}]{Lunardi2007}%
  \BibitemOpen
  \bibfield  {author} {\bibinfo {author} {\bibfnamefont {S.}~\bibnamefont
  {Lunardi}}, \bibinfo {author} {\bibfnamefont {S.~M.}\ \bibnamefont {Lenzi}},
  \bibinfo {author} {\bibfnamefont {F.~D.}\ \bibnamefont {Vedova}}, \bibinfo
  {author} {\bibfnamefont {E.}~\bibnamefont {Farnea}}, \bibinfo {author}
  {\bibfnamefont {A.}~\bibnamefont {Gadea}}, \bibinfo {author} {\bibfnamefont
  {N.}~\bibnamefont {Mărginean}}, \bibinfo {author} {\bibfnamefont
  {D.}~\bibnamefont {Bazzacco}}, \bibinfo {author} {\bibfnamefont
  {S.}~\bibnamefont {Beghini}}, \bibinfo {author} {\bibfnamefont {P.~G.}\
  \bibnamefont {Bizzeti}}, \bibinfo {author} {\bibfnamefont {A.~M.}\
  \bibnamefont {Bizzeti-Sona}}, \bibinfo {author} {\bibfnamefont
  {D.}~\bibnamefont {Bucurescu}}, \bibinfo {author} {\bibfnamefont
  {L.}~\bibnamefont {Corradi}}, \bibinfo {author} {\bibfnamefont {A.~N.}\
  \bibnamefont {Deacon}}, \bibinfo {author} {\bibfnamefont {G.}~\bibnamefont
  {de~Angelis}}, \bibinfo {author} {\bibfnamefont {E.}~\bibnamefont
  {Fioretto}}, \bibinfo {author} {\bibfnamefont {S.~J.}\ \bibnamefont
  {Freeman}}, \bibinfo {author} {\bibfnamefont {M.}~\bibnamefont
  {Ionescu-Bujor}}, \bibinfo {author} {\bibfnamefont {A.}~\bibnamefont
  {Iordachescu}}, \bibinfo {author} {\bibfnamefont {P.}~\bibnamefont {Mason}},
  \bibinfo {author} {\bibfnamefont {D.}~\bibnamefont {Mengoni}}, \bibinfo
  {author} {\bibfnamefont {G.}~\bibnamefont {Montagnoli}}, \bibinfo {author}
  {\bibfnamefont {D.~R.}\ \bibnamefont {Napoli}}, \bibinfo {author}
  {\bibfnamefont {F.}~\bibnamefont {Nowacki}}, \bibinfo {author} {\bibfnamefont
  {R.}~\bibnamefont {Orlandi}}, \bibinfo {author} {\bibfnamefont
  {G.}~\bibnamefont {Pollarolo}}, \bibinfo {author} {\bibfnamefont
  {F.}~\bibnamefont {Recchia}}, \bibinfo {author} {\bibfnamefont
  {F.}~\bibnamefont {Scarlassara}}, \bibinfo {author} {\bibfnamefont {J.~F.}\
  \bibnamefont {Smith}}, \bibinfo {author} {\bibfnamefont {A.~M.}\ \bibnamefont
  {Stefanini}}, \bibinfo {author} {\bibfnamefont {S.}~\bibnamefont {Szilner}},
  \bibinfo {author} {\bibfnamefont {C.~A.}\ \bibnamefont {Ur}}, \bibinfo
  {author} {\bibfnamefont {J.~J.}\ \bibnamefont {Valiente-Dob{\'{o}}n}}, \ and\
  \bibinfo {author} {\bibfnamefont {B.~J.}\ \bibnamefont {Varley}},\ }\href
  {\doibase 10.1103/PhysRevC.76.034303} {\bibfield  {journal} {\bibinfo
  {journal} {Physical Review C}\ }\textbf {\bibinfo {volume} {76}},\ \bibinfo
  {pages} {034303} (\bibinfo {year} {2007})}\BibitemShut {NoStop}%
\bibitem [{\citenamefont {Adrich}\ \emph {et~al.}(2008)\citenamefont {Adrich},
  \citenamefont {Amthor}, \citenamefont {Bazin}, \citenamefont {Bowen},
  \citenamefont {Brown}, \citenamefont {Campbell}, \citenamefont {Cook},
  \citenamefont {Gade}, \citenamefont {Galaviz}, \citenamefont {Glasmacher},
  \citenamefont {McDaniel}, \citenamefont {Miller}, \citenamefont {Obertelli},
  \citenamefont {Shimbara}, \citenamefont {Siwek}, \citenamefont {Tostevin},\
  and\ \citenamefont {Weisshaar}}]{Adrich2008}%
  \BibitemOpen
  \bibfield  {author} {\bibinfo {author} {\bibfnamefont {P.}~\bibnamefont
  {Adrich}}, \bibinfo {author} {\bibfnamefont {A.~M.}\ \bibnamefont {Amthor}},
  \bibinfo {author} {\bibfnamefont {D.}~\bibnamefont {Bazin}}, \bibinfo
  {author} {\bibfnamefont {M.~D.}\ \bibnamefont {Bowen}}, \bibinfo {author}
  {\bibfnamefont {B.~A.}\ \bibnamefont {Brown}}, \bibinfo {author}
  {\bibfnamefont {C.~M.}\ \bibnamefont {Campbell}}, \bibinfo {author}
  {\bibfnamefont {J.~M.}\ \bibnamefont {Cook}}, \bibinfo {author}
  {\bibfnamefont {A.}~\bibnamefont {Gade}}, \bibinfo {author} {\bibfnamefont
  {D.}~\bibnamefont {Galaviz}}, \bibinfo {author} {\bibfnamefont
  {T.}~\bibnamefont {Glasmacher}}, \bibinfo {author} {\bibfnamefont
  {S.}~\bibnamefont {McDaniel}}, \bibinfo {author} {\bibfnamefont
  {D.}~\bibnamefont {Miller}}, \bibinfo {author} {\bibfnamefont
  {A.}~\bibnamefont {Obertelli}}, \bibinfo {author} {\bibfnamefont
  {Y.}~\bibnamefont {Shimbara}}, \bibinfo {author} {\bibfnamefont {K.~P.}\
  \bibnamefont {Siwek}}, \bibinfo {author} {\bibfnamefont {J.~A.}\ \bibnamefont
  {Tostevin}}, \ and\ \bibinfo {author} {\bibfnamefont {D.}~\bibnamefont
  {Weisshaar}},\ }\href {\doibase 10.1103/PhysRevC.77.054306} {\bibfield
  {journal} {\bibinfo  {journal} {Physical Review C}\ }\textbf {\bibinfo
  {volume} {77}},\ \bibinfo {pages} {054306} (\bibinfo {year}
  {2008})}\BibitemShut {NoStop}%
\bibitem [{\citenamefont {Daugas}\ \emph {et~al.}(2011)\citenamefont {Daugas},
  \citenamefont {Matea}, \citenamefont {Delaroche}, \citenamefont
  {Pf{\"{u}}tzner}, \citenamefont {Sawicka}, \citenamefont {Becker},
  \citenamefont {B{\'{e}}lier}, \citenamefont {Bingham}, \citenamefont
  {Borcea}, \citenamefont {Bouchez}, \citenamefont {Buta}, \citenamefont
  {Dragulescu}, \citenamefont {Georgiev}, \citenamefont {Giovinazzo},
  \citenamefont {Girod}, \citenamefont {Grawe}, \citenamefont {Grzywacz},
  \citenamefont {Hammache}, \citenamefont {Ibrahim}, \citenamefont
  {Lewitowicz}, \citenamefont {Libert}, \citenamefont {Mayet}, \citenamefont
  {M{\'{e}}ot}, \citenamefont {Negoita}, \citenamefont {{de Oliveira Santos}},
  \citenamefont {Perru}, \citenamefont {Roig}, \citenamefont {Rykaczewski},
  \citenamefont {Saint-Laurent}, \citenamefont {Sauvestre}, \citenamefont
  {Sorlin}, \citenamefont {Stanoiu}, \citenamefont {Stefan}, \citenamefont
  {Stodel}, \citenamefont {Theisen}, \citenamefont {Verney},\ and\
  \citenamefont {{\.{Z}}ylicz}}]{Daugas2011}%
  \BibitemOpen
  \bibfield  {author} {\bibinfo {author} {\bibfnamefont {J.~M.}\ \bibnamefont
  {Daugas}}, \bibinfo {author} {\bibfnamefont {I.}~\bibnamefont {Matea}},
  \bibinfo {author} {\bibfnamefont {J.-P.}\ \bibnamefont {Delaroche}}, \bibinfo
  {author} {\bibfnamefont {M.}~\bibnamefont {Pf{\"{u}}tzner}}, \bibinfo
  {author} {\bibfnamefont {M.}~\bibnamefont {Sawicka}}, \bibinfo {author}
  {\bibfnamefont {F.}~\bibnamefont {Becker}}, \bibinfo {author} {\bibfnamefont
  {G.}~\bibnamefont {B{\'{e}}lier}}, \bibinfo {author} {\bibfnamefont {C.~R.}\
  \bibnamefont {Bingham}}, \bibinfo {author} {\bibfnamefont {R.}~\bibnamefont
  {Borcea}}, \bibinfo {author} {\bibfnamefont {E.}~\bibnamefont {Bouchez}},
  \bibinfo {author} {\bibfnamefont {A.}~\bibnamefont {Buta}}, \bibinfo {author}
  {\bibfnamefont {E.}~\bibnamefont {Dragulescu}}, \bibinfo {author}
  {\bibfnamefont {G.}~\bibnamefont {Georgiev}}, \bibinfo {author}
  {\bibfnamefont {J.}~\bibnamefont {Giovinazzo}}, \bibinfo {author}
  {\bibfnamefont {M.}~\bibnamefont {Girod}}, \bibinfo {author} {\bibfnamefont
  {H.}~\bibnamefont {Grawe}}, \bibinfo {author} {\bibfnamefont
  {R.}~\bibnamefont {Grzywacz}}, \bibinfo {author} {\bibfnamefont
  {F.}~\bibnamefont {Hammache}}, \bibinfo {author} {\bibfnamefont
  {F.}~\bibnamefont {Ibrahim}}, \bibinfo {author} {\bibfnamefont
  {M.}~\bibnamefont {Lewitowicz}}, \bibinfo {author} {\bibfnamefont
  {J.}~\bibnamefont {Libert}}, \bibinfo {author} {\bibfnamefont
  {P.}~\bibnamefont {Mayet}}, \bibinfo {author} {\bibfnamefont
  {V.}~\bibnamefont {M{\'{e}}ot}}, \bibinfo {author} {\bibfnamefont
  {F.}~\bibnamefont {Negoita}}, \bibinfo {author} {\bibfnamefont
  {F.}~\bibnamefont {{de Oliveira Santos}}}, \bibinfo {author} {\bibfnamefont
  {O.}~\bibnamefont {Perru}}, \bibinfo {author} {\bibfnamefont
  {O.}~\bibnamefont {Roig}}, \bibinfo {author} {\bibfnamefont {K.}~\bibnamefont
  {Rykaczewski}}, \bibinfo {author} {\bibfnamefont {M.~G.}\ \bibnamefont
  {Saint-Laurent}}, \bibinfo {author} {\bibfnamefont {J.~E.}\ \bibnamefont
  {Sauvestre}}, \bibinfo {author} {\bibfnamefont {O.}~\bibnamefont {Sorlin}},
  \bibinfo {author} {\bibfnamefont {M.}~\bibnamefont {Stanoiu}}, \bibinfo
  {author} {\bibfnamefont {I.}~\bibnamefont {Stefan}}, \bibinfo {author}
  {\bibfnamefont {C.}~\bibnamefont {Stodel}}, \bibinfo {author} {\bibfnamefont
  {C.}~\bibnamefont {Theisen}}, \bibinfo {author} {\bibfnamefont
  {D.}~\bibnamefont {Verney}}, \ and\ \bibinfo {author} {\bibfnamefont
  {J.}~\bibnamefont {{\.{Z}}ylicz}},\ }\href {\doibase
  10.1103/PhysRevC.83.054312} {\bibfield  {journal} {\bibinfo  {journal}
  {Physical Review C}\ }\textbf {\bibinfo {volume} {83}},\ \bibinfo {pages}
  {054312} (\bibinfo {year} {2011})}\BibitemShut {NoStop}%
\bibitem [{\citenamefont {Rother}\ \emph {et~al.}(2011)\citenamefont {Rother},
  \citenamefont {Dewald}, \citenamefont {Iwasaki}, \citenamefont {Lenzi},
  \citenamefont {Starosta}, \citenamefont {Bazin}, \citenamefont {Baugher},
  \citenamefont {Brown}, \citenamefont {Crawford}, \citenamefont {Fransen},
  \citenamefont {Gade}, \citenamefont {Ginter}, \citenamefont {Glasmacher},
  \citenamefont {Grinyer}, \citenamefont {Hackstein}, \citenamefont {Ilie},
  \citenamefont {Jolie}, \citenamefont {McDaniel}, \citenamefont {Miller},
  \citenamefont {Petkov}, \citenamefont {Pissulla}, \citenamefont {Ratkiewicz},
  \citenamefont {Ur}, \citenamefont {Voss}, \citenamefont {Walsh},
  \citenamefont {Weisshaar},\ and\ \citenamefont {Zell}}]{Rother2011}%
  \BibitemOpen
  \bibfield  {author} {\bibinfo {author} {\bibfnamefont {W.}~\bibnamefont
  {Rother}}, \bibinfo {author} {\bibfnamefont {A.}~\bibnamefont {Dewald}},
  \bibinfo {author} {\bibfnamefont {H.}~\bibnamefont {Iwasaki}}, \bibinfo
  {author} {\bibfnamefont {S.~M.}\ \bibnamefont {Lenzi}}, \bibinfo {author}
  {\bibfnamefont {K.}~\bibnamefont {Starosta}}, \bibinfo {author}
  {\bibfnamefont {D.}~\bibnamefont {Bazin}}, \bibinfo {author} {\bibfnamefont
  {T.}~\bibnamefont {Baugher}}, \bibinfo {author} {\bibfnamefont {B.~A.}\
  \bibnamefont {Brown}}, \bibinfo {author} {\bibfnamefont {H.~L.}\ \bibnamefont
  {Crawford}}, \bibinfo {author} {\bibfnamefont {C.}~\bibnamefont {Fransen}},
  \bibinfo {author} {\bibfnamefont {A.}~\bibnamefont {Gade}}, \bibinfo {author}
  {\bibfnamefont {T.~N.}\ \bibnamefont {Ginter}}, \bibinfo {author}
  {\bibfnamefont {T.}~\bibnamefont {Glasmacher}}, \bibinfo {author}
  {\bibfnamefont {G.~F.}\ \bibnamefont {Grinyer}}, \bibinfo {author}
  {\bibfnamefont {M.}~\bibnamefont {Hackstein}}, \bibinfo {author}
  {\bibfnamefont {G.}~\bibnamefont {Ilie}}, \bibinfo {author} {\bibfnamefont
  {J.}~\bibnamefont {Jolie}}, \bibinfo {author} {\bibfnamefont
  {S.}~\bibnamefont {McDaniel}}, \bibinfo {author} {\bibfnamefont
  {D.}~\bibnamefont {Miller}}, \bibinfo {author} {\bibfnamefont
  {P.}~\bibnamefont {Petkov}}, \bibinfo {author} {\bibfnamefont
  {T.}~\bibnamefont {Pissulla}}, \bibinfo {author} {\bibfnamefont
  {A.}~\bibnamefont {Ratkiewicz}}, \bibinfo {author} {\bibfnamefont {C.~A.}\
  \bibnamefont {Ur}}, \bibinfo {author} {\bibfnamefont {P.}~\bibnamefont
  {Voss}}, \bibinfo {author} {\bibfnamefont {K.~A.}\ \bibnamefont {Walsh}},
  \bibinfo {author} {\bibfnamefont {D.}~\bibnamefont {Weisshaar}}, \ and\
  \bibinfo {author} {\bibfnamefont {K.~O.}\ \bibnamefont {Zell}},\ }\href
  {\doibase 10.1103/PhysRevLett.106.022502} {\bibfield  {journal} {\bibinfo
  {journal} {Physical Review Letters}\ }\textbf {\bibinfo {volume} {106}},\
  \bibinfo {pages} {022502} (\bibinfo {year} {2011})}\BibitemShut {NoStop}%
\bibitem [{\citenamefont {Wang}\ \emph {et~al.}(2017)\citenamefont {Wang},
  \citenamefont {Audi}, \citenamefont {Kondev}, \citenamefont {Huang},
  \citenamefont {Naimi},\ and\ \citenamefont {Xu}}]{Wang2017}%
  \BibitemOpen
  \bibfield  {author} {\bibinfo {author} {\bibfnamefont {M.}~\bibnamefont
  {Wang}}, \bibinfo {author} {\bibfnamefont {G.}~\bibnamefont {Audi}}, \bibinfo
  {author} {\bibfnamefont {F.~G.}\ \bibnamefont {Kondev}}, \bibinfo {author}
  {\bibfnamefont {W.}~\bibnamefont {Huang}}, \bibinfo {author} {\bibfnamefont
  {S.}~\bibnamefont {Naimi}}, \ and\ \bibinfo {author} {\bibfnamefont
  {X.}~\bibnamefont {Xu}},\ }\href {\doibase 10.1088/1674-1137/41/3/030003}
  {\bibfield  {journal} {\bibinfo  {journal} {Chinese Physics C}\ }\textbf
  {\bibinfo {volume} {41}},\ \bibinfo {pages} {030003} (\bibinfo {year}
  {2017})}\BibitemShut {NoStop}%
\bibitem [{\citenamefont {NNDC}()}]{logft}%
  \BibitemOpen
  \bibfield  {author} {\bibinfo {author} {\bibnamefont {NNDC}},\ }\href@noop {}
  {}\bibinfo {note} {\url{http://www.nndc.bnl.gov/logft/}}\BibitemShut
  {NoStop}%
\bibitem [{\citenamefont {Hardy}\ \emph {et~al.}(1977)\citenamefont {Hardy},
  \citenamefont {Carraz}, \citenamefont {Jonson},\ and\ \citenamefont
  {Hansen}}]{Hardy1977}%
  \BibitemOpen
  \bibfield  {author} {\bibinfo {author} {\bibfnamefont {J.}~\bibnamefont
  {Hardy}}, \bibinfo {author} {\bibfnamefont {L.}~\bibnamefont {Carraz}},
  \bibinfo {author} {\bibfnamefont {B.}~\bibnamefont {Jonson}}, \ and\ \bibinfo
  {author} {\bibfnamefont {P.}~\bibnamefont {Hansen}},\ }\href {\doibase
  10.1016/0370-2693(77)90223-4} {\bibfield  {journal} {\bibinfo  {journal}
  {Physics Letters B}\ }\textbf {\bibinfo {volume} {71}},\ \bibinfo {pages}
  {307} (\bibinfo {year} {1977})}\BibitemShut {NoStop}%
\bibitem [{\citenamefont {Olaizola}\ \emph {et~al.}(2013)\citenamefont
  {Olaizola}, \citenamefont {Fraile}, \citenamefont {Mach}, \citenamefont
  {Aprahamian}, \citenamefont {Briz}, \citenamefont {Cal-Gonz{\'{a}}lez},
  \citenamefont {Ghiţa}, \citenamefont {K{\"{o}}ster}, \citenamefont
  {Kurcewicz}, \citenamefont {Lesher}, \citenamefont {Pauwels}, \citenamefont
  {Picado}, \citenamefont {Poves}, \citenamefont {Radulov}, \citenamefont
  {Simpson},\ and\ \citenamefont {Ud{\'{i}}as}}]{Olaizola2013}%
  \BibitemOpen
  \bibfield  {author} {\bibinfo {author} {\bibfnamefont {B.}~\bibnamefont
  {Olaizola}}, \bibinfo {author} {\bibfnamefont {L.~M.}\ \bibnamefont
  {Fraile}}, \bibinfo {author} {\bibfnamefont {H.}~\bibnamefont {Mach}},
  \bibinfo {author} {\bibfnamefont {A.}~\bibnamefont {Aprahamian}}, \bibinfo
  {author} {\bibfnamefont {J.~A.}\ \bibnamefont {Briz}}, \bibinfo {author}
  {\bibfnamefont {J.}~\bibnamefont {Cal-Gonz{\'{a}}lez}}, \bibinfo {author}
  {\bibfnamefont {D.}~\bibnamefont {Ghiţa}}, \bibinfo {author} {\bibfnamefont
  {U.}~\bibnamefont {K{\"{o}}ster}}, \bibinfo {author} {\bibfnamefont
  {W.}~\bibnamefont {Kurcewicz}}, \bibinfo {author} {\bibfnamefont {S.~R.}\
  \bibnamefont {Lesher}}, \bibinfo {author} {\bibfnamefont {D.}~\bibnamefont
  {Pauwels}}, \bibinfo {author} {\bibfnamefont {E.}~\bibnamefont {Picado}},
  \bibinfo {author} {\bibfnamefont {A.}~\bibnamefont {Poves}}, \bibinfo
  {author} {\bibfnamefont {D.}~\bibnamefont {Radulov}}, \bibinfo {author}
  {\bibfnamefont {G.~S.}\ \bibnamefont {Simpson}}, \ and\ \bibinfo {author}
  {\bibfnamefont {J.~M.}\ \bibnamefont {Ud{\'{i}}as}},\ }\href {\doibase
  10.1103/PhysRevC.88.044306} {\bibfield  {journal} {\bibinfo  {journal}
  {Physical Review C}\ }\textbf {\bibinfo {volume} {88}},\ \bibinfo {pages}
  {044306} (\bibinfo {year} {2013})}\BibitemShut {NoStop}%
\bibitem [{\citenamefont {Georgiev}(2001)}]{Georgiev2001}%
  \BibitemOpen
  \bibfield  {author} {\bibinfo {author} {\bibfnamefont {G.}~\bibnamefont
  {Georgiev}},\ }\emph {\bibinfo {title} {{Magnetic Moments of Isomers and
  Ground States of Exotic Nuclei Produced by Projectile Fragmentation}}},\
  \href
  {https://fys.kuleuven.be/iks/nm/files/thesis/thesis_Georgie_Georgiev.pdf}
  {Ph.D. thesis},\ \bibinfo  {school} {KU Leuven} (\bibinfo {year}
  {2001})\BibitemShut {NoStop}%
\bibitem [{\citenamefont {Daugas}\ \emph {et~al.}(2010)\citenamefont {Daugas},
  \citenamefont {Faul}, \citenamefont {Grawe}, \citenamefont {Pf{\"{u}}tzner},
  \citenamefont {Grzywacz}, \citenamefont {Lewitowicz}, \citenamefont
  {Achouri}, \citenamefont {Ang{\'{e}}lique}, \citenamefont {Baiborodin},
  \citenamefont {Bentida}, \citenamefont {B{\'{e}}raud}, \citenamefont
  {Borcea}, \citenamefont {Bingham}, \citenamefont {Catford}, \citenamefont
  {Emsallem}, \citenamefont {de~France}, \citenamefont {Grzywacz},
  \citenamefont {Lemmon}, \citenamefont {{Lopez Jimenez}}, \citenamefont {{de
  Oliveira Santos}}, \citenamefont {Regan}, \citenamefont {Rykaczewski},
  \citenamefont {Sauvestre}, \citenamefont {Sawicka}, \citenamefont {Stanoiu},
  \citenamefont {Sieja},\ and\ \citenamefont {Nowacki}}]{Daugas2010}%
  \BibitemOpen
  \bibfield  {author} {\bibinfo {author} {\bibfnamefont {J.~M.}\ \bibnamefont
  {Daugas}}, \bibinfo {author} {\bibfnamefont {T.}~\bibnamefont {Faul}},
  \bibinfo {author} {\bibfnamefont {H.}~\bibnamefont {Grawe}}, \bibinfo
  {author} {\bibfnamefont {M.}~\bibnamefont {Pf{\"{u}}tzner}}, \bibinfo
  {author} {\bibfnamefont {R.}~\bibnamefont {Grzywacz}}, \bibinfo {author}
  {\bibfnamefont {M.}~\bibnamefont {Lewitowicz}}, \bibinfo {author}
  {\bibfnamefont {N.~L.}\ \bibnamefont {Achouri}}, \bibinfo {author}
  {\bibfnamefont {J.~C.}\ \bibnamefont {Ang{\'{e}}lique}}, \bibinfo {author}
  {\bibfnamefont {D.}~\bibnamefont {Baiborodin}}, \bibinfo {author}
  {\bibfnamefont {R.}~\bibnamefont {Bentida}}, \bibinfo {author} {\bibfnamefont
  {R.}~\bibnamefont {B{\'{e}}raud}}, \bibinfo {author} {\bibfnamefont
  {C.}~\bibnamefont {Borcea}}, \bibinfo {author} {\bibfnamefont {C.~R.}\
  \bibnamefont {Bingham}}, \bibinfo {author} {\bibfnamefont {W.~N.}\
  \bibnamefont {Catford}}, \bibinfo {author} {\bibfnamefont {A.}~\bibnamefont
  {Emsallem}}, \bibinfo {author} {\bibfnamefont {G.}~\bibnamefont {de~France}},
  \bibinfo {author} {\bibfnamefont {K.~L.}\ \bibnamefont {Grzywacz}}, \bibinfo
  {author} {\bibfnamefont {R.~C.}\ \bibnamefont {Lemmon}}, \bibinfo {author}
  {\bibfnamefont {M.~J.}\ \bibnamefont {{Lopez Jimenez}}}, \bibinfo {author}
  {\bibfnamefont {F.}~\bibnamefont {{de Oliveira Santos}}}, \bibinfo {author}
  {\bibfnamefont {P.~H.}\ \bibnamefont {Regan}}, \bibinfo {author}
  {\bibfnamefont {K.}~\bibnamefont {Rykaczewski}}, \bibinfo {author}
  {\bibfnamefont {J.~E.}\ \bibnamefont {Sauvestre}}, \bibinfo {author}
  {\bibfnamefont {M.}~\bibnamefont {Sawicka}}, \bibinfo {author} {\bibfnamefont
  {M.}~\bibnamefont {Stanoiu}}, \bibinfo {author} {\bibfnamefont
  {K.}~\bibnamefont {Sieja}}, \ and\ \bibinfo {author} {\bibfnamefont
  {F.}~\bibnamefont {Nowacki}},\ }\href {\doibase 10.1103/PhysRevC.81.034304}
  {\bibfield  {journal} {\bibinfo  {journal} {Physical Review C}\ }\textbf
  {\bibinfo {volume} {81}},\ \bibinfo {pages} {034304} (\bibinfo {year}
  {2010})}\BibitemShut {NoStop}%
\bibitem [{\citenamefont {Ivanov}(2007)}]{Ivanov2007}%
  \BibitemOpen
  \bibfield  {author} {\bibinfo {author} {\bibfnamefont {O.~V.}\ \bibnamefont
  {Ivanov}},\ }\emph {\bibinfo {title} {{Decay of 66Fe studied with a new
  $\beta$-$\gamma$-detection set-up at LISOL}}},\ \href
  {https://fys.kuleuven.be/iks/ns/files/thesis/thesisoi.pdf} {Ph.D. thesis},\
  \bibinfo  {school} {KU Leuven} (\bibinfo {year} {2007})\BibitemShut {NoStop}%
\bibitem [{\citenamefont {Pauwels}\ \emph {et~al.}(2009)\citenamefont
  {Pauwels}, \citenamefont {Ivanov}, \citenamefont {Bree}, \citenamefont
  {B{\"{u}}scher}, \citenamefont {Cocolios}, \citenamefont {Huyse},
  \citenamefont {Kudryavtsev}, \citenamefont {Raabe}, \citenamefont {Sawicka},
  \citenamefont {{Van de Walle}}, \citenamefont {{Van Duppen}}, \citenamefont
  {Korgul}, \citenamefont {Stefanescu}, \citenamefont {Hecht}, \citenamefont
  {Hoteling}, \citenamefont {W{\"{o}}hr}, \citenamefont {Walters},
  \citenamefont {Broda}, \citenamefont {Fornal}, \citenamefont {Krolas},
  \citenamefont {Pawlat}, \citenamefont {Wrzesinski}, \citenamefont
  {Carpenter}, \citenamefont {Janssens}, \citenamefont {Lauritsen},
  \citenamefont {Seweryniak}, \citenamefont {Zhu}, \citenamefont {Stone},\ and\
  \citenamefont {Wang}}]{Pauwels2009}%
  \BibitemOpen
  \bibfield  {author} {\bibinfo {author} {\bibfnamefont {D.}~\bibnamefont
  {Pauwels}}, \bibinfo {author} {\bibfnamefont {O.}~\bibnamefont {Ivanov}},
  \bibinfo {author} {\bibfnamefont {N.}~\bibnamefont {Bree}}, \bibinfo {author}
  {\bibfnamefont {J.}~\bibnamefont {B{\"{u}}scher}}, \bibinfo {author}
  {\bibfnamefont {T.~E.}\ \bibnamefont {Cocolios}}, \bibinfo {author}
  {\bibfnamefont {M.}~\bibnamefont {Huyse}}, \bibinfo {author} {\bibfnamefont
  {Y.}~\bibnamefont {Kudryavtsev}}, \bibinfo {author} {\bibfnamefont
  {R.}~\bibnamefont {Raabe}}, \bibinfo {author} {\bibfnamefont
  {M.}~\bibnamefont {Sawicka}}, \bibinfo {author} {\bibfnamefont
  {J.}~\bibnamefont {{Van de Walle}}}, \bibinfo {author} {\bibfnamefont
  {P.}~\bibnamefont {{Van Duppen}}}, \bibinfo {author} {\bibfnamefont
  {A.}~\bibnamefont {Korgul}}, \bibinfo {author} {\bibfnamefont
  {I.}~\bibnamefont {Stefanescu}}, \bibinfo {author} {\bibfnamefont {A.~A.}\
  \bibnamefont {Hecht}}, \bibinfo {author} {\bibfnamefont {N.}~\bibnamefont
  {Hoteling}}, \bibinfo {author} {\bibfnamefont {A.}~\bibnamefont
  {W{\"{o}}hr}}, \bibinfo {author} {\bibfnamefont {W.~B.}\ \bibnamefont
  {Walters}}, \bibinfo {author} {\bibfnamefont {R.}~\bibnamefont {Broda}},
  \bibinfo {author} {\bibfnamefont {B.}~\bibnamefont {Fornal}}, \bibinfo
  {author} {\bibfnamefont {W.}~\bibnamefont {Krolas}}, \bibinfo {author}
  {\bibfnamefont {T.}~\bibnamefont {Pawlat}}, \bibinfo {author} {\bibfnamefont
  {J.}~\bibnamefont {Wrzesinski}}, \bibinfo {author} {\bibfnamefont {M.~P.}\
  \bibnamefont {Carpenter}}, \bibinfo {author} {\bibfnamefont {R.~V.~F.}\
  \bibnamefont {Janssens}}, \bibinfo {author} {\bibfnamefont {T.}~\bibnamefont
  {Lauritsen}}, \bibinfo {author} {\bibfnamefont {D.}~\bibnamefont
  {Seweryniak}}, \bibinfo {author} {\bibfnamefont {S.}~\bibnamefont {Zhu}},
  \bibinfo {author} {\bibfnamefont {J.~R.}\ \bibnamefont {Stone}}, \ and\
  \bibinfo {author} {\bibfnamefont {X.}~\bibnamefont {Wang}},\ }\href {\doibase
  10.1103/PhysRevC.79.044309} {\bibfield  {journal} {\bibinfo  {journal}
  {Physical Review C}\ }\textbf {\bibinfo {volume} {79}},\ \bibinfo {pages}
  {044309} (\bibinfo {year} {2009})}\BibitemShut {NoStop}%
\bibitem [{\citenamefont {Broda}\ \emph {et~al.}(2012)\citenamefont {Broda},
  \citenamefont {Paw{\l}at}, \citenamefont {Kr{\'{o}}las}, \citenamefont
  {Janssens}, \citenamefont {Zhu}, \citenamefont {Walters}, \citenamefont
  {Fornal}, \citenamefont {Chiara}, \citenamefont {Carpenter}, \citenamefont
  {Hoteling}, \citenamefont {Iskra}, \citenamefont {Kondev}, \citenamefont
  {Lauritsen}, \citenamefont {Seweryniak}, \citenamefont {Stefanescu},
  \citenamefont {Wang},\ and\ \citenamefont {Wrzesi{\'{n}}ski}}]{Broda2012}%
  \BibitemOpen
  \bibfield  {author} {\bibinfo {author} {\bibfnamefont {R.}~\bibnamefont
  {Broda}}, \bibinfo {author} {\bibfnamefont {T.}~\bibnamefont {Paw{\l}at}},
  \bibinfo {author} {\bibfnamefont {W.}~\bibnamefont {Kr{\'{o}}las}}, \bibinfo
  {author} {\bibfnamefont {R.~V.~F.}\ \bibnamefont {Janssens}}, \bibinfo
  {author} {\bibfnamefont {S.}~\bibnamefont {Zhu}}, \bibinfo {author}
  {\bibfnamefont {W.~B.}\ \bibnamefont {Walters}}, \bibinfo {author}
  {\bibfnamefont {B.}~\bibnamefont {Fornal}}, \bibinfo {author} {\bibfnamefont
  {C.~J.}\ \bibnamefont {Chiara}}, \bibinfo {author} {\bibfnamefont {M.~P.}\
  \bibnamefont {Carpenter}}, \bibinfo {author} {\bibfnamefont {N.}~\bibnamefont
  {Hoteling}}, \bibinfo {author} {\bibfnamefont {{\L}.~W.}\ \bibnamefont
  {Iskra}}, \bibinfo {author} {\bibfnamefont {F.~G.}\ \bibnamefont {Kondev}},
  \bibinfo {author} {\bibfnamefont {T.}~\bibnamefont {Lauritsen}}, \bibinfo
  {author} {\bibfnamefont {D.}~\bibnamefont {Seweryniak}}, \bibinfo {author}
  {\bibfnamefont {I.}~\bibnamefont {Stefanescu}}, \bibinfo {author}
  {\bibfnamefont {X.}~\bibnamefont {Wang}}, \ and\ \bibinfo {author}
  {\bibfnamefont {J.}~\bibnamefont {Wrzesi{\'{n}}ski}},\ }\href {\doibase
  10.1103/PhysRevC.86.064312} {\bibfield  {journal} {\bibinfo  {journal}
  {Physical Review C}\ }\textbf {\bibinfo {volume} {86}},\ \bibinfo {pages}
  {064312} (\bibinfo {year} {2012})}\BibitemShut {NoStop}%
\bibitem [{\citenamefont {Chiara}\ \emph {et~al.}(2012)\citenamefont {Chiara},
  \citenamefont {Broda}, \citenamefont {Walters}, \citenamefont {Janssens},
  \citenamefont {Albers}, \citenamefont {Alcorta}, \citenamefont {Bertone},
  \citenamefont {Carpenter}, \citenamefont {Hoffman}, \citenamefont
  {Lauritsen}, \citenamefont {Rogers}, \citenamefont {Seweryniak},
  \citenamefont {Zhu}, \citenamefont {Kondev}, \citenamefont {Fornal},
  \citenamefont {Kr{\'{o}}las}, \citenamefont {Wrzesi{\'{n}}ski}, \citenamefont
  {Larson}, \citenamefont {Liddick}, \citenamefont {Prokop}, \citenamefont
  {Suchyta}, \citenamefont {David},\ and\ \citenamefont
  {Doherty}}]{Chiara2012}%
  \BibitemOpen
  \bibfield  {author} {\bibinfo {author} {\bibfnamefont {C.~J.}\ \bibnamefont
  {Chiara}}, \bibinfo {author} {\bibfnamefont {R.}~\bibnamefont {Broda}},
  \bibinfo {author} {\bibfnamefont {W.~B.}\ \bibnamefont {Walters}}, \bibinfo
  {author} {\bibfnamefont {R.~V.~F.}\ \bibnamefont {Janssens}}, \bibinfo
  {author} {\bibfnamefont {M.}~\bibnamefont {Albers}}, \bibinfo {author}
  {\bibfnamefont {M.}~\bibnamefont {Alcorta}}, \bibinfo {author} {\bibfnamefont
  {P.~F.}\ \bibnamefont {Bertone}}, \bibinfo {author} {\bibfnamefont {M.~P.}\
  \bibnamefont {Carpenter}}, \bibinfo {author} {\bibfnamefont {C.~R.}\
  \bibnamefont {Hoffman}}, \bibinfo {author} {\bibfnamefont {T.}~\bibnamefont
  {Lauritsen}}, \bibinfo {author} {\bibfnamefont {A.~M.}\ \bibnamefont
  {Rogers}}, \bibinfo {author} {\bibfnamefont {D.}~\bibnamefont {Seweryniak}},
  \bibinfo {author} {\bibfnamefont {S.}~\bibnamefont {Zhu}}, \bibinfo {author}
  {\bibfnamefont {F.~G.}\ \bibnamefont {Kondev}}, \bibinfo {author}
  {\bibfnamefont {B.}~\bibnamefont {Fornal}}, \bibinfo {author} {\bibfnamefont
  {W.}~\bibnamefont {Kr{\'{o}}las}}, \bibinfo {author} {\bibfnamefont
  {J.}~\bibnamefont {Wrzesi{\'{n}}ski}}, \bibinfo {author} {\bibfnamefont
  {N.}~\bibnamefont {Larson}}, \bibinfo {author} {\bibfnamefont {S.~N.}\
  \bibnamefont {Liddick}}, \bibinfo {author} {\bibfnamefont {C.}~\bibnamefont
  {Prokop}}, \bibinfo {author} {\bibfnamefont {S.}~\bibnamefont {Suchyta}},
  \bibinfo {author} {\bibfnamefont {H.~M.}\ \bibnamefont {David}}, \ and\
  \bibinfo {author} {\bibfnamefont {D.~T.}\ \bibnamefont {Doherty}},\ }\href
  {\doibase 10.1103/PhysRevC.86.041304} {\bibfield  {journal} {\bibinfo
  {journal} {Physical Review C}\ }\textbf {\bibinfo {volume} {86}},\ \bibinfo
  {pages} {041304} (\bibinfo {year} {2012})}\BibitemShut {NoStop}%
\bibitem [{\citenamefont {Chiara}\ \emph {et~al.}(2013)\citenamefont {Chiara},
  \citenamefont {Walters}, \citenamefont {Janssens}, \citenamefont {Broda},
  \citenamefont {Albers}, \citenamefont {Alcorta}, \citenamefont {Bertone},
  \citenamefont {Carpenter}, \citenamefont {Hoffman}, \citenamefont
  {Lauritsen}, \citenamefont {Rogers}, \citenamefont {Seweryniak},
  \citenamefont {Zhu}, \citenamefont {Kondev}, \citenamefont {Fornal},
  \citenamefont {Kr{\'{o}}las}, \citenamefont {Wrzesi{\'{n}}ski}, \citenamefont
  {Larson}, \citenamefont {Liddick}, \citenamefont {Prokop}, \citenamefont
  {Suchyta}, \citenamefont {David},\ and\ \citenamefont
  {Doherty}}]{Chiara2013}%
  \BibitemOpen
  \bibfield  {author} {\bibinfo {author} {\bibfnamefont {C.}~\bibnamefont
  {Chiara}}, \bibinfo {author} {\bibfnamefont {W.}~\bibnamefont {Walters}},
  \bibinfo {author} {\bibfnamefont {R.}~\bibnamefont {Janssens}}, \bibinfo
  {author} {\bibfnamefont {R.}~\bibnamefont {Broda}}, \bibinfo {author}
  {\bibfnamefont {M.}~\bibnamefont {Albers}}, \bibinfo {author} {\bibfnamefont
  {M.}~\bibnamefont {Alcorta}}, \bibinfo {author} {\bibfnamefont
  {P.}~\bibnamefont {Bertone}}, \bibinfo {author} {\bibfnamefont
  {M.}~\bibnamefont {Carpenter}}, \bibinfo {author} {\bibfnamefont
  {C.}~\bibnamefont {Hoffman}}, \bibinfo {author} {\bibfnamefont
  {T.}~\bibnamefont {Lauritsen}}, \bibinfo {author} {\bibfnamefont
  {A.}~\bibnamefont {Rogers}}, \bibinfo {author} {\bibfnamefont
  {D.}~\bibnamefont {Seweryniak}}, \bibinfo {author} {\bibfnamefont
  {S.}~\bibnamefont {Zhu}}, \bibinfo {author} {\bibfnamefont {F.}~\bibnamefont
  {Kondev}}, \bibinfo {author} {\bibfnamefont {B.}~\bibnamefont {Fornal}},
  \bibinfo {author} {\bibfnamefont {W.}~\bibnamefont {Kr{\'{o}}las}}, \bibinfo
  {author} {\bibfnamefont {J.}~\bibnamefont {Wrzesi{\'{n}}ski}}, \bibinfo
  {author} {\bibfnamefont {N.}~\bibnamefont {Larson}}, \bibinfo {author}
  {\bibfnamefont {S.}~\bibnamefont {Liddick}}, \bibinfo {author} {\bibfnamefont
  {C.}~\bibnamefont {Prokop}}, \bibinfo {author} {\bibfnamefont
  {S.}~\bibnamefont {Suchyta}}, \bibinfo {author} {\bibfnamefont
  {H.}~\bibnamefont {David}}, \ and\ \bibinfo {author} {\bibfnamefont
  {D.}~\bibnamefont {Doherty}},\ }\href {\doibase 10.5506/APhysPolB.44.371}
  {\bibfield  {journal} {\bibinfo  {journal} {Acta Physica Polonica B}\
  }\textbf {\bibinfo {volume} {44}},\ \bibinfo {pages} {371} (\bibinfo {year}
  {2013})}\BibitemShut {NoStop}%
\bibitem [{\citenamefont {Chiara}()}]{ChiaraPrivate}%
  \BibitemOpen
  \bibfield  {author} {\bibinfo {author} {\bibfnamefont {C.~J.}\ \bibnamefont
  {Chiara}},\ }\href@noop {} {}\bibinfo {note} {Priv. comm.}\BibitemShut
  {Stop}%
\bibitem [{\citenamefont {Darcey}\ \emph {et~al.}(1971)\citenamefont {Darcey},
  \citenamefont {Chapman},\ and\ \citenamefont {Hinds}}]{Darcey1971}%
  \BibitemOpen
  \bibfield  {author} {\bibinfo {author} {\bibfnamefont {W.}~\bibnamefont
  {Darcey}}, \bibinfo {author} {\bibfnamefont {R.}~\bibnamefont {Chapman}}, \
  and\ \bibinfo {author} {\bibfnamefont {S.}~\bibnamefont {Hinds}},\ }\href
  {\doibase 10.1016/0375-9474(71)90635-X} {\bibfield  {journal} {\bibinfo
  {journal} {Nuclear Physics A}\ }\textbf {\bibinfo {volume} {170}},\ \bibinfo
  {pages} {253} (\bibinfo {year} {1971})}\BibitemShut {NoStop}%
\bibitem [{\citenamefont {Bernas}\ \emph {et~al.}(1981)\citenamefont {Bernas},
  \citenamefont {Peng}, \citenamefont {Doubre}, \citenamefont {Langevin},
  \citenamefont {{Le Vine}}, \citenamefont {Pougheon},\ and\ \citenamefont
  {Roussel}}]{Bernas1981}%
  \BibitemOpen
  \bibfield  {author} {\bibinfo {author} {\bibfnamefont {M.}~\bibnamefont
  {Bernas}}, \bibinfo {author} {\bibfnamefont {J.~C.}\ \bibnamefont {Peng}},
  \bibinfo {author} {\bibfnamefont {H.}~\bibnamefont {Doubre}}, \bibinfo
  {author} {\bibfnamefont {M.}~\bibnamefont {Langevin}}, \bibinfo {author}
  {\bibfnamefont {M.~J.}\ \bibnamefont {{Le Vine}}}, \bibinfo {author}
  {\bibfnamefont {F.}~\bibnamefont {Pougheon}}, \ and\ \bibinfo {author}
  {\bibfnamefont {P.}~\bibnamefont {Roussel}},\ }\href {\doibase
  10.1103/PhysRevC.24.756} {\bibfield  {journal} {\bibinfo  {journal} {Physical
  Review C}\ }\textbf {\bibinfo {volume} {24}},\ \bibinfo {pages} {756}
  (\bibinfo {year} {1981})}\BibitemShut {NoStop}%
\bibitem [{\citenamefont {Bosch}\ \emph {et~al.}(1988)\citenamefont {Bosch},
  \citenamefont {Schmidt-Ott}, \citenamefont {Runte}, \citenamefont
  {Tidemand-Petersson}, \citenamefont {Koschel}, \citenamefont {Meissner},
  \citenamefont {Kirchner}, \citenamefont {Klepper}, \citenamefont {Roeckl},
  \citenamefont {Rykaczewski},\ and\ \citenamefont {Schardt}}]{Bosch1988}%
  \BibitemOpen
  \bibfield  {author} {\bibinfo {author} {\bibfnamefont {U.}~\bibnamefont
  {Bosch}}, \bibinfo {author} {\bibfnamefont {W.-D.}\ \bibnamefont
  {Schmidt-Ott}}, \bibinfo {author} {\bibfnamefont {E.}~\bibnamefont {Runte}},
  \bibinfo {author} {\bibfnamefont {P.}~\bibnamefont {Tidemand-Petersson}},
  \bibinfo {author} {\bibfnamefont {P.}~\bibnamefont {Koschel}}, \bibinfo
  {author} {\bibfnamefont {F.}~\bibnamefont {Meissner}}, \bibinfo {author}
  {\bibfnamefont {R.}~\bibnamefont {Kirchner}}, \bibinfo {author}
  {\bibfnamefont {O.}~\bibnamefont {Klepper}}, \bibinfo {author} {\bibfnamefont
  {E.}~\bibnamefont {Roeckl}}, \bibinfo {author} {\bibfnamefont
  {K.}~\bibnamefont {Rykaczewski}}, \ and\ \bibinfo {author} {\bibfnamefont
  {D.}~\bibnamefont {Schardt}},\ }\href {\doibase 10.1016/0375-9474(88)90362-4}
  {\bibfield  {journal} {\bibinfo  {journal} {Nuclear Physics A}\ }\textbf
  {\bibinfo {volume} {477}},\ \bibinfo {pages} {89} (\bibinfo {year}
  {1988})}\BibitemShut {NoStop}%
\bibitem [{\citenamefont {Fister}\ \emph {et~al.}(1990)\citenamefont {Fister},
  \citenamefont {Jahn}, \citenamefont {von Neumann-Cosel}, \citenamefont
  {Schenk}, \citenamefont {Trelle}, \citenamefont {Wenzel},\ and\ \citenamefont
  {Wienands}}]{Fister1990}%
  \BibitemOpen
  \bibfield  {author} {\bibinfo {author} {\bibfnamefont {U.}~\bibnamefont
  {Fister}}, \bibinfo {author} {\bibfnamefont {R.}~\bibnamefont {Jahn}},
  \bibinfo {author} {\bibfnamefont {P.}~\bibnamefont {von Neumann-Cosel}},
  \bibinfo {author} {\bibfnamefont {P.}~\bibnamefont {Schenk}}, \bibinfo
  {author} {\bibfnamefont {T.~K.}\ \bibnamefont {Trelle}}, \bibinfo {author}
  {\bibfnamefont {D.}~\bibnamefont {Wenzel}}, \ and\ \bibinfo {author}
  {\bibfnamefont {U.}~\bibnamefont {Wienands}},\ }\href {\doibase
  10.1103/PhysRevC.42.2375} {\bibfield  {journal} {\bibinfo  {journal}
  {Physical Review C}\ }\textbf {\bibinfo {volume} {42}},\ \bibinfo {pages}
  {2375} (\bibinfo {year} {1990})}\BibitemShut {NoStop}%
\bibitem [{\citenamefont {Paw{\l}at}\ \emph {et~al.}(1994)\citenamefont
  {Paw{\l}at}, \citenamefont {Broda}, \citenamefont {Kr{\'{o}}las},
  \citenamefont {Maj}, \citenamefont {Ziȩbli{\'{n}}ski}, \citenamefont
  {Grawe}, \citenamefont {Schubart}, \citenamefont {Maier}, \citenamefont
  {Heese}, \citenamefont {Kluge},\ and\ \citenamefont {Schramm}}]{Pawat1994}%
  \BibitemOpen
  \bibfield  {author} {\bibinfo {author} {\bibfnamefont {T.}~\bibnamefont
  {Paw{\l}at}}, \bibinfo {author} {\bibfnamefont {R.}~\bibnamefont {Broda}},
  \bibinfo {author} {\bibfnamefont {W.}~\bibnamefont {Kr{\'{o}}las}}, \bibinfo
  {author} {\bibfnamefont {A.}~\bibnamefont {Maj}}, \bibinfo {author}
  {\bibfnamefont {M.}~\bibnamefont {Ziȩbli{\'{n}}ski}}, \bibinfo {author}
  {\bibfnamefont {H.}~\bibnamefont {Grawe}}, \bibinfo {author} {\bibfnamefont
  {R.}~\bibnamefont {Schubart}}, \bibinfo {author} {\bibfnamefont
  {K.}~\bibnamefont {Maier}}, \bibinfo {author} {\bibfnamefont
  {J.}~\bibnamefont {Heese}}, \bibinfo {author} {\bibfnamefont
  {H.}~\bibnamefont {Kluge}}, \ and\ \bibinfo {author} {\bibfnamefont
  {M.}~\bibnamefont {Schramm}},\ }\href {\doibase 10.1016/0375-9474(94)90247-X}
  {\bibfield  {journal} {\bibinfo  {journal} {Nuclear Physics A}\ }\textbf
  {\bibinfo {volume} {574}},\ \bibinfo {pages} {623} (\bibinfo {year}
  {1994})}\BibitemShut {NoStop}%
\bibitem [{\citenamefont {Ishii}\ \emph {et~al.}(1997)\citenamefont {Ishii},
  \citenamefont {Itoh}, \citenamefont {Ishii}, \citenamefont {Makishima},
  \citenamefont {Ogawa}, \citenamefont {Hossain}, \citenamefont {Hayakawa},\
  and\ \citenamefont {Kohno}}]{Ishii1997}%
  \BibitemOpen
  \bibfield  {author} {\bibinfo {author} {\bibfnamefont {T.}~\bibnamefont
  {Ishii}}, \bibinfo {author} {\bibfnamefont {M.}~\bibnamefont {Itoh}},
  \bibinfo {author} {\bibfnamefont {M.}~\bibnamefont {Ishii}}, \bibinfo
  {author} {\bibfnamefont {A.}~\bibnamefont {Makishima}}, \bibinfo {author}
  {\bibfnamefont {M.}~\bibnamefont {Ogawa}}, \bibinfo {author} {\bibfnamefont
  {I.}~\bibnamefont {Hossain}}, \bibinfo {author} {\bibfnamefont
  {T.}~\bibnamefont {Hayakawa}}, \ and\ \bibinfo {author} {\bibfnamefont
  {T.}~\bibnamefont {Kohno}},\ }\href {\doibase 10.1016/S0168-9002(97)00720-1}
  {\bibfield  {journal} {\bibinfo  {journal} {Nuclear Instruments and Methods
  in Physics Research Section A: Accelerators, Spectrometers, Detectors and
  Associated Equipment}\ }\textbf {\bibinfo {volume} {395}},\ \bibinfo {pages}
  {210} (\bibinfo {year} {1997})}\BibitemShut {NoStop}%
\bibitem [{\citenamefont {Browne}\ and\ \citenamefont
  {Tuli}(2010{\natexlab{a}})}]{Browne2010}%
  \BibitemOpen
  \bibfield  {author} {\bibinfo {author} {\bibfnamefont {E.}~\bibnamefont
  {Browne}}\ and\ \bibinfo {author} {\bibfnamefont {J.}~\bibnamefont {Tuli}},\
  }\href {\doibase 10.1016/j.nds.2010.03.004} {\bibfield  {journal} {\bibinfo
  {journal} {Nuclear Data Sheets}\ }\textbf {\bibinfo {volume} {111}},\
  \bibinfo {pages} {1093} (\bibinfo {year} {2010}{\natexlab{a}})}\BibitemShut
  {NoStop}%
\bibitem [{\citenamefont {Browne}\ and\ \citenamefont
  {Tuli}(2010{\natexlab{b}})}]{Browne2010a}%
  \BibitemOpen
  \bibfield  {author} {\bibinfo {author} {\bibfnamefont {E.}~\bibnamefont
  {Browne}}\ and\ \bibinfo {author} {\bibfnamefont {J.}~\bibnamefont {Tuli}},\
  }\href {\doibase 10.1016/j.nds.2010.09.002} {\bibfield  {journal} {\bibinfo
  {journal} {Nuclear Data Sheets}\ }\textbf {\bibinfo {volume} {111}},\
  \bibinfo {pages} {2425} (\bibinfo {year} {2010}{\natexlab{b}})}\BibitemShut
  {NoStop}%
\bibitem [{\citenamefont {Ameil}\ \emph {et~al.}(1998)\citenamefont {Ameil},
  \citenamefont {Bernas}, \citenamefont {Armbruster}, \citenamefont
  {Czajkowski}, \citenamefont {Dessagne}, \citenamefont {Geissel},
  \citenamefont {Hanelt}, \citenamefont {Kozhuharov}, \citenamefont {Miehe},
  \citenamefont {Donzaud}, \citenamefont {Grewe}, \citenamefont {Heinz},
  \citenamefont {Janas}, \citenamefont {de~Jong}, \citenamefont {Schwab},\ and\
  \citenamefont {Steinh{\"{a}}user}}]{Ameil1998}%
  \BibitemOpen
  \bibfield  {author} {\bibinfo {author} {\bibfnamefont {F.}~\bibnamefont
  {Ameil}}, \bibinfo {author} {\bibfnamefont {M.}~\bibnamefont {Bernas}},
  \bibinfo {author} {\bibfnamefont {P.}~\bibnamefont {Armbruster}}, \bibinfo
  {author} {\bibfnamefont {S.}~\bibnamefont {Czajkowski}}, \bibinfo {author}
  {\bibfnamefont {P.}~\bibnamefont {Dessagne}}, \bibinfo {author}
  {\bibfnamefont {H.}~\bibnamefont {Geissel}}, \bibinfo {author} {\bibfnamefont
  {E.}~\bibnamefont {Hanelt}}, \bibinfo {author} {\bibfnamefont
  {C.}~\bibnamefont {Kozhuharov}}, \bibinfo {author} {\bibfnamefont
  {C.}~\bibnamefont {Miehe}}, \bibinfo {author} {\bibfnamefont
  {C.}~\bibnamefont {Donzaud}}, \bibinfo {author} {\bibfnamefont
  {A.}~\bibnamefont {Grewe}}, \bibinfo {author} {\bibfnamefont
  {A.}~\bibnamefont {Heinz}}, \bibinfo {author} {\bibfnamefont
  {Z.}~\bibnamefont {Janas}}, \bibinfo {author} {\bibfnamefont
  {M.}~\bibnamefont {de~Jong}}, \bibinfo {author} {\bibfnamefont
  {W.}~\bibnamefont {Schwab}}, \ and\ \bibinfo {author} {\bibfnamefont
  {S.}~\bibnamefont {Steinh{\"{a}}user}},\ }\href {\doibase
  10.1007/s100500050062} {\bibfield  {journal} {\bibinfo  {journal} {The
  European Physical Journal A}\ }\textbf {\bibinfo {volume} {1}},\ \bibinfo
  {pages} {275} (\bibinfo {year} {1998})}\BibitemShut {NoStop}%
\bibitem [{\citenamefont {Sorlin}\ \emph {et~al.}(2000)\citenamefont {Sorlin},
  \citenamefont {Donzaud}, \citenamefont {Axelsson}, \citenamefont {Belleguic},
  \citenamefont {Beraud}, \citenamefont {Borcea}, \citenamefont {Canchel},
  \citenamefont {Chabanat}, \citenamefont {Daugas}, \citenamefont {Emsallem},
  \citenamefont {Girod}, \citenamefont {Leenhardt}, \citenamefont {Lewitowicz},
  \citenamefont {Longour}, \citenamefont {Lopez}, \citenamefont {Santos},
  \citenamefont {Petizon}, \citenamefont {Pfeiffer}, \citenamefont {Pougheon},\
  and\ \citenamefont {Sauvestre}}]{Sorlin2000}%
  \BibitemOpen
  \bibfield  {author} {\bibinfo {author} {\bibfnamefont {O.}~\bibnamefont
  {Sorlin}}, \bibinfo {author} {\bibfnamefont {C.}~\bibnamefont {Donzaud}},
  \bibinfo {author} {\bibfnamefont {L.}~\bibnamefont {Axelsson}}, \bibinfo
  {author} {\bibfnamefont {M.}~\bibnamefont {Belleguic}}, \bibinfo {author}
  {\bibfnamefont {R.}~\bibnamefont {Beraud}}, \bibinfo {author} {\bibfnamefont
  {C.}~\bibnamefont {Borcea}}, \bibinfo {author} {\bibfnamefont
  {G.}~\bibnamefont {Canchel}}, \bibinfo {author} {\bibfnamefont
  {E.}~\bibnamefont {Chabanat}}, \bibinfo {author} {\bibfnamefont {J.~M.}\
  \bibnamefont {Daugas}}, \bibinfo {author} {\bibfnamefont {A.}~\bibnamefont
  {Emsallem}}, \bibinfo {author} {\bibfnamefont {M.}~\bibnamefont {Girod}},
  \bibinfo {author} {\bibfnamefont {S.}~\bibnamefont {Leenhardt}}, \bibinfo
  {author} {\bibfnamefont {M.}~\bibnamefont {Lewitowicz}}, \bibinfo {author}
  {\bibfnamefont {C.}~\bibnamefont {Longour}}, \bibinfo {author} {\bibfnamefont
  {M.~J.}\ \bibnamefont {Lopez}}, \bibinfo {author} {\bibfnamefont {F.~D.~O.}\
  \bibnamefont {Santos}}, \bibinfo {author} {\bibfnamefont {L.}~\bibnamefont
  {Petizon}}, \bibinfo {author} {\bibfnamefont {B.}~\bibnamefont {Pfeiffer}},
  \bibinfo {author} {\bibfnamefont {F.}~\bibnamefont {Pougheon}}, \ and\
  \bibinfo {author} {\bibfnamefont {J.~E.}\ \bibnamefont {Sauvestre}},\ }\href
  {\doibase 10.1016/S0375-9474(00)00137-8} {\bibfield  {journal} {\bibinfo
  {journal} {Nuclear Physics A}\ }\textbf {\bibinfo {volume} {669}},\ \bibinfo
  {pages} {351} (\bibinfo {year} {2000})}\BibitemShut {NoStop}%
\bibitem [{\citenamefont {Hannawald}(2000)}]{Hannawald2000}%
  \BibitemOpen
  \bibfield  {author} {\bibinfo {author} {\bibfnamefont {M.~W.}\ \bibnamefont
  {Hannawald}},\ }\emph {\bibinfo {title} {{Kernspektroskopie an N$\simeq$40
  und N$\simeq$82 Nukliden}}},\ \href@noop {} {Ph.D. thesis},\ \bibinfo
  {school} {Johannes Gutenberg-Universit{\"{a}}t Mainz} (\bibinfo {year}
  {2000})\BibitemShut {NoStop}%
\bibitem [{\citenamefont {Singh}()}]{xundlLiddick}%
  \BibitemOpen
  \bibfield  {author} {\bibinfo {author} {\bibfnamefont {B.}~\bibnamefont
  {Singh}},\ }\href@noop {} {}\bibinfo {note} {{XUNDL} compilation of
  \cite{Liddick2012}}\BibitemShut {NoStop}%
\bibitem [{\citenamefont {Brown}\ \emph {et~al.}(2018)\citenamefont {Brown},
  \citenamefont {Dees}, \citenamefont {Adamek}, \citenamefont {Allgeier},
  \citenamefont {Blatnik}, \citenamefont {Bowles}, \citenamefont {Broussard},
  \citenamefont {Carr}, \citenamefont {Clayton}, \citenamefont {Cude-Woods},
  \citenamefont {Currie}, \citenamefont {Ding}, \citenamefont {Filippone},
  \citenamefont {Garc{\'{i}}a}, \citenamefont {Geltenbort}, \citenamefont
  {Hasan}, \citenamefont {Hickerson}, \citenamefont {Hoagland}, \citenamefont
  {Hong}, \citenamefont {Hogan}, \citenamefont {Holley}, \citenamefont {Ito},
  \citenamefont {Knecht}, \citenamefont {Liu}, \citenamefont {Liu},
  \citenamefont {Makela}, \citenamefont {Martin}, \citenamefont {Melconian},
  \citenamefont {Mendenhall}, \citenamefont {Moore}, \citenamefont {Morris},
  \citenamefont {Nepal}, \citenamefont {Nouri}, \citenamefont {Pattie},
  \citenamefont {{P{\'{e}}rez Galv{\'{a}}n}}, \citenamefont {Phillips},
  \citenamefont {Picker}, \citenamefont {Pitt}, \citenamefont {Plaster},
  \citenamefont {Ramsey}, \citenamefont {Rios}, \citenamefont {Salvat},
  \citenamefont {Saunders}, \citenamefont {Sondheim}, \citenamefont {Seestrom},
  \citenamefont {Sjue}, \citenamefont {Slutsky}, \citenamefont {Sun},
  \citenamefont {Swank}, \citenamefont {Swift}, \citenamefont {Tatar},
  \citenamefont {Vogelaar}, \citenamefont {VornDick}, \citenamefont {Wang},
  \citenamefont {Wexler}, \citenamefont {Womack}, \citenamefont {Wrede},
  \citenamefont {Young},\ and\ \citenamefont {Zeck}}]{Brown2018}%
  \BibitemOpen
  \bibfield  {author} {\bibinfo {author} {\bibfnamefont {M.~A.-P.}\
  \bibnamefont {Brown}}, \bibinfo {author} {\bibfnamefont {E.~B.}\ \bibnamefont
  {Dees}}, \bibinfo {author} {\bibfnamefont {E.}~\bibnamefont {Adamek}},
  \bibinfo {author} {\bibfnamefont {B.}~\bibnamefont {Allgeier}}, \bibinfo
  {author} {\bibfnamefont {M.}~\bibnamefont {Blatnik}}, \bibinfo {author}
  {\bibfnamefont {T.~J.}\ \bibnamefont {Bowles}}, \bibinfo {author}
  {\bibfnamefont {L.~J.}\ \bibnamefont {Broussard}}, \bibinfo {author}
  {\bibfnamefont {R.}~\bibnamefont {Carr}}, \bibinfo {author} {\bibfnamefont
  {S.}~\bibnamefont {Clayton}}, \bibinfo {author} {\bibfnamefont
  {C.}~\bibnamefont {Cude-Woods}}, \bibinfo {author} {\bibfnamefont
  {S.}~\bibnamefont {Currie}}, \bibinfo {author} {\bibfnamefont
  {X.}~\bibnamefont {Ding}}, \bibinfo {author} {\bibfnamefont {B.~W.}\
  \bibnamefont {Filippone}}, \bibinfo {author} {\bibfnamefont {A.}~\bibnamefont
  {Garc{\'{i}}a}}, \bibinfo {author} {\bibfnamefont {P.}~\bibnamefont
  {Geltenbort}}, \bibinfo {author} {\bibfnamefont {S.}~\bibnamefont {Hasan}},
  \bibinfo {author} {\bibfnamefont {K.~P.}\ \bibnamefont {Hickerson}}, \bibinfo
  {author} {\bibfnamefont {J.}~\bibnamefont {Hoagland}}, \bibinfo {author}
  {\bibfnamefont {R.}~\bibnamefont {Hong}}, \bibinfo {author} {\bibfnamefont
  {G.~E.}\ \bibnamefont {Hogan}}, \bibinfo {author} {\bibfnamefont {A.~T.}\
  \bibnamefont {Holley}}, \bibinfo {author} {\bibfnamefont {T.~M.}\
  \bibnamefont {Ito}}, \bibinfo {author} {\bibfnamefont {A.}~\bibnamefont
  {Knecht}}, \bibinfo {author} {\bibfnamefont {C.-Y.}\ \bibnamefont {Liu}},
  \bibinfo {author} {\bibfnamefont {J.}~\bibnamefont {Liu}}, \bibinfo {author}
  {\bibfnamefont {M.}~\bibnamefont {Makela}}, \bibinfo {author} {\bibfnamefont
  {J.~W.}\ \bibnamefont {Martin}}, \bibinfo {author} {\bibfnamefont
  {D.}~\bibnamefont {Melconian}}, \bibinfo {author} {\bibfnamefont {M.~P.}\
  \bibnamefont {Mendenhall}}, \bibinfo {author} {\bibfnamefont {S.~D.}\
  \bibnamefont {Moore}}, \bibinfo {author} {\bibfnamefont {C.~L.}\ \bibnamefont
  {Morris}}, \bibinfo {author} {\bibfnamefont {S.}~\bibnamefont {Nepal}},
  \bibinfo {author} {\bibfnamefont {N.}~\bibnamefont {Nouri}}, \bibinfo
  {author} {\bibfnamefont {R.~W.}\ \bibnamefont {Pattie}}, \bibinfo {author}
  {\bibfnamefont {A.}~\bibnamefont {{P{\'{e}}rez Galv{\'{a}}n}}}, \bibinfo
  {author} {\bibfnamefont {D.~G.}\ \bibnamefont {Phillips}}, \bibinfo {author}
  {\bibfnamefont {R.}~\bibnamefont {Picker}}, \bibinfo {author} {\bibfnamefont
  {M.~L.}\ \bibnamefont {Pitt}}, \bibinfo {author} {\bibfnamefont
  {B.}~\bibnamefont {Plaster}}, \bibinfo {author} {\bibfnamefont {J.~C.}\
  \bibnamefont {Ramsey}}, \bibinfo {author} {\bibfnamefont {R.}~\bibnamefont
  {Rios}}, \bibinfo {author} {\bibfnamefont {D.~J.}\ \bibnamefont {Salvat}},
  \bibinfo {author} {\bibfnamefont {A.}~\bibnamefont {Saunders}}, \bibinfo
  {author} {\bibfnamefont {W.}~\bibnamefont {Sondheim}}, \bibinfo {author}
  {\bibfnamefont {S.~J.}\ \bibnamefont {Seestrom}}, \bibinfo {author}
  {\bibfnamefont {S.}~\bibnamefont {Sjue}}, \bibinfo {author} {\bibfnamefont
  {S.}~\bibnamefont {Slutsky}}, \bibinfo {author} {\bibfnamefont
  {X.}~\bibnamefont {Sun}}, \bibinfo {author} {\bibfnamefont {C.}~\bibnamefont
  {Swank}}, \bibinfo {author} {\bibfnamefont {G.}~\bibnamefont {Swift}},
  \bibinfo {author} {\bibfnamefont {E.}~\bibnamefont {Tatar}}, \bibinfo
  {author} {\bibfnamefont {R.~B.}\ \bibnamefont {Vogelaar}}, \bibinfo {author}
  {\bibfnamefont {B.}~\bibnamefont {VornDick}}, \bibinfo {author}
  {\bibfnamefont {Z.}~\bibnamefont {Wang}}, \bibinfo {author} {\bibfnamefont
  {J.}~\bibnamefont {Wexler}}, \bibinfo {author} {\bibfnamefont
  {T.}~\bibnamefont {Womack}}, \bibinfo {author} {\bibfnamefont
  {C.}~\bibnamefont {Wrede}}, \bibinfo {author} {\bibfnamefont {A.~R.}\
  \bibnamefont {Young}}, \ and\ \bibinfo {author} {\bibfnamefont {B.~A.}\
  \bibnamefont {Zeck}},\ }\href {\doibase 10.1103/PhysRevC.97.035505}
  {\bibfield  {journal} {\bibinfo  {journal} {Physical Review C}\ }\textbf
  {\bibinfo {volume} {97}},\ \bibinfo {pages} {035505} (\bibinfo {year}
  {2018})}\BibitemShut {NoStop}%
\bibitem [{\citenamefont {Mart{\'{i}}nez-Pinedo}\ \emph
  {et~al.}(1996)\citenamefont {Mart{\'{i}}nez-Pinedo}, \citenamefont {Poves},
  \citenamefont {Caurier},\ and\ \citenamefont {Zuker}}]{Martinez-Pinedo1996}%
  \BibitemOpen
  \bibfield  {author} {\bibinfo {author} {\bibfnamefont {G.}~\bibnamefont
  {Mart{\'{i}}nez-Pinedo}}, \bibinfo {author} {\bibfnamefont {A.}~\bibnamefont
  {Poves}}, \bibinfo {author} {\bibfnamefont {E.}~\bibnamefont {Caurier}}, \
  and\ \bibinfo {author} {\bibfnamefont {A.~P.}\ \bibnamefont {Zuker}},\ }\href
  {\doibase 10.1103/PhysRevC.53.R2602} {\bibfield  {journal} {\bibinfo
  {journal} {Physical Review C}\ }\textbf {\bibinfo {volume} {53}},\ \bibinfo
  {pages} {R2602} (\bibinfo {year} {1996})}\BibitemShut {NoStop}%
\bibitem [{\citenamefont {Morales}\ \emph {et~al.}(2017)\citenamefont
  {Morales}, \citenamefont {Benzoni}, \citenamefont {Watanabe}, \citenamefont
  {Tsunoda}, \citenamefont {Otsuka}, \citenamefont {Nishimura}, \citenamefont
  {Browne}, \citenamefont {Daido}, \citenamefont {Doornenbal}, \citenamefont
  {Fang}, \citenamefont {Lorusso}, \citenamefont {Patel}, \citenamefont {Rice},
  \citenamefont {Sinclair}, \citenamefont {S{\"{o}}derstr{\"{o}}m},
  \citenamefont {Sumikama}, \citenamefont {Wu}, \citenamefont {Xu},
  \citenamefont {Yagi}, \citenamefont {Yokoyama}, \citenamefont {Baba},
  \citenamefont {Avigo}, \citenamefont {{Bello Garrote}}, \citenamefont
  {Blasi}, \citenamefont {Bracco}, \citenamefont {Camera}, \citenamefont
  {Ceruti}, \citenamefont {Crespi}, \citenamefont {de~Angelis}, \citenamefont
  {Delattre}, \citenamefont {Dombradi}, \citenamefont {Gottardo}, \citenamefont
  {Isobe}, \citenamefont {Kojouharov}, \citenamefont {Kurz}, \citenamefont
  {Kuti}, \citenamefont {Matsui}, \citenamefont {Melon}, \citenamefont
  {Mengoni}, \citenamefont {Miyazaki}, \citenamefont {Modamio-Hoybjor},
  \citenamefont {Momiyama}, \citenamefont {Napoli}, \citenamefont {Niikura},
  \citenamefont {Orlandi}, \citenamefont {Sakurai}, \citenamefont {Sahin},
  \citenamefont {Sohler}, \citenamefont {Schaffner}, \citenamefont {Taniuchi},
  \citenamefont {Taprogge}, \citenamefont {Vajta}, \citenamefont
  {Valiente-Dob{\'{o}}n}, \citenamefont {Wieland},\ and\ \citenamefont
  {Yalcinkaya}}]{Morales2017}%
  \BibitemOpen
  \bibfield  {author} {\bibinfo {author} {\bibfnamefont {A.}~\bibnamefont
  {Morales}}, \bibinfo {author} {\bibfnamefont {G.}~\bibnamefont {Benzoni}},
  \bibinfo {author} {\bibfnamefont {H.}~\bibnamefont {Watanabe}}, \bibinfo
  {author} {\bibfnamefont {Y.}~\bibnamefont {Tsunoda}}, \bibinfo {author}
  {\bibfnamefont {T.}~\bibnamefont {Otsuka}}, \bibinfo {author} {\bibfnamefont
  {S.}~\bibnamefont {Nishimura}}, \bibinfo {author} {\bibfnamefont
  {F.}~\bibnamefont {Browne}}, \bibinfo {author} {\bibfnamefont
  {R.}~\bibnamefont {Daido}}, \bibinfo {author} {\bibfnamefont
  {P.}~\bibnamefont {Doornenbal}}, \bibinfo {author} {\bibfnamefont
  {Y.}~\bibnamefont {Fang}}, \bibinfo {author} {\bibfnamefont {G.}~\bibnamefont
  {Lorusso}}, \bibinfo {author} {\bibfnamefont {Z.}~\bibnamefont {Patel}},
  \bibinfo {author} {\bibfnamefont {S.}~\bibnamefont {Rice}}, \bibinfo {author}
  {\bibfnamefont {L.}~\bibnamefont {Sinclair}}, \bibinfo {author}
  {\bibfnamefont {P.-A.}\ \bibnamefont {S{\"{o}}derstr{\"{o}}m}}, \bibinfo
  {author} {\bibfnamefont {T.}~\bibnamefont {Sumikama}}, \bibinfo {author}
  {\bibfnamefont {J.}~\bibnamefont {Wu}}, \bibinfo {author} {\bibfnamefont
  {Z.}~\bibnamefont {Xu}}, \bibinfo {author} {\bibfnamefont {A.}~\bibnamefont
  {Yagi}}, \bibinfo {author} {\bibfnamefont {R.}~\bibnamefont {Yokoyama}},
  \bibinfo {author} {\bibfnamefont {H.}~\bibnamefont {Baba}}, \bibinfo {author}
  {\bibfnamefont {R.}~\bibnamefont {Avigo}}, \bibinfo {author} {\bibfnamefont
  {F.}~\bibnamefont {{Bello Garrote}}}, \bibinfo {author} {\bibfnamefont
  {N.}~\bibnamefont {Blasi}}, \bibinfo {author} {\bibfnamefont
  {A.}~\bibnamefont {Bracco}}, \bibinfo {author} {\bibfnamefont
  {F.}~\bibnamefont {Camera}}, \bibinfo {author} {\bibfnamefont
  {S.}~\bibnamefont {Ceruti}}, \bibinfo {author} {\bibfnamefont
  {F.}~\bibnamefont {Crespi}}, \bibinfo {author} {\bibfnamefont
  {G.}~\bibnamefont {de~Angelis}}, \bibinfo {author} {\bibfnamefont {M.-C.}\
  \bibnamefont {Delattre}}, \bibinfo {author} {\bibfnamefont {Z.}~\bibnamefont
  {Dombradi}}, \bibinfo {author} {\bibfnamefont {A.}~\bibnamefont {Gottardo}},
  \bibinfo {author} {\bibfnamefont {T.}~\bibnamefont {Isobe}}, \bibinfo
  {author} {\bibfnamefont {I.}~\bibnamefont {Kojouharov}}, \bibinfo {author}
  {\bibfnamefont {N.}~\bibnamefont {Kurz}}, \bibinfo {author} {\bibfnamefont
  {I.}~\bibnamefont {Kuti}}, \bibinfo {author} {\bibfnamefont {K.}~\bibnamefont
  {Matsui}}, \bibinfo {author} {\bibfnamefont {B.}~\bibnamefont {Melon}},
  \bibinfo {author} {\bibfnamefont {D.}~\bibnamefont {Mengoni}}, \bibinfo
  {author} {\bibfnamefont {T.}~\bibnamefont {Miyazaki}}, \bibinfo {author}
  {\bibfnamefont {V.}~\bibnamefont {Modamio-Hoybjor}}, \bibinfo {author}
  {\bibfnamefont {S.}~\bibnamefont {Momiyama}}, \bibinfo {author}
  {\bibfnamefont {D.}~\bibnamefont {Napoli}}, \bibinfo {author} {\bibfnamefont
  {M.}~\bibnamefont {Niikura}}, \bibinfo {author} {\bibfnamefont
  {R.}~\bibnamefont {Orlandi}}, \bibinfo {author} {\bibfnamefont
  {H.}~\bibnamefont {Sakurai}}, \bibinfo {author} {\bibfnamefont
  {E.}~\bibnamefont {Sahin}}, \bibinfo {author} {\bibfnamefont
  {D.}~\bibnamefont {Sohler}}, \bibinfo {author} {\bibfnamefont
  {H.}~\bibnamefont {Schaffner}}, \bibinfo {author} {\bibfnamefont
  {R.}~\bibnamefont {Taniuchi}}, \bibinfo {author} {\bibfnamefont
  {J.}~\bibnamefont {Taprogge}}, \bibinfo {author} {\bibfnamefont
  {Z.}~\bibnamefont {Vajta}}, \bibinfo {author} {\bibfnamefont
  {J.}~\bibnamefont {Valiente-Dob{\'{o}}n}}, \bibinfo {author} {\bibfnamefont
  {O.}~\bibnamefont {Wieland}}, \ and\ \bibinfo {author} {\bibfnamefont
  {M.}~\bibnamefont {Yalcinkaya}},\ }\href {\doibase
  10.1016/j.physletb.2016.12.025} {\bibfield  {journal} {\bibinfo  {journal}
  {Physics Letters B}\ }\textbf {\bibinfo {volume} {765}},\ \bibinfo {pages}
  {328} (\bibinfo {year} {2017})}\BibitemShut {NoStop}%
\bibitem [{\citenamefont {Wraith}\ \emph {et~al.}(2017)\citenamefont {Wraith},
  \citenamefont {Yang}, \citenamefont {Xie}, \citenamefont {Babcock},
  \citenamefont {Biero{\'{n}}}, \citenamefont {Billowes}, \citenamefont
  {Bissell}, \citenamefont {Blaum}, \citenamefont {Cheal}, \citenamefont
  {Filippin}, \citenamefont {{Garcia Ruiz}}, \citenamefont {Gins},
  \citenamefont {Grob}, \citenamefont {Gaigalas}, \citenamefont {Godefroid},
  \citenamefont {Gorges}, \citenamefont {Heylen}, \citenamefont {Honma},
  \citenamefont {J{\"{o}}nsson}, \citenamefont {Kaufmann}, \citenamefont
  {Kowalska}, \citenamefont {Kr{\"{a}}mer}, \citenamefont
  {Malbrunot-Ettenauer}, \citenamefont {Neugart}, \citenamefont {Neyens},
  \citenamefont {N{\"{o}}rtersh{\"{a}}user}, \citenamefont {Nowacki},
  \citenamefont {Otsuka}, \citenamefont {Papuga}, \citenamefont
  {S{\'{a}}nchez}, \citenamefont {Tsunoda},\ and\ \citenamefont
  {Yordanov}}]{Wraith2017}%
  \BibitemOpen
  \bibfield  {author} {\bibinfo {author} {\bibfnamefont {C.}~\bibnamefont
  {Wraith}}, \bibinfo {author} {\bibfnamefont {X.}~\bibnamefont {Yang}},
  \bibinfo {author} {\bibfnamefont {L.}~\bibnamefont {Xie}}, \bibinfo {author}
  {\bibfnamefont {C.}~\bibnamefont {Babcock}}, \bibinfo {author} {\bibfnamefont
  {J.}~\bibnamefont {Biero{\'{n}}}}, \bibinfo {author} {\bibfnamefont
  {J.}~\bibnamefont {Billowes}}, \bibinfo {author} {\bibfnamefont
  {M.}~\bibnamefont {Bissell}}, \bibinfo {author} {\bibfnamefont
  {K.}~\bibnamefont {Blaum}}, \bibinfo {author} {\bibfnamefont
  {B.}~\bibnamefont {Cheal}}, \bibinfo {author} {\bibfnamefont
  {L.}~\bibnamefont {Filippin}}, \bibinfo {author} {\bibfnamefont
  {R.}~\bibnamefont {{Garcia Ruiz}}}, \bibinfo {author} {\bibfnamefont
  {W.}~\bibnamefont {Gins}}, \bibinfo {author} {\bibfnamefont {L.}~\bibnamefont
  {Grob}}, \bibinfo {author} {\bibfnamefont {G.}~\bibnamefont {Gaigalas}},
  \bibinfo {author} {\bibfnamefont {M.}~\bibnamefont {Godefroid}}, \bibinfo
  {author} {\bibfnamefont {C.}~\bibnamefont {Gorges}}, \bibinfo {author}
  {\bibfnamefont {H.}~\bibnamefont {Heylen}}, \bibinfo {author} {\bibfnamefont
  {M.}~\bibnamefont {Honma}}, \bibinfo {author} {\bibfnamefont
  {P.}~\bibnamefont {J{\"{o}}nsson}}, \bibinfo {author} {\bibfnamefont
  {S.}~\bibnamefont {Kaufmann}}, \bibinfo {author} {\bibfnamefont
  {M.}~\bibnamefont {Kowalska}}, \bibinfo {author} {\bibfnamefont
  {J.}~\bibnamefont {Kr{\"{a}}mer}}, \bibinfo {author} {\bibfnamefont
  {S.}~\bibnamefont {Malbrunot-Ettenauer}}, \bibinfo {author} {\bibfnamefont
  {R.}~\bibnamefont {Neugart}}, \bibinfo {author} {\bibfnamefont
  {G.}~\bibnamefont {Neyens}}, \bibinfo {author} {\bibfnamefont
  {W.}~\bibnamefont {N{\"{o}}rtersh{\"{a}}user}}, \bibinfo {author}
  {\bibfnamefont {F.}~\bibnamefont {Nowacki}}, \bibinfo {author} {\bibfnamefont
  {T.}~\bibnamefont {Otsuka}}, \bibinfo {author} {\bibfnamefont
  {J.}~\bibnamefont {Papuga}}, \bibinfo {author} {\bibfnamefont
  {R.}~\bibnamefont {S{\'{a}}nchez}}, \bibinfo {author} {\bibfnamefont
  {Y.}~\bibnamefont {Tsunoda}}, \ and\ \bibinfo {author} {\bibfnamefont
  {D.}~\bibnamefont {Yordanov}},\ }\href {\doibase
  10.1016/j.physletb.2017.05.085} {\bibfield  {journal} {\bibinfo  {journal}
  {Physics Letters B}\ }\textbf {\bibinfo {volume} {771}},\ \bibinfo {pages}
  {385} (\bibinfo {year} {2017})}\BibitemShut {NoStop}%
\bibitem [{\citenamefont {Shimizu}\ \emph {et~al.}(2012)\citenamefont
  {Shimizu}, \citenamefont {Abe}, \citenamefont {Tsunoda}, \citenamefont
  {Utsuno}, \citenamefont {Yoshida}, \citenamefont {Mizusaki}, \citenamefont
  {Honma},\ and\ \citenamefont {Otsuka}}]{Shimizu2012}%
  \BibitemOpen
  \bibfield  {author} {\bibinfo {author} {\bibfnamefont {N.}~\bibnamefont
  {Shimizu}}, \bibinfo {author} {\bibfnamefont {T.}~\bibnamefont {Abe}},
  \bibinfo {author} {\bibfnamefont {Y.}~\bibnamefont {Tsunoda}}, \bibinfo
  {author} {\bibfnamefont {Y.}~\bibnamefont {Utsuno}}, \bibinfo {author}
  {\bibfnamefont {T.}~\bibnamefont {Yoshida}}, \bibinfo {author} {\bibfnamefont
  {T.}~\bibnamefont {Mizusaki}}, \bibinfo {author} {\bibfnamefont
  {M.}~\bibnamefont {Honma}}, \ and\ \bibinfo {author} {\bibfnamefont
  {T.}~\bibnamefont {Otsuka}},\ }\href {\doibase 10.1093/ptep/pts012}
  {\bibfield  {journal} {\bibinfo  {journal} {Progress of Theoretical and
  Experimental Physics}\ }\textbf {\bibinfo {volume} {2012}},\ \bibinfo {pages}
  {1} (\bibinfo {year} {2012})}\BibitemShut {NoStop}%
\bibitem [{\citenamefont {Shimizu}\ \emph {et~al.}(2017)\citenamefont
  {Shimizu}, \citenamefont {Abe}, \citenamefont {Honma}, \citenamefont
  {Otsuka}, \citenamefont {Togashi}, \citenamefont {Tsunoda}, \citenamefont
  {Utsuno},\ and\ \citenamefont {Yoshida}}]{Shimizu2017}%
  \BibitemOpen
  \bibfield  {author} {\bibinfo {author} {\bibfnamefont {N.}~\bibnamefont
  {Shimizu}}, \bibinfo {author} {\bibfnamefont {T.}~\bibnamefont {Abe}},
  \bibinfo {author} {\bibfnamefont {M.}~\bibnamefont {Honma}}, \bibinfo
  {author} {\bibfnamefont {T.}~\bibnamefont {Otsuka}}, \bibinfo {author}
  {\bibfnamefont {T.}~\bibnamefont {Togashi}}, \bibinfo {author} {\bibfnamefont
  {Y.}~\bibnamefont {Tsunoda}}, \bibinfo {author} {\bibfnamefont
  {Y.}~\bibnamefont {Utsuno}}, \ and\ \bibinfo {author} {\bibfnamefont
  {T.}~\bibnamefont {Yoshida}},\ }\href {\doibase 10.1088/1402-4896/aa65e4}
  {\bibfield  {journal} {\bibinfo  {journal} {Physica Scripta}\ }\textbf
  {\bibinfo {volume} {92}},\ \bibinfo {pages} {063001} (\bibinfo {year}
  {2017})}\BibitemShut {NoStop}%
\bibitem [{\citenamefont {Liddick}\ \emph {et~al.}(2011)\citenamefont
  {Liddick}, \citenamefont {Suchyta}, \citenamefont {Abromeit}, \citenamefont
  {Ayres}, \citenamefont {Bey}, \citenamefont {Bingham}, \citenamefont {Bolla},
  \citenamefont {Carpenter}, \citenamefont {Cartegni}, \citenamefont {Chiara},
  \citenamefont {Crawford}, \citenamefont {Darby}, \citenamefont {Grzywacz},
  \citenamefont {G{\"{u}}rdal}, \citenamefont {Ilyushkin}, \citenamefont
  {Larson}, \citenamefont {Madurga}, \citenamefont {McCutchan}, \citenamefont
  {Miller}, \citenamefont {Padgett}, \citenamefont {Paulauskas}, \citenamefont
  {Pereira}, \citenamefont {Rajabali}, \citenamefont {Rykaczewski},
  \citenamefont {Vinnikova}, \citenamefont {Walters},\ and\ \citenamefont
  {Zhu}}]{Liddick2011}%
  \BibitemOpen
  \bibfield  {author} {\bibinfo {author} {\bibfnamefont {S.~N.}\ \bibnamefont
  {Liddick}}, \bibinfo {author} {\bibfnamefont {S.}~\bibnamefont {Suchyta}},
  \bibinfo {author} {\bibfnamefont {B.}~\bibnamefont {Abromeit}}, \bibinfo
  {author} {\bibfnamefont {A.}~\bibnamefont {Ayres}}, \bibinfo {author}
  {\bibfnamefont {A.}~\bibnamefont {Bey}}, \bibinfo {author} {\bibfnamefont
  {C.~R.}\ \bibnamefont {Bingham}}, \bibinfo {author} {\bibfnamefont
  {M.}~\bibnamefont {Bolla}}, \bibinfo {author} {\bibfnamefont {M.~P.}\
  \bibnamefont {Carpenter}}, \bibinfo {author} {\bibfnamefont {L.}~\bibnamefont
  {Cartegni}}, \bibinfo {author} {\bibfnamefont {C.~J.}\ \bibnamefont
  {Chiara}}, \bibinfo {author} {\bibfnamefont {H.~L.}\ \bibnamefont
  {Crawford}}, \bibinfo {author} {\bibfnamefont {I.~G.}\ \bibnamefont {Darby}},
  \bibinfo {author} {\bibfnamefont {R.}~\bibnamefont {Grzywacz}}, \bibinfo
  {author} {\bibfnamefont {G.}~\bibnamefont {G{\"{u}}rdal}}, \bibinfo {author}
  {\bibfnamefont {S.}~\bibnamefont {Ilyushkin}}, \bibinfo {author}
  {\bibfnamefont {N.}~\bibnamefont {Larson}}, \bibinfo {author} {\bibfnamefont
  {M.}~\bibnamefont {Madurga}}, \bibinfo {author} {\bibfnamefont {E.~A.}\
  \bibnamefont {McCutchan}}, \bibinfo {author} {\bibfnamefont {D.}~\bibnamefont
  {Miller}}, \bibinfo {author} {\bibfnamefont {S.}~\bibnamefont {Padgett}},
  \bibinfo {author} {\bibfnamefont {S.~V.}\ \bibnamefont {Paulauskas}},
  \bibinfo {author} {\bibfnamefont {J.}~\bibnamefont {Pereira}}, \bibinfo
  {author} {\bibfnamefont {M.~M.}\ \bibnamefont {Rajabali}}, \bibinfo {author}
  {\bibfnamefont {K.}~\bibnamefont {Rykaczewski}}, \bibinfo {author}
  {\bibfnamefont {S.}~\bibnamefont {Vinnikova}}, \bibinfo {author}
  {\bibfnamefont {W.~B.}\ \bibnamefont {Walters}}, \ and\ \bibinfo {author}
  {\bibfnamefont {S.}~\bibnamefont {Zhu}},\ }\href {\doibase
  10.1103/PhysRevC.84.061305} {\bibfield  {journal} {\bibinfo  {journal}
  {Physical Review C}\ }\textbf {\bibinfo {volume} {84}},\ \bibinfo {pages}
  {061305} (\bibinfo {year} {2011})}\BibitemShut {NoStop}%
\bibitem [{\citenamefont {Jordan}\ \emph {et~al.}(2013)\citenamefont {Jordan},
  \citenamefont {Algora}, \citenamefont {Ta{\'{i}}n}, \citenamefont {Rubio},
  \citenamefont {Agramunt}, \citenamefont {Perez-Cerdan}, \citenamefont
  {Molina}, \citenamefont {Caballero}, \citenamefont {N{\'{a}}cher},
  \citenamefont {Krasznahorkay}, \citenamefont {Hunyadi}, \citenamefont
  {Guly{\'{a}}s}, \citenamefont {Vit{\'{e}}z}, \citenamefont {Csatl{\'{o}}s},
  \citenamefont {Csige}, \citenamefont {{\"{A}}ysto}, \citenamefont
  {Penttil{\"{a}}}, \citenamefont {Moore}, \citenamefont {Eronen},
  \citenamefont {Jokinen}, \citenamefont {Nieminen}, \citenamefont {Hakala},
  \citenamefont {Karvonen}, \citenamefont {Kankainen}, \citenamefont
  {Saastamoinen}, \citenamefont {Rissanen}, \citenamefont {Kessler},
  \citenamefont {Weber}, \citenamefont {Ronkainen}, \citenamefont {Rahaman},
  \citenamefont {Elomaa}, \citenamefont {Hager}, \citenamefont {Rinta-Antila},
  \citenamefont {Sonoda}, \citenamefont {Burkard}, \citenamefont
  {H{\"{u}}ller}, \citenamefont {Batist}, \citenamefont {Gelletly},
  \citenamefont {Nichols}, \citenamefont {Yoshida}, \citenamefont {Sonzogni},
  \citenamefont {Per{\"{a}}j{\"{a}}rvi}, \citenamefont {Petrovici},
  \citenamefont {Schmid},\ and\ \citenamefont {Faessler}}]{Jordan2013}%
  \BibitemOpen
  \bibfield  {author} {\bibinfo {author} {\bibfnamefont {D.}~\bibnamefont
  {Jordan}}, \bibinfo {author} {\bibfnamefont {A.}~\bibnamefont {Algora}},
  \bibinfo {author} {\bibfnamefont {J.~L.}\ \bibnamefont {Ta{\'{i}}n}},
  \bibinfo {author} {\bibfnamefont {B.}~\bibnamefont {Rubio}}, \bibinfo
  {author} {\bibfnamefont {J.}~\bibnamefont {Agramunt}}, \bibinfo {author}
  {\bibfnamefont {A.~B.}\ \bibnamefont {Perez-Cerdan}}, \bibinfo {author}
  {\bibfnamefont {F.}~\bibnamefont {Molina}}, \bibinfo {author} {\bibfnamefont
  {L.}~\bibnamefont {Caballero}}, \bibinfo {author} {\bibfnamefont
  {E.}~\bibnamefont {N{\'{a}}cher}}, \bibinfo {author} {\bibfnamefont
  {A.}~\bibnamefont {Krasznahorkay}}, \bibinfo {author} {\bibfnamefont {M.~D.}\
  \bibnamefont {Hunyadi}}, \bibinfo {author} {\bibfnamefont {J.}~\bibnamefont
  {Guly{\'{a}}s}}, \bibinfo {author} {\bibfnamefont {A.}~\bibnamefont
  {Vit{\'{e}}z}}, \bibinfo {author} {\bibfnamefont {M.}~\bibnamefont
  {Csatl{\'{o}}s}}, \bibinfo {author} {\bibfnamefont {L.}~\bibnamefont
  {Csige}}, \bibinfo {author} {\bibfnamefont {J.}~\bibnamefont {{\"{A}}ysto}},
  \bibinfo {author} {\bibfnamefont {H.}~\bibnamefont {Penttil{\"{a}}}},
  \bibinfo {author} {\bibfnamefont {I.~D.}\ \bibnamefont {Moore}}, \bibinfo
  {author} {\bibfnamefont {T.}~\bibnamefont {Eronen}}, \bibinfo {author}
  {\bibfnamefont {A.}~\bibnamefont {Jokinen}}, \bibinfo {author} {\bibfnamefont
  {A.}~\bibnamefont {Nieminen}}, \bibinfo {author} {\bibfnamefont
  {J.}~\bibnamefont {Hakala}}, \bibinfo {author} {\bibfnamefont
  {P.}~\bibnamefont {Karvonen}}, \bibinfo {author} {\bibfnamefont
  {A.}~\bibnamefont {Kankainen}}, \bibinfo {author} {\bibfnamefont
  {A.}~\bibnamefont {Saastamoinen}}, \bibinfo {author} {\bibfnamefont
  {J.}~\bibnamefont {Rissanen}}, \bibinfo {author} {\bibfnamefont
  {T.}~\bibnamefont {Kessler}}, \bibinfo {author} {\bibfnamefont
  {C.}~\bibnamefont {Weber}}, \bibinfo {author} {\bibfnamefont
  {J.}~\bibnamefont {Ronkainen}}, \bibinfo {author} {\bibfnamefont
  {S.}~\bibnamefont {Rahaman}}, \bibinfo {author} {\bibfnamefont
  {V.}~\bibnamefont {Elomaa}}, \bibinfo {author} {\bibfnamefont
  {U.}~\bibnamefont {Hager}}, \bibinfo {author} {\bibfnamefont
  {S.}~\bibnamefont {Rinta-Antila}}, \bibinfo {author} {\bibfnamefont
  {T.}~\bibnamefont {Sonoda}}, \bibinfo {author} {\bibfnamefont
  {K.}~\bibnamefont {Burkard}}, \bibinfo {author} {\bibfnamefont
  {W.}~\bibnamefont {H{\"{u}}ller}}, \bibinfo {author} {\bibfnamefont
  {L.}~\bibnamefont {Batist}}, \bibinfo {author} {\bibfnamefont
  {W.}~\bibnamefont {Gelletly}}, \bibinfo {author} {\bibfnamefont {A.~L.}\
  \bibnamefont {Nichols}}, \bibinfo {author} {\bibfnamefont {T.}~\bibnamefont
  {Yoshida}}, \bibinfo {author} {\bibfnamefont {A.~A.}\ \bibnamefont
  {Sonzogni}}, \bibinfo {author} {\bibfnamefont {K.}~\bibnamefont
  {Per{\"{a}}j{\"{a}}rvi}}, \bibinfo {author} {\bibfnamefont {A.}~\bibnamefont
  {Petrovici}}, \bibinfo {author} {\bibfnamefont {K.~W.}\ \bibnamefont
  {Schmid}}, \ and\ \bibinfo {author} {\bibfnamefont {A.}~\bibnamefont
  {Faessler}},\ }\href {\doibase 10.1103/PhysRevC.87.044318} {\bibfield
  {journal} {\bibinfo  {journal} {Physical Review C}\ }\textbf {\bibinfo
  {volume} {87}},\ \bibinfo {pages} {044318} (\bibinfo {year}
  {2013})}\BibitemShut {NoStop}%
\bibitem [{\citenamefont {Tain}\ \emph {et~al.}(2015)\citenamefont {Tain},
  \citenamefont {Valencia}, \citenamefont {Algora}, \citenamefont {Agramunt},
  \citenamefont {Rubio}, \citenamefont {Rice}, \citenamefont {Gelletly},
  \citenamefont {Regan}, \citenamefont {Zakari-Issoufou}, \citenamefont
  {Fallot}, \citenamefont {Porta}, \citenamefont {Rissanen}, \citenamefont
  {Eronen}, \citenamefont {{\"{A}}yst{\"{o}}}, \citenamefont {Batist},
  \citenamefont {Bowry}, \citenamefont {Bui}, \citenamefont {Caballero-Folch},
  \citenamefont {Cano-Ott}, \citenamefont {Elomaa}, \citenamefont {Estevez},
  \citenamefont {Farrelly}, \citenamefont {Garcia}, \citenamefont
  {Gomez-Hornillos}, \citenamefont {Gorlychev}, \citenamefont {Hakala},
  \citenamefont {Jordan}, \citenamefont {Jokinen}, \citenamefont {Kolhinen},
  \citenamefont {Kondev}, \citenamefont {Mart{\'{i}}nez}, \citenamefont
  {Mendoza}, \citenamefont {Moore}, \citenamefont {Penttil{\"{a}}},
  \citenamefont {Podoly{\'{a}}k}, \citenamefont {Reponen}, \citenamefont
  {Sonnenschein},\ and\ \citenamefont {Sonzogni}}]{Tain2015}%
  \BibitemOpen
  \bibfield  {author} {\bibinfo {author} {\bibfnamefont {J.~L.}\ \bibnamefont
  {Tain}}, \bibinfo {author} {\bibfnamefont {E.}~\bibnamefont {Valencia}},
  \bibinfo {author} {\bibfnamefont {A.}~\bibnamefont {Algora}}, \bibinfo
  {author} {\bibfnamefont {J.}~\bibnamefont {Agramunt}}, \bibinfo {author}
  {\bibfnamefont {B.}~\bibnamefont {Rubio}}, \bibinfo {author} {\bibfnamefont
  {S.}~\bibnamefont {Rice}}, \bibinfo {author} {\bibfnamefont {W.}~\bibnamefont
  {Gelletly}}, \bibinfo {author} {\bibfnamefont {P.}~\bibnamefont {Regan}},
  \bibinfo {author} {\bibfnamefont {A.-A.}\ \bibnamefont {Zakari-Issoufou}},
  \bibinfo {author} {\bibfnamefont {M.}~\bibnamefont {Fallot}}, \bibinfo
  {author} {\bibfnamefont {A.}~\bibnamefont {Porta}}, \bibinfo {author}
  {\bibfnamefont {J.}~\bibnamefont {Rissanen}}, \bibinfo {author}
  {\bibfnamefont {T.}~\bibnamefont {Eronen}}, \bibinfo {author} {\bibfnamefont
  {J.}~\bibnamefont {{\"{A}}yst{\"{o}}}}, \bibinfo {author} {\bibfnamefont
  {L.}~\bibnamefont {Batist}}, \bibinfo {author} {\bibfnamefont
  {M.}~\bibnamefont {Bowry}}, \bibinfo {author} {\bibfnamefont {V.~M.}\
  \bibnamefont {Bui}}, \bibinfo {author} {\bibfnamefont {R.}~\bibnamefont
  {Caballero-Folch}}, \bibinfo {author} {\bibfnamefont {D.}~\bibnamefont
  {Cano-Ott}}, \bibinfo {author} {\bibfnamefont {V.-V.}\ \bibnamefont
  {Elomaa}}, \bibinfo {author} {\bibfnamefont {E.}~\bibnamefont {Estevez}},
  \bibinfo {author} {\bibfnamefont {G.~F.}\ \bibnamefont {Farrelly}}, \bibinfo
  {author} {\bibfnamefont {A.~R.}\ \bibnamefont {Garcia}}, \bibinfo {author}
  {\bibfnamefont {B.}~\bibnamefont {Gomez-Hornillos}}, \bibinfo {author}
  {\bibfnamefont {V.}~\bibnamefont {Gorlychev}}, \bibinfo {author}
  {\bibfnamefont {J.}~\bibnamefont {Hakala}}, \bibinfo {author} {\bibfnamefont
  {M.~D.}\ \bibnamefont {Jordan}}, \bibinfo {author} {\bibfnamefont
  {A.}~\bibnamefont {Jokinen}}, \bibinfo {author} {\bibfnamefont {V.~S.}\
  \bibnamefont {Kolhinen}}, \bibinfo {author} {\bibfnamefont {F.~G.}\
  \bibnamefont {Kondev}}, \bibinfo {author} {\bibfnamefont {T.}~\bibnamefont
  {Mart{\'{i}}nez}}, \bibinfo {author} {\bibfnamefont {E.}~\bibnamefont
  {Mendoza}}, \bibinfo {author} {\bibfnamefont {I.}~\bibnamefont {Moore}},
  \bibinfo {author} {\bibfnamefont {H.}~\bibnamefont {Penttil{\"{a}}}},
  \bibinfo {author} {\bibfnamefont {Z.}~\bibnamefont {Podoly{\'{a}}k}},
  \bibinfo {author} {\bibfnamefont {M.}~\bibnamefont {Reponen}}, \bibinfo
  {author} {\bibfnamefont {V.}~\bibnamefont {Sonnenschein}}, \ and\ \bibinfo
  {author} {\bibfnamefont {A.~A.}\ \bibnamefont {Sonzogni}},\ }\href {\doibase
  10.1103/PhysRevLett.115.062502} {\bibfield  {journal} {\bibinfo  {journal}
  {Physical Review Letters}\ }\textbf {\bibinfo {volume} {115}},\ \bibinfo
  {pages} {062502} (\bibinfo {year} {2015})}\BibitemShut {NoStop}%
\bibitem [{\citenamefont {Rasco}\ \emph {et~al.}(2016)\citenamefont {Rasco},
  \citenamefont {Woli{\'{n}}ska-Cichocka}, \citenamefont {Fija{\l}kowska},
  \citenamefont {Rykaczewski}, \citenamefont {Karny}, \citenamefont {Grzywacz},
  \citenamefont {Goetz}, \citenamefont {Gross}, \citenamefont {Stracener},
  \citenamefont {Zganjar}, \citenamefont {Batchelder}, \citenamefont
  {Blackmon}, \citenamefont {Brewer}, \citenamefont {Go}, \citenamefont
  {Heffron}, \citenamefont {King}, \citenamefont {Matta}, \citenamefont
  {Miernik}, \citenamefont {Nesaraja}, \citenamefont {Paulauskas},
  \citenamefont {Rajabali}, \citenamefont {Wang}, \citenamefont {Winger},
  \citenamefont {Xiao},\ and\ \citenamefont {Zachary}}]{Rasco2016}%
  \BibitemOpen
  \bibfield  {author} {\bibinfo {author} {\bibfnamefont {B.~C.}\ \bibnamefont
  {Rasco}}, \bibinfo {author} {\bibfnamefont {M.}~\bibnamefont
  {Woli{\'{n}}ska-Cichocka}}, \bibinfo {author} {\bibfnamefont
  {A.}~\bibnamefont {Fija{\l}kowska}}, \bibinfo {author} {\bibfnamefont
  {K.~P.}\ \bibnamefont {Rykaczewski}}, \bibinfo {author} {\bibfnamefont
  {M.}~\bibnamefont {Karny}}, \bibinfo {author} {\bibfnamefont {R.~K.}\
  \bibnamefont {Grzywacz}}, \bibinfo {author} {\bibfnamefont {K.~C.}\
  \bibnamefont {Goetz}}, \bibinfo {author} {\bibfnamefont {C.~J.}\ \bibnamefont
  {Gross}}, \bibinfo {author} {\bibfnamefont {D.~W.}\ \bibnamefont
  {Stracener}}, \bibinfo {author} {\bibfnamefont {E.~F.}\ \bibnamefont
  {Zganjar}}, \bibinfo {author} {\bibfnamefont {J.~C.}\ \bibnamefont
  {Batchelder}}, \bibinfo {author} {\bibfnamefont {J.~C.}\ \bibnamefont
  {Blackmon}}, \bibinfo {author} {\bibfnamefont {N.~T.}\ \bibnamefont
  {Brewer}}, \bibinfo {author} {\bibfnamefont {S.}~\bibnamefont {Go}}, \bibinfo
  {author} {\bibfnamefont {B.}~\bibnamefont {Heffron}}, \bibinfo {author}
  {\bibfnamefont {T.}~\bibnamefont {King}}, \bibinfo {author} {\bibfnamefont
  {J.~T.}\ \bibnamefont {Matta}}, \bibinfo {author} {\bibfnamefont
  {K.}~\bibnamefont {Miernik}}, \bibinfo {author} {\bibfnamefont {C.~D.}\
  \bibnamefont {Nesaraja}}, \bibinfo {author} {\bibfnamefont {S.~V.}\
  \bibnamefont {Paulauskas}}, \bibinfo {author} {\bibfnamefont {M.~M.}\
  \bibnamefont {Rajabali}}, \bibinfo {author} {\bibfnamefont {E.~H.}\
  \bibnamefont {Wang}}, \bibinfo {author} {\bibfnamefont {J.~A.}\ \bibnamefont
  {Winger}}, \bibinfo {author} {\bibfnamefont {Y.}~\bibnamefont {Xiao}}, \ and\
  \bibinfo {author} {\bibfnamefont {C.~J.}\ \bibnamefont {Zachary}},\ }\href
  {\doibase 10.1103/PhysRevLett.117.092501} {\bibfield  {journal} {\bibinfo
  {journal} {Physical Review Letters}\ }\textbf {\bibinfo {volume} {117}},\
  \bibinfo {pages} {092501} (\bibinfo {year} {2016})}\BibitemShut {NoStop}%
\bibitem [{\citenamefont {Rasco}\ \emph {et~al.}(2017)\citenamefont {Rasco},
  \citenamefont {Rykaczewski}, \citenamefont {Fija{\l}kowska}, \citenamefont
  {Karny}, \citenamefont {Woli{\'{n}}ska-Cichocka}, \citenamefont {Grzywacz},
  \citenamefont {Gross}, \citenamefont {Stracener}, \citenamefont {Zganjar},
  \citenamefont {Blackmon}, \citenamefont {Brewer}, \citenamefont {Goetz},
  \citenamefont {Johnson}, \citenamefont {Jost}, \citenamefont {Hamilton},
  \citenamefont {Miernik}, \citenamefont {Madurga}, \citenamefont {Miller},
  \citenamefont {Padgett}, \citenamefont {Paulauskas}, \citenamefont
  {Ramayya},\ and\ \citenamefont {Spejewski}}]{Rasco2017}%
  \BibitemOpen
  \bibfield  {author} {\bibinfo {author} {\bibfnamefont {B.~C.}\ \bibnamefont
  {Rasco}}, \bibinfo {author} {\bibfnamefont {K.~P.}\ \bibnamefont
  {Rykaczewski}}, \bibinfo {author} {\bibfnamefont {A.}~\bibnamefont
  {Fija{\l}kowska}}, \bibinfo {author} {\bibfnamefont {M.}~\bibnamefont
  {Karny}}, \bibinfo {author} {\bibfnamefont {M.}~\bibnamefont
  {Woli{\'{n}}ska-Cichocka}}, \bibinfo {author} {\bibfnamefont {R.~K.}\
  \bibnamefont {Grzywacz}}, \bibinfo {author} {\bibfnamefont {C.~J.}\
  \bibnamefont {Gross}}, \bibinfo {author} {\bibfnamefont {D.~W.}\ \bibnamefont
  {Stracener}}, \bibinfo {author} {\bibfnamefont {E.~F.}\ \bibnamefont
  {Zganjar}}, \bibinfo {author} {\bibfnamefont {J.~C.}\ \bibnamefont
  {Blackmon}}, \bibinfo {author} {\bibfnamefont {N.~T.}\ \bibnamefont
  {Brewer}}, \bibinfo {author} {\bibfnamefont {K.~C.}\ \bibnamefont {Goetz}},
  \bibinfo {author} {\bibfnamefont {J.~W.}\ \bibnamefont {Johnson}}, \bibinfo
  {author} {\bibfnamefont {C.~U.}\ \bibnamefont {Jost}}, \bibinfo {author}
  {\bibfnamefont {J.~H.}\ \bibnamefont {Hamilton}}, \bibinfo {author}
  {\bibfnamefont {K.}~\bibnamefont {Miernik}}, \bibinfo {author} {\bibfnamefont
  {M.}~\bibnamefont {Madurga}}, \bibinfo {author} {\bibfnamefont
  {D.}~\bibnamefont {Miller}}, \bibinfo {author} {\bibfnamefont
  {S.}~\bibnamefont {Padgett}}, \bibinfo {author} {\bibfnamefont {S.~V.}\
  \bibnamefont {Paulauskas}}, \bibinfo {author} {\bibfnamefont {A.~V.}\
  \bibnamefont {Ramayya}}, \ and\ \bibinfo {author} {\bibfnamefont {E.~H.}\
  \bibnamefont {Spejewski}},\ }\href {\doibase 10.1103/PhysRevC.95.054328}
  {\bibfield  {journal} {\bibinfo  {journal} {Physical Review C}\ }\textbf
  {\bibinfo {volume} {95}},\ \bibinfo {pages} {054328} (\bibinfo {year}
  {2017})}\BibitemShut {NoStop}%
\bibitem [{\citenamefont {Rice}\ \emph {et~al.}(2017)\citenamefont {Rice},
  \citenamefont {Algora}, \citenamefont {Tain}, \citenamefont {Valencia},
  \citenamefont {Agramunt}, \citenamefont {Rubio}, \citenamefont {Gelletly},
  \citenamefont {Regan}, \citenamefont {Zakari-Issoufou}, \citenamefont
  {Fallot}, \citenamefont {Porta}, \citenamefont {Rissanen}, \citenamefont
  {Eronen}, \citenamefont {{\"{A}}yst{\"{o}}}, \citenamefont {Batist},
  \citenamefont {Bowry}, \citenamefont {Bui}, \citenamefont {Caballero-Folch},
  \citenamefont {Cano-Ott}, \citenamefont {Elomaa}, \citenamefont {Estevez},
  \citenamefont {Farrelly}, \citenamefont {Garcia}, \citenamefont
  {Gomez-Hornillos}, \citenamefont {Gorlychev}, \citenamefont {Hakala},
  \citenamefont {Jordan}, \citenamefont {Jokinen}, \citenamefont {Kolhinen},
  \citenamefont {Kondev}, \citenamefont {Mart{\'{i}}nez}, \citenamefont
  {Mason}, \citenamefont {Mendoza}, \citenamefont {Moore}, \citenamefont
  {Penttil{\"{a}}}, \citenamefont {Podoly{\'{a}}k}, \citenamefont {Reponen},
  \citenamefont {Sonnenschein}, \citenamefont {Sonzogni},\ and\ \citenamefont
  {Sarriguren}}]{Rice2017}%
  \BibitemOpen
  \bibfield  {author} {\bibinfo {author} {\bibfnamefont {S.}~\bibnamefont
  {Rice}}, \bibinfo {author} {\bibfnamefont {A.}~\bibnamefont {Algora}},
  \bibinfo {author} {\bibfnamefont {J.~L.}\ \bibnamefont {Tain}}, \bibinfo
  {author} {\bibfnamefont {E.}~\bibnamefont {Valencia}}, \bibinfo {author}
  {\bibfnamefont {J.}~\bibnamefont {Agramunt}}, \bibinfo {author}
  {\bibfnamefont {B.}~\bibnamefont {Rubio}}, \bibinfo {author} {\bibfnamefont
  {W.}~\bibnamefont {Gelletly}}, \bibinfo {author} {\bibfnamefont {P.~H.}\
  \bibnamefont {Regan}}, \bibinfo {author} {\bibfnamefont {A.-A.}\ \bibnamefont
  {Zakari-Issoufou}}, \bibinfo {author} {\bibfnamefont {M.}~\bibnamefont
  {Fallot}}, \bibinfo {author} {\bibfnamefont {A.}~\bibnamefont {Porta}},
  \bibinfo {author} {\bibfnamefont {J.}~\bibnamefont {Rissanen}}, \bibinfo
  {author} {\bibfnamefont {T.}~\bibnamefont {Eronen}}, \bibinfo {author}
  {\bibfnamefont {J.}~\bibnamefont {{\"{A}}yst{\"{o}}}}, \bibinfo {author}
  {\bibfnamefont {L.}~\bibnamefont {Batist}}, \bibinfo {author} {\bibfnamefont
  {M.}~\bibnamefont {Bowry}}, \bibinfo {author} {\bibfnamefont {V.~M.}\
  \bibnamefont {Bui}}, \bibinfo {author} {\bibfnamefont {R.}~\bibnamefont
  {Caballero-Folch}}, \bibinfo {author} {\bibfnamefont {D.}~\bibnamefont
  {Cano-Ott}}, \bibinfo {author} {\bibfnamefont {V.-V.}\ \bibnamefont
  {Elomaa}}, \bibinfo {author} {\bibfnamefont {E.}~\bibnamefont {Estevez}},
  \bibinfo {author} {\bibfnamefont {G.~F.}\ \bibnamefont {Farrelly}}, \bibinfo
  {author} {\bibfnamefont {A.~R.}\ \bibnamefont {Garcia}}, \bibinfo {author}
  {\bibfnamefont {B.}~\bibnamefont {Gomez-Hornillos}}, \bibinfo {author}
  {\bibfnamefont {V.}~\bibnamefont {Gorlychev}}, \bibinfo {author}
  {\bibfnamefont {J.}~\bibnamefont {Hakala}}, \bibinfo {author} {\bibfnamefont
  {M.~D.}\ \bibnamefont {Jordan}}, \bibinfo {author} {\bibfnamefont
  {A.}~\bibnamefont {Jokinen}}, \bibinfo {author} {\bibfnamefont {V.~S.}\
  \bibnamefont {Kolhinen}}, \bibinfo {author} {\bibfnamefont {F.~G.}\
  \bibnamefont {Kondev}}, \bibinfo {author} {\bibfnamefont {T.}~\bibnamefont
  {Mart{\'{i}}nez}}, \bibinfo {author} {\bibfnamefont {P.}~\bibnamefont
  {Mason}}, \bibinfo {author} {\bibfnamefont {E.}~\bibnamefont {Mendoza}},
  \bibinfo {author} {\bibfnamefont {I.}~\bibnamefont {Moore}}, \bibinfo
  {author} {\bibfnamefont {H.}~\bibnamefont {Penttil{\"{a}}}}, \bibinfo
  {author} {\bibfnamefont {Z.}~\bibnamefont {Podoly{\'{a}}k}}, \bibinfo
  {author} {\bibfnamefont {M.}~\bibnamefont {Reponen}}, \bibinfo {author}
  {\bibfnamefont {V.}~\bibnamefont {Sonnenschein}}, \bibinfo {author}
  {\bibfnamefont {A.~A.}\ \bibnamefont {Sonzogni}}, \ and\ \bibinfo {author}
  {\bibfnamefont {P.}~\bibnamefont {Sarriguren}},\ }\href {\doibase
  10.1103/PhysRevC.96.014320} {\bibfield  {journal} {\bibinfo  {journal}
  {Physical Review C}\ }\textbf {\bibinfo {volume} {96}},\ \bibinfo {pages}
  {014320} (\bibinfo {year} {2017})}\BibitemShut {NoStop}%
\bibitem [{\citenamefont {Valencia}\ \emph {et~al.}(2017)\citenamefont
  {Valencia}, \citenamefont {Tain}, \citenamefont {Algora}, \citenamefont
  {Agramunt}, \citenamefont {Estevez}, \citenamefont {Jordan}, \citenamefont
  {Rubio}, \citenamefont {Rice}, \citenamefont {Regan}, \citenamefont
  {Gelletly}, \citenamefont {Podoly{\'{a}}k}, \citenamefont {Bowry},
  \citenamefont {Mason}, \citenamefont {Farrelly}, \citenamefont
  {Zakari-Issoufou}, \citenamefont {Fallot}, \citenamefont {Porta},
  \citenamefont {Bui}, \citenamefont {Rissanen}, \citenamefont {Eronen},
  \citenamefont {Moore}, \citenamefont {Penttil{\"{a}}}, \citenamefont
  {{\"{A}}yst{\"{o}}}, \citenamefont {Elomaa}, \citenamefont {Hakala},
  \citenamefont {Jokinen}, \citenamefont {Kolhinen}, \citenamefont {Reponen},
  \citenamefont {Sonnenschein}, \citenamefont {Cano-Ott}, \citenamefont
  {Garcia}, \citenamefont {Mart{\'{i}}nez}, \citenamefont {Mendoza},
  \citenamefont {Caballero-Folch}, \citenamefont {Gomez-Hornillos},
  \citenamefont {Gorlichev}, \citenamefont {Kondev}, \citenamefont {Sonzogni},\
  and\ \citenamefont {Batist}}]{Valencia2017}%
  \BibitemOpen
  \bibfield  {author} {\bibinfo {author} {\bibfnamefont {E.}~\bibnamefont
  {Valencia}}, \bibinfo {author} {\bibfnamefont {J.~L.}\ \bibnamefont {Tain}},
  \bibinfo {author} {\bibfnamefont {A.}~\bibnamefont {Algora}}, \bibinfo
  {author} {\bibfnamefont {J.}~\bibnamefont {Agramunt}}, \bibinfo {author}
  {\bibfnamefont {E.}~\bibnamefont {Estevez}}, \bibinfo {author} {\bibfnamefont
  {M.~D.}\ \bibnamefont {Jordan}}, \bibinfo {author} {\bibfnamefont
  {B.}~\bibnamefont {Rubio}}, \bibinfo {author} {\bibfnamefont
  {S.}~\bibnamefont {Rice}}, \bibinfo {author} {\bibfnamefont {P.}~\bibnamefont
  {Regan}}, \bibinfo {author} {\bibfnamefont {W.}~\bibnamefont {Gelletly}},
  \bibinfo {author} {\bibfnamefont {Z.}~\bibnamefont {Podoly{\'{a}}k}},
  \bibinfo {author} {\bibfnamefont {M.}~\bibnamefont {Bowry}}, \bibinfo
  {author} {\bibfnamefont {P.}~\bibnamefont {Mason}}, \bibinfo {author}
  {\bibfnamefont {G.~F.}\ \bibnamefont {Farrelly}}, \bibinfo {author}
  {\bibfnamefont {A.}~\bibnamefont {Zakari-Issoufou}}, \bibinfo {author}
  {\bibfnamefont {M.}~\bibnamefont {Fallot}}, \bibinfo {author} {\bibfnamefont
  {A.}~\bibnamefont {Porta}}, \bibinfo {author} {\bibfnamefont {V.~M.}\
  \bibnamefont {Bui}}, \bibinfo {author} {\bibfnamefont {J.}~\bibnamefont
  {Rissanen}}, \bibinfo {author} {\bibfnamefont {T.}~\bibnamefont {Eronen}},
  \bibinfo {author} {\bibfnamefont {I.}~\bibnamefont {Moore}}, \bibinfo
  {author} {\bibfnamefont {H.}~\bibnamefont {Penttil{\"{a}}}}, \bibinfo
  {author} {\bibfnamefont {J.}~\bibnamefont {{\"{A}}yst{\"{o}}}}, \bibinfo
  {author} {\bibfnamefont {V.-V.}\ \bibnamefont {Elomaa}}, \bibinfo {author}
  {\bibfnamefont {J.}~\bibnamefont {Hakala}}, \bibinfo {author} {\bibfnamefont
  {A.}~\bibnamefont {Jokinen}}, \bibinfo {author} {\bibfnamefont {V.~S.}\
  \bibnamefont {Kolhinen}}, \bibinfo {author} {\bibfnamefont {M.}~\bibnamefont
  {Reponen}}, \bibinfo {author} {\bibfnamefont {V.}~\bibnamefont
  {Sonnenschein}}, \bibinfo {author} {\bibfnamefont {D.}~\bibnamefont
  {Cano-Ott}}, \bibinfo {author} {\bibfnamefont {A.~R.}\ \bibnamefont
  {Garcia}}, \bibinfo {author} {\bibfnamefont {T.}~\bibnamefont
  {Mart{\'{i}}nez}}, \bibinfo {author} {\bibfnamefont {E.}~\bibnamefont
  {Mendoza}}, \bibinfo {author} {\bibfnamefont {R.}~\bibnamefont
  {Caballero-Folch}}, \bibinfo {author} {\bibfnamefont {B.}~\bibnamefont
  {Gomez-Hornillos}}, \bibinfo {author} {\bibfnamefont {V.}~\bibnamefont
  {Gorlichev}}, \bibinfo {author} {\bibfnamefont {F.~G.}\ \bibnamefont
  {Kondev}}, \bibinfo {author} {\bibfnamefont {A.~A.}\ \bibnamefont
  {Sonzogni}}, \ and\ \bibinfo {author} {\bibfnamefont {L.}~\bibnamefont
  {Batist}},\ }\href {\doibase 10.1103/PhysRevC.95.024320} {\bibfield
  {journal} {\bibinfo  {journal} {Physical Review C}\ }\textbf {\bibinfo
  {volume} {95}},\ \bibinfo {pages} {024320} (\bibinfo {year}
  {2017})}\BibitemShut {NoStop}%
\bibitem [{\citenamefont {Fija{\l}kowska}\ \emph {et~al.}(2017)\citenamefont
  {Fija{\l}kowska}, \citenamefont {Karny}, \citenamefont {Rykaczewski},
  \citenamefont {Rasco}, \citenamefont {Grzywacz}, \citenamefont {Gross},
  \citenamefont {Woli{\'{n}}ska-Cichocka}, \citenamefont {Goetz}, \citenamefont
  {Stracener}, \citenamefont {Bielewski}, \citenamefont {Goans}, \citenamefont
  {Hamilton}, \citenamefont {Johnson}, \citenamefont {Jost}, \citenamefont
  {Madurga}, \citenamefont {Miernik}, \citenamefont {Miller}, \citenamefont
  {Padgett}, \citenamefont {Paulauskas}, \citenamefont {Ramayya},\ and\
  \citenamefont {Zganjar}}]{Fijakowska2017}%
  \BibitemOpen
  \bibfield  {author} {\bibinfo {author} {\bibfnamefont {A.}~\bibnamefont
  {Fija{\l}kowska}}, \bibinfo {author} {\bibfnamefont {M.}~\bibnamefont
  {Karny}}, \bibinfo {author} {\bibfnamefont {K.~P.}\ \bibnamefont
  {Rykaczewski}}, \bibinfo {author} {\bibfnamefont {B.~C.}\ \bibnamefont
  {Rasco}}, \bibinfo {author} {\bibfnamefont {R.}~\bibnamefont {Grzywacz}},
  \bibinfo {author} {\bibfnamefont {C.~J.}\ \bibnamefont {Gross}}, \bibinfo
  {author} {\bibfnamefont {M.}~\bibnamefont {Woli{\'{n}}ska-Cichocka}},
  \bibinfo {author} {\bibfnamefont {K.~C.}\ \bibnamefont {Goetz}}, \bibinfo
  {author} {\bibfnamefont {D.~W.}\ \bibnamefont {Stracener}}, \bibinfo {author}
  {\bibfnamefont {W.}~\bibnamefont {Bielewski}}, \bibinfo {author}
  {\bibfnamefont {R.}~\bibnamefont {Goans}}, \bibinfo {author} {\bibfnamefont
  {J.~H.}\ \bibnamefont {Hamilton}}, \bibinfo {author} {\bibfnamefont {J.~W.}\
  \bibnamefont {Johnson}}, \bibinfo {author} {\bibfnamefont {C.}~\bibnamefont
  {Jost}}, \bibinfo {author} {\bibfnamefont {M.}~\bibnamefont {Madurga}},
  \bibinfo {author} {\bibfnamefont {K.}~\bibnamefont {Miernik}}, \bibinfo
  {author} {\bibfnamefont {D.}~\bibnamefont {Miller}}, \bibinfo {author}
  {\bibfnamefont {S.~W.}\ \bibnamefont {Padgett}}, \bibinfo {author}
  {\bibfnamefont {S.~V.}\ \bibnamefont {Paulauskas}}, \bibinfo {author}
  {\bibfnamefont {A.~V.}\ \bibnamefont {Ramayya}}, \ and\ \bibinfo {author}
  {\bibfnamefont {E.~F.}\ \bibnamefont {Zganjar}},\ }\href {\doibase
  10.1103/PhysRevLett.119.052503} {\bibfield  {journal} {\bibinfo  {journal}
  {Physical Review Letters}\ }\textbf {\bibinfo {volume} {119}},\ \bibinfo
  {pages} {052503} (\bibinfo {year} {2017})}\BibitemShut {NoStop}%
\bibitem [{\citenamefont {Tsunoda}\ \emph {et~al.}(2013)\citenamefont
  {Tsunoda}, \citenamefont {Otsuka}, \citenamefont {Shimizu}, \citenamefont
  {Honma},\ and\ \citenamefont {Utsuno}}]{Tsunoda2013}%
  \BibitemOpen
  \bibfield  {author} {\bibinfo {author} {\bibfnamefont {Y.}~\bibnamefont
  {Tsunoda}}, \bibinfo {author} {\bibfnamefont {T.}~\bibnamefont {Otsuka}},
  \bibinfo {author} {\bibfnamefont {N.}~\bibnamefont {Shimizu}}, \bibinfo
  {author} {\bibfnamefont {M.}~\bibnamefont {Honma}}, \ and\ \bibinfo {author}
  {\bibfnamefont {Y.}~\bibnamefont {Utsuno}},\ }\href {\doibase
  10.1088/1742-6596/445/1/012028} {\bibfield  {journal} {\bibinfo  {journal}
  {Journal of Physics: Conference Series}\ }\textbf {\bibinfo {volume} {445}},\
  \bibinfo {pages} {0} (\bibinfo {year} {2013})}\BibitemShut {NoStop}%
\bibitem [{\citenamefont {Perdrisat}(1966)}]{Perdrisat1966}%
  \BibitemOpen
  \bibfield  {author} {\bibinfo {author} {\bibfnamefont {C.~F.}\ \bibnamefont
  {Perdrisat}},\ }\href {\doibase 10.1103/RevModPhys.38.41} {\bibfield
  {journal} {\bibinfo  {journal} {Reviews of Modern Physics}\ }\textbf
  {\bibinfo {volume} {38}},\ \bibinfo {pages} {41} (\bibinfo {year}
  {1966})}\BibitemShut {NoStop}%
\end{thebibliography}%
\end{document}